\documentclass[preprint,twocolumn]{aastex63}

\usepackage{float, bm, graphicx, amsmath, morefloats}
\usepackage[caption=false]{subfig}
\shorttitle{Low Luminosity Type IIP Supernovae}
\shortauthors{Das et al.}

\usepackage{float,graphicx,amsmath,tabularx,booktabs,natbib,threeparttable}
\usepackage{float, bm, graphicx, amsmath, morefloats}
\usepackage[caption=false]{subfig}
\usepackage{float,graphicx,amsmath,tabularx,booktabs,natbib,threeparttable}
\usepackage{subfig}
\usepackage[para]{footmisc}
\usepackage{tablefootnote}

\usepackage{graphicx}
\usepackage[caption=false]{subfig}
\usepackage{lipsum} 
\usepackage{placeins}

\usepackage{comment}
\definecolor{dark-red}{rgb}{0.4,0.15,0.15}
\definecolor{dark-blue}{rgb}{0.15,0.15,0.4}
\definecolor{medium-blue}{rgb}{0,0,0.5}
\hypersetup{
    colorlinks, linkcolor={dark-red},
    citecolor={dark-blue}, urlcolor={medium-blue}
}

\newcommand{\beqa}{\begin{eqnarray}} 
\newcommand{\eeqa}{\end{eqnarray}}

\newcommand{\bsub}{\begin{subequations}}
\newcommand{\esub}{\end{subequations}}
\newcommand{\beal}{\begin{align}}
\newcommand{\ealn}{\end{align}}

\newcommand{\Msun}{{\ensuremath{\mathrm{M}_{\odot}}}}

\graphicspath{{./}{figures/}}

\begin{document}

\title{Low-Luminosity Type IIP Supernovae from the Zwicky Transient Facility Census of the Local Universe. II: Lightcurve Analysis}

\author[0000-0001-8372-997X]{Kaustav~K.~Das}\thanks{E-mail: kdas@astro.caltech.edu}\affiliation{Cahill Center for Astrophysics, California Institute of Technology, MC 249-17, 
1200 E California Boulevard, Pasadena, CA, 91125, USA}

\author[0000-0002-5619-4938]{Mansi~M.~Kasliwal}
\affiliation{Cahill Center for Astrophysics, 
California Institute of Technology, MC 249-17, 
1200 E California Boulevard, Pasadena, CA, 91125, USA}

\author[0000-0003-1546-6615]{Jesper Sollerman}
\affiliation{The Oskar Klein Centre, Department of Astronomy, Stockholm University, AlbaNova, SE-10691 Stockholm, Sweden}

\author[0000-0002-4223-103X]{Christoffer Fremling}
\affil{Caltech Optical Observatories, California Institute of Technology, Pasadena, CA 91125, USA}

\author{Takashi J. Moriya}
\affiliation{Astronomical Science Program, Graduate Institute for Advanced Studies, SOKENDAI, 2-21-1 Osawa, Mitaka, Tokyo 181-8588, Japan}
\affiliation{National Astronomical Observatory of Japan, National Institutes of Natural Sciences, 2-21-1 Osawa, Mitaka, Tokyo 181-8588, Japan}
\affiliation{School of Physics and Astronomy, Monash University, Clayton, VIC 3800, Australia}

\author{K-Ryan Hinds}
\affiliation{Astrophysics Research Institute, Liverpool John Moores University, IC2,  Liverpool L3 5RF, UK}

\author{Daniel A. Perley}
\affiliation{Astrophysics Research Institute, Liverpool John Moores University, IC2,  Liverpool L3 5RF, UK}

\author[0000-0001-8018-5348]{Eric C. Bellm}
\affiliation{DIRAC Institute, Department of Astronomy, University of Washington, 3910 15th Avenue NE, Seattle, WA 98195, USA}

\author[0000-0001-9152-6224]{Tracy X. Chen}
\affiliation{IPAC, California Institute of Technology, 1200 E. California
             Blvd, Pasadena, CA 91125, USA}

\author{Evan P. O'Connor}
\affiliation{The Oskar Klein Centre, Department of Astronomy, Stockholm University, AlbaNova, SE-10691 Stockholm, Sweden}

\author[0000-0002-8262-2924]{Michael W. Coughlin}
\affiliation{School of Physics and Astronomy, University of Minnesota, Minneapolis, MN 55455, USA}             
\author[0000-0002-3934-2644]{W. V. Jacobson-Galán}
\altaffiliation{NASA Hubble Fellow}
\affiliation{Cahill Center for Astrophysics, 
California Institute of Technology, MC 249-17, 
1200 E California Boulevard, Pasadena, CA, 91125, USA}

\author{Anjasha Gangopadhyay}
\affiliation{The Oskar Klein Centre, Department of Astronomy, Stockholm University, AlbaNova, SE-10691 Stockholm, Sweden}

\author[0000-0002-3168-0139]{Matthew Graham}
\affiliation{Cahill Center for Astrophysics, 
California Institute of Technology, MC 249-17, 
1200 E California Boulevard, Pasadena, CA, 91125, USA}

\author[0000-0001-5390-8563]{S.~R.~Kulkarni}
\affiliation{Cahill Center for Astrophysics, 
California Institute of Technology, MC 249-17, 
1200 E California Boulevard, Pasadena, CA, 91125, USA}

\author{Josiah Purdum}
\affiliation{Caltech Optical Observatories, California Institute of Technology, Pasadena, CA 91125, USA}

\author[0000-0003-2700-1030]{Nikhil Sarin}
\affiliation{The Oskar Klein Centre, Department of Astronomy, Stockholm University, AlbaNova, SE-10691 Stockholm, Sweden}

\author[0000-0001-6797-1889]{Steve Schulze}
\affiliation{Center for Interdisciplinary Exploration and Research in Astrophysics (CIERA), 1800 Sherman Ave., Evanston, IL 60201, USA}

\author[0000-0003-2091-622X]{Avinash Singh}
\affiliation{The Oskar Klein Centre, Department of Astronomy, Stockholm University, AlbaNova, SE-10691 Stockholm, Sweden}

\author[0000-0002-6347-3089]{Daichi Tsuna}
\affiliation{TAPIR, Mailcode 350-17, California Institute of Technology, Pasadena, CA 91125, USA}

\author[0000-0002-9998-6732]{Avery Wold}
\affiliation{IPAC, California Institute of Technology, 1200 E. California
             Blvd, Pasadena, CA 91125, USA}


\begin{abstract}

The Zwicky Transient Facility Census of the Local Universe survey yielded a sample of 330 Type IIP supernovae (SNe) with well-constrained peak luminosities. In paper I \citep{Das2025}, we measured their luminosity function and volumetric rate. Here (paper II), we present the largest systematic study of lightcurve properties for Type IIP SNe from a volume-limited survey, analyzing a selected subset of 129 events, including 16 low-luminosity Type IIP (LLIIP) SNe with M$_{r,peak} \geq ‑16$ mag. We find that plateau slope correlates with peak brightness, with many LLIIP SNe showing positive slopes---suggesting smaller progenitor radii and distinct density profiles compared to brighter Type IIP SNe. The plateau duration shows only a weak dependence on peak brightness, likely suggesting binary interaction. One SN exhibits a plateau-to-tail drop of $>3.5$ mag, consistent with an electron-capture or failed SN with very low or zero nickel mass. We derive explosion and progenitor parameters of the entire Type IIP SN sample using semi-analytical and radiation-hydrodynamical models. Based on radiation-hydrodynamical model fitting, LLIIP SNe are characterized by low nickel masses ($0.001–0.025$~\Msun), low explosion energies ($0.1–0.28 \times 10^{51}$~erg), low ejecta masses ($8.1^{+0.8}_{-1.7}$~\Msun), and ZAMS masses below $11$~\Msun. In comparison, the full Type IIP SN sample spans a wider range with nickel masses ($0.001$–$0.222$ \Msun), explosion energies ($0.10$–$4.43 \times 10^{51}$erg), ejecta masses ($5.4–24.8$~\Msun), and ZAMS masses ($9.3$–$16.7$~\Msun). We find strong correlations between peak brightness, explosion energy, and nickel mass that extend to the low-luminosity end. We conclude that LLIIP SNe represent the faint, low-energy end of the Type IIP population and originate from the lowest-mass core-collapse progenitors.

\end{abstract}


\section{Introduction} \label{sec:intro}



Core-collapse supernovae (CCSNe) are the explosive deaths of massive stars that play crucial roles in galactic chemical evolution, the formation of new stars, and the creation of compact neutron stars and black holes. Moreover, their luminosities, rates, and association with massive star formation make them essential tools for probing key astrophysical parameters across vast cosmic volumes. The most common type of core-collapse SNe are Type II SNe \citep[e.g.,][]{Li11, Shivvers2017}. They arise from the explosions of stars that retain significant portions of their hydrogen envelopes.

Low-luminosity Type IIP supernovae (LLIIP SNe), defined here as those with $M_{\textrm{r,peak}} \ge -16$ mag, as motivated by \textcolor{black}{prior literature \citep[e.g.,][]{Pastorello2004, Spiro2014, Reguitti2021}, are particularly intriguing because they are thought to originate from progenitor stars with just enough mass to undergo core collapse.} These events likely arise from 8–12 \Msun\ progenitors, as supported by pre-SN progenitor imaging \citep{Maund2005, Li2006, Mattila2008, ONeill2019, VanDyk2023}, light curve simulations \citep{Pumo2017, Fraser2011}, evolutionary models of low-mass red supergiants (RSGs) \citep{Lisakov2017}, and nebular spectroscopy \citep{Jerkstrand2018}. They provide valuable insight into the lower mass threshold for core collapse, which defines the boundary between stars that form white dwarfs and those that form neutron stars or black holes. LLIIP SNe also serve as key candidates for probing the upper mass limit of super-asymptotic giant branch (sAGB) star formation, \textcolor{black}{exploring whether they can explode as electron-capture supernovae (ECSN) \citep{Nomoto1984, Kitaura2006, Janka2008, Takahashi2013, JOnes2013, Hiramatsu2021b, Wang2025}, and understanding} their role in shaping elemental abundances in subsequent generations of stars \citep{Tsujimoto1999, Siess2006}. Furthermore, a typical initial mass function (IMF) predicts that more than 40\% of CCSN progenitors fall within the 8–12 \Msun\ range, with 25\% within the 8–10 \Msun\ subset \citep{Sukhbold2016}. It has also been suggested that LLIIP SNe may instead originate from more \textcolor{black}{massive RSGs with significant fallback \citep[e.g.,][]{Woosley1995, Zampieri2003, Moriya2010}.} Thus, systematic studies of LLIIP SNe are essential for improving our understanding of the progenitors, explosion properties, and the dust and chemical enrichment contributed by nearly half of all massive stars that undergo core-collapse.

\textcolor{black}{
However, to date, no study of the lightcurve properties of LLIIP SNe has been conducted using data from a systematic SN survey. Due to their low luminosity, they are relatively difficult to detect and classify. There are only about a dozen such objects presented in the literature, focused on single-object studies or small heterogeneous samples. Early examples such as SN~1997D, SN~1999eu and
SN~1999br \citep{Turatto1998, Pastorello2004} revealed the defining characteristics of
the class: very low plateau luminosities ($M_{V} \approx $$-15$~mag), unusually
narrow P-Cygni profiles indicating expansion velocities of only
$\sim$1000--2500~km\,s$^{-1}$, extremely small $^{56}$Ni masses ($10^{-3}$--$10^{-2}$~M$_\odot$),
and low explosion energies. Subsequent studies of nearby events such as SN~2005cs and
SN~2008bk \citep{Pastorello2009, Mattila2008, VanDyk2012} connected these faint
light curves to low-mass red supergiant progenitors ($\sim$8--12~M$_\odot$) through
direct pre-explosion imaging and hydrodynamical modeling. \citet{Spiro2014} modeled a set of underluminous Type~IIP SNe and confirmed uniformly low explosion energies; \citet{Nakaoka2018} and
\citet{Reguitti2021} found similarly low $^{56}$Ni yields and slow expansion velocities
for SN~2016bkv and SN~2018hwm, respectively; and studies such as
\citet{Jager2020, Muller2020, Sheng2021, Valerin2022, Kozyreva2022, Bostroem2023, Teja2024,
Dastidar2025} have recently reported additional LLIIP SN candidates with low luminosities and
energetics consistent with either low-mass RSG/sAGB explosions or fallback-suppressed
$^{56}$Ni production. In contrast to these low-luminosity events, canonical Type~II SNe typically reach plateau
absolute magnitudes of $M_r \simeq -16$ to $-18$~mag, have plateau durations of
$\sim$80--120~days, ejecta masses of $\sim$8--18~M$_\odot$, explosion energies of order
$0.3$--$1\times10^{51}$~erg, and $^{56}$Ni yields of $\sim$0.01--0.1~M$_\odot$
\citep[e.g.,][]{Hamuy2003, Anderson2014, Sanders2015, Valenti2016, Martinez2022, Fang2025}. These
properties are generally consistent with explosions of RSG progenitors with
initial masses $\sim$12--18~M$_\odot$ \citep[e.g., see ][]{VanDyk2025}}.

In Part I of the series of three papers on LLIIP SNe \citep{Das2025}, we measured the luminosity function and volumetric rate of the largest sample to date of 36 LLIIP SNe (and 330 Type IIP SNe) from the Zwicky Transient Facility Census of the Local Universe \citep[ZTF CLU;][]{Bellm2019, Graham2019, Dekany20, Masci2019, De2020} survey. This paper delves into their lightcurve properties. In this paper, we analyze the observed properties of the multi-band lightcurves of a selected subset of 129 Type II SNe and 16 LLIIP SNe. \textcolor{black}{It is well established that the morphology of Type~II SN lightcurves depends on key progenitor properties (e.g., final mass, envelope mass, and radius) and on explosion parameters such as the deposited energy and the amount and distribution of synthesized $^{56}$Ni \citep[e.g.,][]{Kasen2009, Goldberg2019}. Here, we measure observed lightcurve parametrs  and infer physical and explosion parameters using semi-analytical and radiation-hydrodynamical models that link SN observables to progenitor properties \citep[e.g.,][]{Litvinova1983, Hamuy2003, Kasen2009, Goldberg2019, Morozova2018, Moriya2023}. Such models have been previously applied to both individual events and larger samples of Type II SNe \citep{Hamuy2003, Arcavi2011, Anderson2014, Spiro2014, Sanders2015, Valenti2016, Gutierrez2017, Morozova2018, Martinez2022, Martinez2022b, Subrayan2023, Silva2024, Fang2025}. } This paper is structured as follows: in Section \ref{sec:sample}, we define the sample selection criteria. The data and extinction correction methods are described in Sections \ref{sec:data} and \ref{sec:extinction}, respectively. Sections \ref{sec:analysis} and \ref{sec:results} detail the fitting methods and results. We discuss the implications of the derived physical properties in Section \ref{sec:discussion} and conclude in Section \ref{sec:conclusion}.

\section{Sample Selection}
\label{sec:sample}




\textcolor{black}{From the ZTF CLU experiment, we selected Type~IIP SNe with well-sampled $r$-band light curves and a clearly defined plateau phase. The selection required (i) a GP-fit peak
apparent magnitude of $m_{r,\rm peak}<20$~mag, (ii) sufficient temporal coverage around peak and during the plateau (more than ten total detections and at least one detection on both sides of peak, or an equivalent set of conditions ensuring a constrained rise), and (iii) a plateau lasting at least 40~days with $<1$~mag decline. 
We refer to Paper I \citep{Das2025} for the complete sample selection criteria to select 330 Type IIP SNe.}
To select the sample for lightcurve analysis, we require that the end of the plateau phase is observed in the $r$-band lightcurve. This is to ensure we can extract reliable progenitor and explosion parameters from the lightcurves. The epoch of the end of the plateau ($t_{\rm plateau,end}$) is determined as the epoch where the derivative of the slope of the $r$-band lightcurve ($\Delta_{\rm slope}$) reach a local minimum, and its value is  $< -0.003$ mag day$^{-2}$, indicating a sharp transition (see Section \ref{sec:obsanalysis} and Figure \ref{fig:GPfits} for details). This marks the end of the plateau phase in the lightcurve evolution. Using this criterion, we identified 129 candidates that fulfill this selection criterion. Out of these, 68 objects have $\geq 2$ $r$-band data points in the radioactive tail phase, while 61 SNe do not have an observed radioactive tail.

%


\section{Data}
\label{sec:data}

In this section, we describe the photometric and spectroscopic data used.

\subsection{Optical photometry}
\label{section:photdata}

\begin{figure*}
    \centering
    \includegraphics[width=0.49\textwidth]{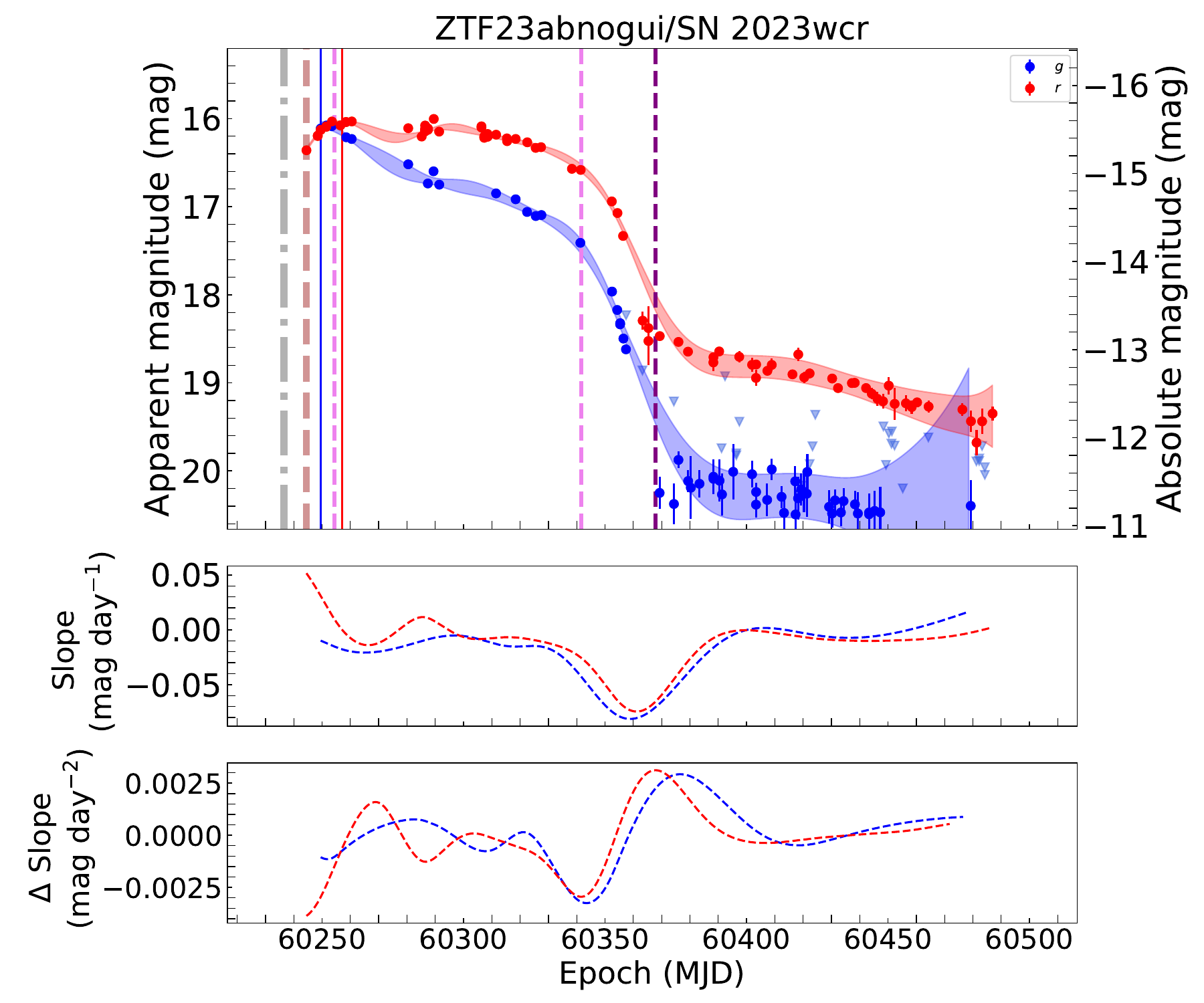}\includegraphics[width=0.49\textwidth]{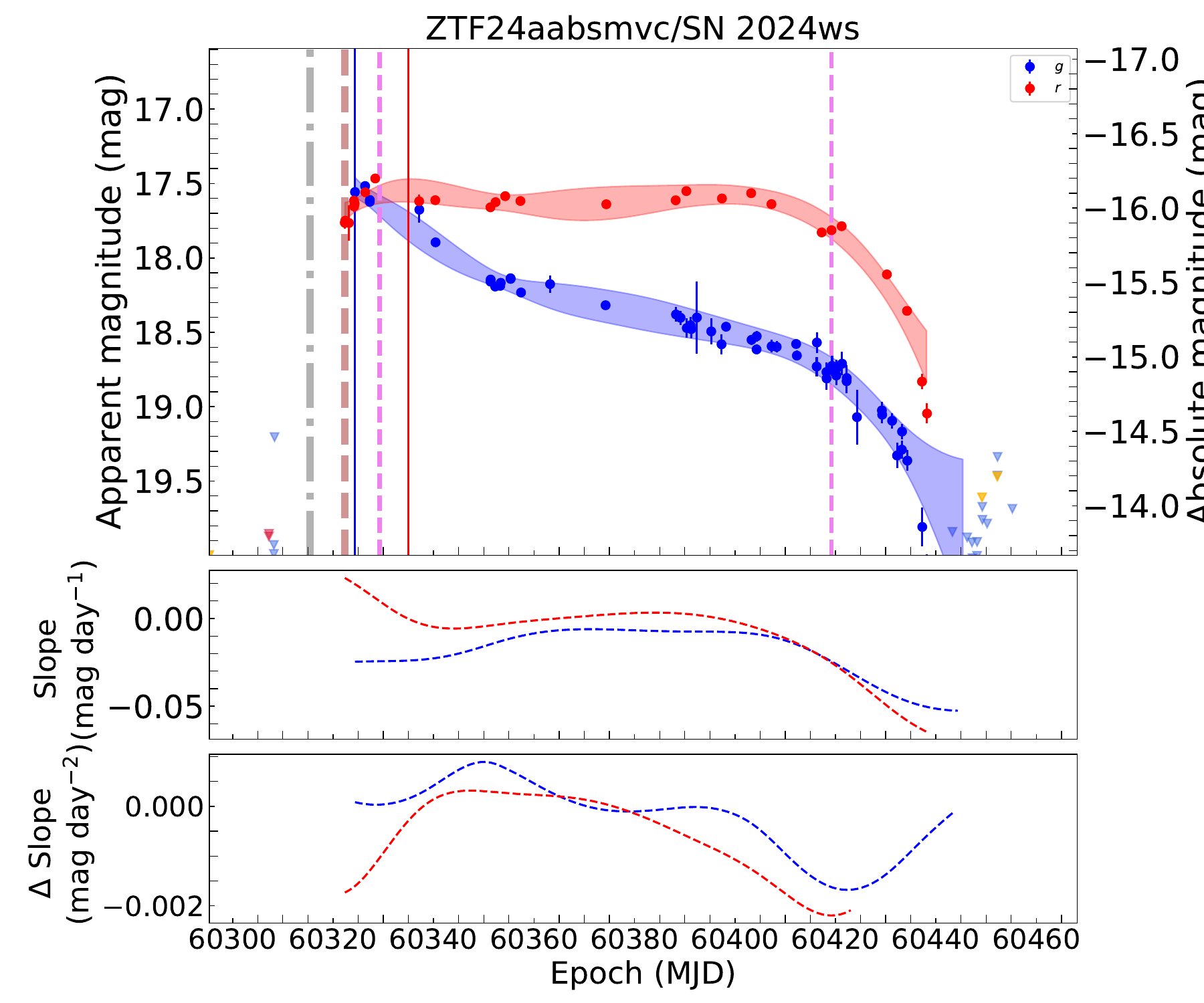}
    \includegraphics[width=0.49\textwidth]{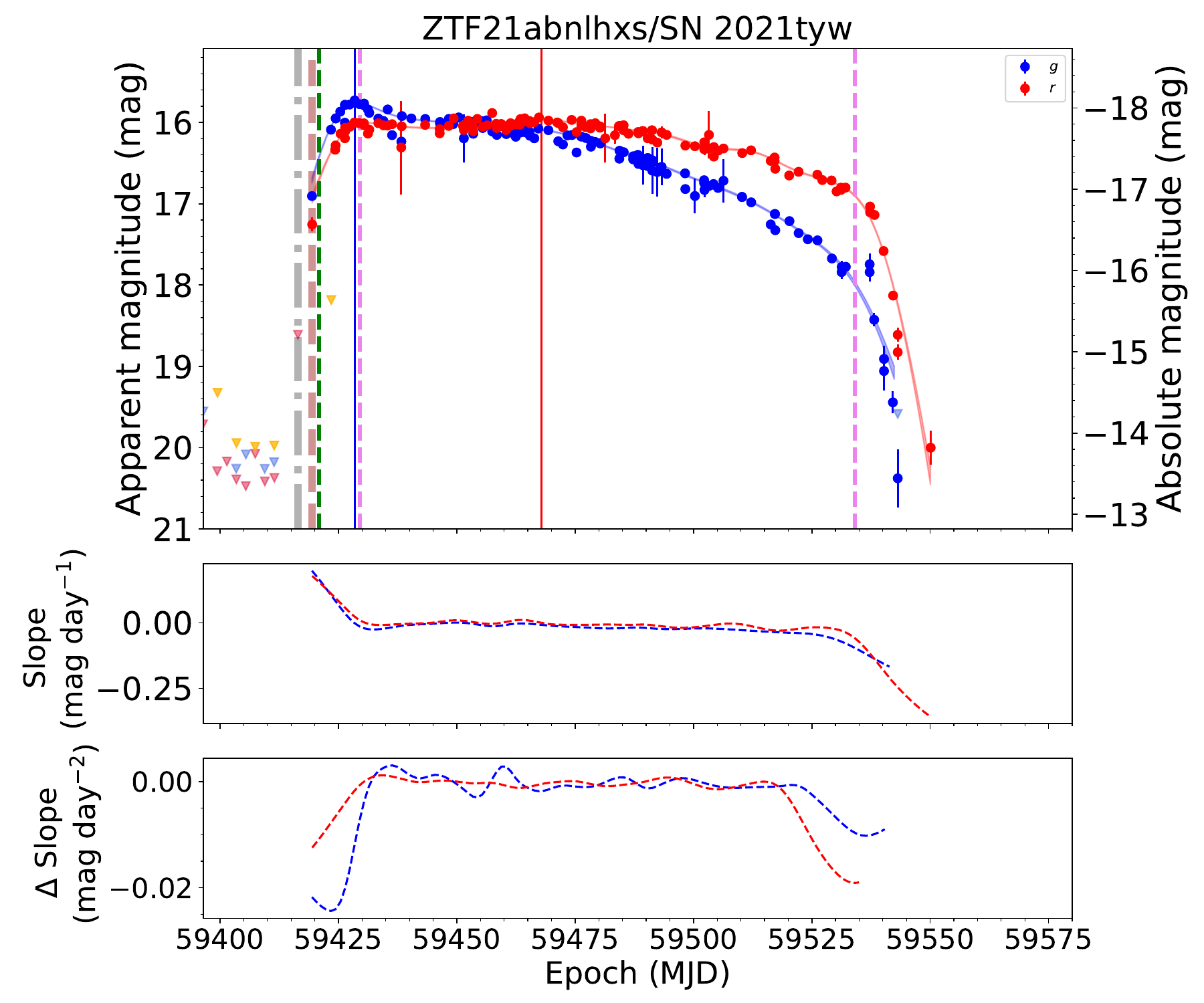}\includegraphics[width=0.49\textwidth]{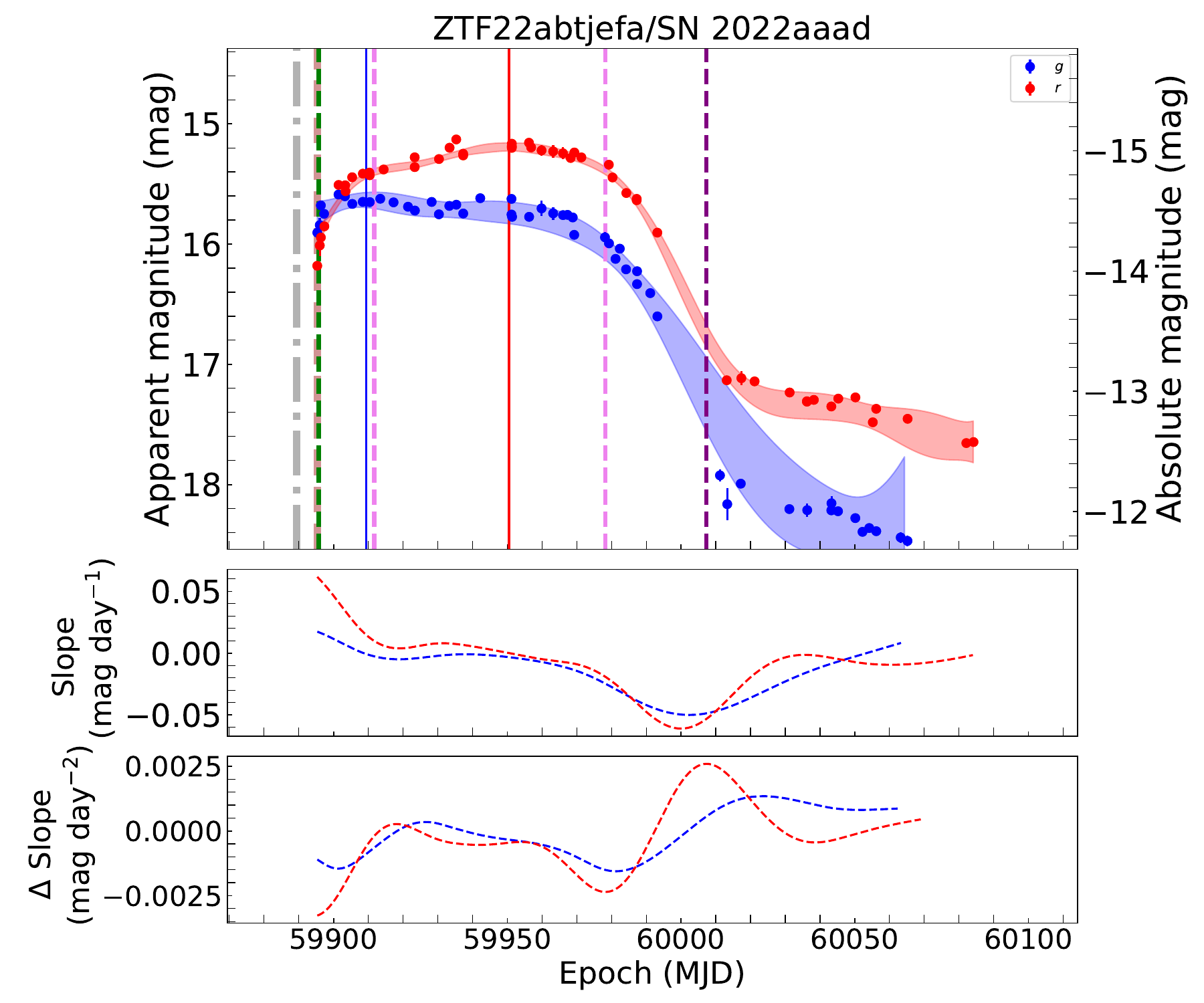} 
    \includegraphics[width=0.49\textwidth]{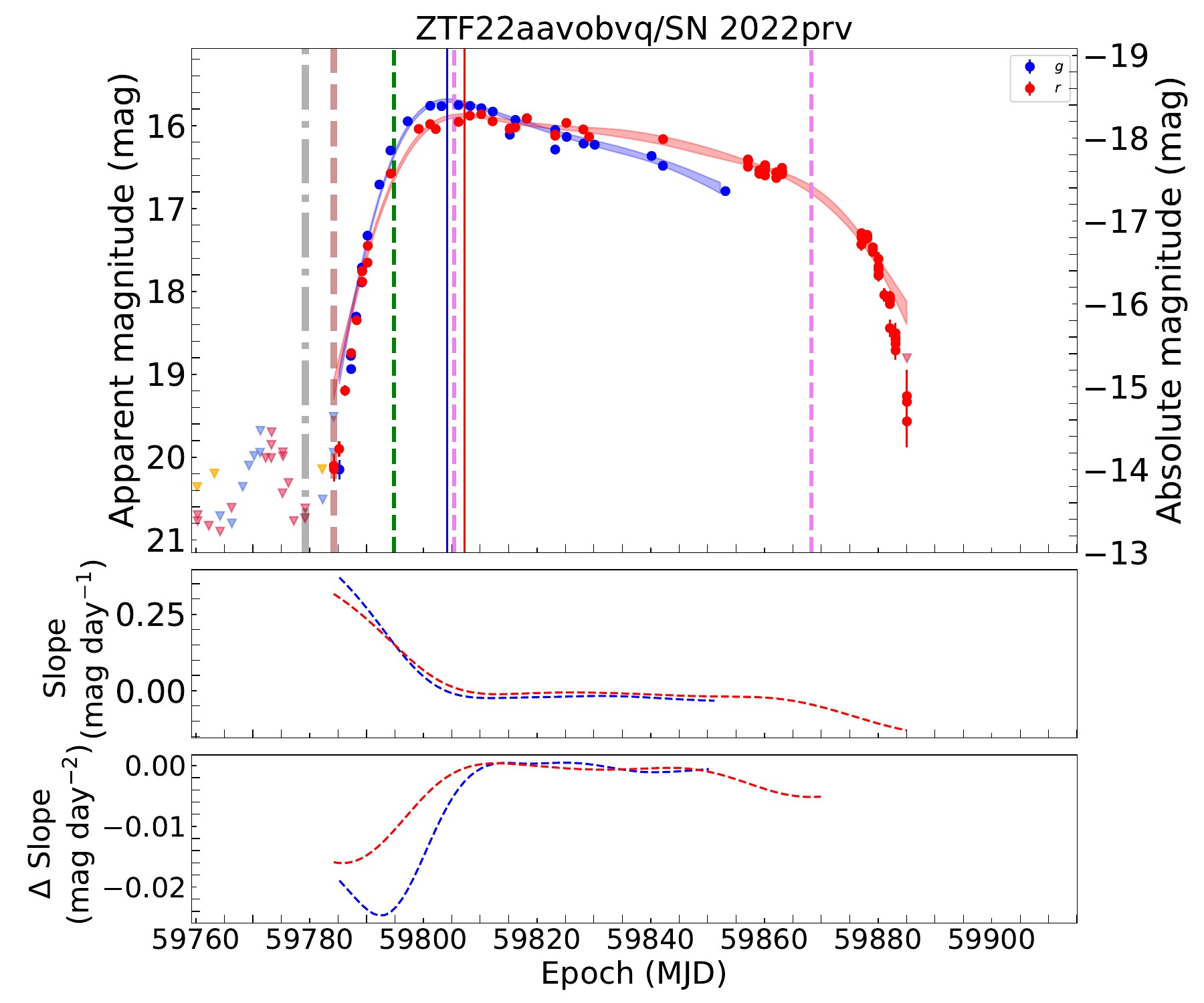}\includegraphics[width=0.49\textwidth]{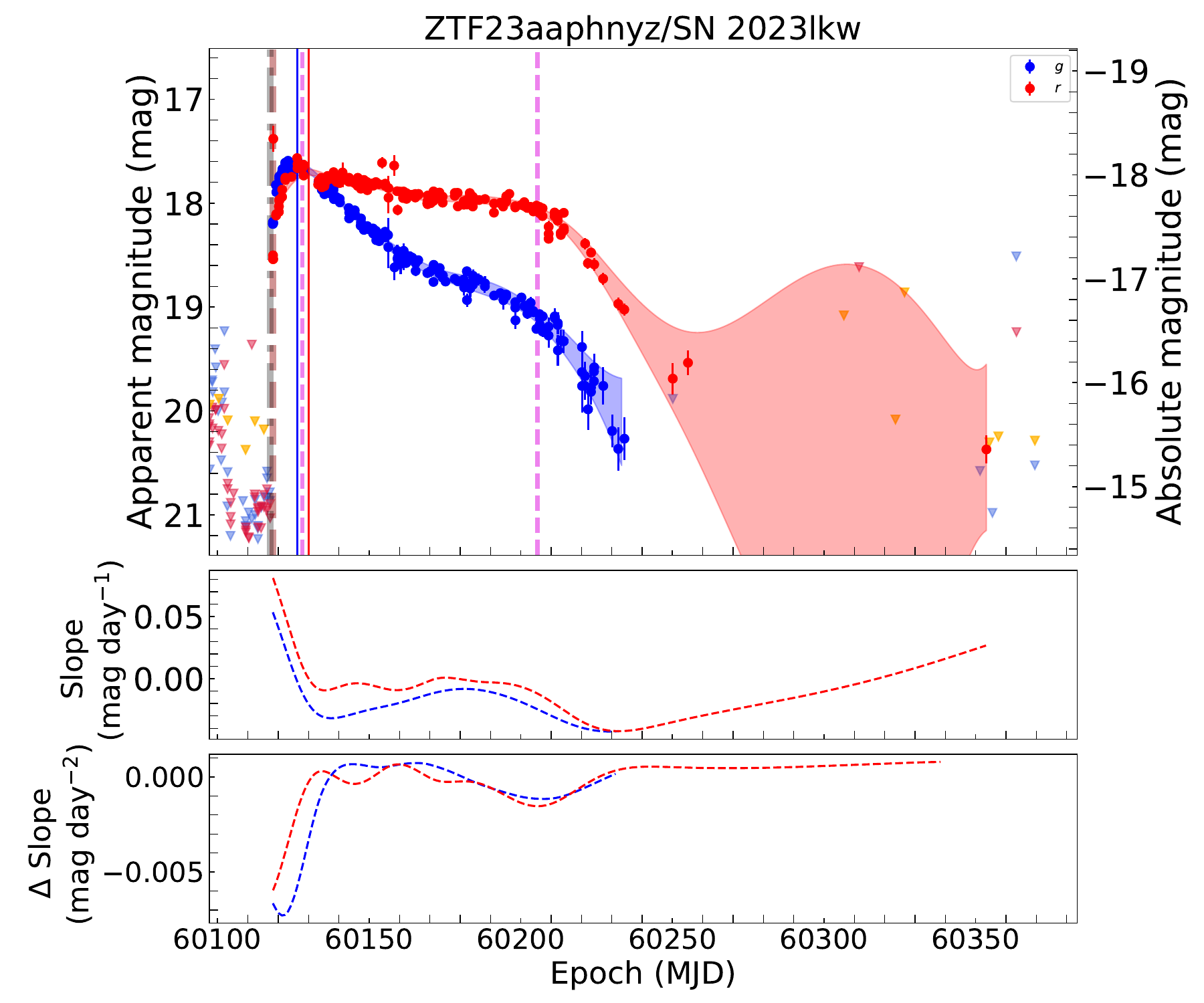} 
    
   \caption{The upper panels show Gaussian Process fits to the lightcurves (shaded regions). Vertical dashed red lines indicate the epochs of first detection; vertical dashed grey lines mark the estimated explosion epochs. Vertical solid red and blue lines show the epochs of peak $r$- and $g$-band magnitudes, respectively. Vertical solid violet lines denote the start and end of the $r$-band plateau phase. Green dashed lines mark the epoch when the flux reaches 50\% of the peak, used to compute the rise time. The lower panels show the light curve slope in units of mag day$^{-1}$ and the change in slope ($\Delta$ slope) in units of mag day$^{-2}$.
}

    \label{fig:GPfits}
\end{figure*}

We perform forced photometry on the ZTF difference images using the ZTF forced-photometry service developed by \citet{Masci2019} in $g$, $r$ and $i$ bands. For this work, we consider
anything less than a 3$\sigma$ detection an upper limit. 
In the tail phase, for a few SNe we also took photometry in the $g$, $r$, $i$ bands with the Optical Imager (IO:O) at the 2.0 m robotic Liverpool Telescope \citep{Steele2004} and the Alhambra Faint Object Spectrograph and Camera at the Nordic Optical Telescope \citep{Djupvik2010}. The P60 and LT data sets were processed with the FPipe image subtraction pipeline \citep{Fremling2016}, using Sloan Digital Sky Survey \citep[SDSS;][]{Ahn2012a} and PanSTARRS \citep[PS1;][]{Chambers2016} reference frames. See \citet{Das2025} for details on photometric observations and reductions. Figure \ref{fig:GPfits} shows a few representative lightcurves as examples. The photometry data and lightcurve plots for all the SNe in the sample will be made available on \href{https://zenodo.org/records/15717884?token=eyJhbGciOiJIUzUxMiJ9.eyJpZCI6IjNhYjE3NjYwLTU2NDEtNDBkZi1iYmI5LTQ1YzQxY2EwYjllNyIsImRhdGEiOnt9LCJyYW5kb20iOiJhNjA2ZTkzNDFjYjU5NGM3ZGIzMmExMTRlOGY1NzU5MCJ9.X0jCmJz6FJbXwmcY_6ZZ1kYCcRHwrABpoZ2epNXLlpIjVIwrFkEEOGD0Z7Cfk7luRPlHvOCwJdWBwB40vX2JoQ}{\texttt{Zenodo}}.

\subsection{Optical spectroscopy}
\label{section:spectra}

For each transient, at least one spectrum is usually obtained close to the peak luminosity to establish an initial spectroscopic classification. We primarily utilize the Double Beam Spectrograph \citep[DBSP;][]{Oke1982} on the Palomar 200-inch Hale telescope and the Spectral Energy Distribution Machine \citep[SEDM;][]{Blagorodnova2018, Rigault2019, Kim2022} on the Palomar 60-inch telescope for this purpose. Other instruments used to obtain classification spectra include: the Alhambra Faint Object Spectrograph and Camera at the Nordic Optical Telescope \citep[NOT;][]{Djupvik2010}, the Spectrograph for the Rapid Acquisition of Transients \citep[SPRAT;][]{Piascik2014} on the Liverpool Telescope, the Low-Resolution Imaging Spectrometer \citep[LRIS;][]{Oke1995} on the Keck I telescope. See \citet{Das2025} for details on spectral observations and reductions. \textcolor{black}{We employ the \texttt{SuperNova Identification} \citep[\texttt{SNID};][]{Blondin2007} and \texttt{superfit} \citep{Howell2005} codes for classifications.}



\begin{figure*}
    \centering
    \includegraphics[width=0.32\textwidth]{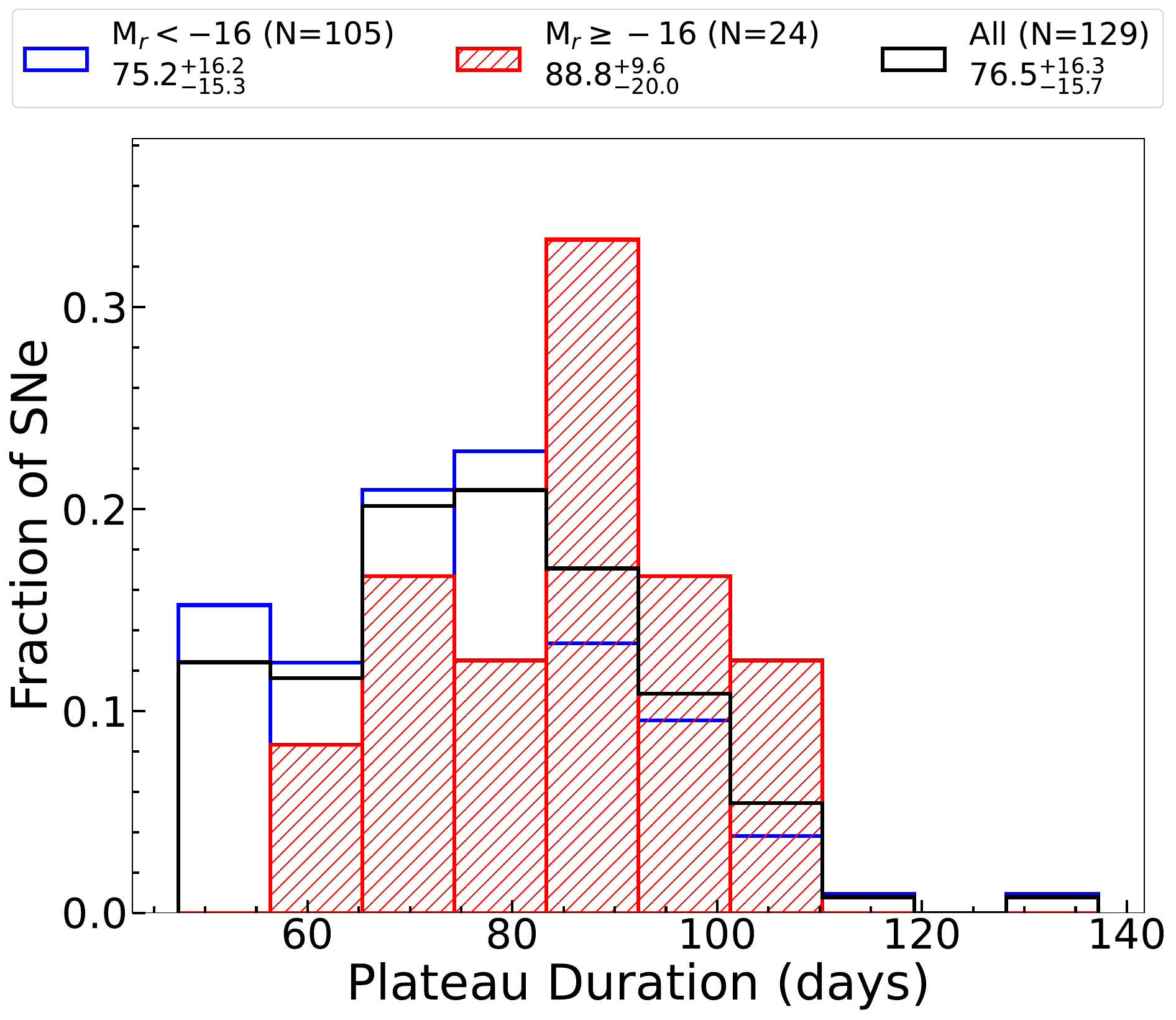}\includegraphics[width=0.32\textwidth]{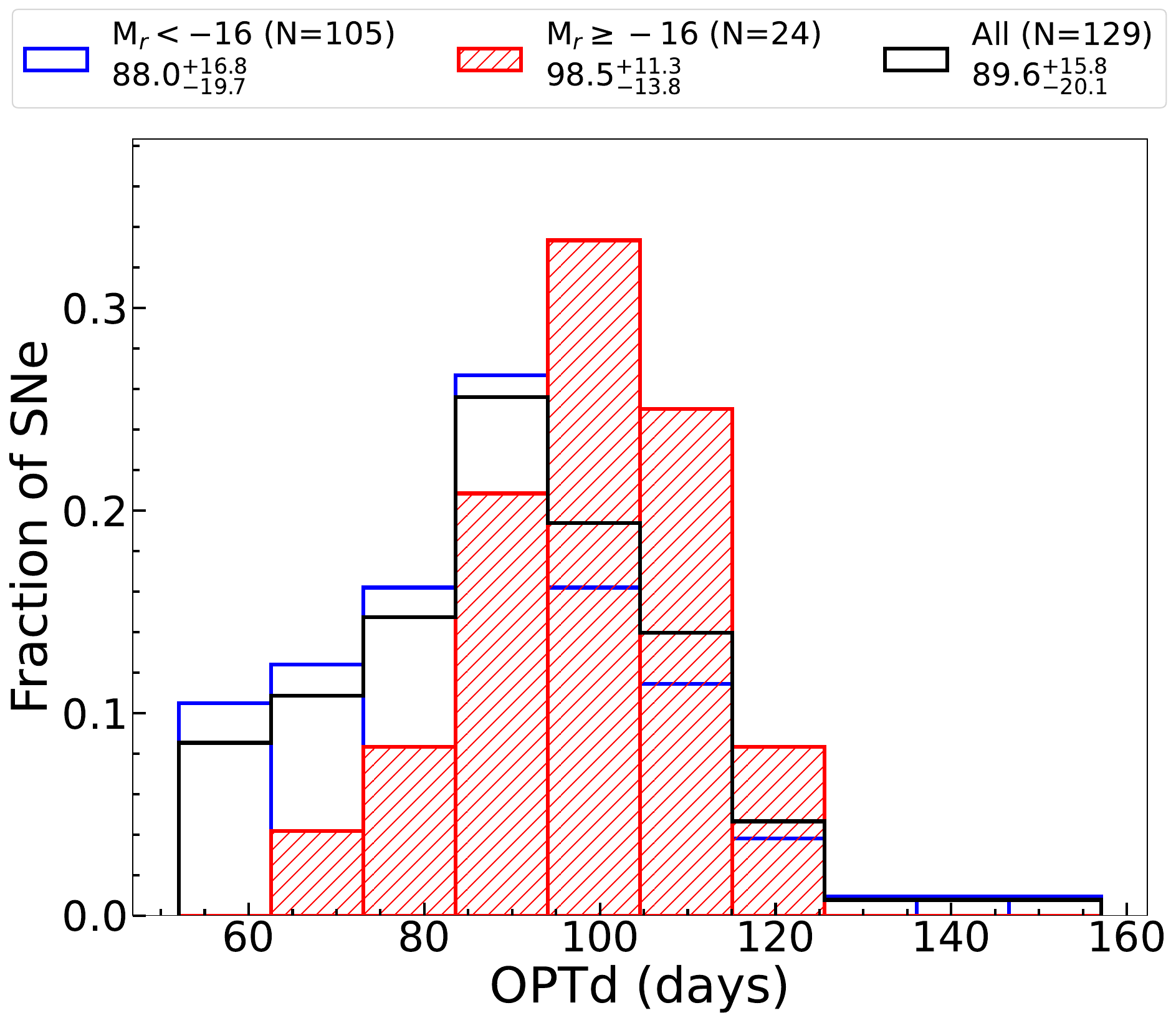}\includegraphics[width=0.32\textwidth]{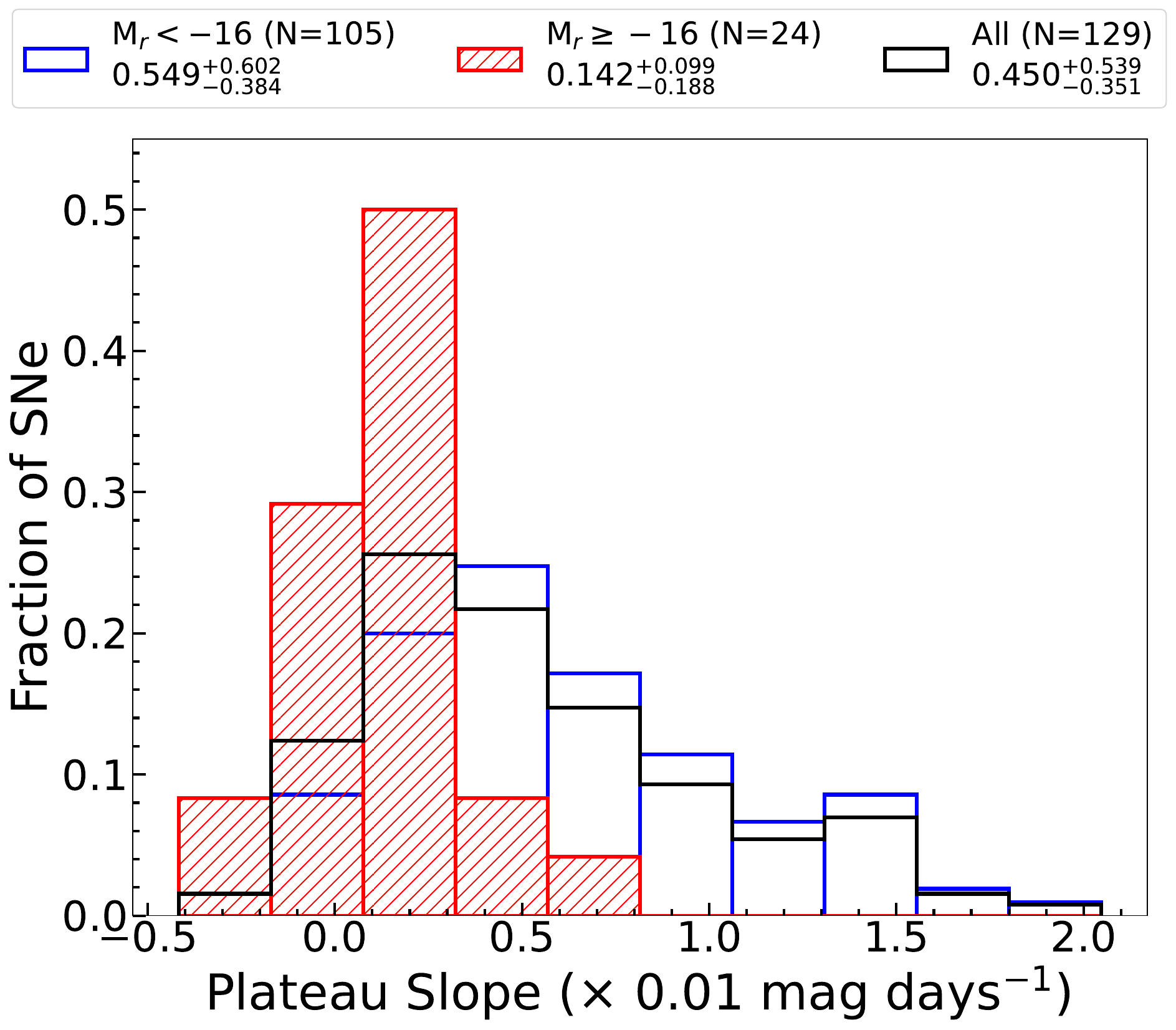}

    \includegraphics[width=0.32\textwidth]{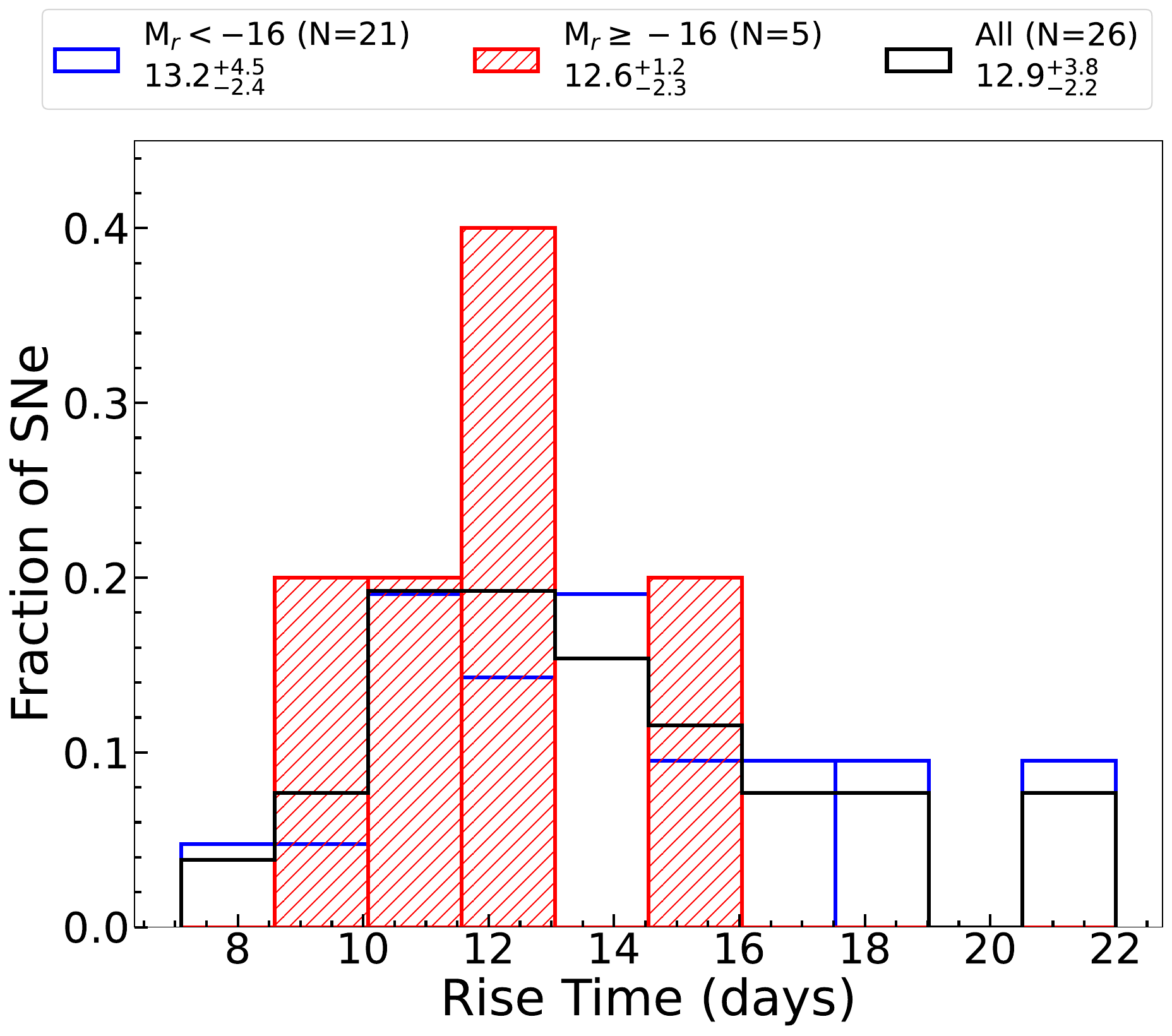}\includegraphics[width=0.32\textwidth]{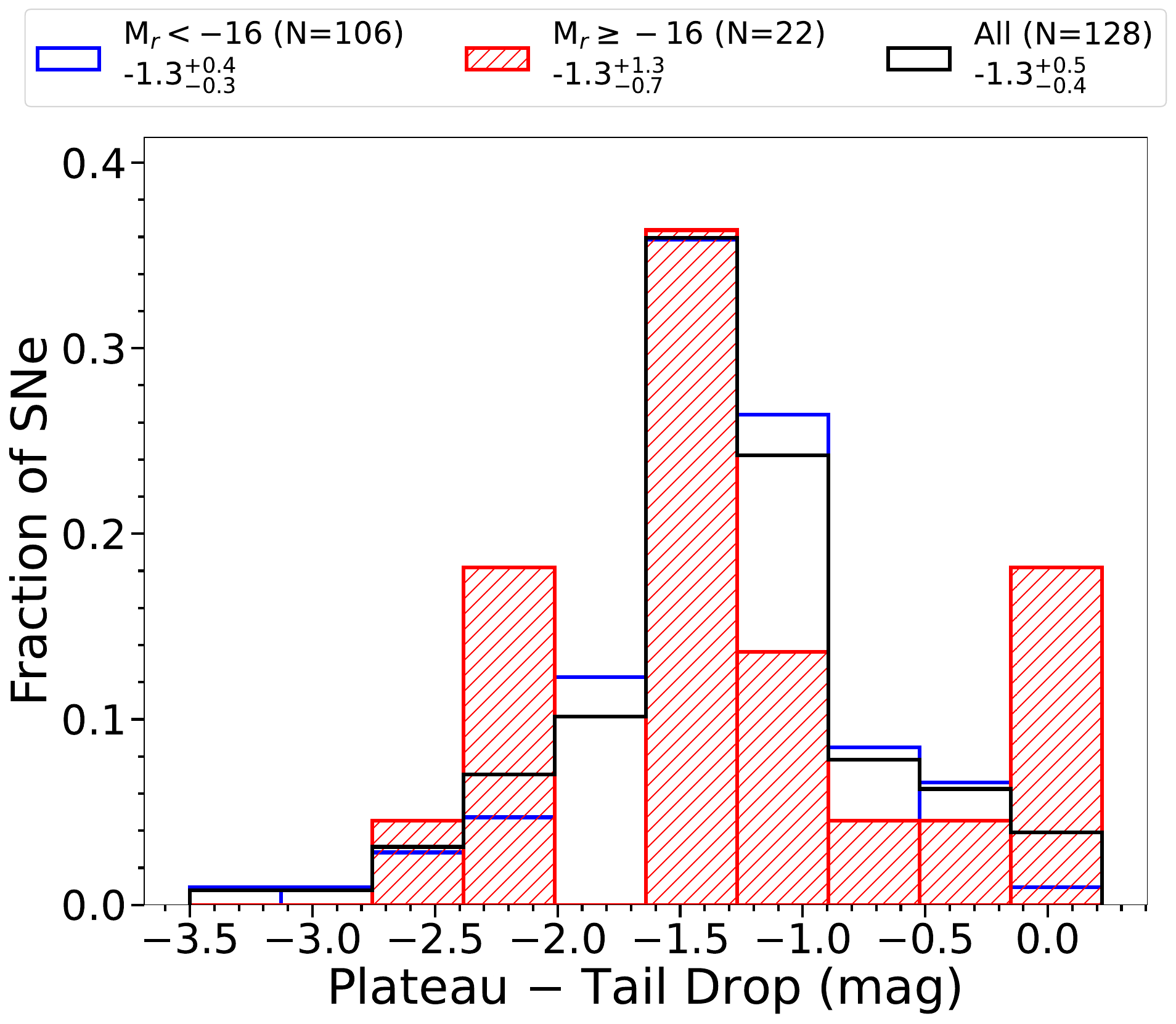}
\caption{
Distribution of observable properties of the $r$-band light curves for the SN sample. From left to right and top to bottom, the panels show the distributions of: (1) plateau duration, (2) optically thick phase duration (OPTd), (3) plateau slope, (4) rise time to peak, and (5) magnitude drop between the plateau and the tail phase. These observables are measured from the Gaussian Process fits described in Section~\ref{sec:obsanalysis}. LLIIP SNe with $M_r \geq -16$ are shown in hatched red, those with $M_r < -16$ in blue, and the full sample is shown in black. The median (50th percentile) and the 16th and 84th percentiles of the distributions are indicated as $\text{median}^{+\text{84th}-\text{50th}}_{-\text{50th}-\text{16th}}$ in the legend.
}

\label{fig:observables}
\end{figure*}

\begin{figure*}
    \centering
    \includegraphics[width=0.32\textwidth]{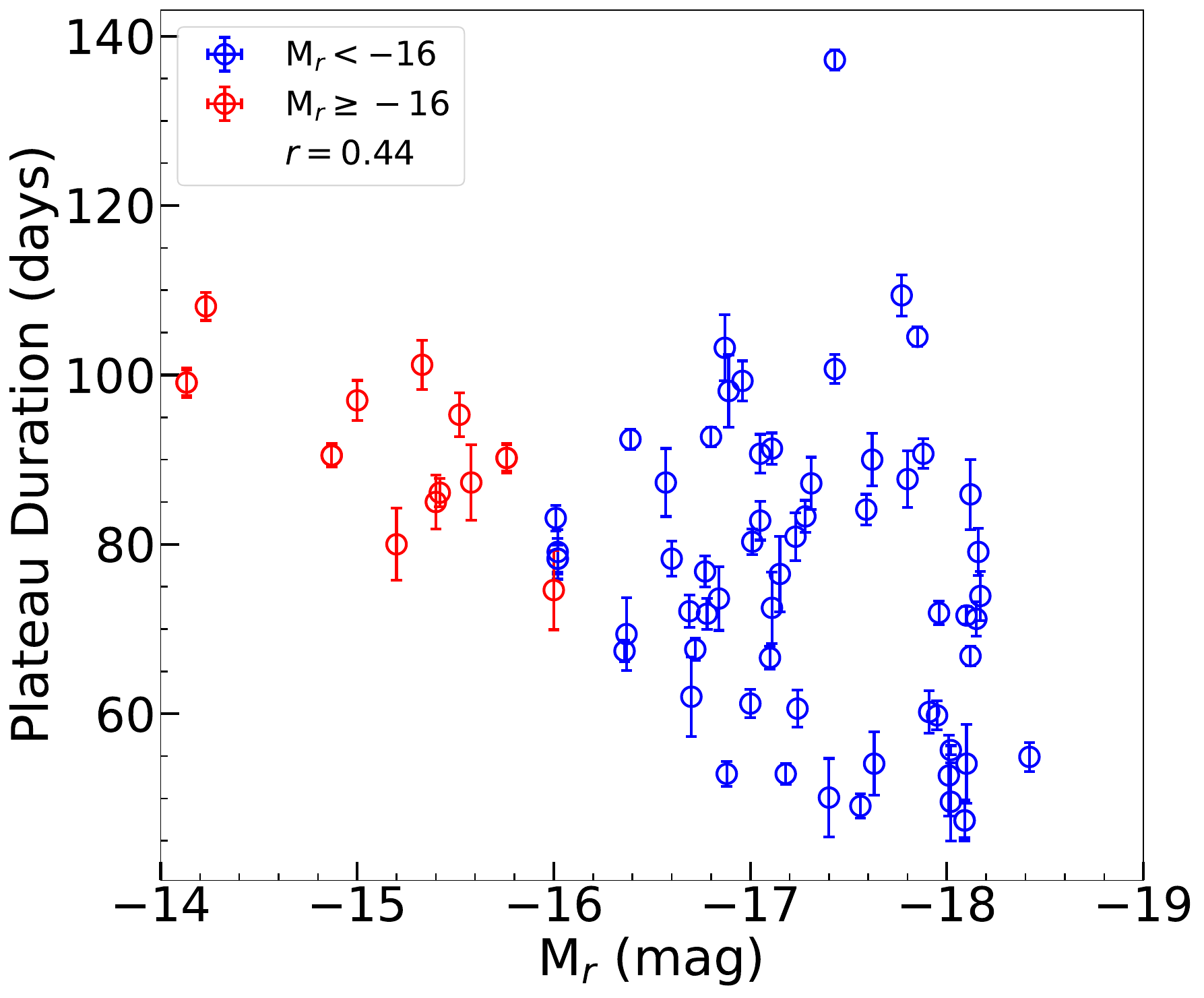}\includegraphics[width=0.32\textwidth]{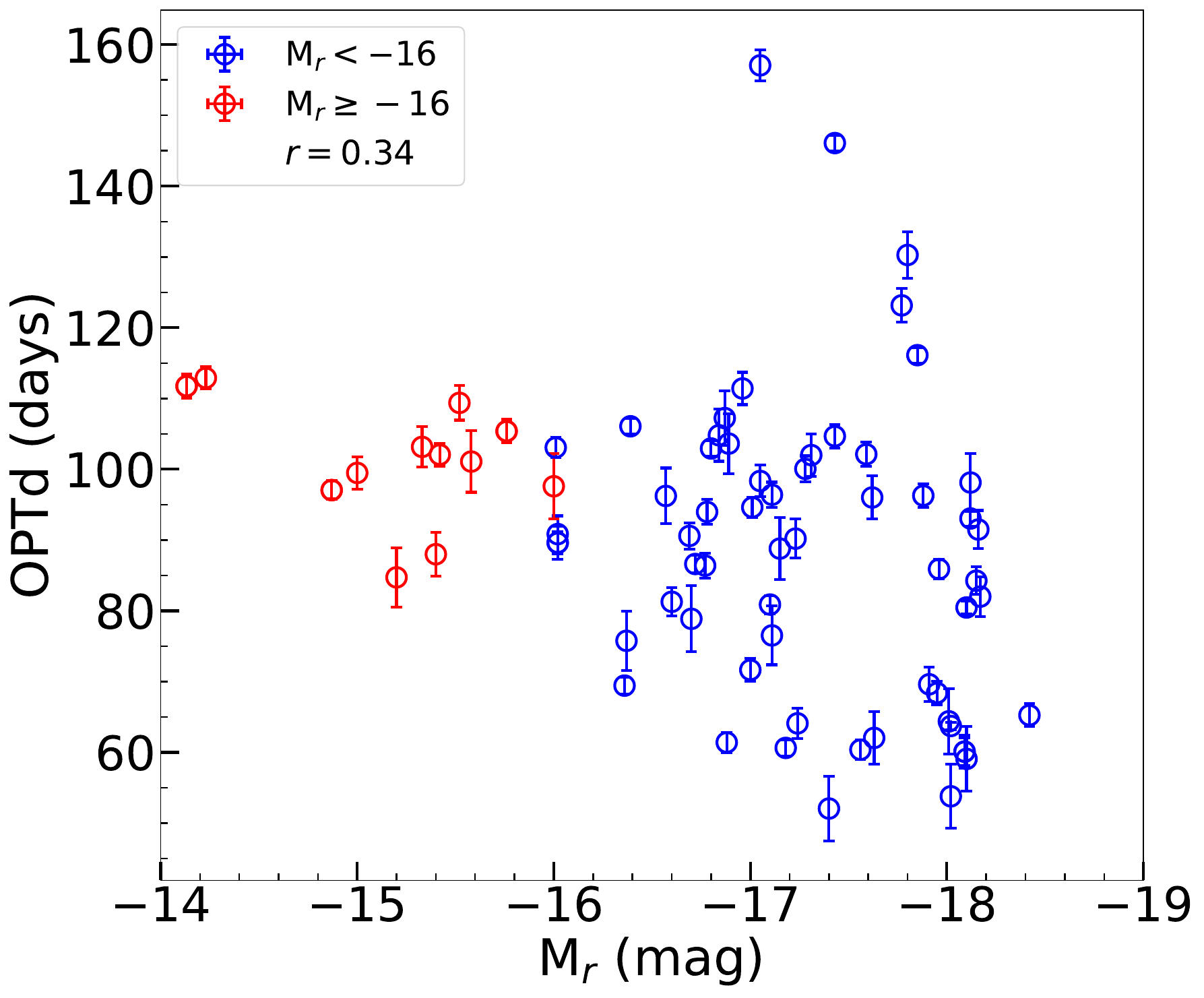}\includegraphics[width=0.32\textwidth]{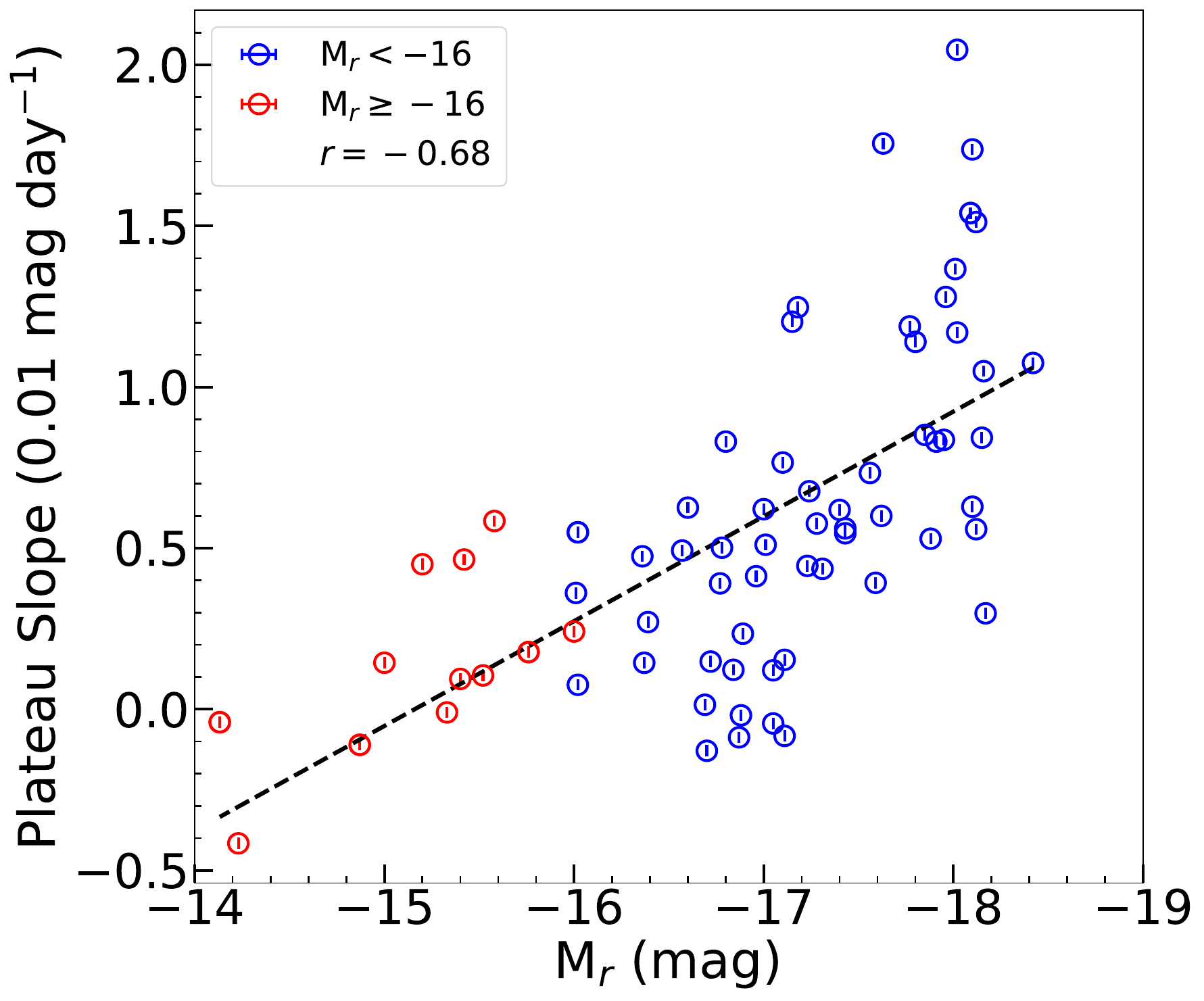}

    \includegraphics[width=0.32\textwidth]{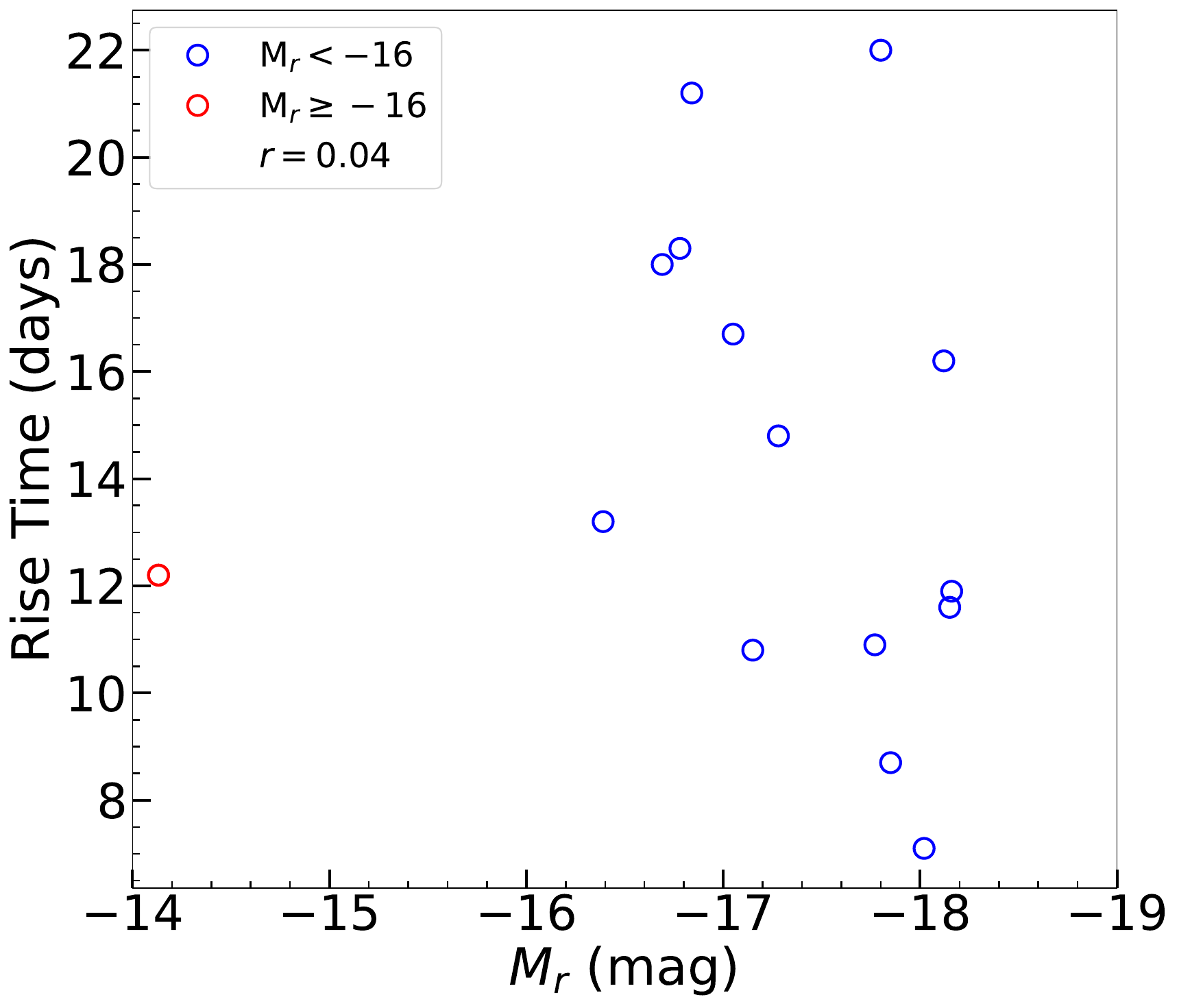}\includegraphics[width=0.32\textwidth]{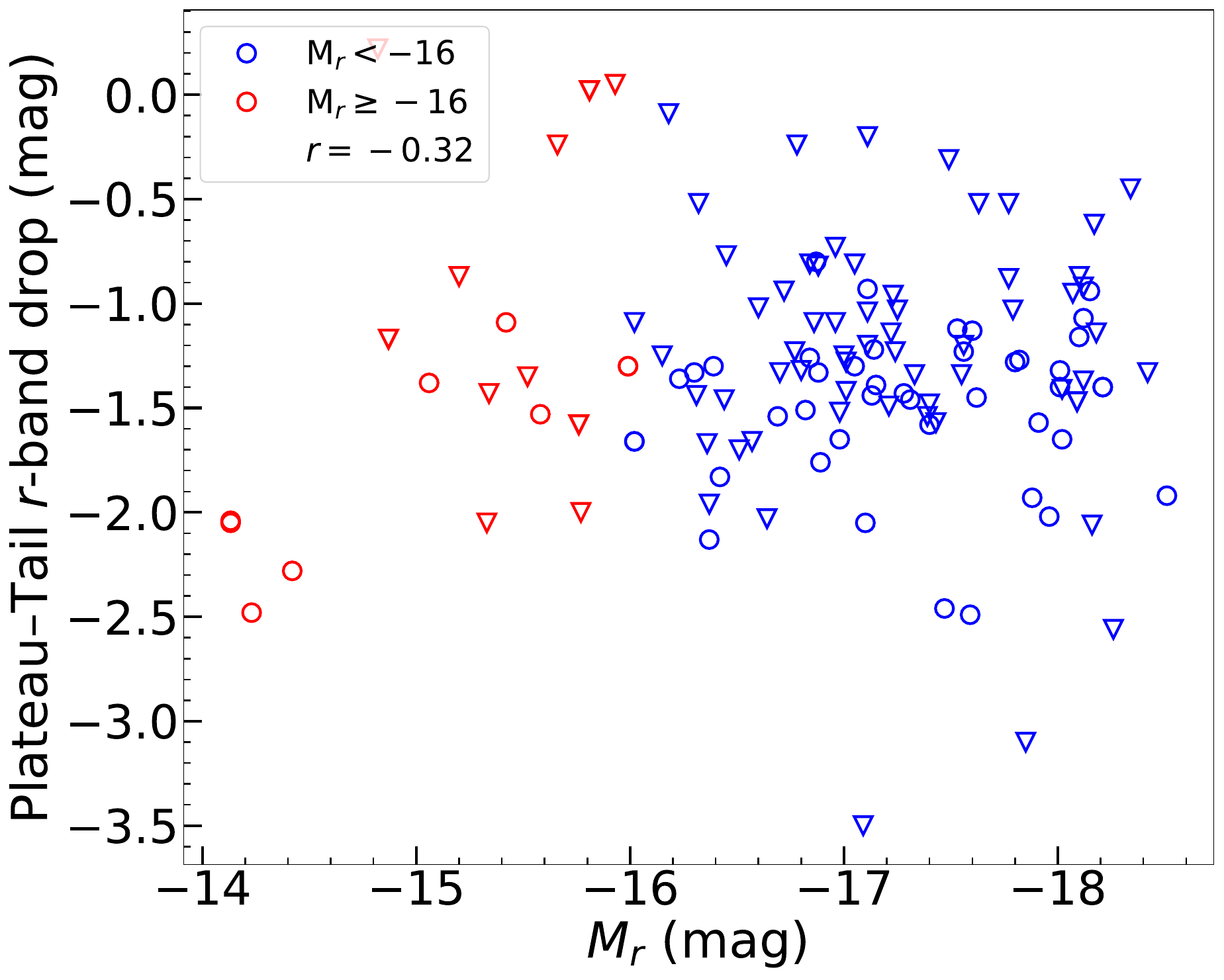}
\caption{
Observable light curve parameters as a function of $r$-band peak absolute magnitude ($M_r$). From left to right and top to bottom, the panels show: (1) plateau duration vs. $M_r$, (2) optically thick phase duration (OPTd) vs. $M_r$, (3) plateau slope vs. $M_r$, (4) rise time vs. $M_r$, and (5) the magnitude drop between the plateau and tail vs. $M_r$. In the last panel, upper limits are indicated by V-shaped triangles for SNe without a detected tail phase. LLIIP SNe with $M_r \geq -16$ are shown in red, while SNe with $M_r < -16$ are shown in blue. See details in Section~\ref{sec:obsresults}.
}

\label{fig:obsvspeak}
\end{figure*}

\begin{figure*}
    \centering
    \includegraphics[width=0.5\textwidth]{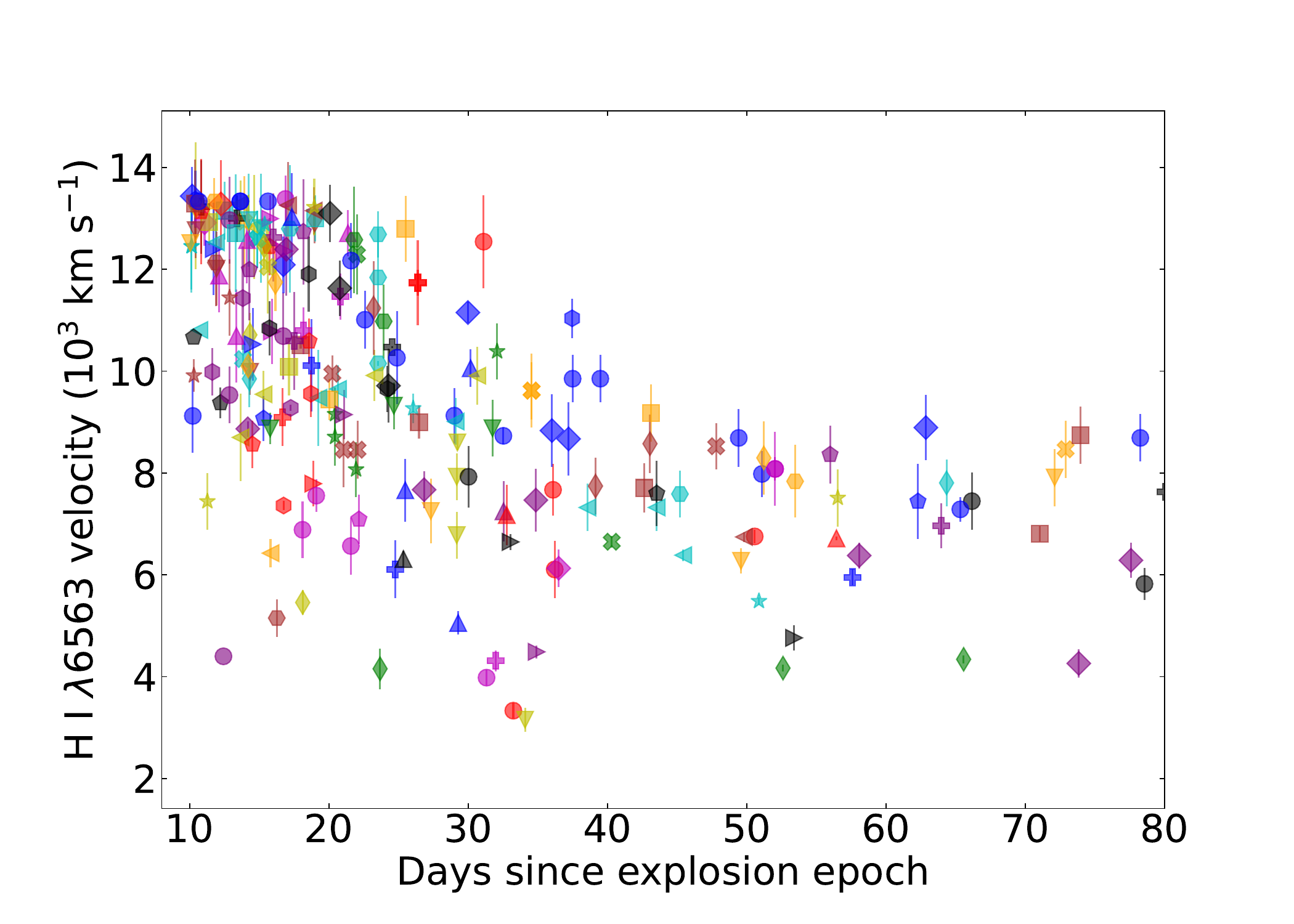}\includegraphics[width=0.45\textwidth]{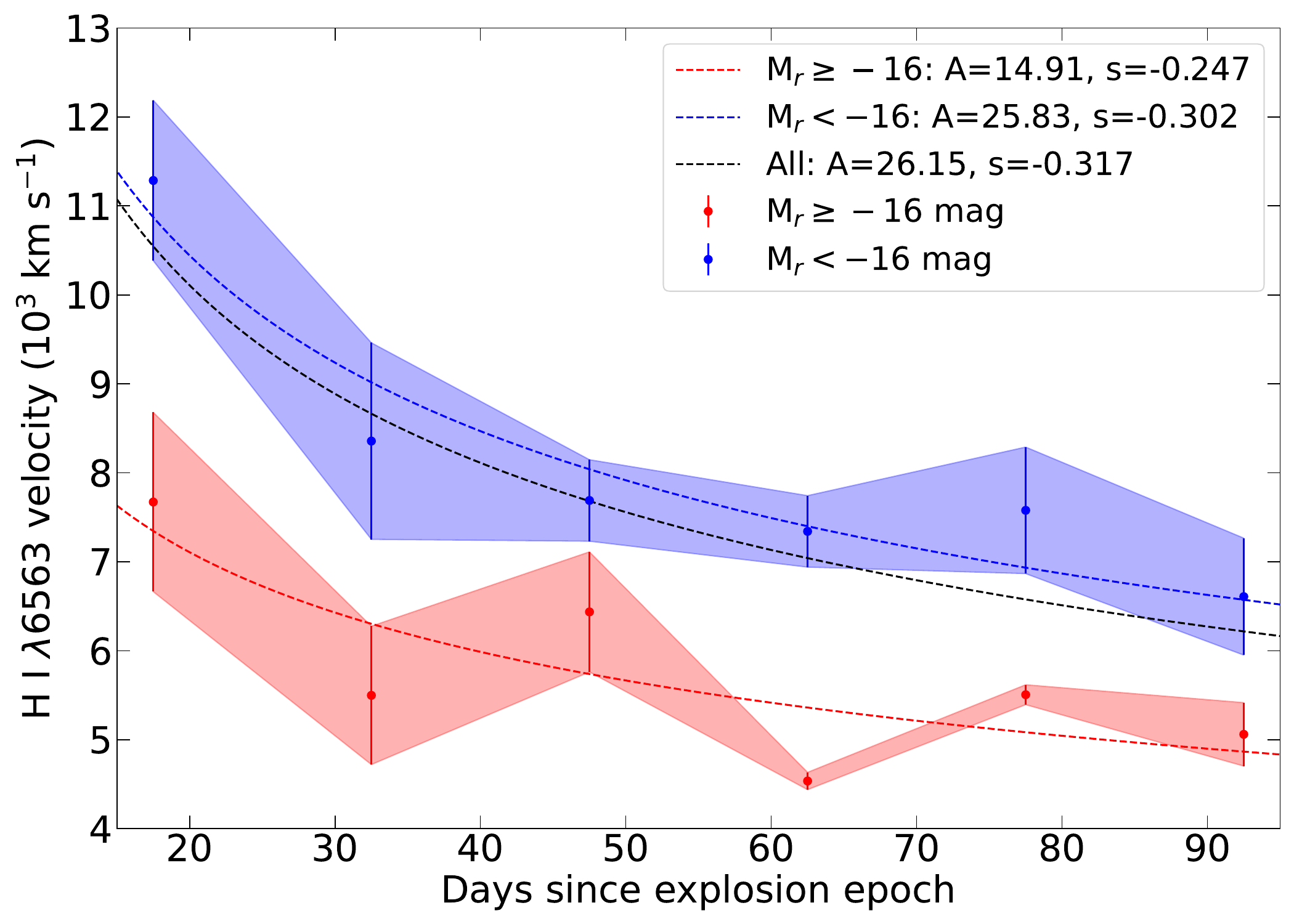}
\caption{
Left: Expansion velocities measured from the absorption minima of the H~I $\lambda6563$ P-Cygni line for all SNe in our sample. Each marker and color represents a unique SN. 
Right: Mean H~I $\lambda6563$ velocities with standard deviation for the full sample (black), LLIIP SNe with $M_r \geq -16$ (red), and brighter SNe with $M_r < -16$ (blue).
}
\label{fig:Hvel}
\end{figure*}

\begin{figure*}
    \centering
    \includegraphics[width=0.47\textwidth]{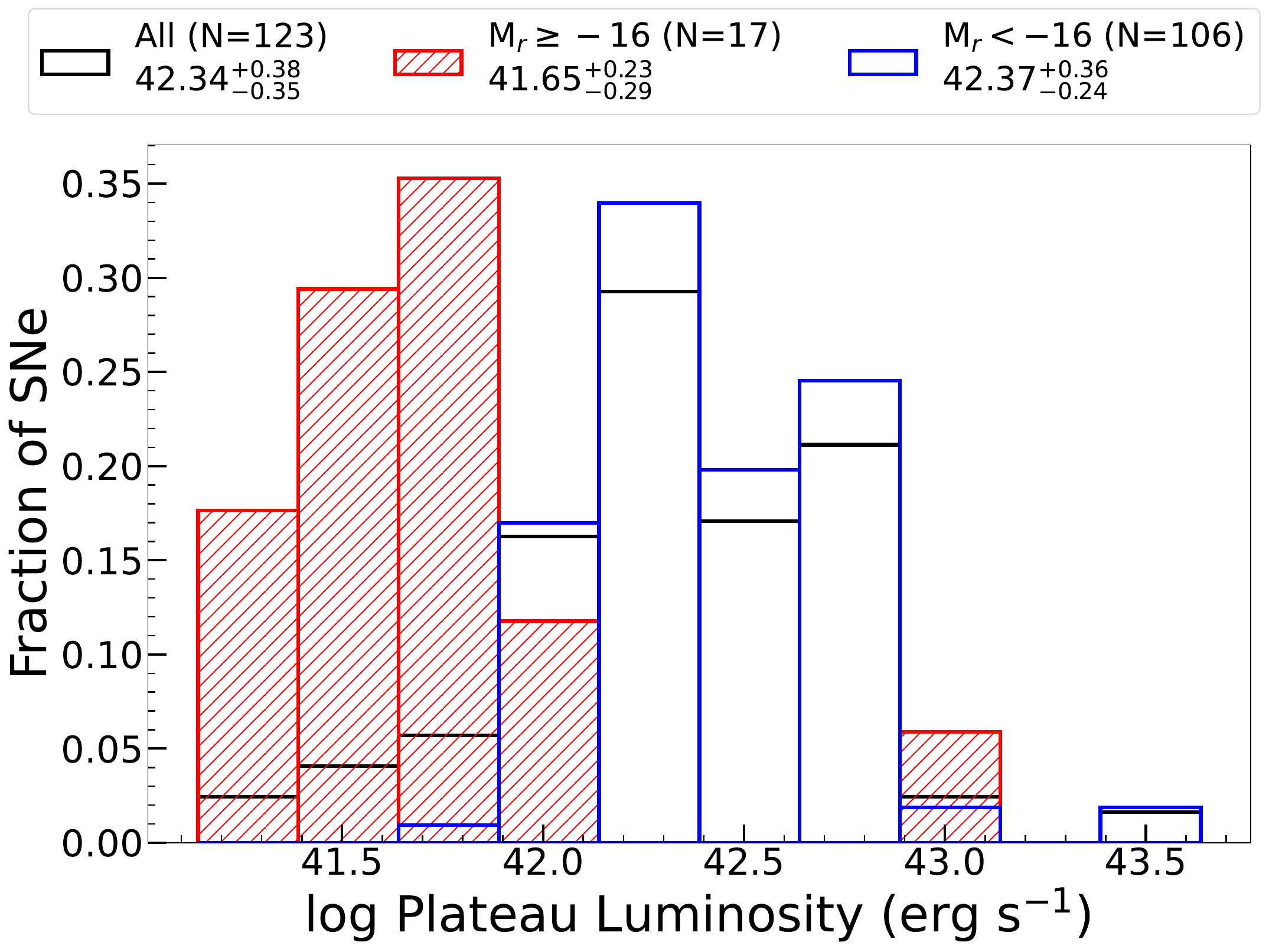}
    \includegraphics[width=0.5\textwidth]{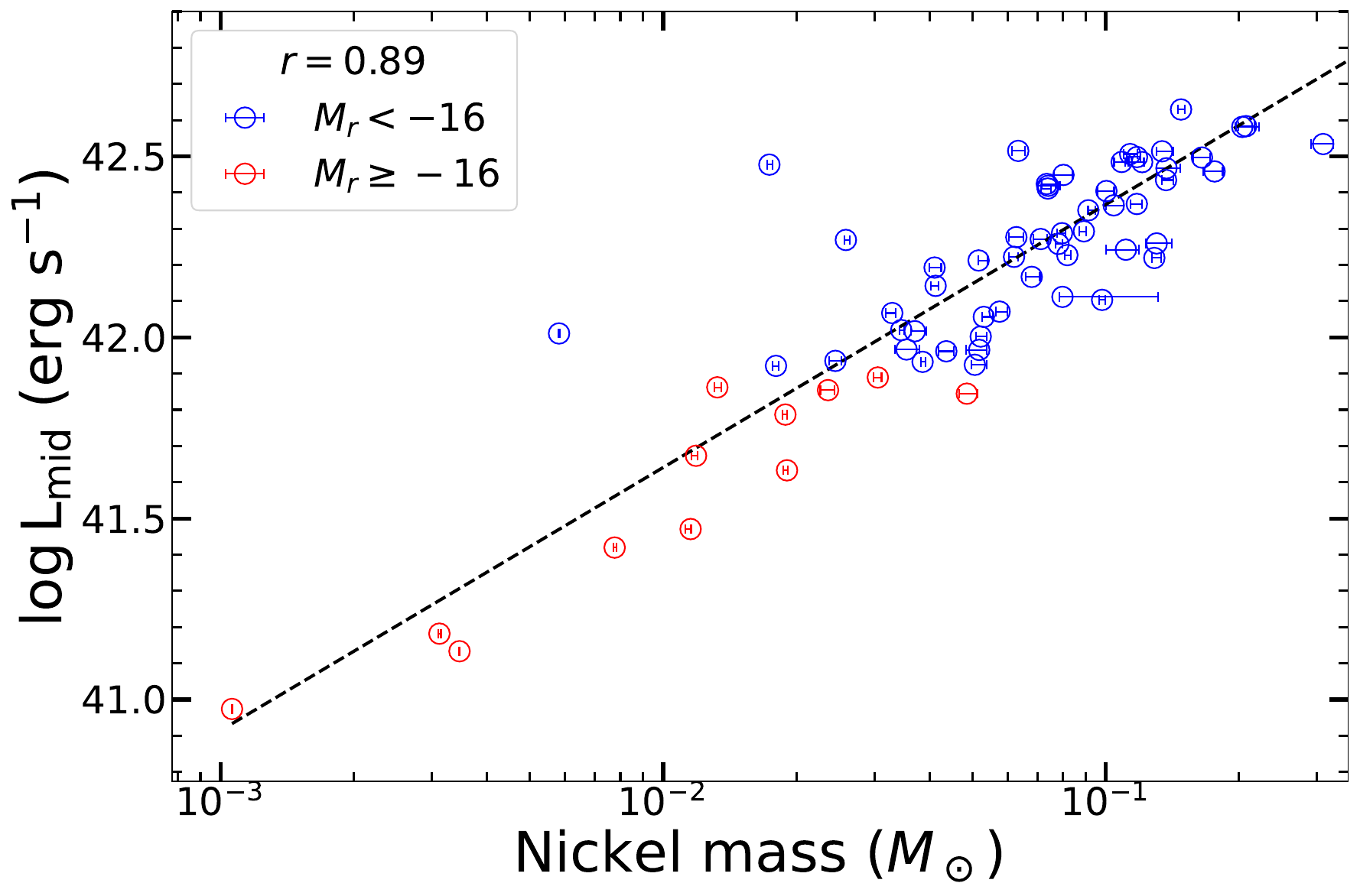}

\caption{
Left: Fractional distribution of bolometric luminosity at the mid-plateau phase. LLIIP SNe with $M_r \geq -16$ are shown in hatched red, those with $M_r < -16$ in blue, and the full sample is shown in black. Right: Correlation between mid-plateau bolometric luminosity and nickel mass inferred from the bolometric lightcurve tail. LLIIP SNe with $M_r \geq -16$ are shown in red, and SNe with $M_r < -16$ are shown in blue.
}

    \label{fig:bolcomp}
\end{figure*}


\section{Extinction Correction}
\label{sec:extinction}

We corrected for Galactic extinction using the maps provided by \citet{Schlafly11}. For the 
corrections, we use the extinction law described by \citet{Cardelli1989} with a value of $R_V = 3.1$.

To estimate host-galaxy extinction, we employ the empirical $g-r$ color template 
constructed in Paper~I \citep{Das2025} from a subset of Type~IIP SNe with negligible host reddening (identified via  weak or absent host \ion{Na}{1}~D absorption). The intrinsic template has a 1$\sigma$ scatter of 
$\sim$0.35\,mag. For each SN in our sample, we measure the deviation of its observed $g-r$ color from this 
template and attribute any excess beyond the 1$\sigma$ intrinsic range to host-galaxy reddening, converting 
it to $A_{V,\mathrm{host}}$ assuming the same extinction law. The template is available on \href{https://zenodo.org/records/14538857}{Zenodo}. See \citet{Das2025} for further details.

\section{Analysis}
\label{sec:analysis}

\subsection{Observed lightcurve properties}
\label{sec:obsanalysis}
\begin{table*}[ht!]
\centering
\caption{Definition of lightcurve shape parameters based on lightcurve phase, slope (mag day$^{-1}$), and rate of change of slope $\Delta_{\rm slope}$ (mag day$^{-2}$).}
\begin{tabular}{lll}
\hline
Parameter & Phase / Slope condition & $\Delta_{\rm slope}$ condition \\
\hline
$t_{\rm plateau,start}$ & Phase $<$ 60 days; earlier of slope $<$ 0.01 mag day$^{-1}$ or max. $\Delta_{\rm slope}$ & Max. $\Delta_{\rm slope}$ \\

$t_{\rm plateau,end}$ & Phase $>$ $t_{\rm plateau,start} + 40$ days; slope $< -0.03$ mag day$^{-1}$ & Min. $\Delta_{\rm slope} < -0.003$ mag day$^{-2}$ \\

$t_{\rm tail}$ & Phase $>$ $t_{\rm plateau,end}$ and $<$ $t_{\rm plateau,end} + 50$ days; slope $<$ 0 & Max. $\Delta_{\rm slope} \geq 0.002$ mag day$^{-2}$ \\

$t_{\rm rise}$ & Time between half-max. brightness ($( \mathrm{M}_r + 0.75 )$ mag) and peak epoch & Not applicable \\
\hline
\end{tabular}
\label{tab:lcparams}
\end{table*}

We employed a Gaussian Process (GP) algorithm\footnote{https://george.readthedocs.io/} \citep{Ambikasaran2015} to \textcolor{black}{fit} the
$r$-band and $g$-band lightcurves. The top panel of Figure~\ref{fig:GPfits} displays the interpolated $r$-band lightcurve, the middle panel shows its first derivative with respect to time (i.e., the slope, in mag day$^{-1}$), and the bottom panel shows the time derivative of the slope, denoted as $\Delta_{\rm slope}$ (in mag day$^{-2}$).

The $r$-band lightcurve parameters are measured using the following criteria:

\begin{itemize}
    \item Plateau onset ($t_{\rm plateau,start}$): Defined as the earlier of (i) the epoch when the slope becomes shallower than 0.01 mag day$^{-1}$, or (ii) the epoch when the change in slope with respect to time ($\Delta_{\rm slope}$) reaches its maximum value, measured within 60 days after first detection.
    
    \item Plateau end ($t_{\rm plateau,end}$): The epoch of plateau end ($t_{\rm plateau,end}$) is identified after 40 days, when the lightcurve begins to steeply decline. This point is determined as the epoch where $\Delta_{\rm slope}$ reaches a minimum below $-0.003$ mag day$^{-2}$, indicating a sharp transition.
    
    \item Tail onset ($t_{\rm tail}$): Defined as the epoch post-plateau when $\Delta_{\rm slope}$ reaches a maximum exceeding $0.002$ mag day$^{-2}$, provided this occurs within 50 days after the plateau ends.
    
    \item Rise time ($t_{\rm rise}$): Defined as the time interval between the epoch when the $r$-band magnitude is 0.75 mag fainter than peak (half-maximum) and the epoch of maximum light.

\end{itemize}

To estimate uncertainties in these observables, we adopt a Monte Carlo approach. We simulate 50 realizations of each lightcurve based on Gaussian noise scaled to the flux errors. For each realization, we recompute the observables using the same criteria. The standard deviation of the resulting distribution of values is used as the uncertainty on the measurements of these observables. The plateau duration is measured as $t_{\rm plateau,end} - t_{\rm plateau,start}$. The optically thick plateau duration (OPTd) is measured as the duration between the explosion epoch and the midpoint of the plateau end and tail onset epochs. The distributions of these observed parameters are shown in Figure \ref{fig:observables}. \textcolor{black}{A sharp luminosity drop between the plateau and the radioactive tail is a key prediction for both ECSNe and failed SNe with very low or negligible $^{56}$Ni production \citep[e.g.,][]{Tominaga2013, Turatto1998}. 
To identify such ECSN and failed-SN candidates in our sample, we measure the drop in the $r$-band magnitude from the end of the plateau to the first epoch on the tail, whenever the tail is detected.}
If a tail does not exist, this is marked as ``Limit = yes'' in Table~\ref{tab:lightcurveparams} or indicated as a V-shaped triangle in Figure~\ref{fig:obsvspeak}. \textcolor{black}{
These photometric definitions follow the same physical interpretation as those adopted
by \citet{Anderson2014}, who also identify the plateau phase through changes in the
light-curve slope and characterize the transition to the radioactive tail using the
steep post-plateau decline. Our implementation differs in that we use the derivatives
of the lightcurve to define these transitions in a fully automated and
objective way, whereas \citet{Anderson2014} relied on parametric spline fits and visually guided
measurements for some phases. The data and code used for these measurements will be made publicly available on \href{https://zenodo.org/records/15717884?token=eyJhbGciOiJIUzUxMiJ9.eyJpZCI6IjNhYjE3NjYwLTU2NDEtNDBkZi1iYmI5LTQ1YzQxY2EwYjllNyIsImRhdGEiOnt9LCJyYW5kb20iOiJhNjA2ZTkzNDFjYjU5NGM3ZGIzMmExMTRlOGY1NzU5MCJ9.X0jCmJz6FJbXwmcY_6ZZ1kYCcRHwrABpoZ2epNXLlpIjVIwrFkEEOGD0Z7Cfk7luRPlHvOCwJdWBwB40vX2JoQ}{\texttt{Zenodo}} and \href{https://github.com}{\texttt{GitHub}} upon publication. The criteria are summarized in Table~\ref{tab:lcparams}. }

\subsection{Measuring H~I velocities}

We derive expansion velocities of the H~I $\lambda6563$ line by measuring the minima of
the absorption troughs in the P-Cygni profiles. A third-order polynomial is fit to the
region surrounding the absorption minimum to estimate the velocity. For spectra with low
resolution or strong host-galaxy contamination, we manually inspect and identify the
minimum. Uncertainties are estimated via a Monte Carlo method. We first compute a noise
spectrum by subtracting a heavily smoothed version of the original spectrum; the
standard deviation of this residual defines the spectral noise. We then generate
perturbed realizations by adding Gaussian noise of equal standard deviation to the
smoothed spectrum and remeasure the velocity for each realization. The 1$\sigma$ spread
in the resulting velocity distribution is adopted as the measurement uncertainty. Only
measurements with a standard deviation less than 50\% of their velocity are considered
reliable and included. The measured velocities and their errors are documented in
Table~\ref{table_vel}.

We find that LLIIP SNe exhibit systematically lower H~I $\lambda 6563$ velocities compared to their higher-luminosity counterparts at all phases Figure~\ref{fig:Hvel}. 
\textcolor{black}{We fit a power-law of the form $V(t) = A\,t^{\,-s}$ to the binned velocity measurements. 
Photospheric velocities in Type~II SNe are expected to follow 
a power-law decline during the plateau phase due to homologous expansion and the recession of the recombination front 
(e.g., \citealt{Kasen2009}). 
The best-fit relations are shown in Figure~\ref{fig:Hvel}. 
For the LLIIP sample, we obtain $V(t) = (14.9 \pm 1.3)\, t^{-0.25 \pm 0.05}$, and for the full sample 
$V(t) = (26.2 \pm 2.1)\, t^{-0.32 \pm 0.05}$, where $V$ is in units of $10^{3}$~km~s$^{-1}$ and 
$t$ is the phase in days since explosion. }

\textcolor{black}{In principle, Fe~II $\lambda5169$ is a better tracer of the photospheric velocity than
H~I (e.g., \citealt{Dessart:aa}). However, most of our spectra were obtained with SEDM at $R\sim100$, and we do not have sufficient resolution and SNR to robustly measure the Fe~II $\lambda5169$ absorption profile in most cases. We
therefore use the H$\alpha$ velocity as a practical proxy. When needed for comparison
with studies that quote Fe~II velocities, we employ the empirical scaling relation
between H$\alpha$ and Fe~II derived by} \citet{Faran2014},
\begin{equation*}
v_{\rm Fe\,II} = (0.855 \pm 0.006)\,v_{\rm H\alpha} - (1499 \pm 87)\ {\rm km\ s^{-1}},
\end{equation*} 
\textcolor{black}{to estimate the corresponding Fe~II velocity. These measured slopes ($s \approx 0.2$--$0.4$) lie within the range of shallow power-law declines inferred for Type~II SN photospheric velocities in previous observational
studies \citep[e.g.,][]{Faran2014, deJaeger2019, Lin2024}. While both LLIIP SNe and brighter IIP SNe show comparable decline rates, LLIIP SNe exhibit lower velocities. At 50 days post-explosion, the average velocity inferred from these fits is approximately $5900 \pm 1200$~km~s$^{-1}$ for LLIIP SNe and $8440 \pm 2100$~km~s$^{-1}$ for SNe with $M_r \geq -16$ mag. These velocities are broadly consistent with those reported for Type~II SNe in the literature \citep[][]{Faran2014, deJaeger2019, Lin2024}.}

\subsection{Bolometric Luminosity}
First, we measure the bolometric luminosity of the SNe in our sample. Since we only have the $g$- and $r$-band magnitudes available for most of the SNe in our sample, we estimate the bolometric luminosities using the bolometric correction (BC) approach from \citet{Lyman2014b} to estimate the bolometric magnitudes:
\[
\text{BC}_g = 0.053 - 0.089 \times (g - r) - 0.736 \times (g - r)^2
\]
$M_{bol} = BC_g + M_g$,  where $M_{bol}$ is the bolometric magnitude, and $M_g$ is the absolute magnitude of the SN in $g$-band filter. \citet{Jager2020} and \citet{Sheng2021} showed that the bolometric luminosity of LLIIP SNe measured by bolometric correction approach from \citet{Lyman2014b} is similar to that obtained by fitting a blackbody function to multiband photometry, which supports the validity of the empirical correlation for the bolometric correction given by \citet{Lyman2014b}.

We bin the data into intervals of 2 days and use $g$- and $r$-band photometry to get the bolometric luminosity for that epoch. If only one band of photometry is available in a bin, then we use the color from the nearest epoch where photometry data for both bands is available within 10 days to calculate the bolometric luminosity. 


The distribution of the plateau bolometric luminosity is shown in the left panel of Figure~\ref{fig:bolcomp}. The full sample has a median peak luminosity of $\log(L_{\rm peak}/{\rm erg\ s^{-1}}) = 42.33^{+0.38}_{-0.39}$.  
LLIIP SNe have lower peak luminosities with a median of $41.65^{+0.24}_{-0.21}$. We find a strong correlation between the mid-plateau bolometric luminosity ($L_{\rm mid}$) and the nickel mass (Figure~\ref{fig:bolcomp}). The nickel mass is measured by fitting the radioactive tail (see Section~\ref{sec:semianalysis}). The best-fit relation is
\[\log L_{\rm mid} = (0.73 \pm 0.05)\,\log M_{\rm Ni} + (43.09 \pm 0.06),\]

with a Pearson correlation coefficient of $r=0.89$ and a $p$-value $<10^{-5}$.  \textcolor{black}{
These luminosities are consistent with previous measurements for both normal and 
low-luminosity Type~II SNe. Large-sample studies typically find mid-plateau 
bolometric luminosities of $\log L_{\rm mid} \sim 42.3$--42.7~erg~s$^{-1}$ for 
ordinary SNe~IIP \citep[e.g.,][]{Bersten2009, Pejcha2015, Valenti2016}, 
whereas underluminous events such as SN~1997D, SN~2005cs, and SN~2008bk exhibit 
significantly fainter plateaus with $\log L_{\rm mid} \sim 41.4$--41.7~erg~s$^{-1}$ 
\citep{Spiro2014}. The tight correlation we find  between $L_{\rm mid}$ and $M_{\rm Ni}$ is also consistent with earlier work showing  that fainter plateau luminosities correspond to smaller $^{56}$Ni yields  \citep{Hamuy2003, Pejcha2015, Muller2017, Martinez2022}.}


\subsection{Estimating physical and explosion parameters}

In order to estimate progenitor and explosion parameters of the Type IIP sample, we use semi-analytic models \citet{Nagy2016} and radiation hydrodynamical models from \citet{Moriya2023}. We also use scaling relations from \cite{Goldberg2019} to constrain the explosion parameters of LLIIP SNe.

\subsubsection{Lightcurve fitting to Semi-analytical models}
\label{sec:semianalysis}
 The model in \citet{Nagy2016} is based on a two-component configuration consisting of a uniform dense stellar core and an extended low-mass envelope where the density decreases as an exponential function. \citet{Nagy2016, Jager2020} claim that the results from the two-component semi-analytic LC model are consistent with current state-of-the-art calculations for Type II SNe. They are useful to derive estimates or constraints of basic parameters like the explosion energy, ejected mass, and initial radius of the progenitor as well as the amount of synthesized radioactive nickel. \textcolor{black}{The model assumes spherical symmetry and do not adequately model the initial transient behavior at early stages (t\,$<$\,20 days).} Thus, we only fit for lightcurve data 20 days after the explosion. We use the MCMC routine in \citet{Jager2020} to fit these semi-analytical expressions. The Python code used for these fits will be made available on \href{https://github.com/kaustavkdas/LLIIP_lightcurve}{GitHub} after publication. The nickel mass is measured from the lightcurve tail by fitting for both the nickel mass and the gamma-ray trapping efficiency parameter as free parameters. The priors used for the fits are shown in Table \ref{tab:semipriors}. 
 We fit the semi-analytic models to Type IIP SNe. After marginalization, we have estimates with confidence intervals (1$\sigma$) for each of these parameters. The best-fit values for each SN are listed in Table \ref{tab:nagyfit}. The sample statistics are summarized in Table \ref{tab:semianalytic_summary}.

\subsubsection{Lightcurve fitting to radiation-hydrodynamical Type II model grids}

\citet{Moriya2023} presented a comprehensive model grid containing 228,016 synthetic Type II SN light curves, based on calculations using the radiation-hydrodynamical code STELLA. This set of model grids has previously been used in \citet{Subrayan2023} and \citet{Silva2024} to infer explosion parameters for a sample of 45 and 186 Type II SNe, respectively. Here we also include previously unpublished models for lower-mass progenitors ($9, 10$ \Msun) with low explosion energies (1.0, 2.0, 3.0, 4.0, $\times 10^{50} \mathrm{erg}$). The overall parameter space covered by the model grid is progenitor masses (9, 10, $12$, $14$, $16$, and $18~\Msun$ at the zero-age main sequence (ZAMS), solar metallicity), explosion energies ($0.5$, $1.0$, $1.5$, $2.0$, $2.5$, $3.0$, $3.5$, $4.0$, $4.5$, and $5.0\times 10^{51}\ \mathrm{erg}$),  nickel  masses ($0.001$, $0.01$, $0.02$, $0.04$, $0.06$, $0.08$, $0.1$, $0.2$, and $0.3~\Msun$),  mass-loss rates ($10^{-5.0}$, $10^{-4.5}$, $10^{-4.0}$, $10^{-3.5}$, $10^{-3.0}$, $10^{-2.5}$, $10^{-2.0}$, $10^{-1.5}$, and $10^{-1.0}$ \Msun~yr$^{-1}$ with a wind velocity of 10 km~s$^{-1}$), circumstellar matter radii ($1,2,4,6,8,$ and $10\times 10^{14}
\mathrm{cm}$), and ten circumstellar structures ($\beta=0.5,1.0,1.5,2.0,2.5,3.0,3.5,4.0,4.5,$ and $5.0$), where $\beta$ is a wind structure parameter determined by the efficiency of wind acceleration \citep{Moriya2023}. The radioactive nickel is assumed to be uniformly mixed up to half the mass of the hydrogen-rich envelope.

We use this model grid to measure the explosion parameters of the CLU Type IIP SNe. Using Bayesian Inference methods requires the models to be finely sampled within the parameter space. However, because the model grid is discrete and non-uniform, we extrapolate the model parameter by doing a weighted average of the model magnitudes weighted by the separation of the vector space of the nearest-neighbour models from the model vector to be sampled.
For a given parameter vector $\vec{\theta}$, the method finds the closest models $\vec{\theta}_{close}$ and weighs them appropriately using 
\begin{equation}
    m(t, \vec{\theta}) = \sum_{\vec{\theta}_{\it{i}} \in  \vec{\theta}_{\textit{close}} }\hat{w}^{-1}(\vec{\theta},\vec{\theta}_{\it{i}}) m (t, \vec{\theta}_{\it{i}}),
    \label{EQ1}
\end{equation}
where $m(t, t_{\text{exp}}, \vec{\theta})$ is the magnitude for a given $\vec{\theta}$ at time $t$, where
\begin{equation}
    w(\vec{\theta},\vec{\theta}_{\it{i}}) = \Big( \prod_{\it{j}} \big|\theta^{\it{j}} - \theta_{\it{i}}^{\it{j}} \big| \Big) \Big( \big|\theta^{\it{j}} - \theta_{\it{i}}^{\it{j}} \big|_\textrm{average} \Big)^{-1}.
    \label{EQ3}
\end{equation}

The priors used for the fits are shown in Table \ref{tab:moriyapriors}. The Python code used for these fits will be made available on \href{https://github.com/kaustavkdas/LLIIP_lightcurve}{GitHub} after publication. The best-fit values for each SN are listed in Table \ref{moriyafit}. The sample statistics are summarized in Table \ref{tab:radhydro_summary}.

\begin{figure*}
    \centering
    \includegraphics[width=0.33\textwidth]{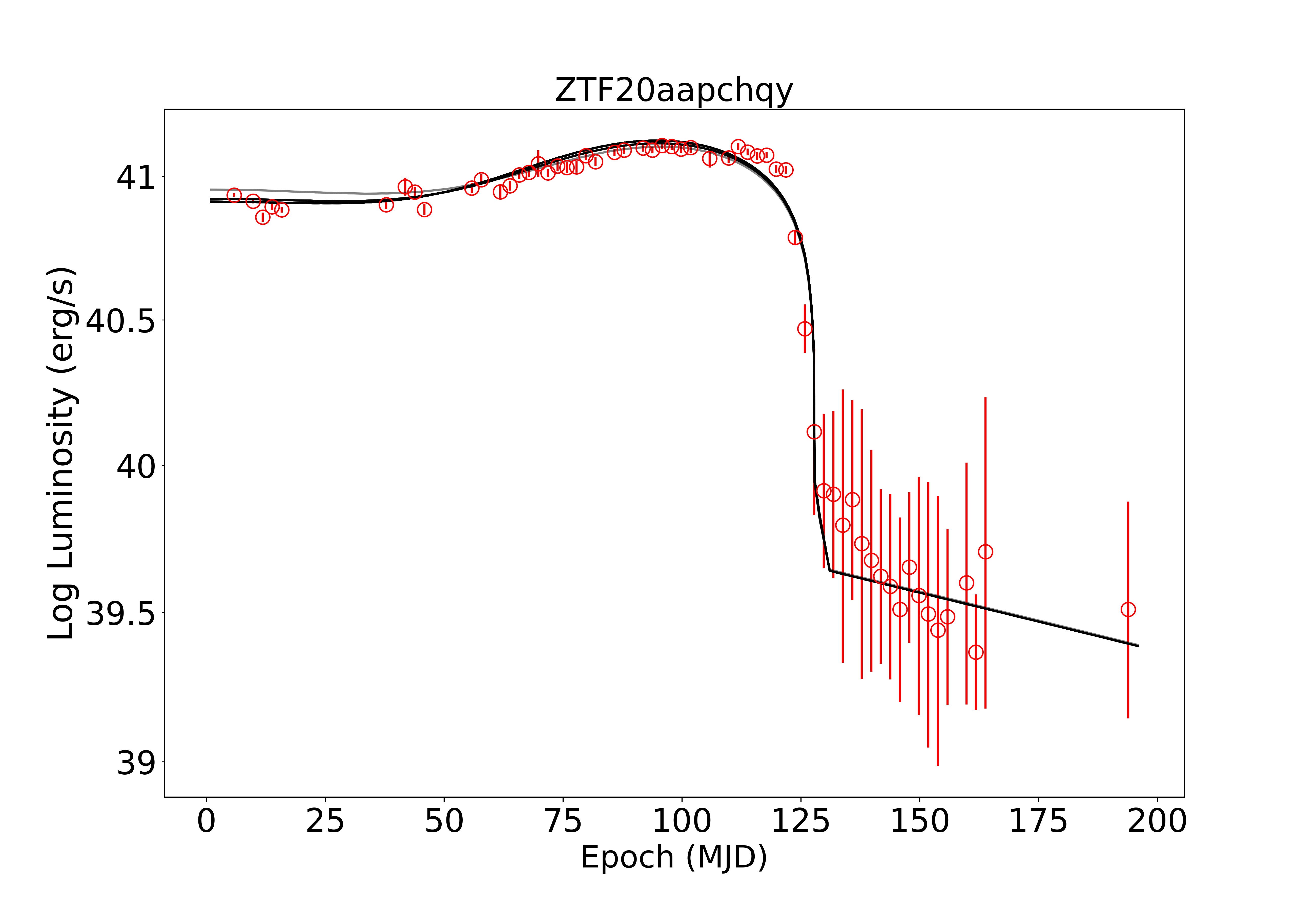}\includegraphics[width=0.33\textwidth]{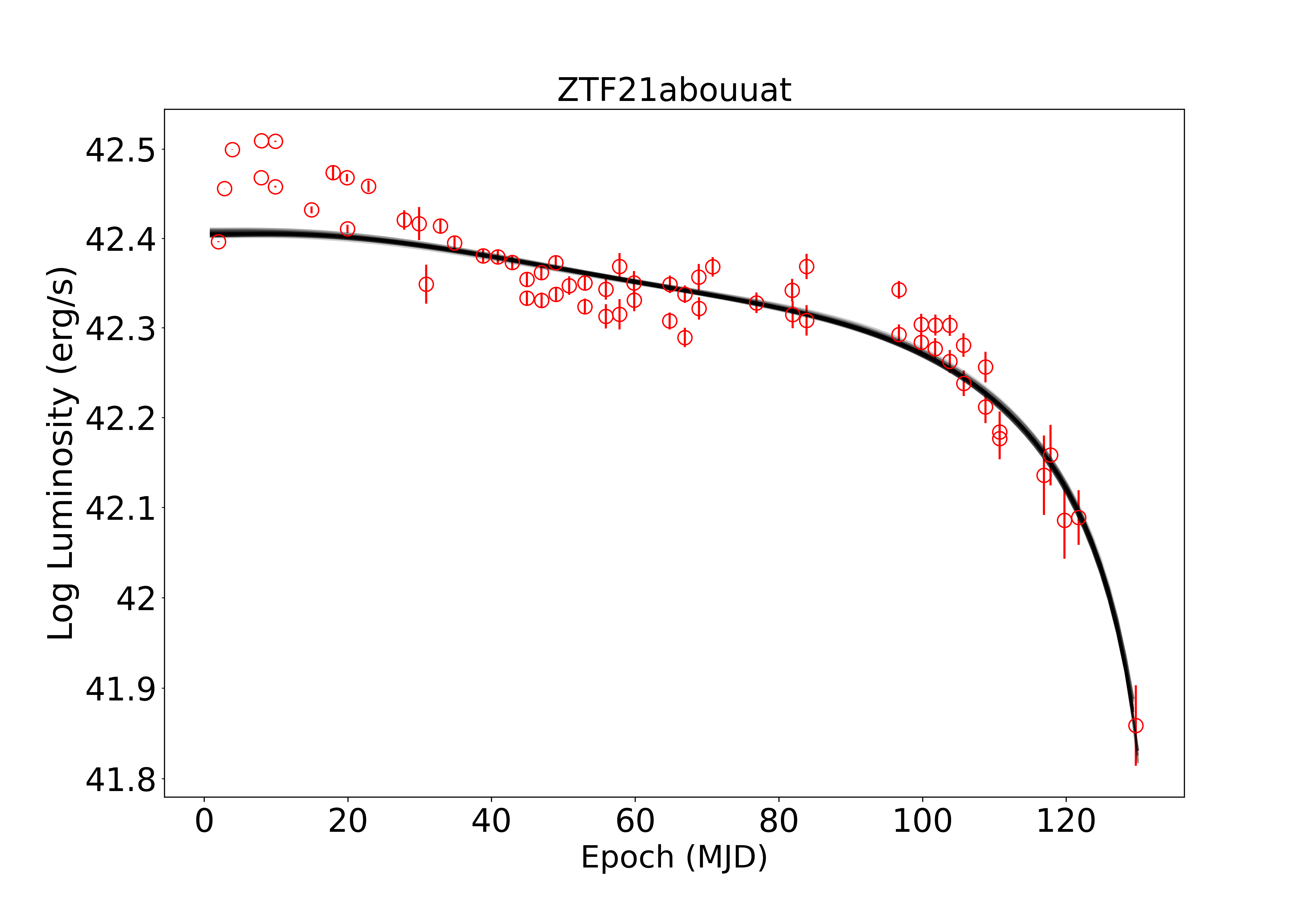}\includegraphics[width=0.33\textwidth]{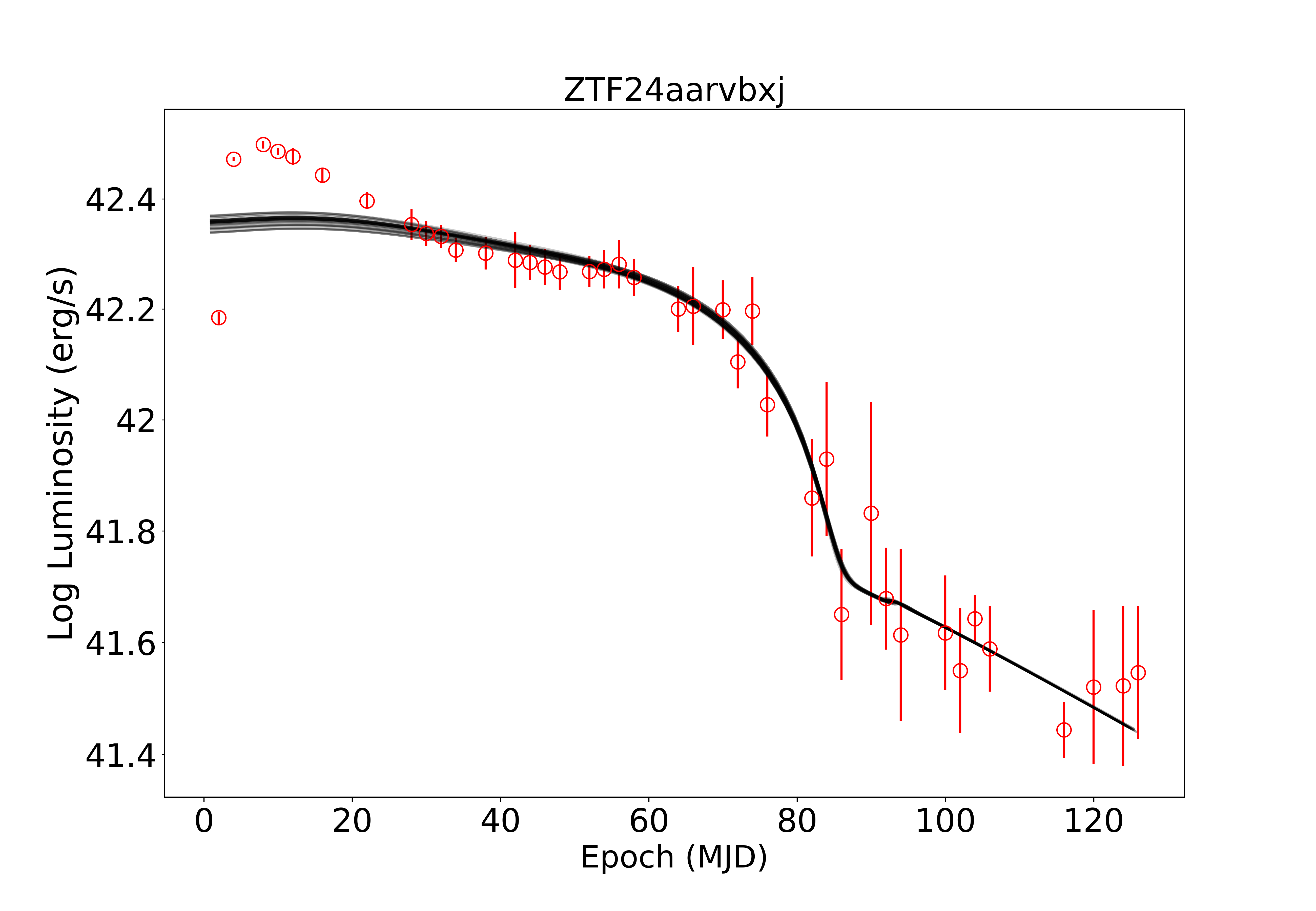}
    
    \includegraphics[width=0.33\textwidth]{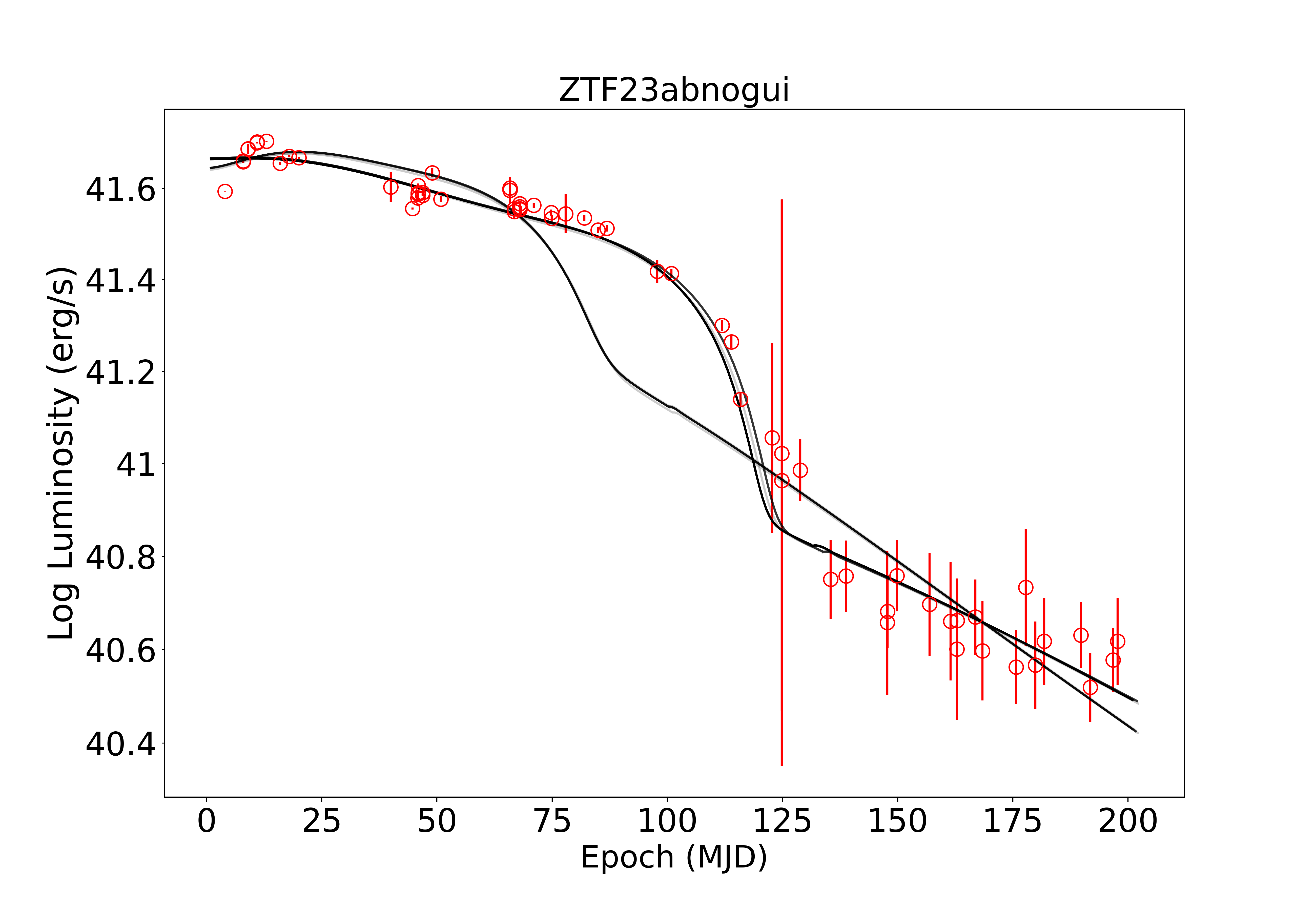}\includegraphics[width=0.33\textwidth]{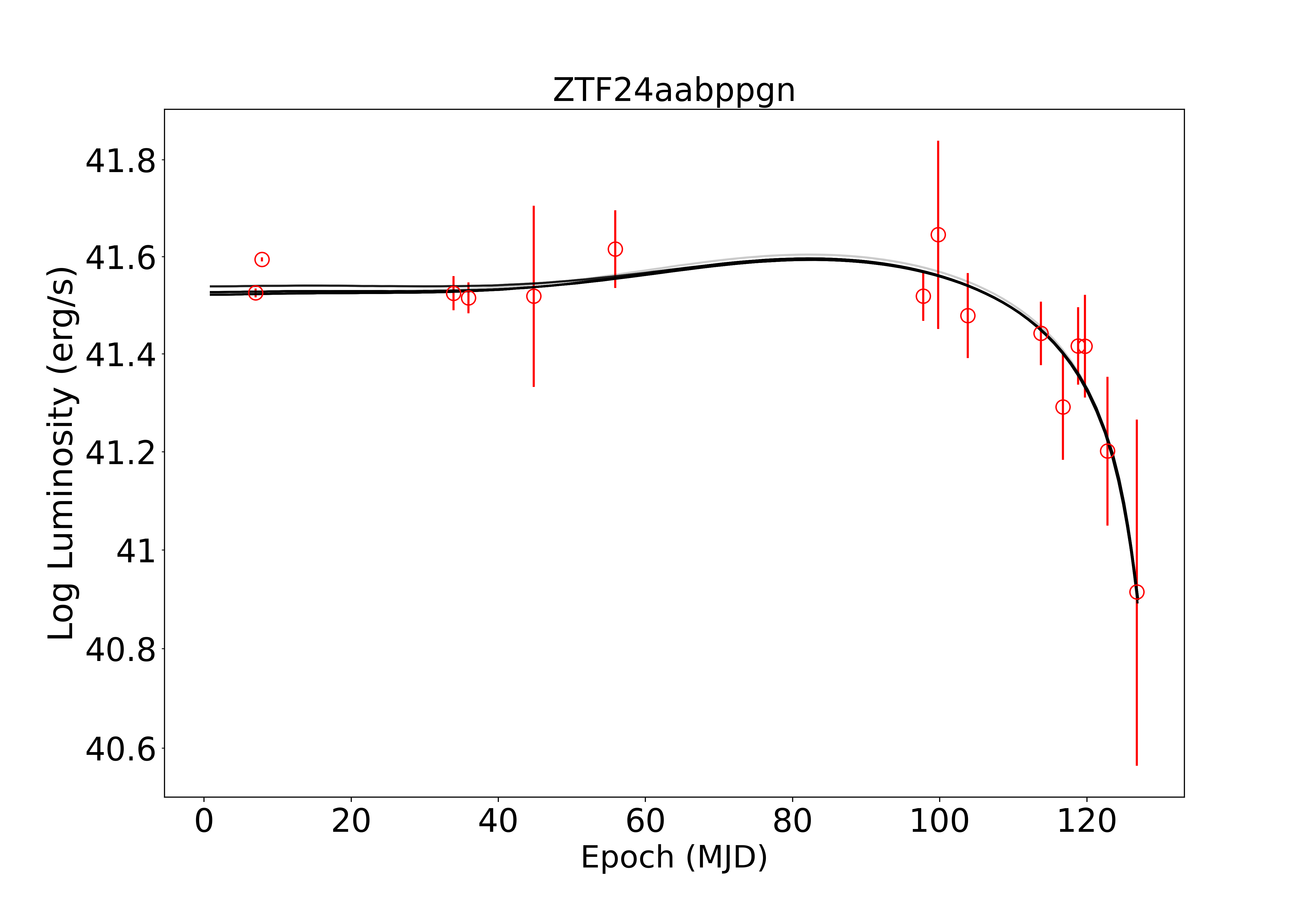}\includegraphics[width=0.33\textwidth]{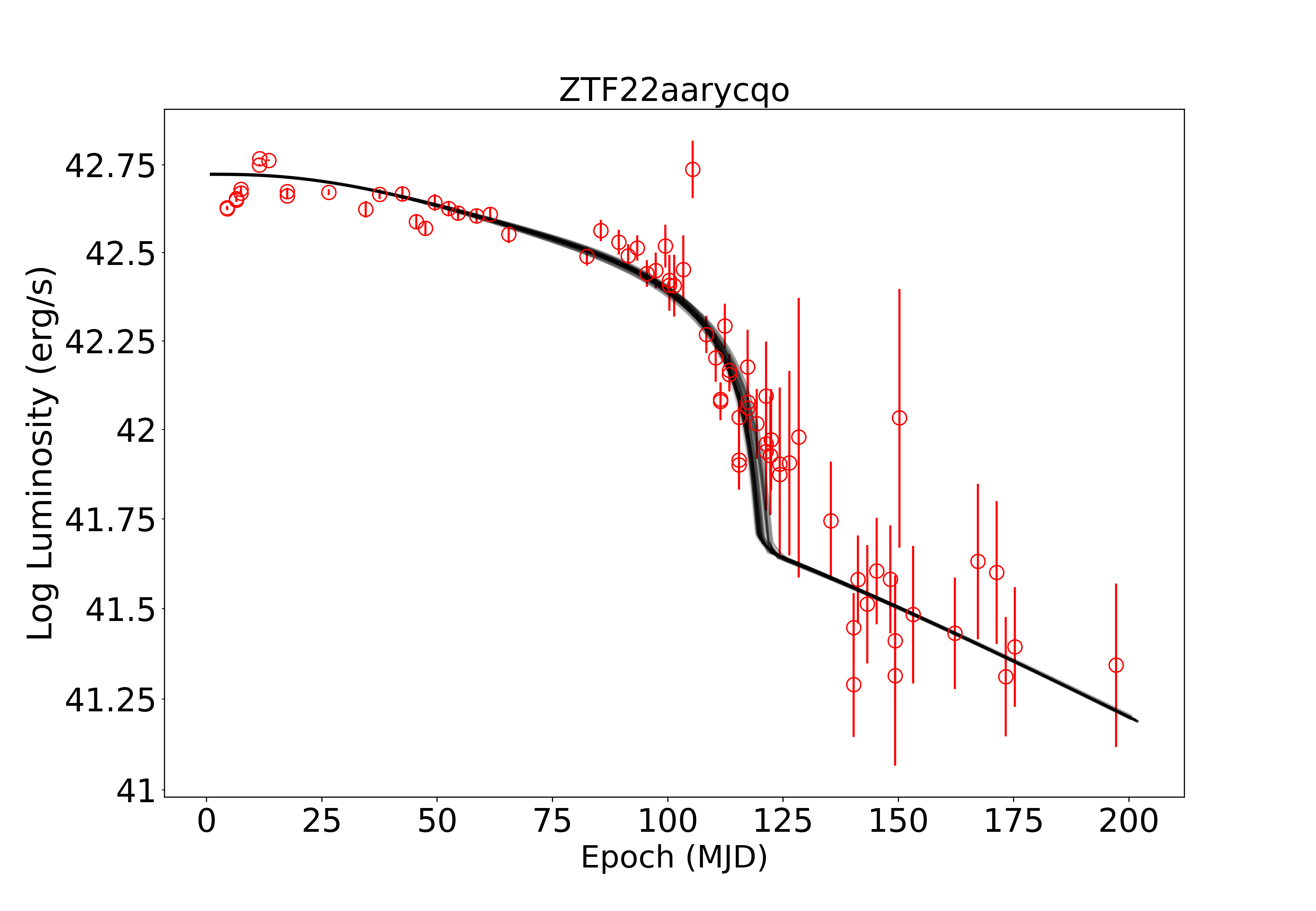}
    
\caption{
Bolometric light curve fits using semi-analytical models from \citet{Nagy2016}. Red dots show the observed bolometric luminosities, while black lines represent the best-fit model lightcurves. The panels show (left to right, top to bottom): ZTF20aapchqy (SN~2020cxd), ZTF21abouuat (SN~2021ucg), ZTF24aarvbxj (SN~2024lby), ZTF23abnogui (SN~2023wcr), ZTF24aabppgn (SN~2024wp), and ZTF22aarycqo (SN~2022ojo).
}
    
    \label{fig:semifits}
\end{figure*}

\begin{figure*}
    \centering
    \includegraphics[width=0.33\textwidth]{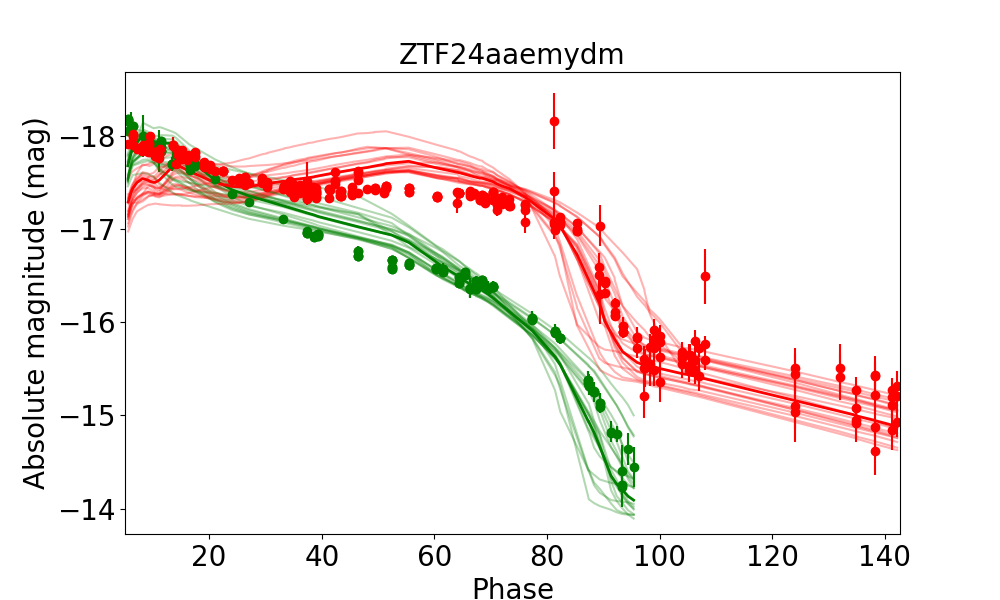}\includegraphics[width=0.33\textwidth]{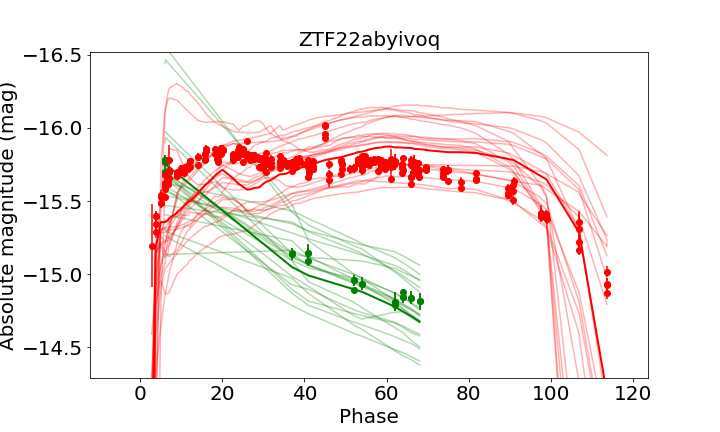}\includegraphics[width=0.33\textwidth]{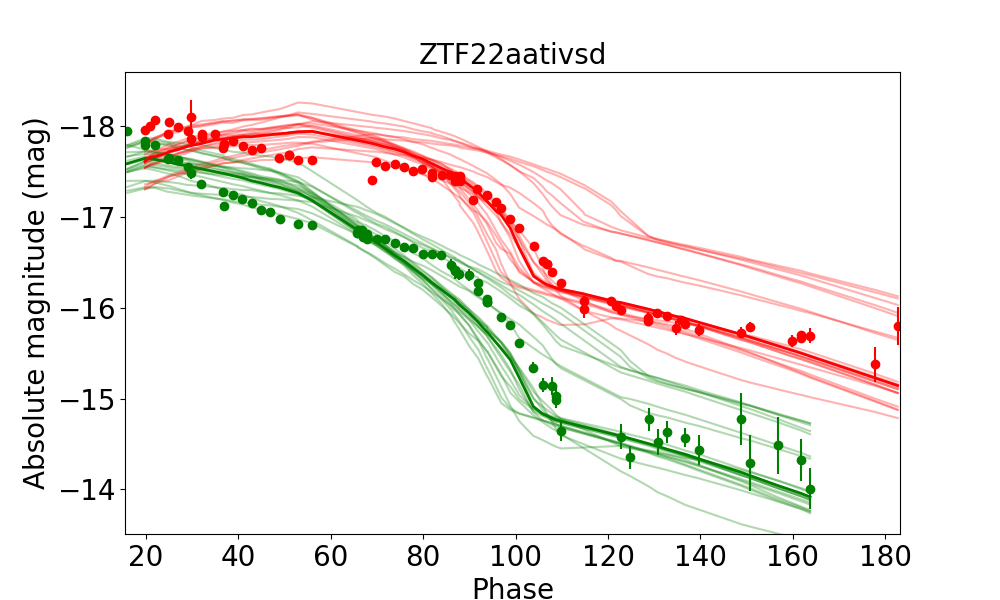}
    
    \includegraphics[width=0.33\textwidth]{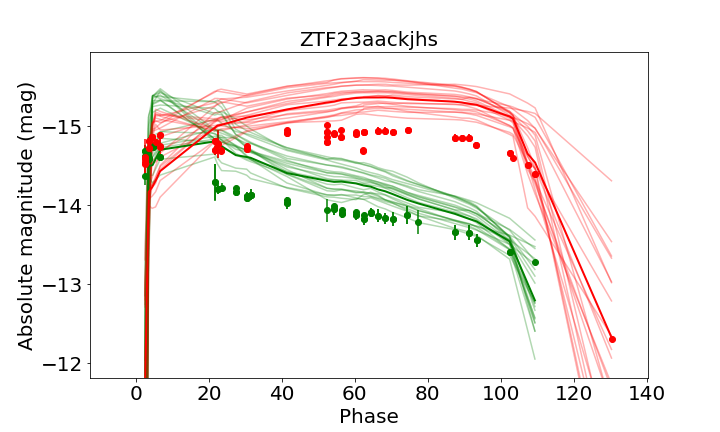}\includegraphics[width=0.33\textwidth]{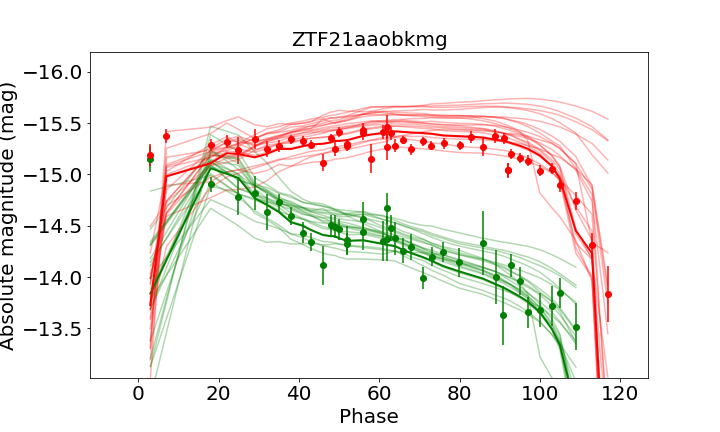}\includegraphics[width=0.33\textwidth]{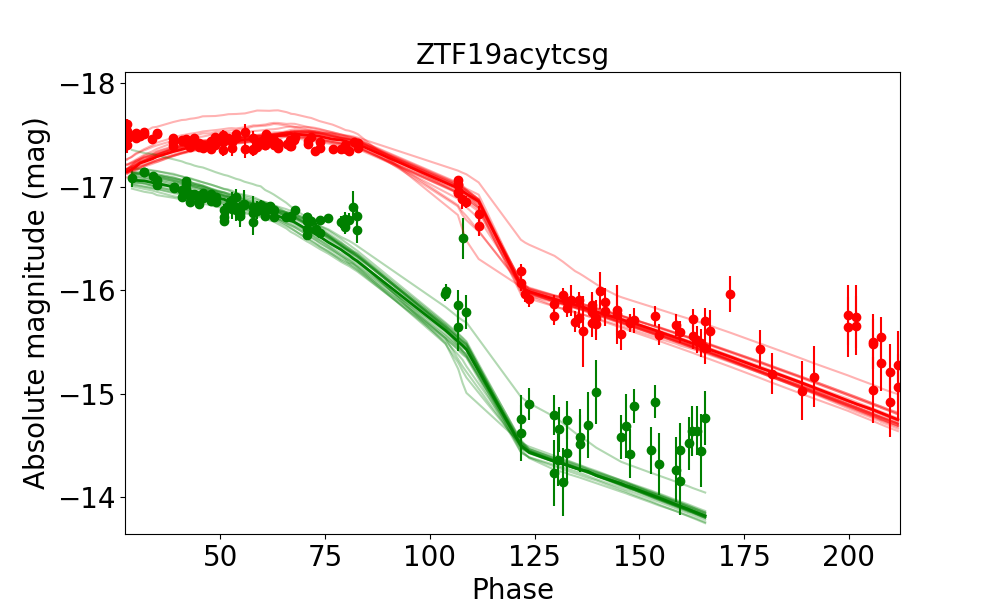}

\caption{
Radiation-hydrodynamical model fits to the $r$- and $g$-band light curves using the model grid from \citet{Moriya2023}. The model grid in \citet{Moriya2023} has been extended to include 9~$M_\odot$ and 10~$M_\odot$ progenitors with explosion energies in the range $(1$--$5) \times 10^{50}$ erg. The panels show (left to right, top to bottom): ZTF24aaemydm (SN~2024chx), ZTF22abyivoq (SN~2022acko),  ZTF22aativsd (SN~2022ovb), ZTF23aackjhs (SN~2023bvj), ZTF21aaobkmg (SN~2021eui) and ZTF19acytcsg (SN~2019wvz).
}

    \label{fig:moriyafits}
\end{figure*}

\begin{figure*}
    \centering
    \includegraphics[width=0.33\textwidth]{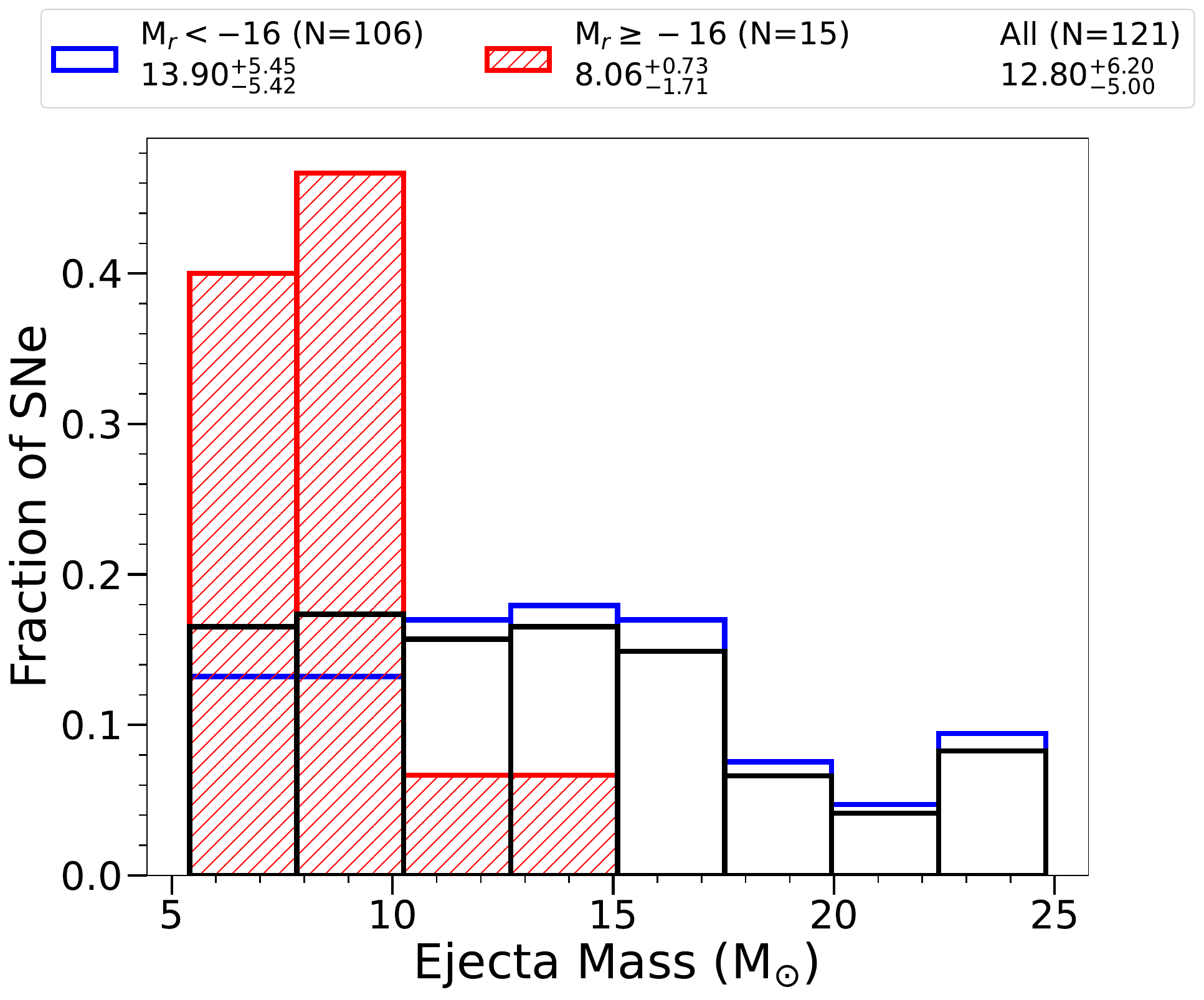}\includegraphics[width=0.33\textwidth]{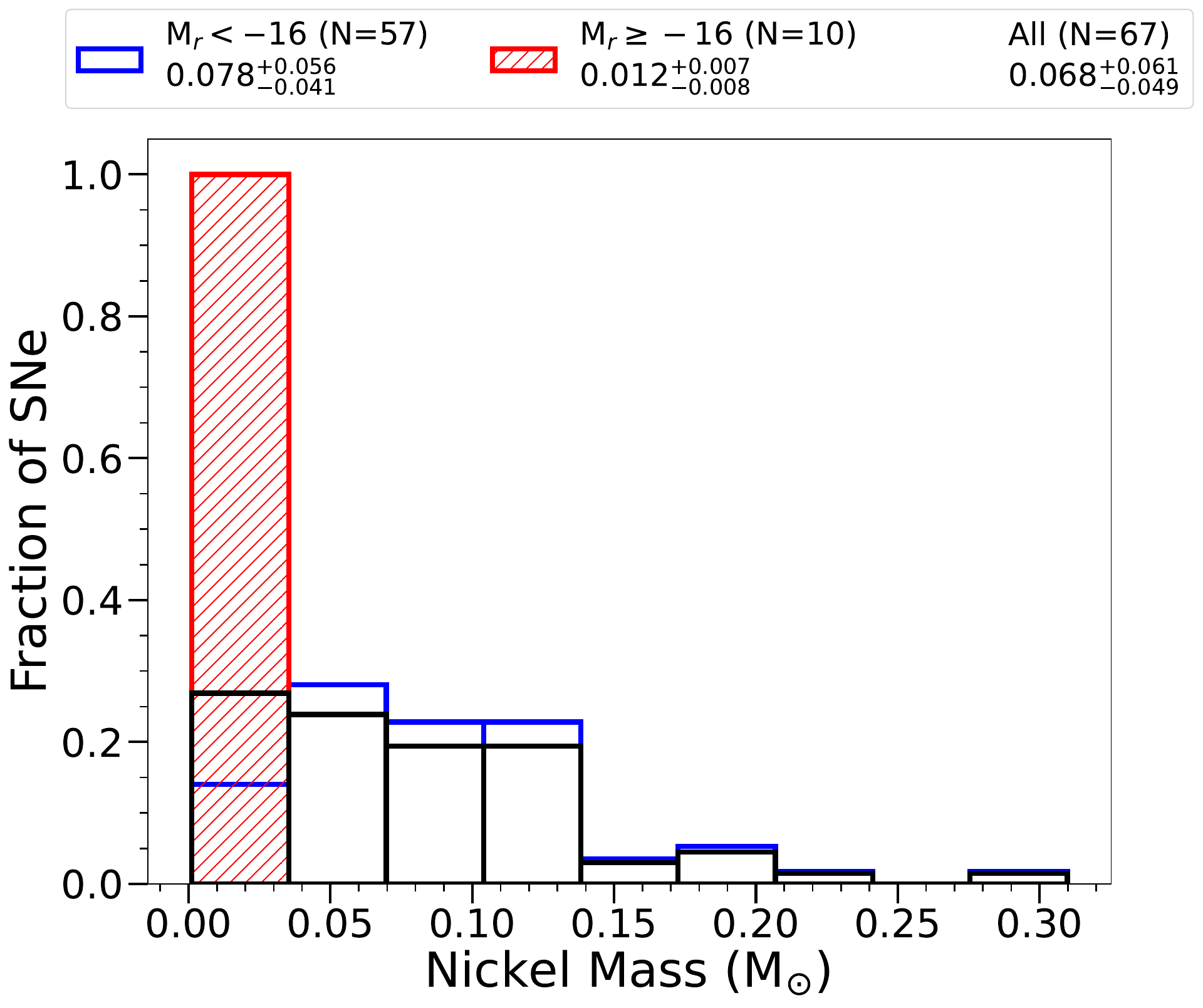}

    \includegraphics[width=0.33\textwidth]{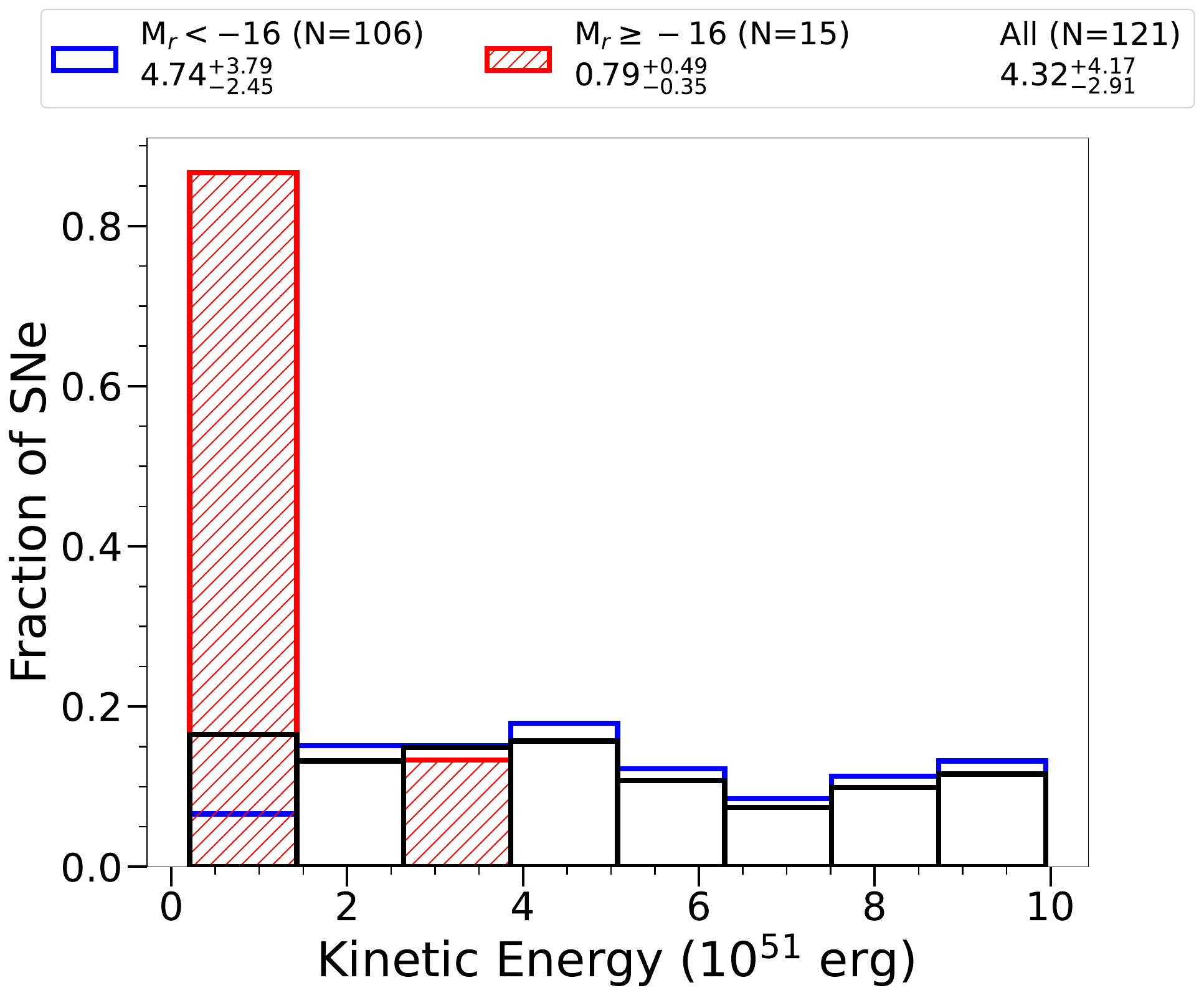}\includegraphics[width=0.33\textwidth]{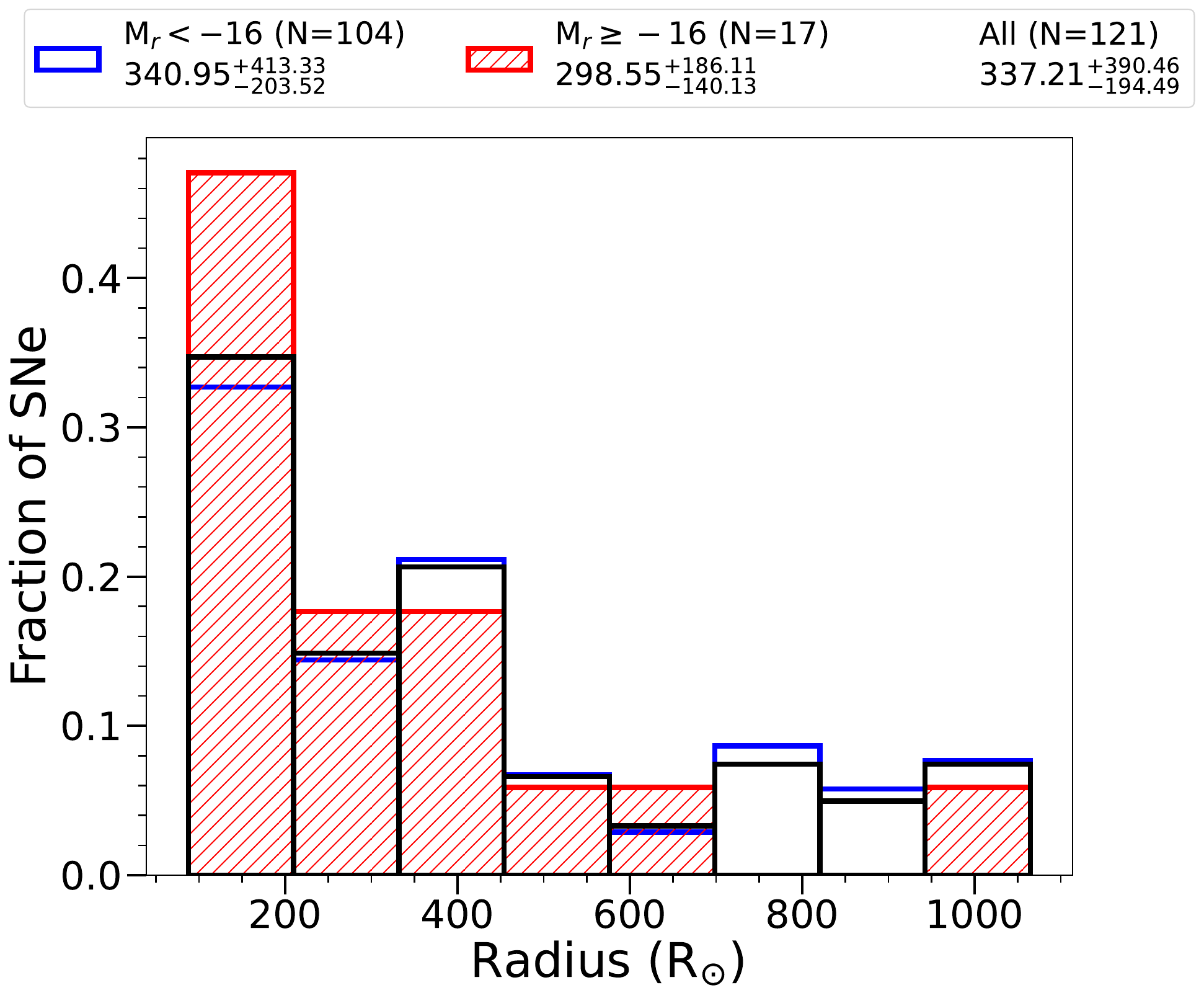}

\caption{Distributions of ejecta mass, nickel mass, explosion energy, and radius using semi-analytical models from \citet{Nagy2016}. LLIIP SNe with $M_r \geq -16$ are shown in hatched red, SNe IIP with $M_r < -16$ in blue, and the full sample is shown in black. }

    \label{fig:semiparameters}
\end{figure*}

\begin{figure*}
    \centering
    \includegraphics[width=0.33\textwidth]{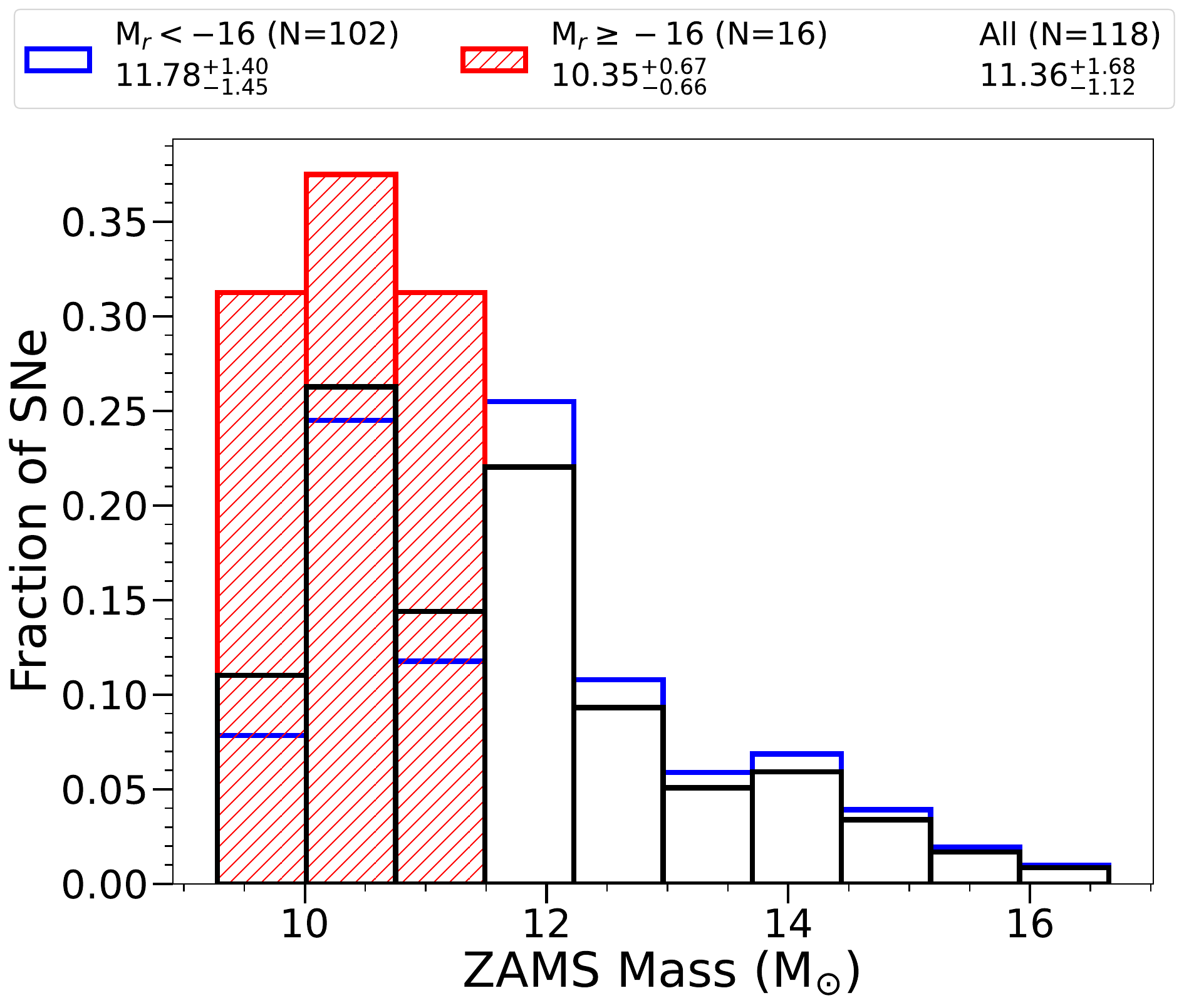}\includegraphics[width=0.33\textwidth]{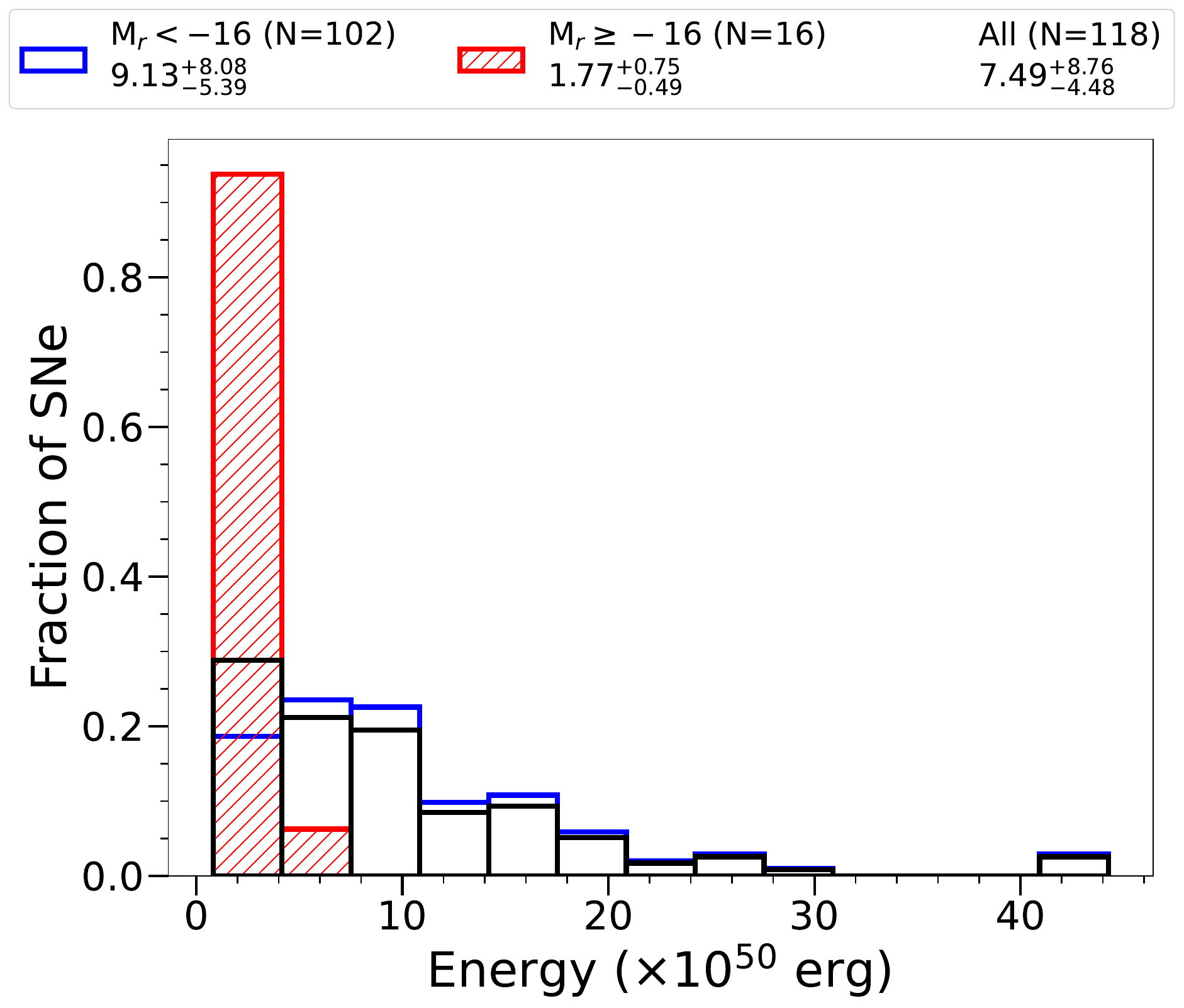}\includegraphics[width=0.33\textwidth]{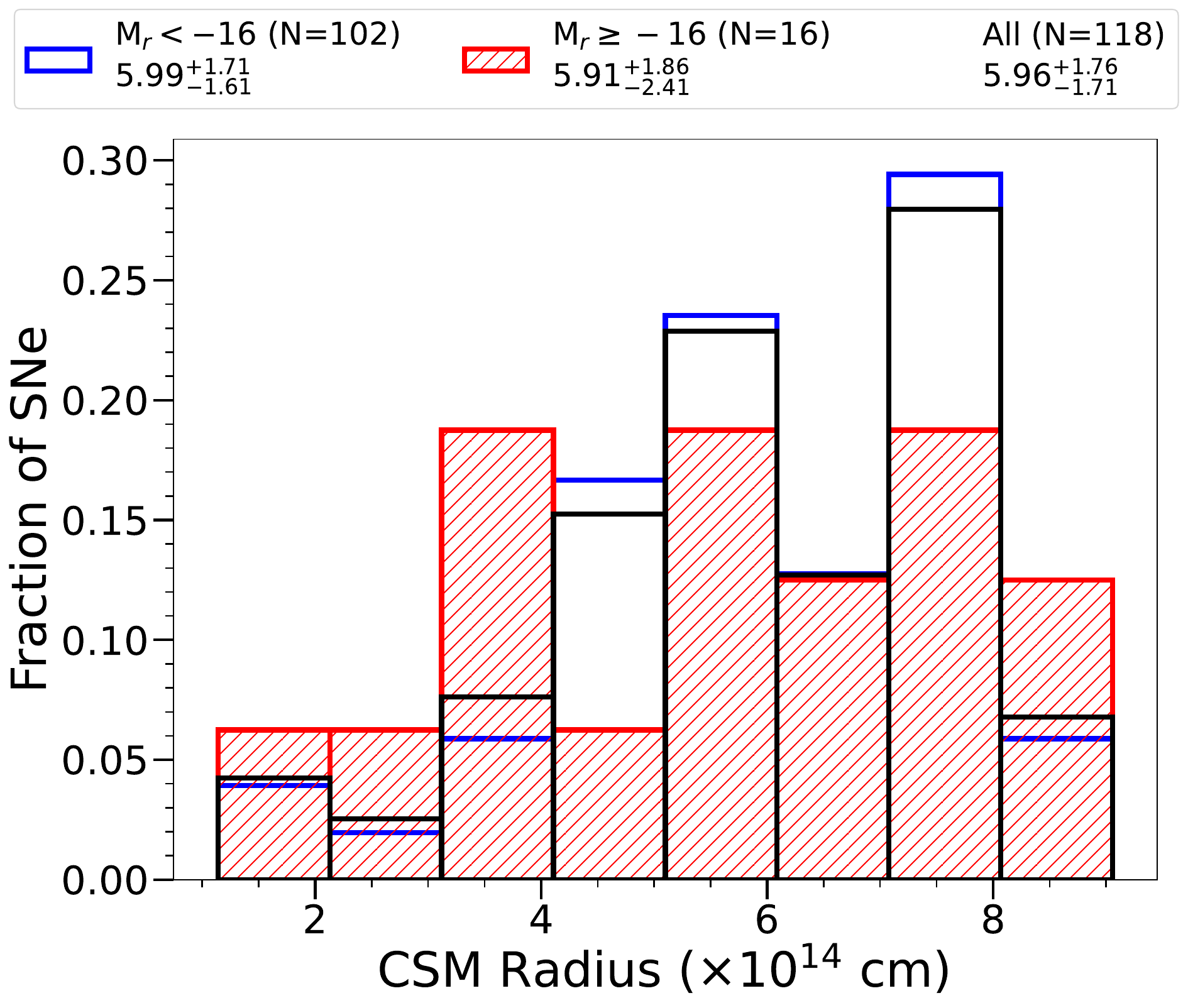}
    
    \includegraphics[width=0.33\textwidth]{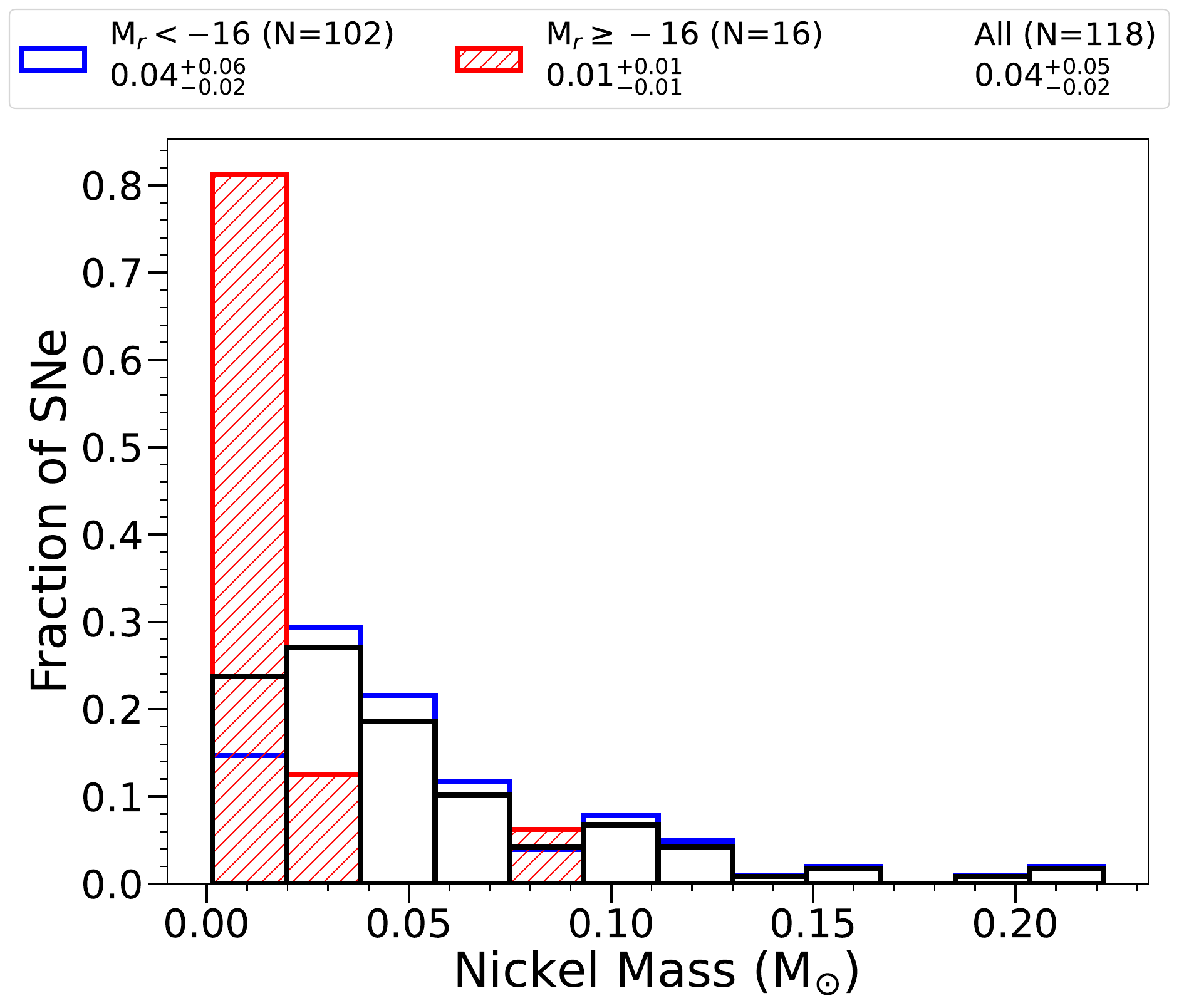}\includegraphics[width=0.33\textwidth]{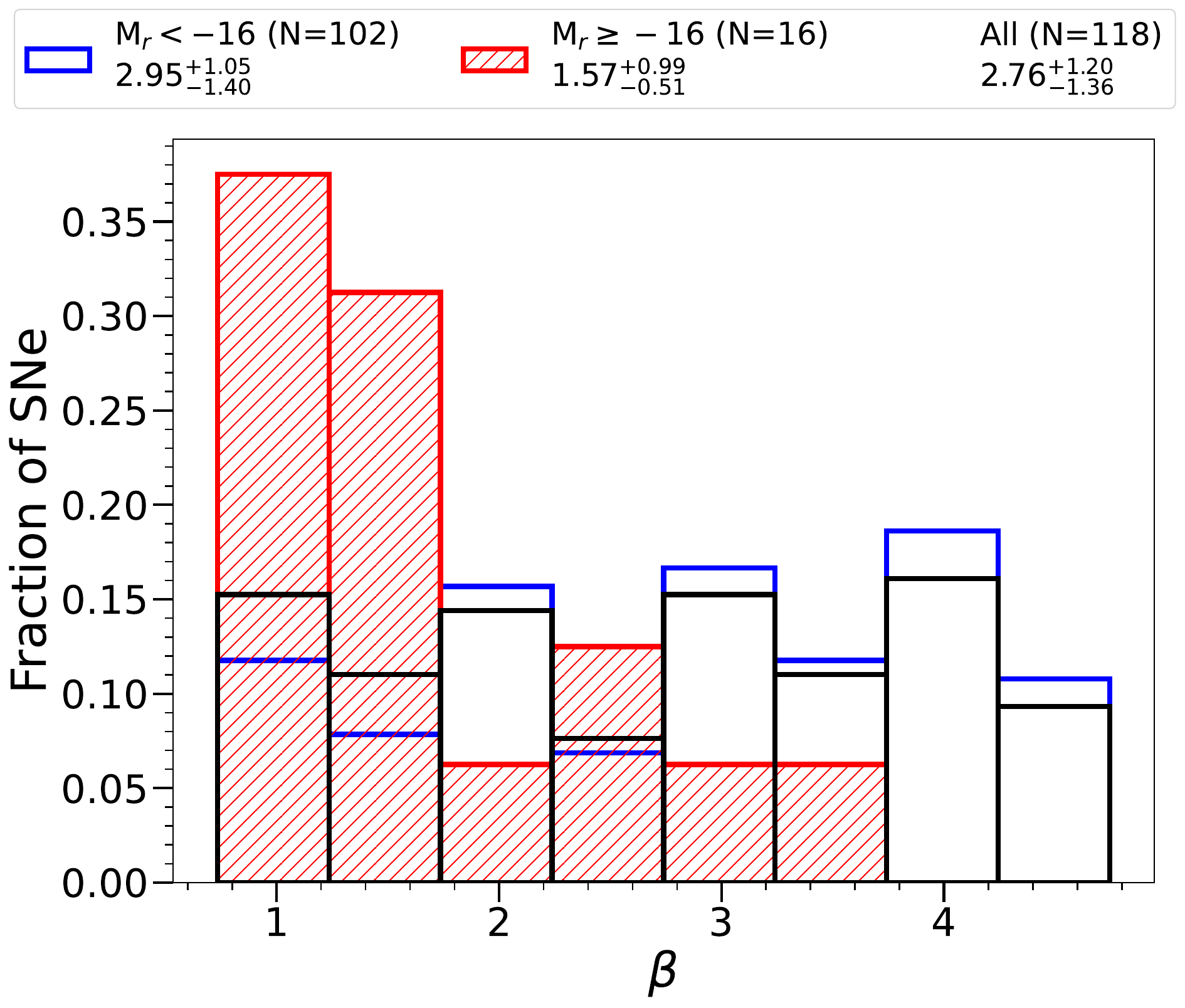}\includegraphics[width=0.33\textwidth]{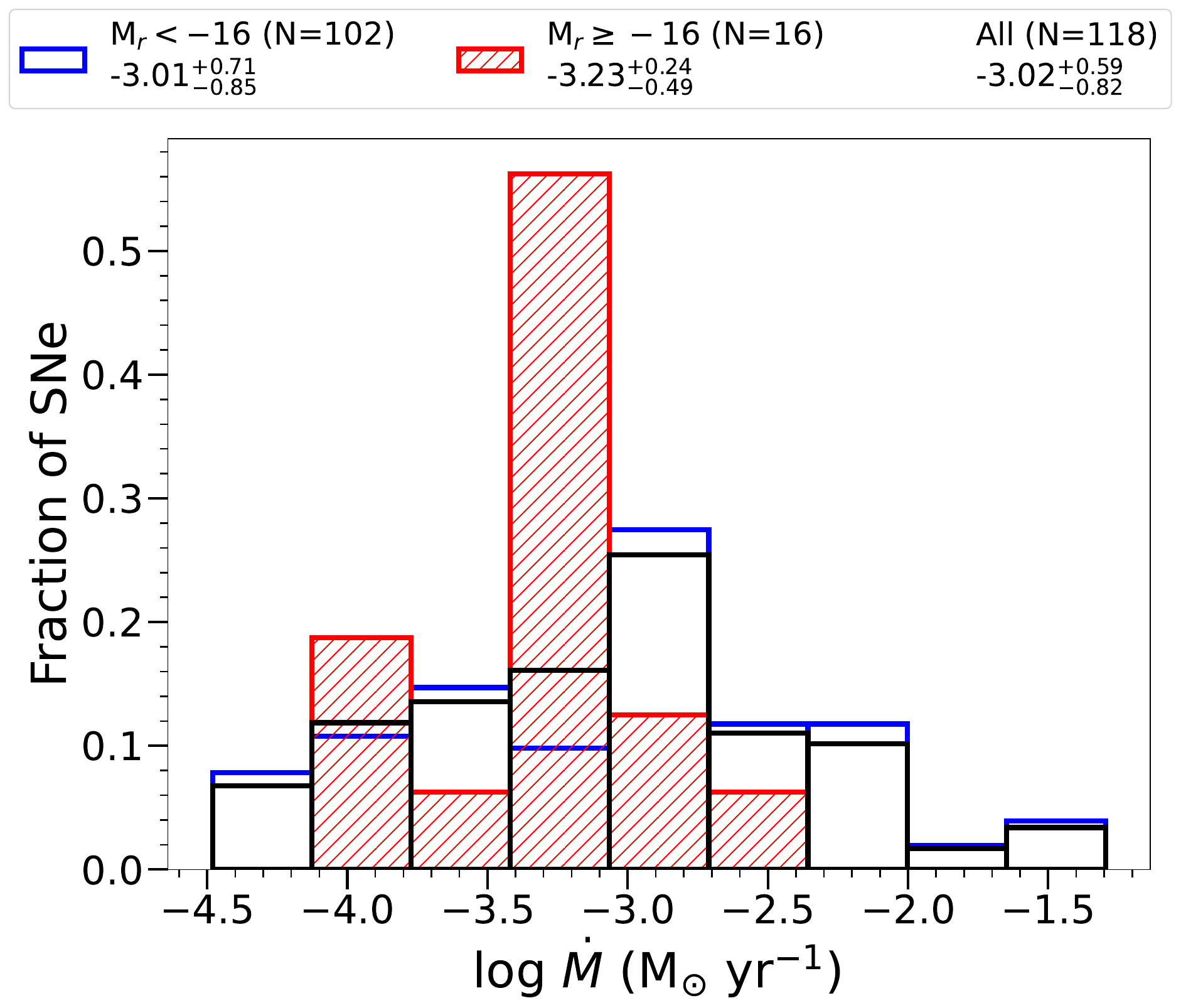}
    \caption{Distributions of the ZAMS mass,  explosion energy,  CSM radius, nickel mass, CSM structure parameter, and mass loss rates using radiation-hydrodynamical models from \citet{Moriya2023}. LLIIP SNe with $M_r \geq -16$ are shown in hatched red, those with $M_r < -16$ in blue, and the full sample is shown in black.}
    \label{fig:moriyaparameters}

\end{figure*}

The best-fit model lightcurves are shown in Figures \ref{fig:semifits} and \ref{fig:moriyafits}. The distribution of the best-fit parameters are shown in Figures \ref{fig:semiparameters} and \ref{fig:moriyaparameters}. 

\subsubsection{Scaling Relations}

Scaling relations have been commonly used to infer explosion parameters from lightcurve observables \citep{Popov1993, Kasen2009}. Ejecta mass ($M_{\rm ej}$), progenitor radius ($R$), and explosion energy ($E$) are commonly used as the primary independent variables in scaling relations.  In this paper, we use scaling relations derived by \citet{Goldberg2019}. We use the following relations to constrain the ejected mass ($M_{10} \equiv M_{\rm ej}/10\,M_\odot$) and explosion energy ($E_{51} \equiv E_{\rm exp}/10^{51}\,\mathrm{erg}$) as a function of progenitor radius ($R_{500} \equiv R/500\,R_\odot$), via the following relations:
\begin{align}
\log(E_{51}) &= -0.728 + 2.148\log(L_{42}) - 0.280\log(M_{\rm Ni}) \nonumber \\
             &\quad + 2.091\log(t_{\rm p,2}) - 1.632\log(R_{500}), \\
\log(M_{10}) &= -0.947 + 1.474\log(L_{42}) - 0.518\log(M_{\rm Ni}) \nonumber \\
             &\quad + 3.867\log(t_{\rm p,2}) - 1.120\log(R_{500}),
\label{eq:scaling_relations}
\end{align}
where $M_{\rm Ni}$ is in units of $M_\odot$, $L_{42} = L_{50d}/10^{42}\,{\rm erg\,s^{-1}}$, $L_{50d}$ is the plateau luminosity 50 days after explosion and $t_{\rm p,2} = t_{\rm p}/100\,\mathrm{days}$.  Here, $t_{\rm p}$ is measured from the parametric fit to the bolometric lightcurve using the following equation used in  \citet{Goldberg2019, Valenti2016}:
\begin{equation}
\log(L_{\mathrm{bol}}(t)) = \frac{-A_0}{1 + \exp\left( \frac{t - t_{\rm p}}{W_0} \right)} + P_0 t + M_0,
\label{eq:valenti}
\end{equation}
where $t_{\rm p}$ represents the inflection point of the lightcurve and corresponds to the midpoint of the plateau-to-tail transition.

The constraints for LLIIP SNe thus obtained are discussed in Section \ref{sec:results}.

\section{Results}
\label{sec:results}

In this section, we summarize the results obtained using the analysis techniques
described above. The best-fit radiation-hydrodynamical values for each SN are listed in Table~\ref{moriyafit}, and the sample statistics are
summarized in Table~\ref{tab:radhydro_summary}. Similarly, the best-fit values from the
semi-analytical modeling are given in Table~\ref{tab:nagyfit}, with the corresponding
sample statistics summarized in Table~\ref{tab:semianalytic_summary}.

\subsection{Observed Parameters}
\label{sec:obsresults}

\begin{table*}[ht]
\centering
\begin{tabular}{l l c c}
\hline
Parameter & Sample & Median & Range \\
\hline
Plateau Duration (days)        & M$_r \geq -16$ & 89$^{+10}_{-20}$ & 64--108 \\
                               & M$_r < -16$ & 75$^{+16}_{-15}$ & 47--137 \\
                               & All     & 76$^{+16}_{-16}$ & 47--137 \\
\hline
OPTd (days)                    & M$_r \geq -16$ & 99$^{+11}_{-14}$ & 67--116 \\
                               & M$_r < -16$ & 88$^{+17}_{-20}$ & 52--157 \\
                               & All     & 90$^{+16}_{-20}$ & 52--157 \\
\hline
Rise Time (days)               & M$_r \geq -16$ & 13$^{+1}_{-2}$ & 10--16 \\
                               & M$_r < -16$ & 13$^{+5}_{-2}$ & 7--22 \\
                               & All     & 13$^{+4}_{-2}$ & 7--22 \\
\hline
Plateau Slope ($\times 0.01$ mag day$^{-1}$) & M$_r \geq -16$ & 0.1$^{+0.1}_{-0.2}$ & -0.4--0.6 \\
                               & M$_r < -16$ & 0.5$^{+0.6}_{-0.4}$ & -0.1--2.0 \\
                               & All     & 0.4$^{+0.5}_{-0.4}$ & -0.4--2.0 \\
\hline
Plateau - Tail Drop (mag)      & M$_r \geq -16$ & -1.33$^{+1.25}_{-0.72}$ & -2.48--0.22 \\
                               & M$_r < -16$ & -1.33$^{+0.42}_{-0.35}$ & -3.50---0.09 \\
                               & All     & -1.33$^{+0.46}_{-0.41}$ & -3.50--0.22 \\
\hline
\end{tabular}
\caption{Summary of $r$-band lightcurve observables, divided into faint ($M_r \geq -16$), bright ($M_r < -16$), and the full sample. Median values with uncertainties and full observed ranges are shown. Plateau slopes are scaled by a factor of $10^{-2}$ mag day$^{-1}$.}
\label{tab:observables}
\end{table*}


The observed parameters in the $r$-band and their errors are shown in Table \ref{tab:lightcurveparams}. The sample statistics are shown in Table \ref{tab:observables}. The lower and upper uncertainties correspond to the 16th and 84th percentiles, respectively. The distributions of these observed parameters are shown in Figure \ref{fig:observables}.

\begin{figure}
    \centering
    \includegraphics[width=1.0\columnwidth]{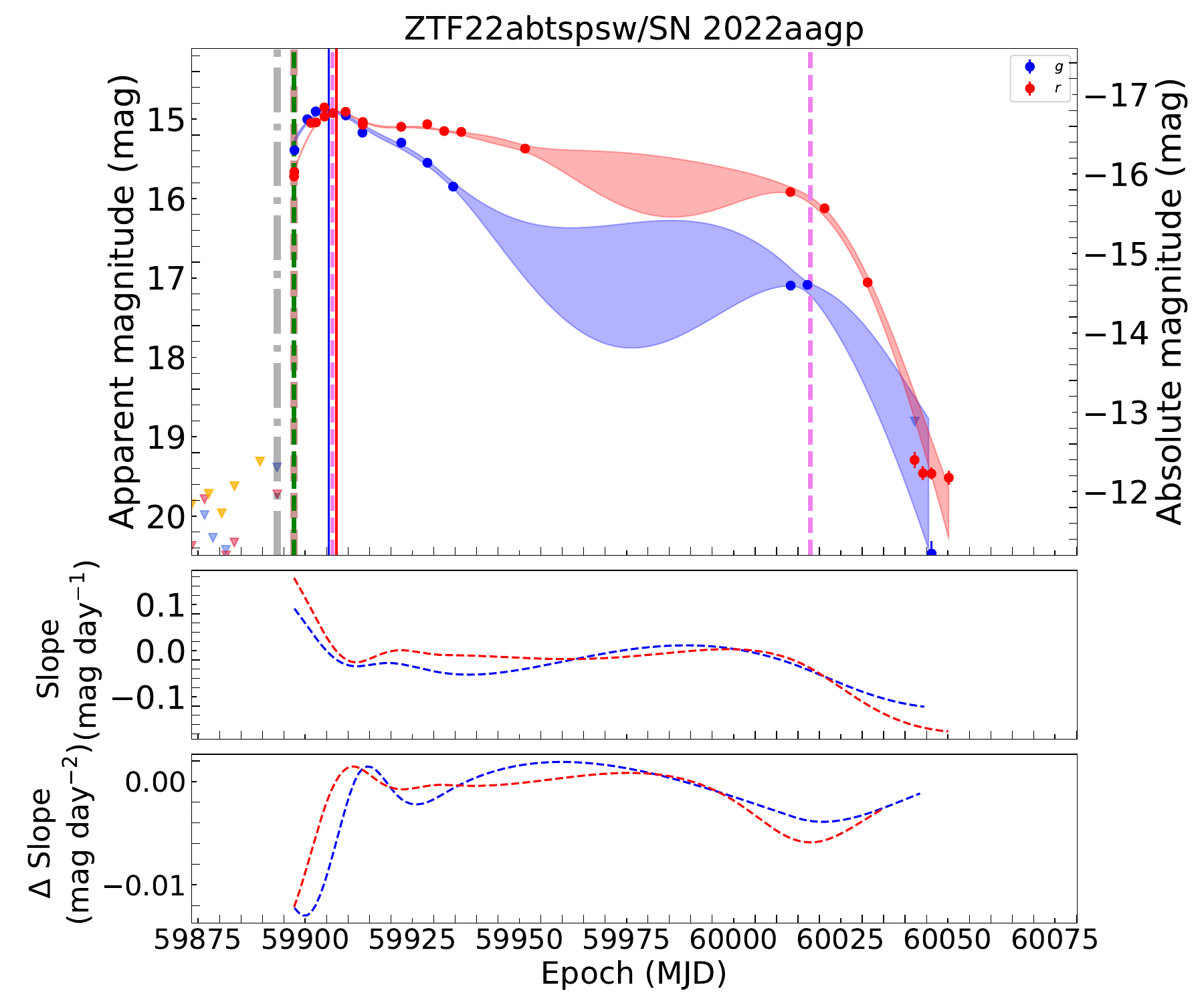}
\caption
{The $r$- and $g$-band lightcurve of ZTF22abtspsw/SN~2022aagp with GP fits (shaded regions). Vertical lines mark key epochs: first detection (dashed red), explosion epoch (dashed grey), $r$- and $g$-band peaks (solid red and blue), plateau start and end (solid violet), and 50\% peak flux (dashed green). The lower panels show the light curve slope (mag day$^{-1}$) and its change ($\Delta$ slope, mag day$^{-2}$). The steepest observed drop of $>3.5$~mag was seen in this SN, consistent with predictions for ECSN and failed SNe with little or no nickel mass.
}
\label{fig:SN2022aagp}
\end{figure}

For LLIIP SNe ($M_r > -16$), the median plateau duration was $89^{+10}_{-20}$ days, with a range of 64 to 108 days. The median duration of the optically thick phase (OPTd) was $99^{+11}_{-14}$ days, spanning from 67 to 116 days. The median rise time was $13^{+1}_{-2}$ days, with a range from 10 to 16 days. The median plateau slope was $0.1^{+0.1}_{-0.2} \times 10^{-2}$ mag day$^{-1}$, ranging from $-0.4$ to $0.6 \times 10^{-2}$ mag day$^{-1}$. Finally, the change in magnitude between the plateau and the radioactive tail had a median value of $-1.33^{+1.25}_{-0.72}$ mag, with observed values ranging from $-2.48$ to $-0.22$ mag. For the full Type IIP SN sample, the median values were $76^{+16}_{-16}$ days for the plateau duration, $90^{+16}_{-20}$ days for OPTd, $13^{+4}_{-2}$ days for the rise time, $0.4^{+0.5}_{-0.4} \times 10^{-2}$ mag day$^{-1}$ for the plateau slope, and $-1.33^{+0.46}_{-0.41}$ mag for the plateau-to-tail drop. The steepest drop of $>3.50$ mag was seen in SN~2022aagp, shown in Figure \ref{fig:SN2022aagp}. A sharp decline from the plateau to the tail is predicted for ECSNe and failed SNe with very low or zero nickel mass \citep[see e.g.,][]{Tominaga2013, Turatto1998}. Other objects with sharp drops include SN~2021tyw ($> 3.10$ mag), SN~2022prv ($2.56$ mag), SN~2023vog ($2.49$ mag) and SN~2020cxd ($2.48$ mag). 

We further investigated how the lightcurve observables correlate with peak magnitude. As shown in Figure~\ref{fig:obsvspeak}, the plateau slope shows a strong correlation with peak brightness, with a Pearson correlation coefficient of $r = -0.68$ ($p < 10^{-5}$). The best-fit line is given by:
\begin{align*}
\text{Plateau slope (0.01 mag day$^{-1}$)} =\; & (-0.33 \pm 0.04)\,M_r \\
& +\,(-4.93 \pm 0.70)
\end{align*}
Brighter SNe tend to have steeper plateau slopes, indicating more rapid declines during the plateau phase. We also note that a major fraction of LLIIP SNe have positive slope. The plateau duration shows only a weak correlation with peak magnitude ($r = 0.42$). The plateau durations for faint and bright SNe span a broad range and overlap significantly. The physical implications for these results are discussed in Section \ref{sec:obsdiscuss}.

\subsection{Progenitor and explosion parameters}

\subsubsection{Nickel mass}
\textcolor{black}{For the semi-analytical model fits, the nickel mass is measured only for SNe
with a distinct radioactive tail. Within the LLIIP subsample,
the median nickel mass is $0.0132^{+0.0089}_{-0.0100}$~M$_\odot$, spanning
0.0009--0.0420~M$_\odot$.  The lowest value in our sample is that of SN~2020cxd, with $M_{\rm Ni}=0.0009$~M$_\odot$. For the full Type~IIP sample, the semi-analytical fits yield a median nickel mass of $0.065^{+0.068}_{-0.047}$~M$_\odot$ (range 0.0009--0.3098~M$_\odot$). The radiation-hydrodynamical fits of \citet{Moriya2023} give broadly consistent values, with a median of $0.038^{+0.059}_{-0.024}$~M$_\odot$ for the full sample and $0.013^{+0.003}_{-0.010}$~M$_\odot$ for the LLIIP subsample. \citet{Rodriguez2021} studied 110 events and found $^{56}$Ni masses between
0.005--0.177~M$_\odot$ (average $0.037\pm0.005$~M$_\odot$). \citet{Martinez2022} reported
0.006--0.069~M$_\odot$ (median $\sim0.036$~M$_\odot$) from 17 SNe, while \citet{Pejcha2015} and
\citet{Muller2017} obtained similar medians ($\sim$0.03~M$_\odot$) with maxima reaching
0.28~M$_\odot$. \citet{Subrayan2023} measured Ni masses for 45 SNe, with only a single
object as low as 0.01~M$_\odot$, while most lie between 0.02 and 0.1~M$_\odot$.
\citet{Anderson2019} found a broad distribution (0.001--0.360~M$_\odot$), and
\citet{Hamuy2003} reported 0.0016--0.26~M$_\odot$ from 21 SNe. In comparison with these literature samples, the LLIIP SNe in our sample clearly occupy
the low end of the nickel-mass distribution. Their very small $^{56}$Ni yields are
consistent with expectations for explosions of low-mass red supergiant progenitors or
events experiencing significant fallback, as discussed in Section~\ref{sec:progenitor}.
}

\subsubsection{Explosion energy}

\textcolor{black}{From the \citet{Moriya2023} radiation-hydrodynamical fits, the LLIIP SNe show
low explosion energies, with a median of $0.17^{+0.07}_{-0.03}\times10^{51}$~erg and a
range of 0.10--0.28$\times10^{51}$~erg. For the full Type~IIP sample, the median energy
is $0.76^{+0.89}_{-0.44}\times10^{51}$~erg. In contrast, the semi-analytical
\citet{Nagy2016} models yield systematically higher values, with a median of
$0.79^{+0.49}_{-0.35}\times10^{51}$~erg for the LLIIP subsample (range
0.21--3.64$\times10^{51}$~erg) and $4.32^{+4.17}_{-2.91}\times10^{51}$~erg for the full
sample (range 0.21--9.94$\times10^{51}$~erg). We note that both the ejecta masses and
explosion energies from the semi-analytical fits tend to be higher, likely due to the two-component density profile assumed in these models; the caveats are discussed further in Section~\ref{sec:caveats}.  In the literature, \citet{Hamuy2003} measured 0.6--5.5$\times10^{51}$~erg for 21 Type~II SNe, while \citet{Utrobin2019} and \citet{Pumo2017} obtained up to
1.4--2.0$\times10^{51}$~erg. \citet{Martinez2022} report a median of 0.61$\times10^{51}$~erg (range 0.30--1.01$\times10^{51}$~erg). The LLIIP SNe in our sample clearly occupy the low-energy end, clustering just above $10^{50}$~erg, and may in some cases fall below the model grid minimum of $0.1\times10^{51}$~erg. This supports interpretations in which LLIIP SNe arise from weak explosions of low-mass cores (see Section~\ref{sec:progenitor}).
}

\subsubsection{Ejecta mass/ZAMS mass}

\textcolor{black}{
The semi-analytical models yield ejecta masses of 5.73--12.89~M$_\odot$ for the LLIIP
sample, with a median of $8.06^{+0.73}_{-1.71}$~M$_\odot$. For the full Type~IIP sample,
the corresponding median ejecta mass is $12.80^{+6.20}_{-5.00}$~M$_\odot$, spanning
5.41--24.79~M$_\odot$. In earlier studies, \citet{Martinez2022} report a typical ejecta
mass of $\sim$9.2~M$_\odot$ (16--84 percentile range 8.2--12.7~M$_\odot$), while
\citet{Utrobin2019} infer somewhat higher masses up to $\sim$14.8~M$_\odot$. Even larger
values were found by \citet{Hamuy2003}, who estimated ejecta masses in the range
14--56~M$_\odot$ for some Type~II SNe. LLIIP SNe lie at the lower end of these
distributions, consistent with expectations for low-mass RSG or sAGB progenitors, or for
progenitors that have experienced significant envelope stripping prior to core collapse. ZAMS masses inferred in \citet{Subrayan2023} and \citet{Silva2024} span 12--16~M$_\odot$,
with a steep IMF placing most objects near $\sim$12~M$_\odot$. From our
radiation-hydrodynamical fits, the LLIIP SNe are inferred to arise from relatively
low-mass progenitors, with ZAMS masses typically $<12$~M$_\odot$ and a median of
$10.5^{+0.4}_{-0.7}$~M$_\odot$. For the full Type~II sample, the median ZAMS mass is
$11.5^{+1.6}_{-1.2}$~M$_\odot$. Thus, the LLIIP subsample lies toward the low-mass end
of the progenitor-mass distribution, reinforcing the picture that the faintest events
originate from the lowest-mass core-collapse progenitors.}

\subsubsection{Progenitor radius}

\textcolor{black}{The progenitor radii inferred from the semi-analytical fits for the
LLIIP SNe span 138--1024~R$_\odot$, with a median of $299^{+219}_{-142}$~R$_\odot$. For
the full Type~IIP sample, the corresponding median is $337^{+390}_{-194}$~R$_\odot$.
These radii are smaller than those typically inferred for normal Type~II SNe in
the literature. For example, \citet{Martinez2022} report progenitor radii of
460--610~R$_\odot$, with a median of $\sim$495~R$_\odot$, while \citet{Hamuy2003} found a
broad range of 80--600~R$_\odot$ using hydrodynamical modeling. More recently,
\citet{Irani2024} identified a bimodal distribution in breakout radii for 34 SNe~II,
with roughly half of the sample preferring compact radii ($<10^{14}$~cm) and the other
half favoring larger radii ($>10^{14}$~cm), consistent with substantial diversity in
outer-envelope structure among RSG progenitors. For additional context, the
radius distribution of 74 Galactic RSGs spans 100--2135~R$_\odot$ with a
median of $625^{+393}_{-275}$~R$_\odot$ \citep{Levesque2005}. Thus, these results suggest that LLIIP progenitors are either intrinsically more compact RSGs, or that their outer hydrogen envelopes have been partially stripped through binary interaction.
}
\subsubsection{Progenitor mass-loss}

\textcolor{black}{
The radiation-hydrodynamical modeling yields median pre-explosion mass-loss rates of
$\log\,\dot{M} = -3.01^{+0.83}_{-0.60}$~M$_\odot$\,yr$^{-1}$ for the full sample, and
$\log\,\dot{M} = -3.32^{+0.37}_{-0.26}$ for the LLIIP subsample. Our results agree well
with those of \citet{Subrayan2023} and \citet{Silva2024}, who report mass-loss rates in
the range $10^{-4}$--$10^{-2}$~M$_\odot$\,yr$^{-1}$ for samples of Type~II SNe. The
systematically smaller mass-loss rates of the LLIIP subsample may reflect weaker winds
or shorter-lived mass-loss episodes prior to explosion. The corresponding CSM radii inferred from the radiation-hydrodynamical fits span $1.1$--$8.6\times10^{14}$~cm, with a median of $5.8^{+1.9}_{-1.6}\times10^{14}$~cm for
the full sample and $4.3^{+3.0}_{-1.0}\times10^{14}$~cm for the LLIIP events, indicating
an extended circumstellar environment that can influence early-time lightcurve
and spectra \citep[e.g.,][]{Galbany2016, Valenti2016, Forster2018,
Morozova2018, Bruch2021, Bruch2023, Galan2024, Hinds2025, Galan2025}.}

\begin{table}[ht]
\centering
\begin{tabular}{l l c c}
\hline
Parameter & Sample & Median & Range \\
\hline
$M_\mathrm{Ni}$ ($0.01\ \mathrm{M_\odot}$) & M$_r \geq -16$ & 1.32$^{+0.89}_{-1.00}$ & 0.09--4.20 \\
                                    & M$_r < -16$    & 7.36$^{+6.36}_{-4.80}$ & 0.56--30.98 \\
                                    & All            & 6.52$^{+6.80}_{-4.71}$ & 0.09--30.98 \\
\hline
$M_\mathrm{ej}$ ($\mathrm{M_\odot}$) & M$_r \geq -16$ & 8.06$^{+0.73}_{-1.71}$ & 5.73--12.89 \\
                                    & M$_r < -16$    & 13.90$^{+5.45}_{-5.42}$ & 5.41--24.79 \\
                                    & All            & 12.80$^{+6.20}_{-5.00}$ & 5.41--24.79 \\
\hline
$E_\mathrm{kin}$ ($10^{51}$ erg)    & M$_r \geq -16$ & 0.79$^{+0.49}_{-0.35}$ & 0.21--3.64 \\
                                    & M$_r < -16$    & 4.74$^{+3.79}_{-2.45}$ & 1.04--9.94 \\
                                    & All            & 4.32$^{+4.17}_{-2.91}$ & 0.21--9.94 \\
\hline
$R$ ($\mathrm{R_\odot}$)            & M$_r \geq -16$ & 299$^{+219}_{-142}$ & 138--1024 \\
                                    & M$_r < -16$    & 340$^{+407}_{-202}$ & 88--1065 \\
                                    & All            & 337$^{+390}_{-194}$ & 88--1065 \\
\hline
\end{tabular}
\caption{Best-fit physical parameters derived from semi-analytical model fits \citep{Nagy2016} for faint ($M_r \geq -16$), bright ($M_r < -16$), and all Type IIP SNe. Reported values include the median and 16th/84th percentile uncertainties, along with the full observed range. Nickel masses are scaled by a factor of $10^{-2}$ $\mathrm{M_\odot}$.}
\label{tab:semianalytic_summary}
\end{table}


\begin{table}[ht]
\centering
\begin{tabular}{l l c c}
\hline
Parameter & Sample & Median & Range \\
\hline
$E_\mathrm{exp}$ ($10^{51}$ erg)              & M$_r \geq -16$  & 0.17$^{+0.07}_{-0.03}$ & 0.10--0.28 \\
                                              & M$_r < -16$     & 0.91$^{+0.81}_{-0.54}$ & 0.16--4.43 \\
                                              & All             & 0.76$^{+0.89}_{-0.44}$ & 0.10--4.43 \\
\hline
$R_\mathrm{CSM}$ ($10^{14}$ cm)               & M$_r \geq -16$  & 4.3$^{+3.0}_{-1.0}$ & 1.8--8.6 \\
                                              & M$_r < -16$     & 5.8$^{+1.9}_{-1.4}$ & 1.1--8.4 \\
                                              & All             & 5.8$^{+1.9}_{-1.6}$ & 1.1--8.6 \\
\hline
$M_\mathrm{Ni}$ ($0.01\,\mathrm{M_\odot}$)    & M$_r \geq -16$  & 1.3$^{+0.3}_{-1.0}$ & 0.1--2.5 \\
                                              & M$_r < -16$     & 4.1$^{+5.9}_{-2.1}$ & 0.2--22.2 \\
                                              & All             & 3.8$^{+5.9}_{-2.4}$ & 0.1--22.2 \\
\hline
$M_\mathrm{ZAMS}$ ($\mathrm{M_\odot}$)        & M$_r \geq -16$  & 10.5$^{+0.4}_{-0.7}$ & 9.4--11.0 \\
                                              & M$_r < -16$     & 11.8$^{+1.4}_{-1.4}$ & 9.3--16.7 \\
                                              & All             & 11.5$^{+1.6}_{-1.2}$ & 9.3--16.7 \\
\hline
$-$ $\log\ \dot{M}\ (\mathrm{M_\odot\ yr^{-1}}$) & M$_r \geq -16$  & 3.32$^{+0.37}_{-0.26}$ & 2.86--4.37 \\
                                              & M$_r < -16$     & 3.00$^{+0.85}_{-0.72}$ & 1.29--4.48 \\
                                              & All             & 3.01$^{+0.83}_{-0.60}$ & 1.29--4.48 \\
\hline
\end{tabular}
\caption{Best-fit physical parameters derived from radiation-hydrodynamical model fits \citep{Moriya2023} for faint ($M_r \geq -16$), bright ($M_r < -16$), and all Type IIP SNe. Reported values include the median and 16th/84th percentile uncertainties, along with the full observed range.}
\label{tab:radhydro_summary}
\end{table}




\subsection{Scaling Relations}
\label{sec:scaling}

\begin{figure}
    \centering
    \includegraphics[width=0.47\textwidth]{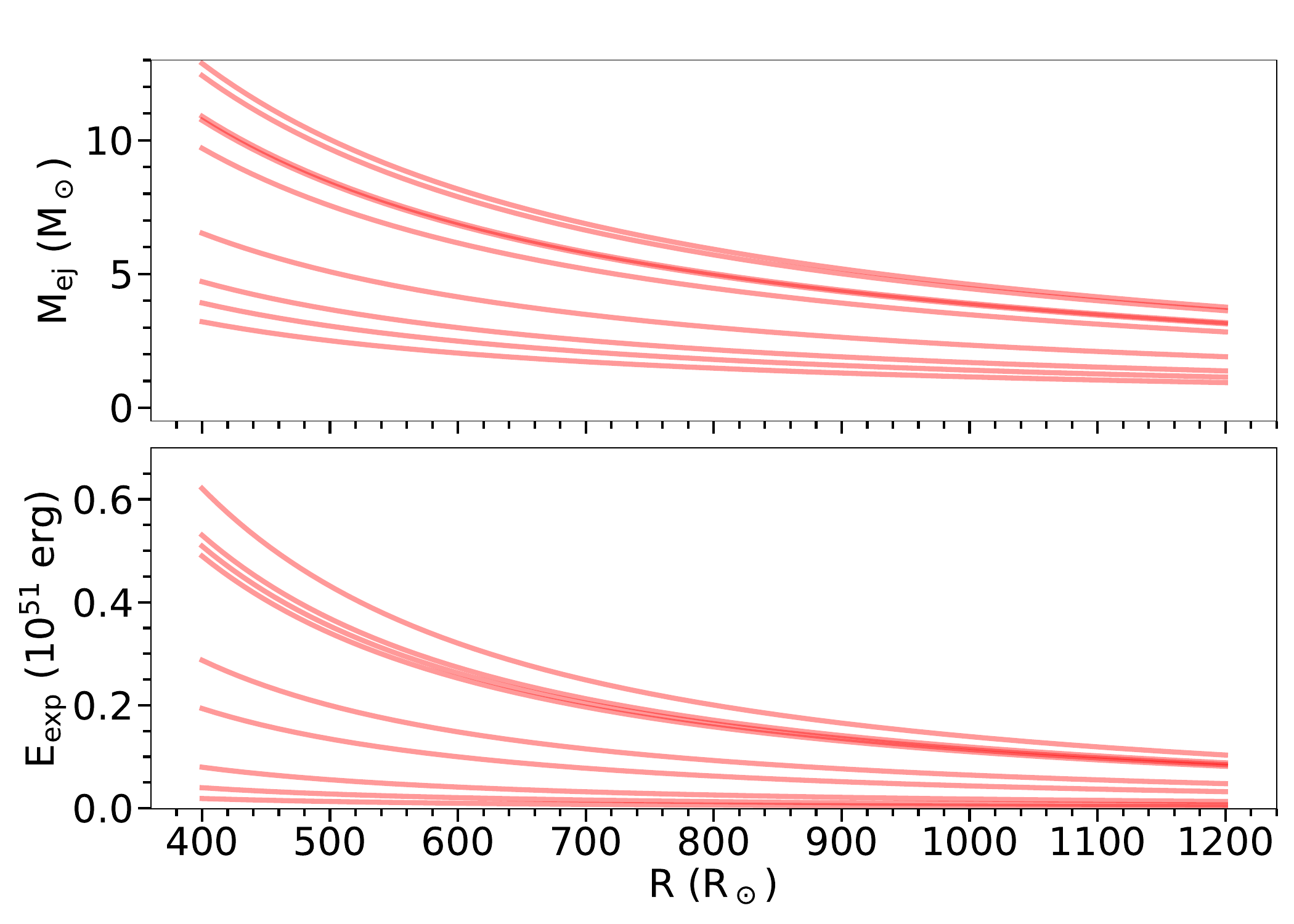}
 
    \caption{
Ejecta mass (top) and explosion energy (bottom) versus progenitor radius, derived using the scaling relations from \citet{Goldberg2019}. Red lines denote SNe with $M_r > -16$
}

    \label{fig:scaling}
\end{figure}

We investigate the physical properties of the low-luminosity Type IIP (LLIIP) SNe in our sample using scaling relations between plateau luminosity, nickel mass, duration, and progenitor radius. The constraints for LLIIP SNe thus obtained are shown in Figure \ref{fig:scaling}. 
At a representative radius of $500$ R$_\odot$, we find that the LLIIP SNe are characterized by relatively low ejecta masses ranging from $\sim2.1$ to $\sim6.5$ M$_\odot$ and low explosion energies between $\sim0.02$ and $\sim0.20$ $\times 10^{51}$ erg.  These results reinforce the picture that LLIIP SNe arise from low-mass progenitors undergoing weak explosions, consistent with those derived from semi-analytical and radiation-hydrodynamical analysis.

\section{Discussion}
\label{sec:discussion}
\label{sec:obsdiscuss}

\subsection{LLIIP SNe exhibit less steep gradients, with most displaying upward slopes}

We first attempt to explain the positive slope seen in LLIIP SNe through basic order-of-magnitude arguments. In standard Type IIP models such as \citet{Popov1993}, the onset of the plateau is set by hydrogen recombination, where the luminosity reaches
\begin{equation}
    L_{\rm pl} = 4\pi R_{\rm ej}^2 \sigma T_I^4,
\end{equation}
with $T_I \approx 5000$~K being the hydrogen recombination temperature. For low-luminosity plateaus, this implies a smaller ejecta radius at the onset of the plateau phase, since
\begin{equation}
    R_{\rm ej} \propto L_{\rm pl}^{1/2}.
\end{equation}
This is consistent with the lower expansion velocities observed in LLIIP SNe compared to more luminous events, resulting in smaller ejecta radius.

The photon diffusion time at this stage is given by
\begin{equation}
    t_{\rm diff, onset} \sim \frac{\kappa M_{\rm ej}}{4\pi R_{\rm ej} c} \propto M_{\rm ej} L_{\rm pl}^{-1/2},
\end{equation}
and can be estimated as
\begin{align}
    t_{\rm diff, onset} \sim 200~{\rm days} 
    &\left(\frac{\kappa}{0.34~{\rm cm}^2~{\rm g}^{-1}}\right)
    \left(\frac{M_{\rm ej}}{8~M_\odot}\right) \notag \\
    &\times \left(\frac{L_{\rm pl}}{3 \times 10^{41}~{\rm erg}~{\rm s}^{-1}}\right)^{-1/2}.
\end{align}
The diffusion time $t_{\rm diff}$ decreases with time as the ejecta expands, owing to both the dilution of density and the recombination-driven reduction in opacity.

Following the explosion, the ejecta cool both adiabatically (on the dynamical timescale $t_{\rm dyn}$) and radiatively (on the diffusion timescale $t_{\rm diff}$). The plateau slope is governed by the ratio of these timescales. When $t_{\rm diff} \gg t_{\rm dyn}$, the internal energy (dominated by radiation) evolves as
\begin{equation}
    \frac{dE_{\rm int}}{dt} = -\frac{E_{\rm int}}{t} \quad \Rightarrow \quad E_{\rm int} \propto t^{-1}.
\end{equation}

As the photosphere recedes, the emergent flux remains approximately blackbody at $T_I$, with the luminosity given by
\begin{equation}
    4\pi R_{\rm ph}^2 \sigma T_I^4 = \frac{E_{\rm int}}{t_{\rm diff}} \approx \frac{E_{\rm int}}{\kappa M_{\rm ej}(r<R_{\rm ph}) / (4\pi R_{\rm ph} c)}.
\end{equation}
Assuming a uniform density profile in the inner ejecta, the mass enclosed within radius \( R_{\rm ph} \) evolves as
\begin{equation}
    M_{\rm ej}(r < R_{\rm ph}) 
    \propto M_{\rm ej} \left( \frac{R_{\rm ph}}{t} \right)^3.
\end{equation}
Combining the above, the photospheric radius evolves as
\begin{equation}
    R_{\rm ph}(t) \propto \left( \frac{E_{\rm int}(t)\,t^3}{M_{\rm ej}} \right)^{1/4}.
\end{equation}
Using \( E_{\rm int}(t) \propto t^{-1} \), this simplifies to
\begin{equation}
    R_{\rm ph}(t) \propto t^{1/2} M_{\rm ej}^{-1/4},
\end{equation}
and the luminosity becomes
\begin{equation}
    L_{\rm pl}(t) = 4\pi R_{\rm ph}^2 \sigma T_I^4 
    \propto t M_{\rm ej}^{-1/2}.
\end{equation}
Thus, rising plateaus can naturally emerge in the low-luminosity regime.

\begin{figure*}
    \centering
    \includegraphics[width=0.33\textwidth]{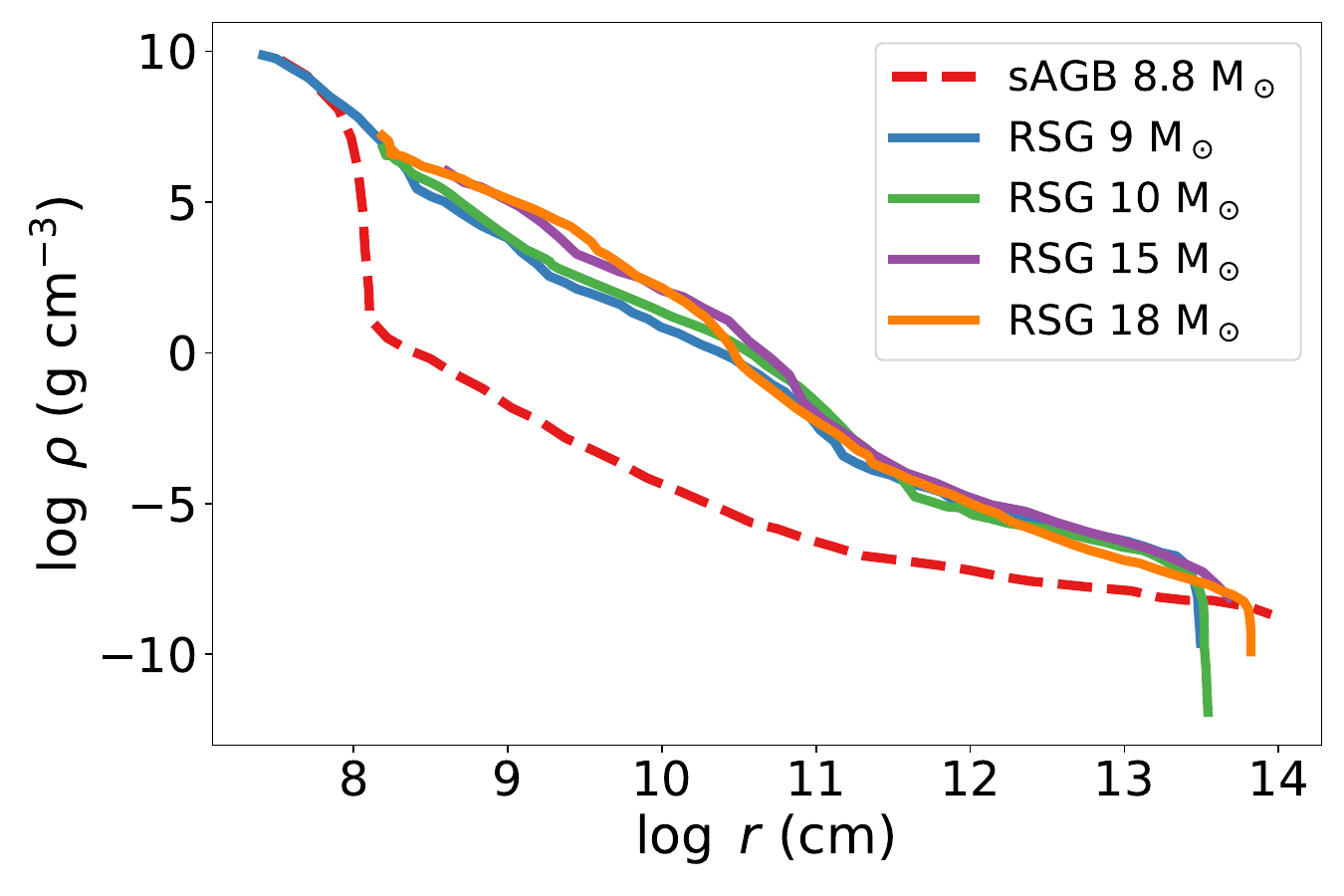}\includegraphics[width=0.33\textwidth]{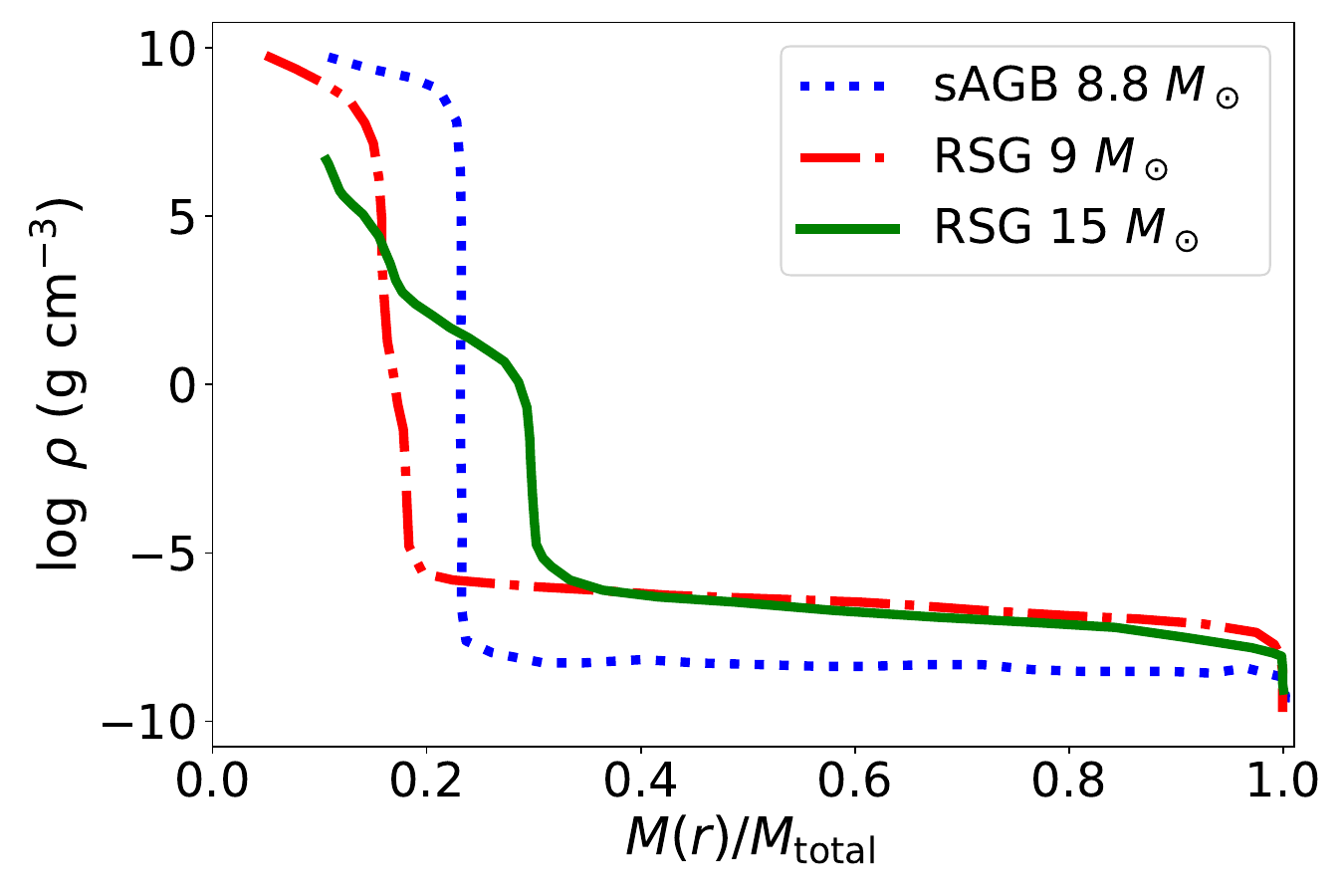}\includegraphics[width=0.33\textwidth]{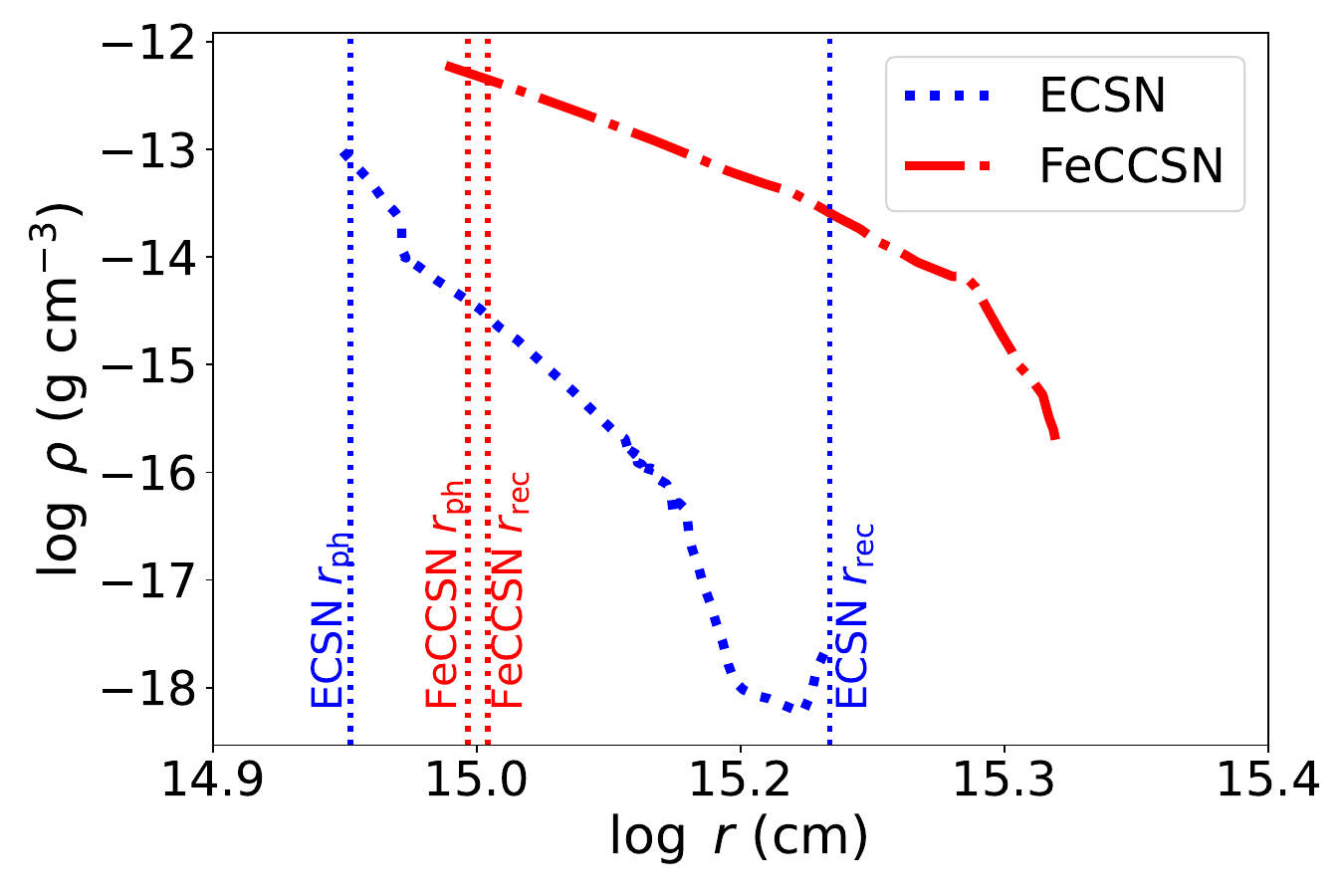}
\caption{
Left: Radial density profile of the $8-10$ \Msun\ stellar progenitor models used in \citet{Utrobin2017,Kozyreva2021,Sato2024} 
Center: Density profile as a function of mass coordinate of the $8-10$ \Msun\ stellar progenitor models used in \citet{Utrobin2017,Kozyreva2021}
Right: Density profile of FeCCSN and ECSN models used in \citet{Sato2024}. The radii of the photosphere ($r_{ph}$) and the recombination front ($r_{rec}$) at 40 days after explosion are indicated by vertical lines.
}
\label{fig:progenitor}
\end{figure*}

The notably flatter or slightly rising slopes in LLIIP SNe can also be explained by the contribution of $^{56}$Ni decay to the plateau luminosity. In Figure~\ref{fig:progenitor}, we show the radial density and mass profiles of $8$--$10$~\Msun\ RSG and sAGB progenitors from the literature. The steeper mass density profiles of low-mass progenitors imply that the photospheric radius remains closer to the core of the star throughout the plateau phase \citep{Sato2024}. As a result, $^{56}$Ni decay likely makes a stronger contribution to the overall plateau luminosity. \citet{Nakar2016} demonstrated that $^{56}$Ni decay can contribute significantly ($\sim10\%$) to the total luminosity during the photospheric phase, influencing the plateau slope. Although their models do not explicitly predict rising slopes, a combination of plateau extension and slope flattening could lead to an apparent increase in brightness over time as seen in SN~2016gfy \citep{Singh2019}. \citet{Kozyreva2019} supports this interpretation, showing that both cooling-envelope emission and nickel distribution shape the late-time behavior of the plateau. However, we note that such $^{56}$Ni-driven flattening or rising would not occur unless the progenitor radii were small. For larger progenitors, the recombination-powered luminosity would dominate and naturally decline with time, overpowering the modest nickel contribution. Thus, in the low-luminosity regime, the small radii and low explosion energies of LLIIP progenitors are necessary for $^{56}$Ni decay to leave a visible imprint on the plateau slope. Furthermore, \citet{Kozyreva2021} find that ECSNe and low-mass CCSNe from $8$--$10$~\Msun\ progenitors exhibit rising light curves in the $V$ and redder bands during the first 50 days, followed by a plateau and a subsequent drop to the radioactive tail. This is consistent with the steadily rising $r$-band light curves observed in many LLIIP SNe.

Another possible contributor in the rise of the of the lightcurves could be explosion asymmetry. 3D simulations for a 9 \Msun\ by \citet{Stockinger2020} show that large nickel-rich plumes can distort the ejecta and drive faster expansion in some directions, which lifts the light curve during the initial decline to the plateau and the early phase of the plateau. However, for a large sample, such viewing-angle-dependent effects are expected to average out.

\subsection{The plateau duration and peak magnitude are not correlated}

Single-star models predict a strong correlation between peak brightness and plateau duration in Type II SNe \citep[e.g.,][]{Kasen2009}. However, the observed lack of such a correlation suggests that the stars undergo substantial pre-supernova mass loss, likely requiring additional processes such as binary interaction to explain the observed diversity, especially at the low-mass end.
\citet{Eldridge2018,Hiramatsu2021, Dessart2024} also find that binary stripping can explain the duration of plateau seen in Type II SNe lightcurves. This aligns with the increasing observational consensus that a significant number of massive stars are found in binary systems, with many of them undergoing interaction \textcolor{black}{\citep[e.g.,][]{Sana2012, Duchene2013, Kobulnicky2014, Sana2014, Gravity2018, Bordier2022, Guo2022}.} \textcolor{black}{Also, \citet{Zapartas2019} estimate that $33$--$50\%$ of all Type~II SN progenitors undergo binary interaction, based on analytical estimates and population-synthesis simulations that track which progenitors exchange mass with a companion prior to explosion.}
Inferring the explosion and progenitor properties from the lightcurves using models that incorporate self-consistent progenitor evolution with binary effects will constrain the binary fraction and extent of stripping among massive stars that explode as core-collapse SNe. 


\subsection{Correlation of physical parameters}

\begin{figure*}
    \centering
    \includegraphics[width=0.33\textwidth]{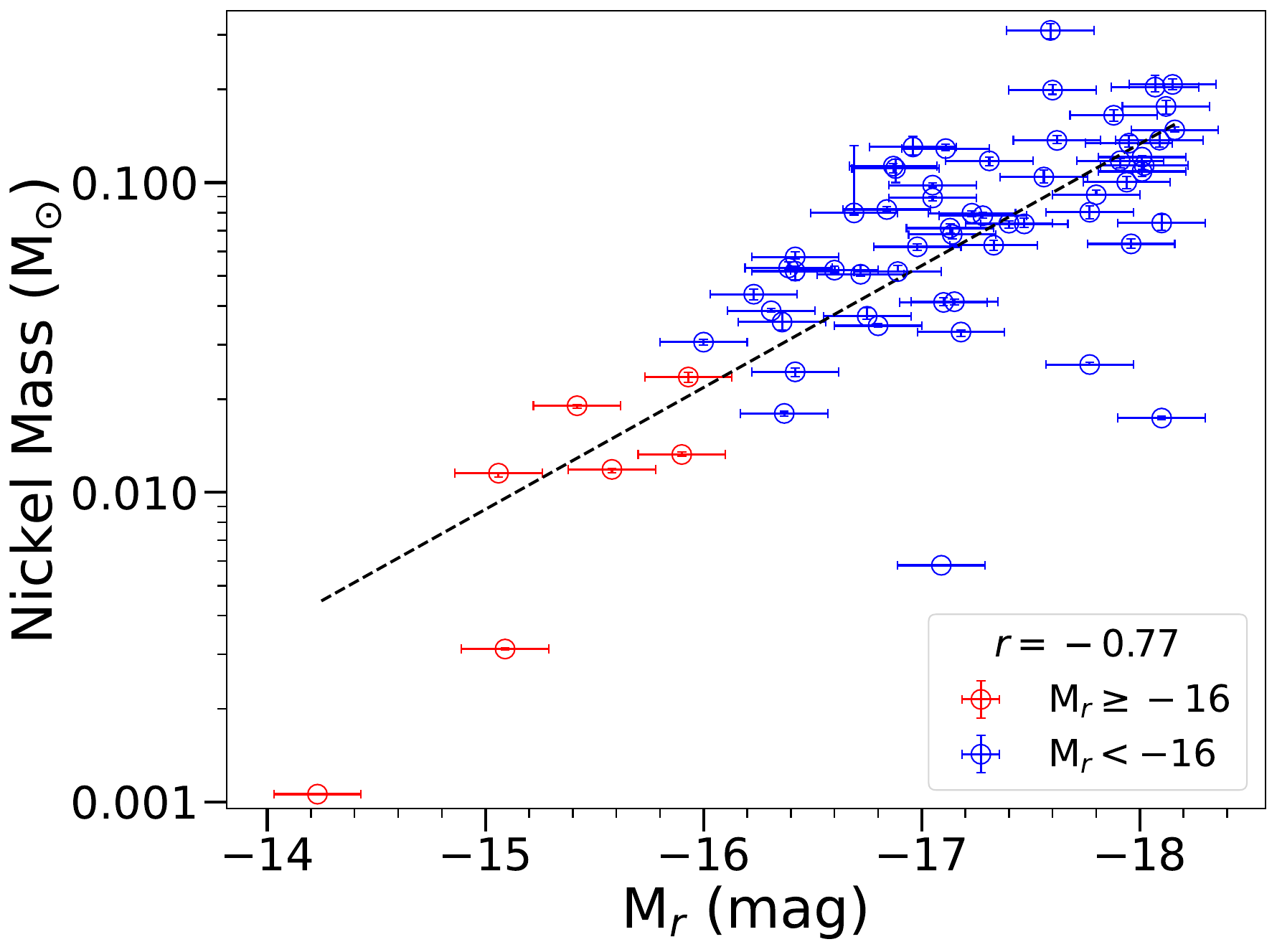}\includegraphics[width=0.33\textwidth]{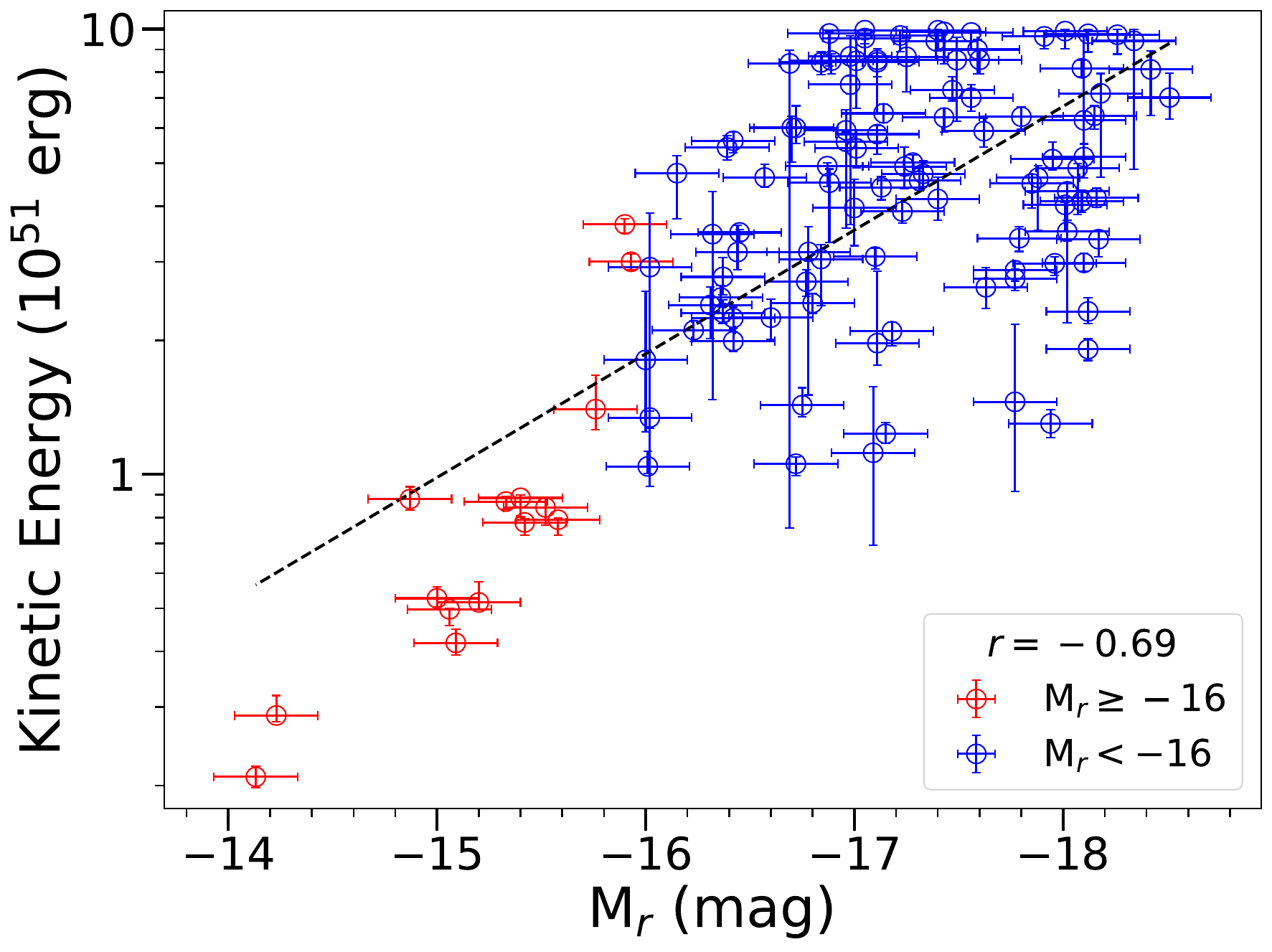}\includegraphics[width=0.33\textwidth]{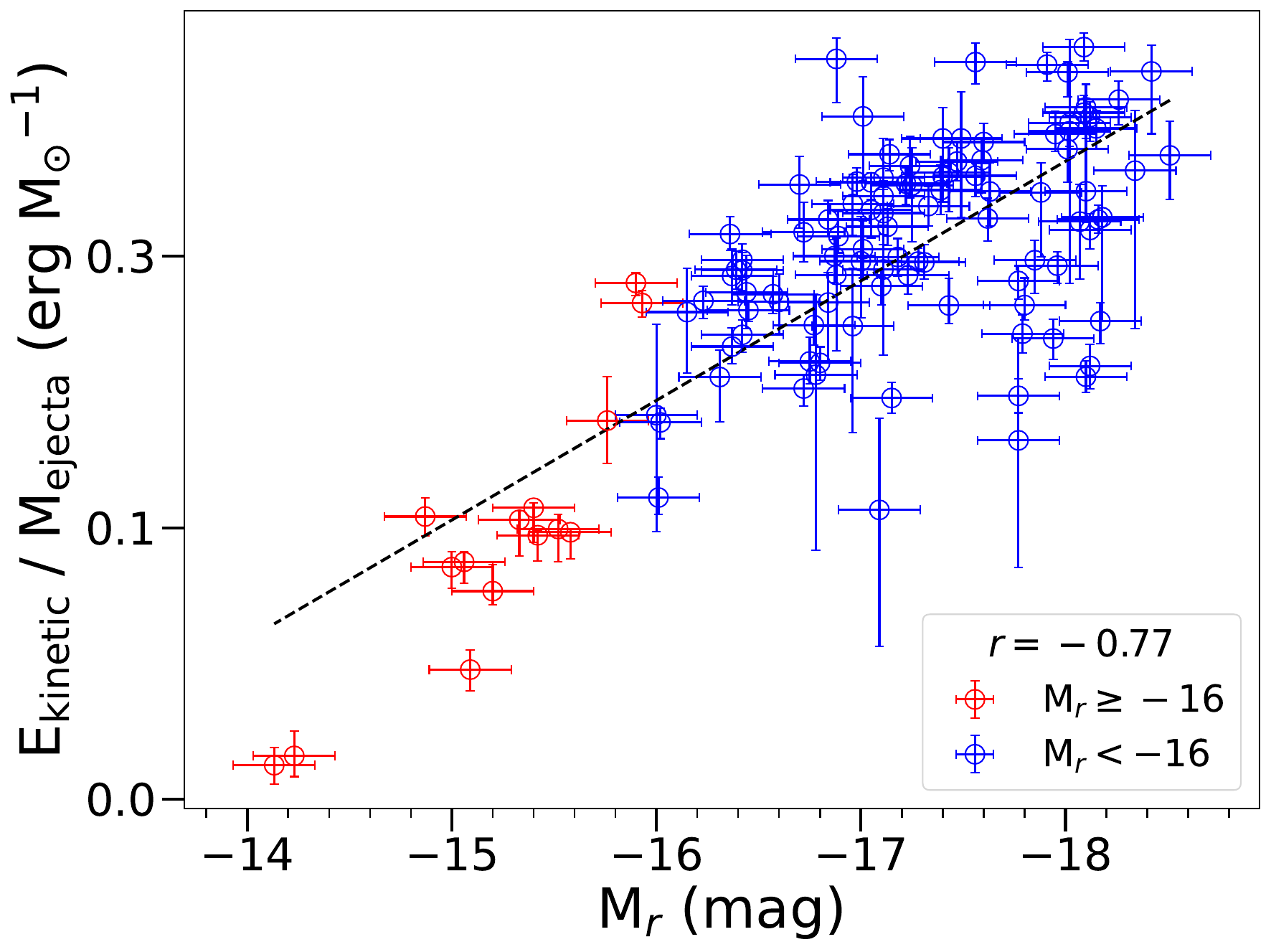} 
    
    \includegraphics[width=0.33\textwidth]{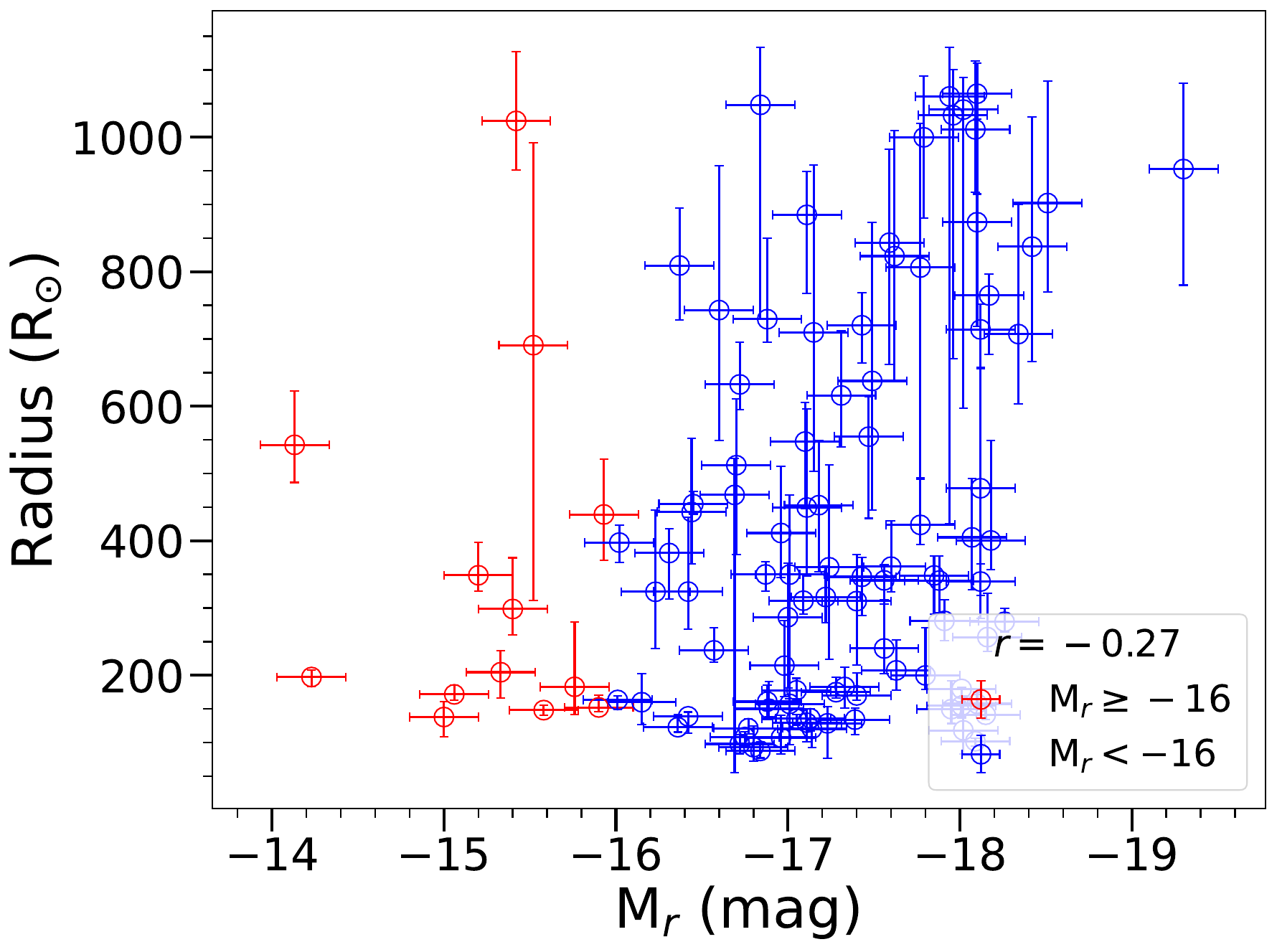}\includegraphics[width=0.33\textwidth]{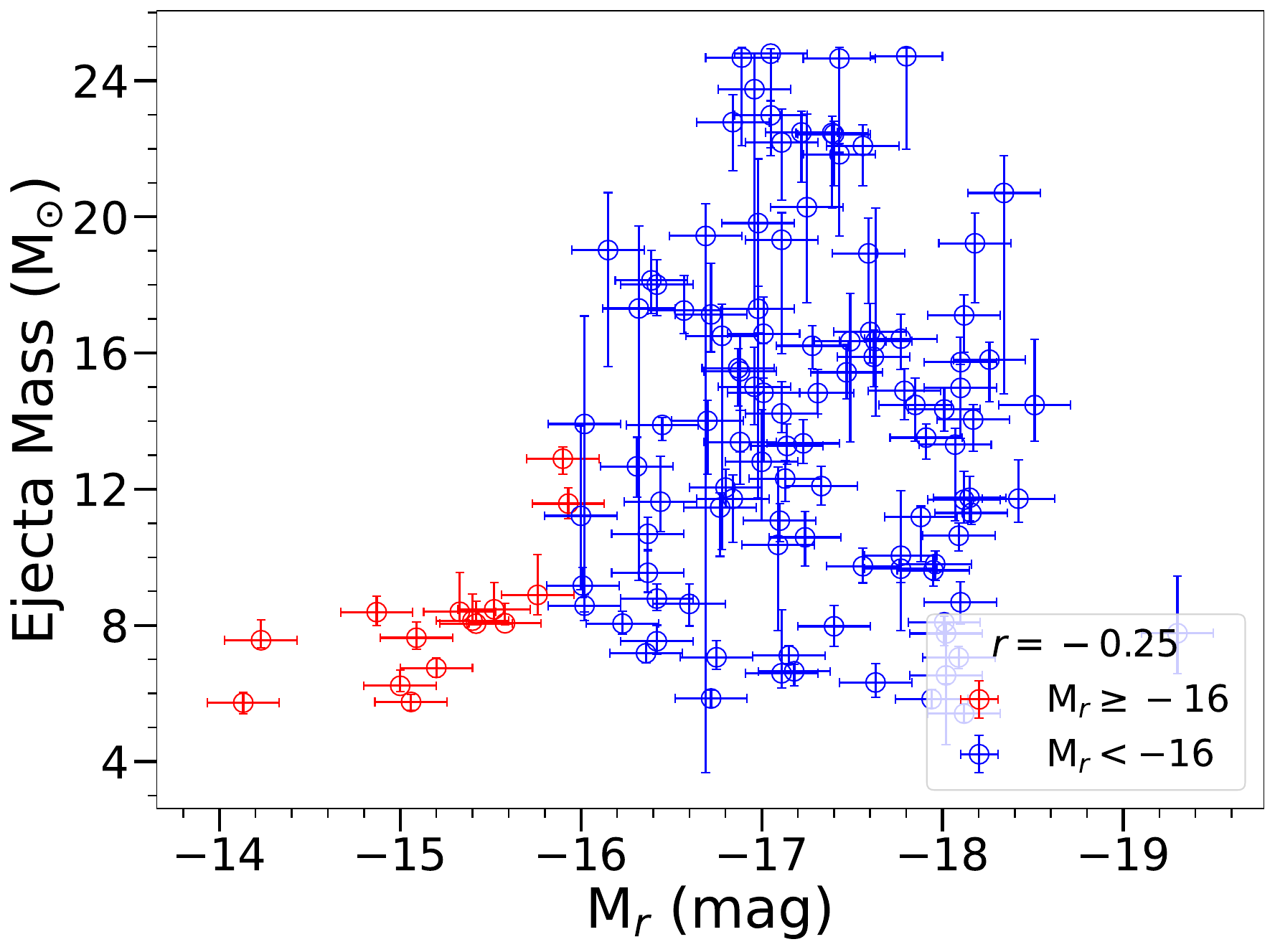}\includegraphics[width=0.33\textwidth]{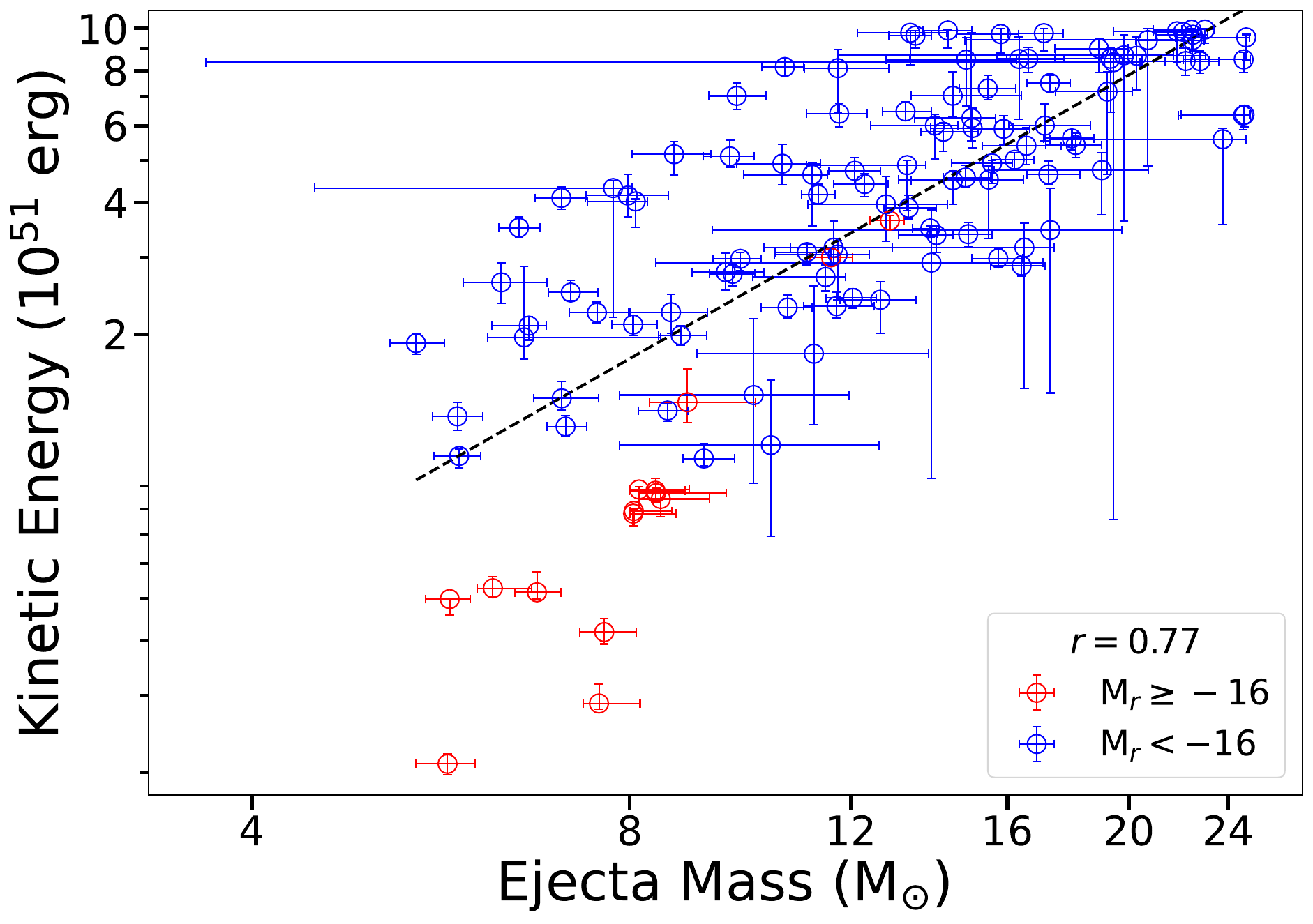}
    
    \includegraphics[width=0.33\textwidth]{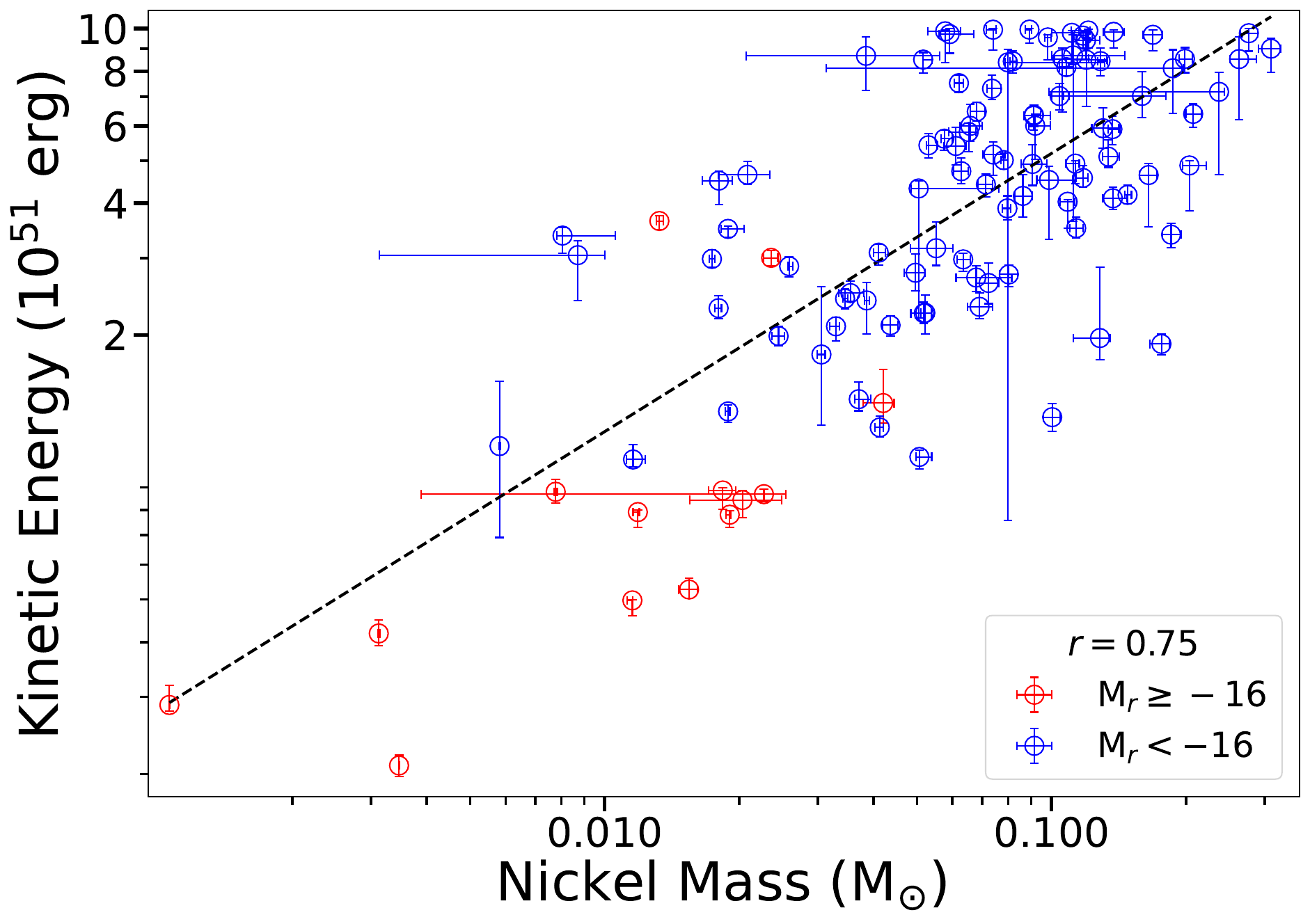}

\caption{Correlations between peak $r$-band magnitude and physical parameters (nickel mass, explosion energy, energy per unit mass, radius, ejecta mass), and among the physical parameters themselves, based on semi-analytical models from \citet{Nagy2016}. LLIIP SNe with $M_r \geq -16$ are shown in red, SNe IIP with $M_r < -16$ in blue.}

    \label{fig:semicorrelations}
\end{figure*}

\begin{figure*}
    \centering
    \includegraphics[width=0.333\textwidth]{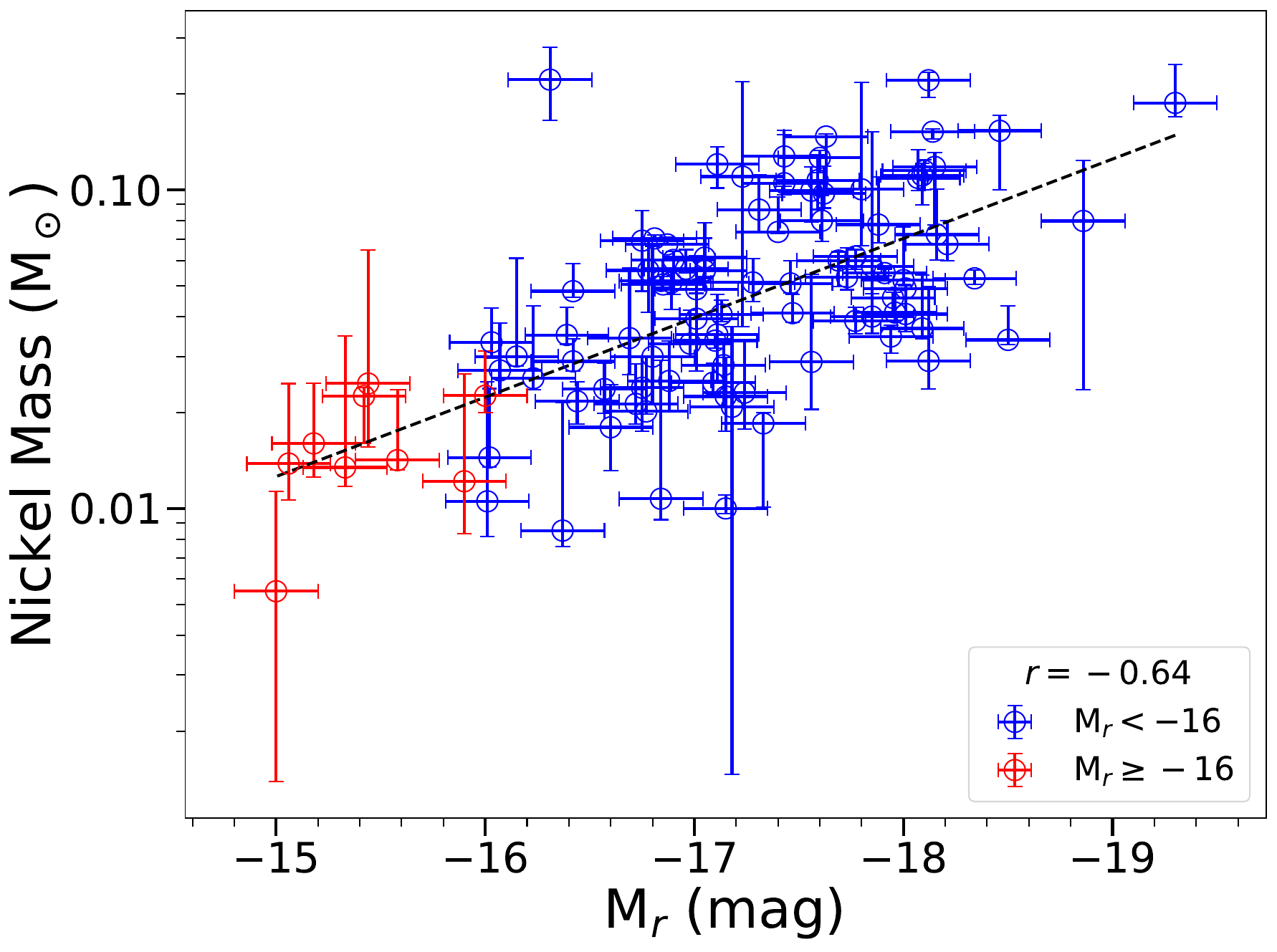}\includegraphics[width=0.33\textwidth]{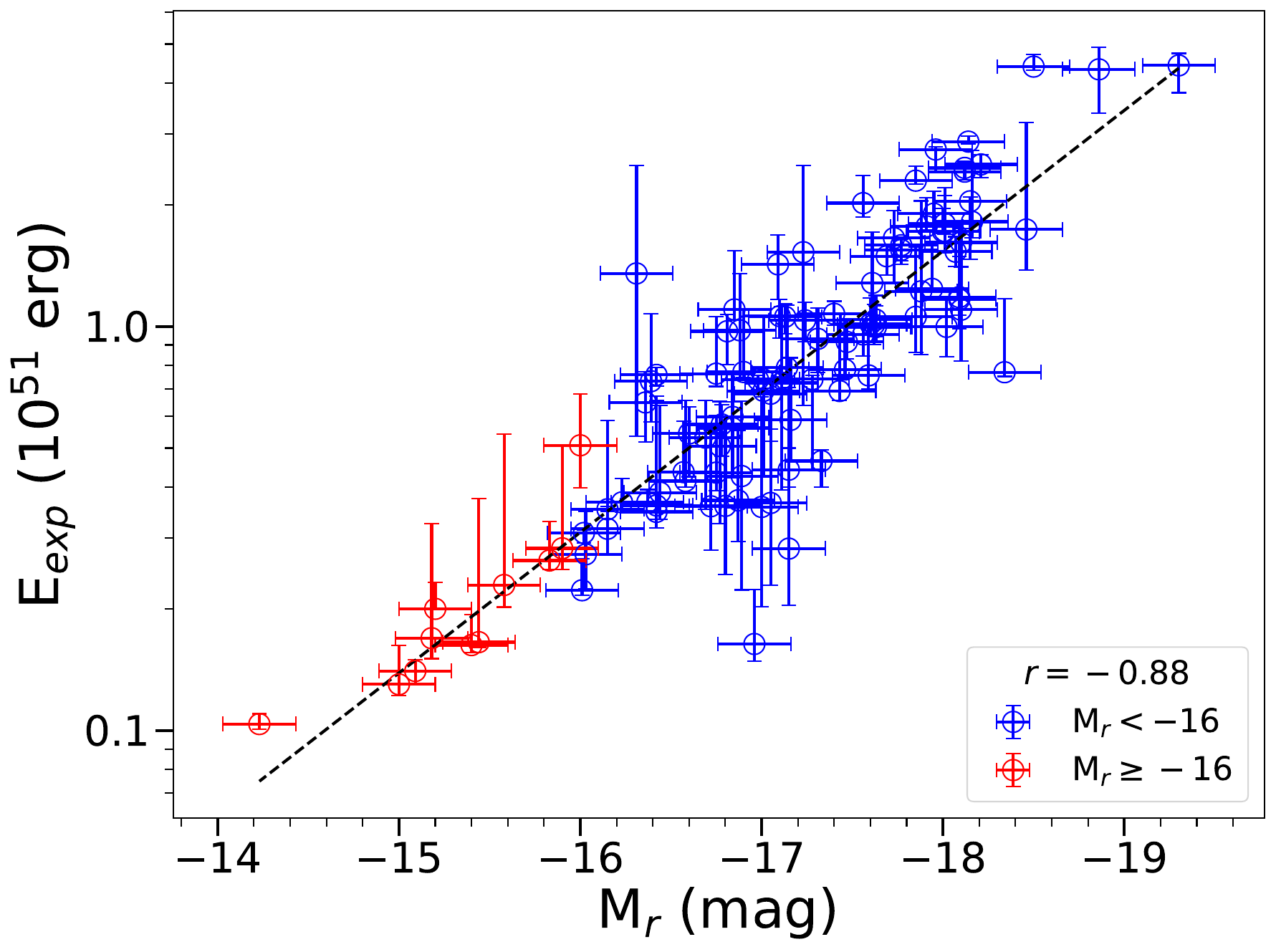}\includegraphics[width=0.33\textwidth]{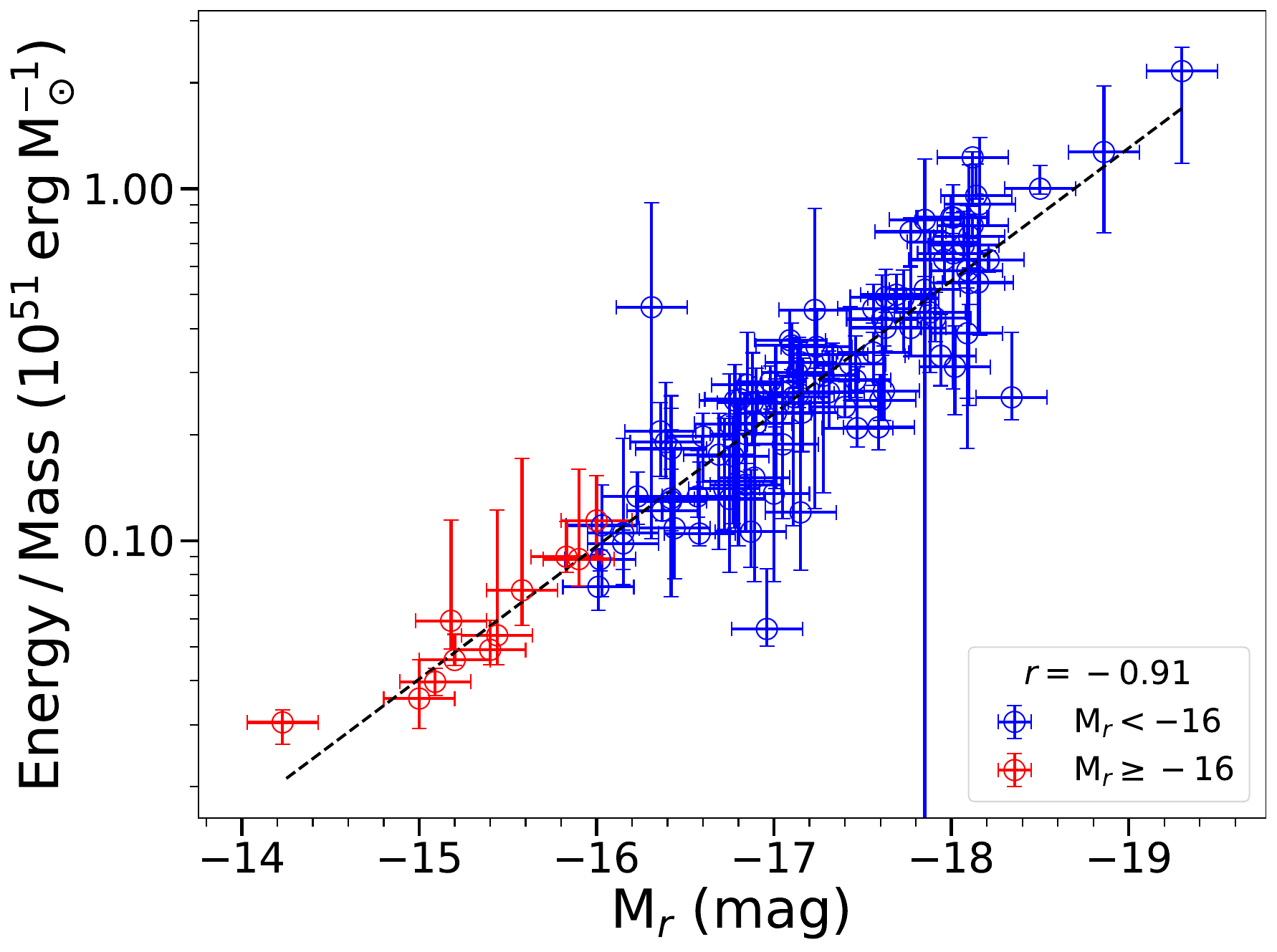}

    \includegraphics[width=0.33\textwidth]{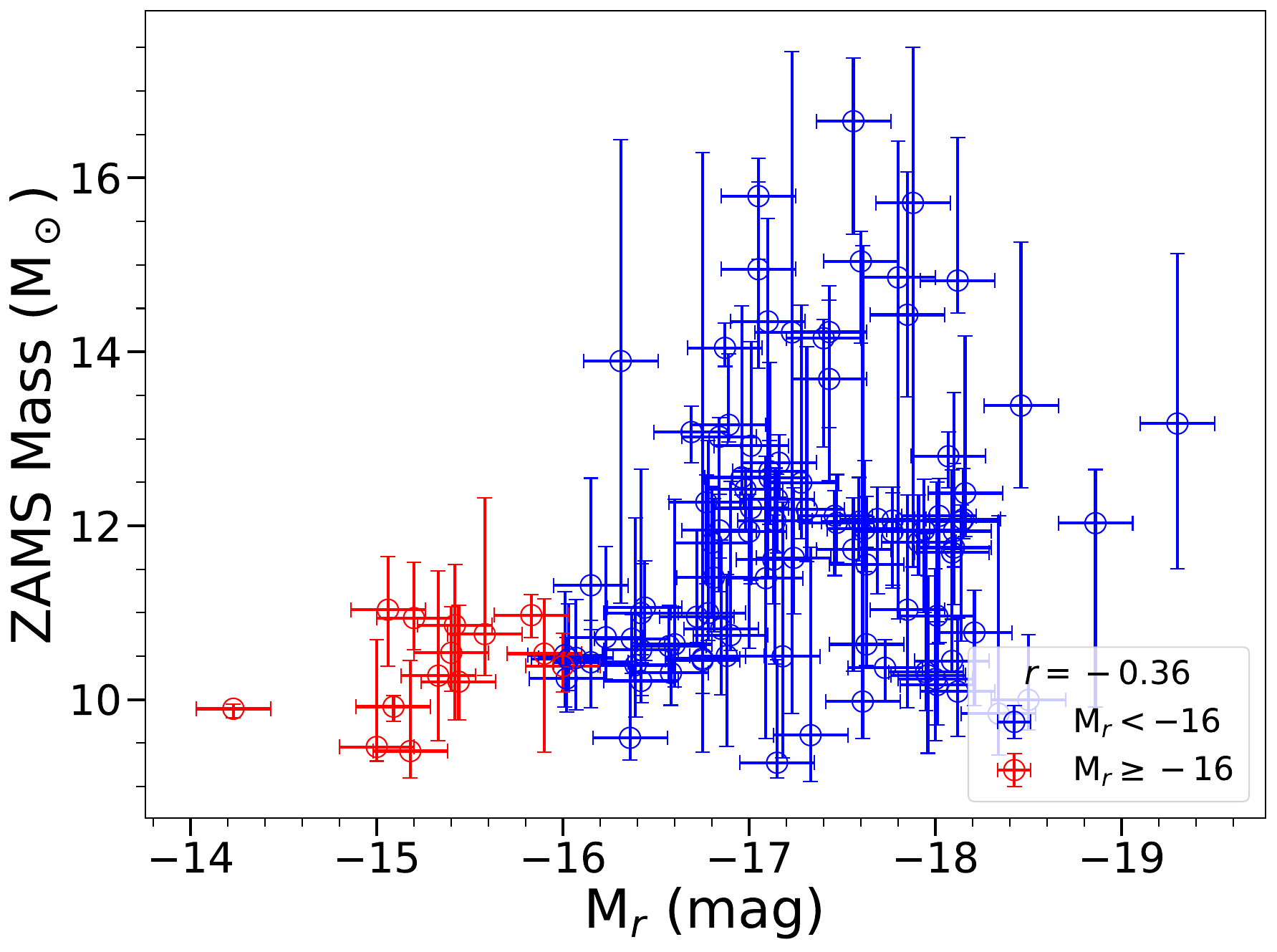}\includegraphics[width=0.33\textwidth]{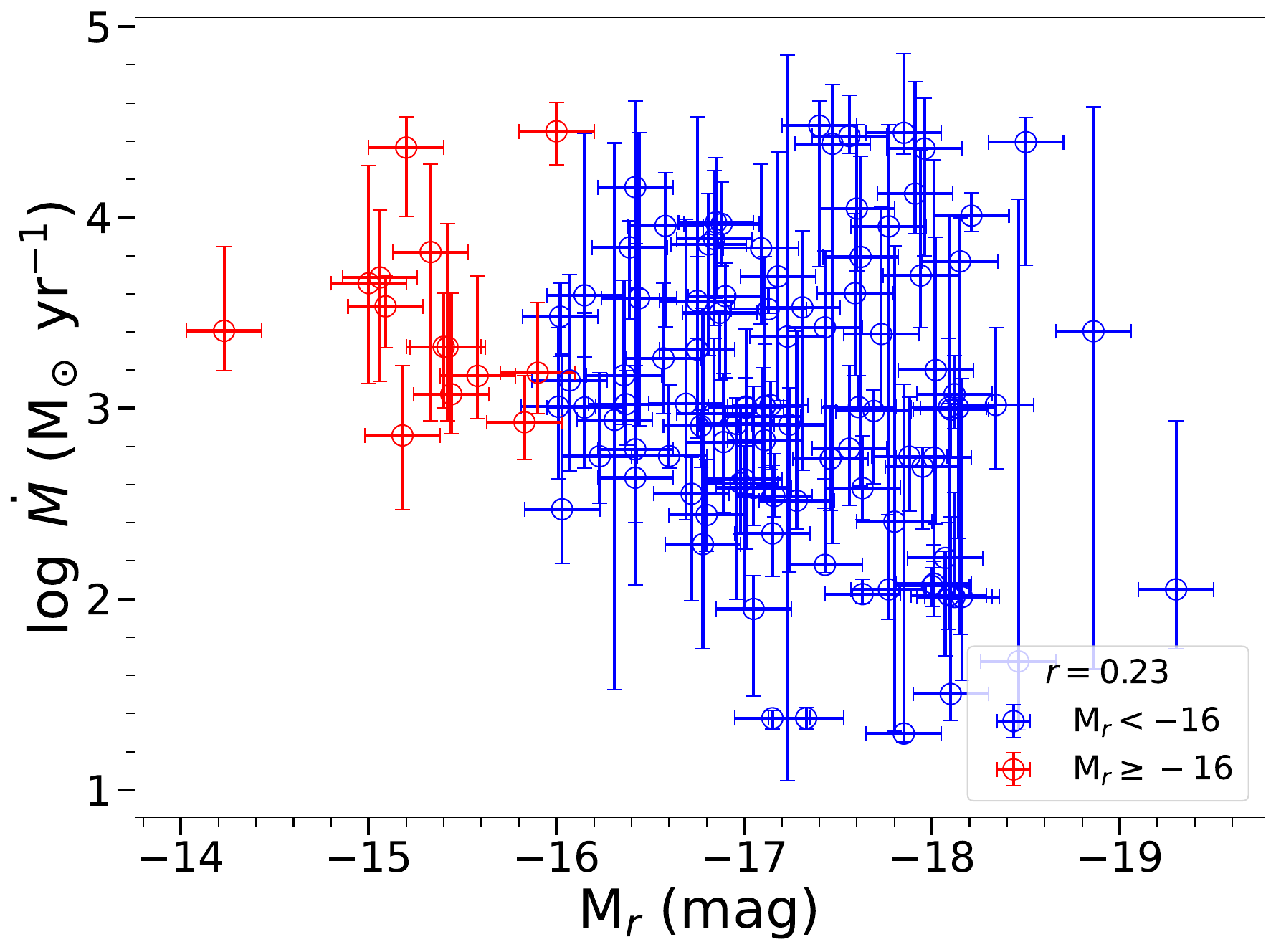}\includegraphics[width=0.33\textwidth]{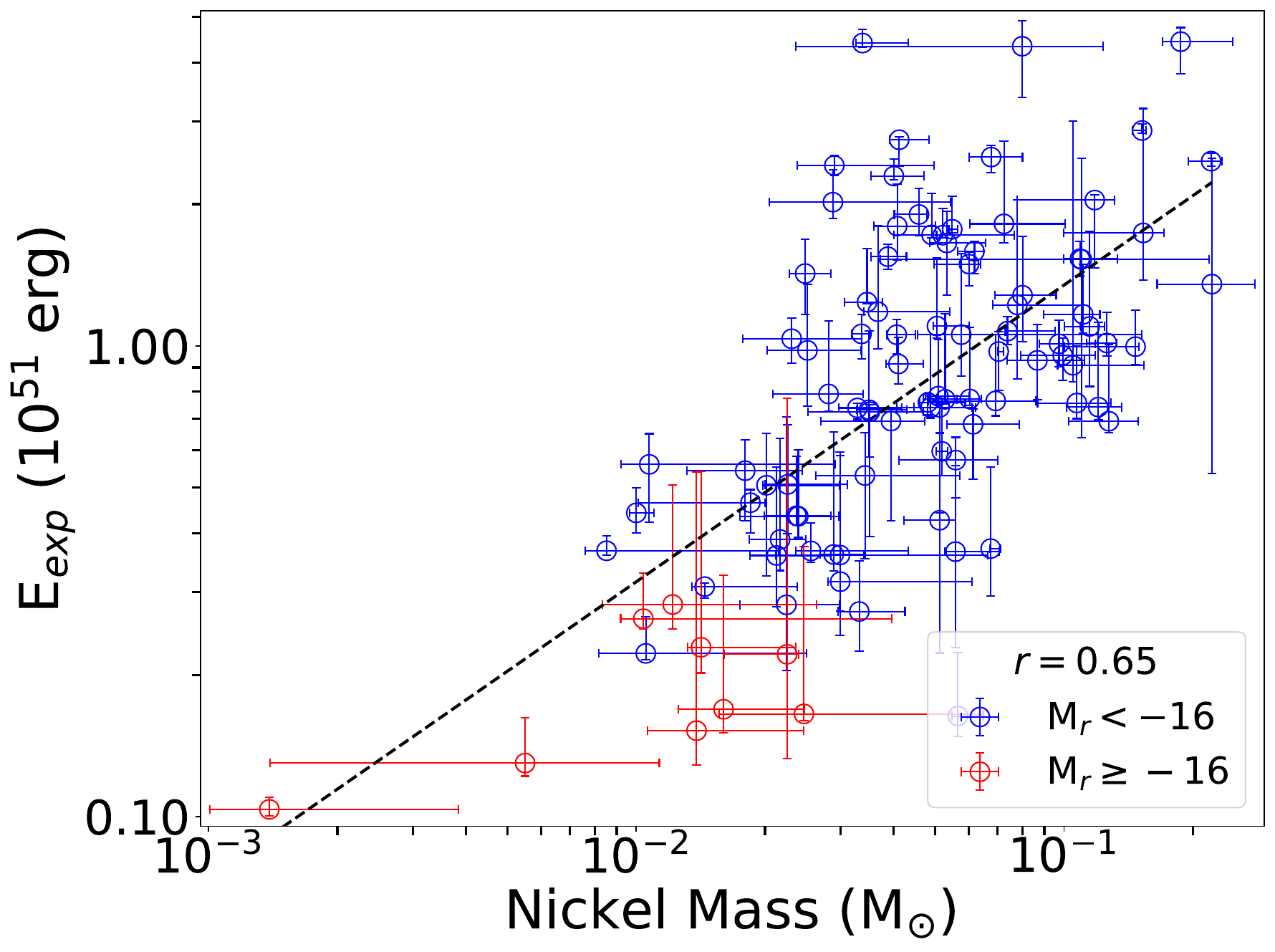}

\caption{Correlations between peak $r$-band magnitude and physical parameters (nickel mass, explosion energy, energy per unit mass, radius, ejecta mass), and among the physical parameters themselves, based on radiation-hydrodynamical model fits from \citet{Moriya2023}. LLIIP SNe with $M_r \geq -16$ are shown in red, those with $M_r < -16$ in blue.}    
    
    \label{fig:moriyacorrelations}
\end{figure*}

\textcolor{black}{The physical parameters inferred for the sample exhibit a number of clear and physically 
informative correlations. Figures~\ref{fig:semicorrelations} and~\ref{fig:moriyacorrelations} 
summarize the trends between peak luminosity and the key explosion properties. For the 
\citet{Nagy2016} semi-analytical fits, nickel mass, kinetic energy, and the energy-to-mass 
ratio all show strong negative correlations with peak absolute magnitude: fainter SNe 
systematically eject less $^{56}$Ni, have lower kinetic energies, and have smaller specific 
energies. The trends follow the relations:}

\[
\log_{10}\!\left(\frac{M_{\rm Ni}}{M_\odot}\right)
= (-0.39 \pm 0.04) M_r + (-7.95 \pm 0.71), 
\]
\[
\log_{10}\!\left(\frac{E_{\rm kin}}{10^{51}\,{\rm erg}}\right)
= (-0.28 \pm 0.03) M_r + (-4.17 \pm 0.46),
\]
\[
\log_{10}\!\left(\frac{E_{\rm kin}}{M_{\rm ej}} \right)
= (-0.22 \pm 0.02) M_r + (-4.29 \pm 0.29),
\]
\textcolor{black}{with Pearson coefficients $|r| \approx 0.7$--0.8. These scalings extend the well-known 
luminosity–$^{56}$Ni relation \citep{Hamuy2003, Pejcha2015, Muller2017, Martinez2022} to the faint end and show that LLIIP SNe occupy the faint, low-energy end of a continuous distribution.}

\begin{figure*}
    \centering
    \includegraphics[width=0.49\textwidth]{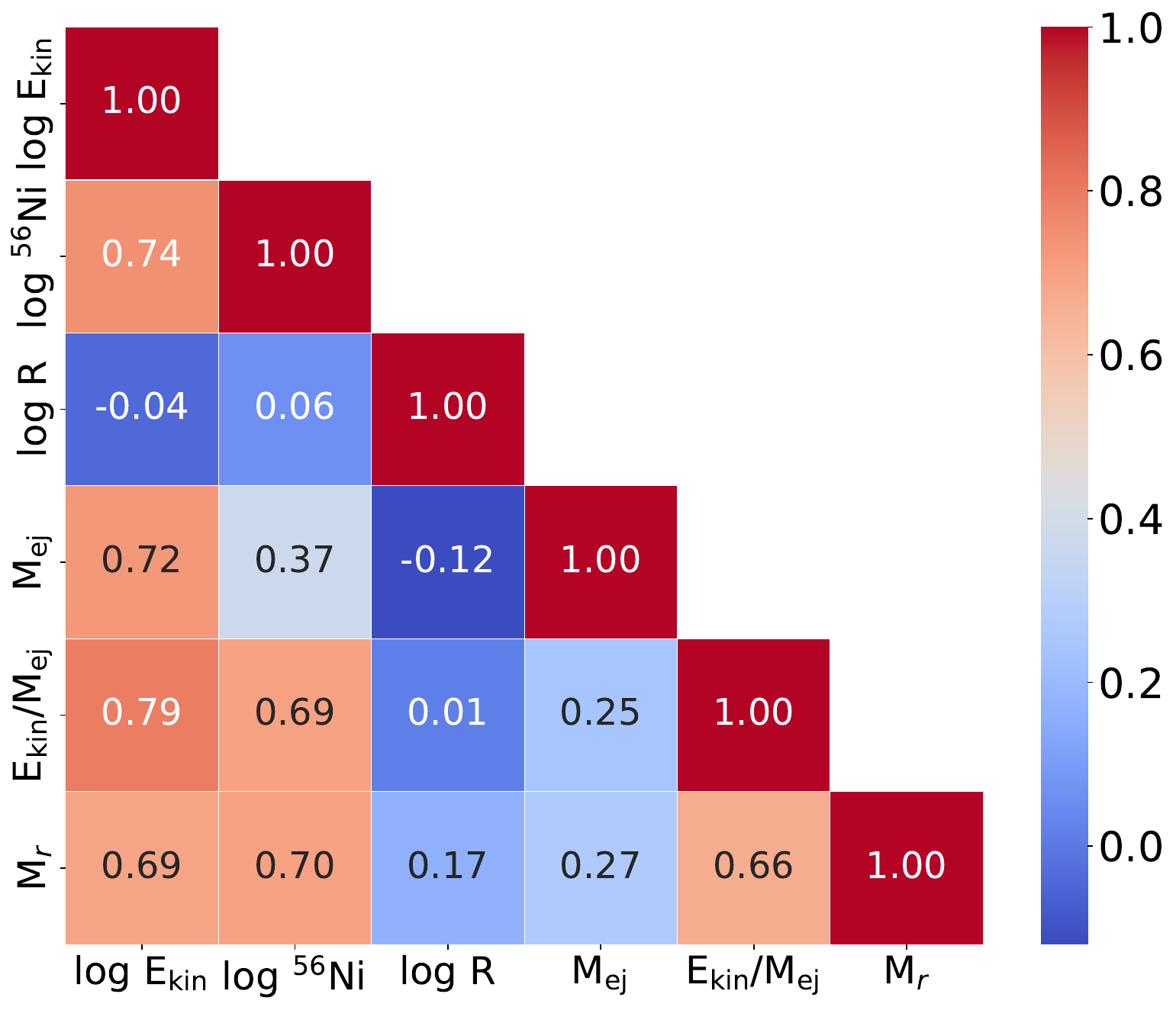}\includegraphics[width=0.49\textwidth]{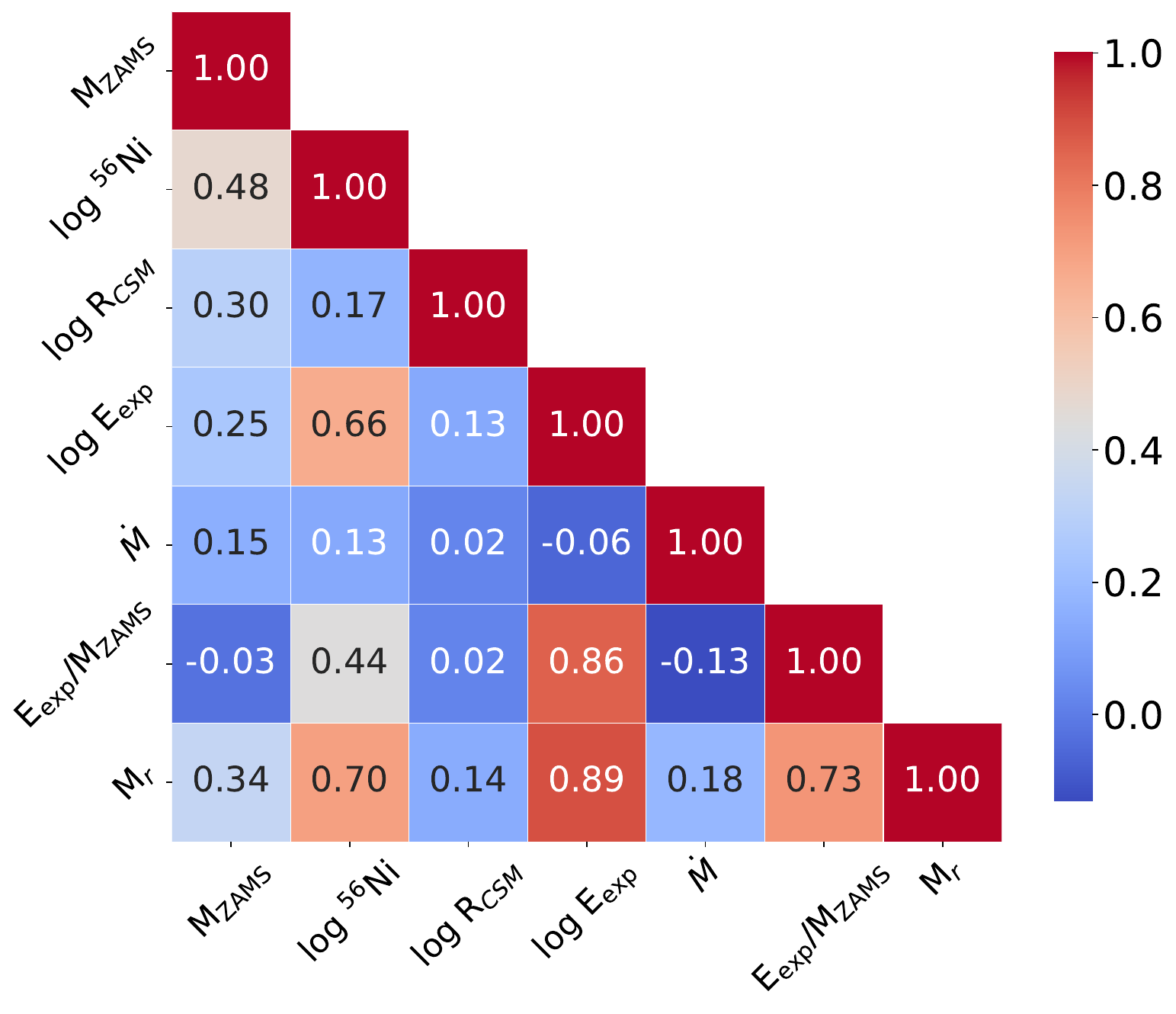}\caption{
Correlation matrices of physical parameters derived from semi-analytical model fits (left) and radiation-hydrodynamical model fits (right). 
}
\label{fig:correlationmatrix}
\end{figure*}

The semi-analytical parameters also correlate with each other. The correlations among physical parameters inferred from the \citet{Nagy2016} models are shown in Figure~\ref{fig:correlationmatrix}. Nickel mass and kinetic 
energy are tightly linked ($r = 0.75$), consistent with the idea that more energetic 
explosions synthesize more $^{56}$Ni. The empirical relation,

\[
E_{\rm kin} \,[10^{51}\,\mathrm{erg}] 
= (22.37 \pm 3.70)\,
\left(\frac{M_{\rm Ni}}{M_\odot}\right)^{0.63 \pm 0.05},
\]

\begin{figure*}
    \centering
    \includegraphics[width=0.99\textwidth]{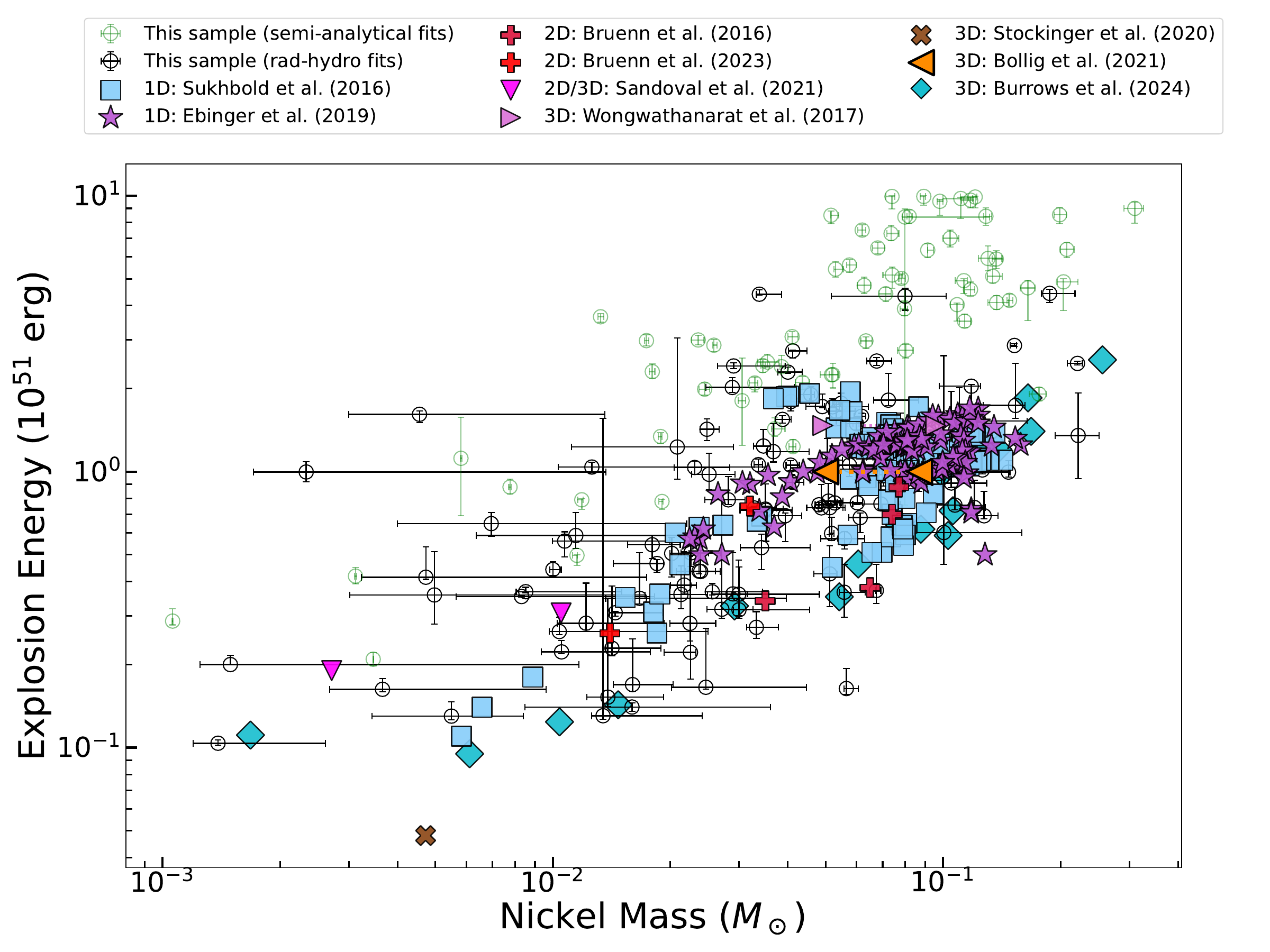}\caption{Nickel mass versus explosion energy for a range of core-collapse SN models, compared to our observations. Theoretical models include 1D, 2D and 3D simulations \citep{Ebinger2019, Sukhbold2016, Bruenn2016, Bruenn2023, Sandoval2021, Bollig2021, Burrows2024}. The observational sample includes semi-analytical fits (green open circles) and radiation-hydrodynamical model fits (black) from this work.}
\label{fig:eknimodel}
\end{figure*}

\textcolor{black}{is consistent with predictions in neutrino-driven explosion models 
\citep{Ebinger2019, Sukhbold2016, Bruenn2016, Bruenn2023, Sandoval2021, Bollig2021, Burrows2024}, as shown in Figure~\ref{fig:eknimodel}.}
Kinetic energy also correlates strongly with ejecta mass ($r = 0.77$), following

\[
E_{\rm kin} \,[10^{51}\,\mathrm{erg}] 
= (0.06 \pm 0.02)\,
\left(\frac{M_{\rm ej}}{M_\odot}\right)^{1.63 \pm 0.13},
\]

indicating that explosions with more massive ejecta tend to be more energetic. In contrast, 
radius and mass-loss rate show only weak or no correlations, implying greater diversity 
in envelope structure and pre-SN winds.

The correlations inferred from the \citet{Moriya2023} radiation-hydrodynamical models 
reveal the same trends, but with tighter relationships. Nickel mass and peak magnitude correlate with 
$r = -0.64$,

\[
\log_{10}\!\left(\frac{M_{\rm Ni}}{M_\odot}\right)
= (-0.25 \pm 0.03)\,M_r - (5.64 \pm 0.51),
\]

and explosion energy shows an even stronger dependence on luminosity ($r = -0.88$):

\[
\log_{10}\!\left(\frac{E_{\rm exp}}{\mathrm{erg}}\right)
= (-0.35 \pm 0.02)\,M_r - (6.08 \pm 0.31).
\]

The specific energy correlates most strongly with luminosity ($r = -0.91$),

\[
\log_{10}\!\left(\frac{E_{\rm exp}}{M_{\rm ej}}\right)
= (-0.38 \pm 0.02)\,M_r - (7.05 \pm 0.29).
\]

As in the semi-analytical case, explosion energy and nickel mass are tightly linked 
($r = 0.65$),

\[
E_{\rm exp} = (5.7 \pm 1.4)\times 
\left(\frac{M_{\rm Ni}}{M_\odot}\right)^{0.63 \pm 0.08},
\]

\textcolor{black}{consistent with neutrino-driven simulations. Mass-loss rate shows no significant correlations ($r < 0.20$), 
suggesting that pre-SN winds and envelope stripping vary independently of the core properties that set the SN 
explosion physics. The overall distribution of the physical parameter correlations is summarized in Figures~\ref{fig:semicorrelations}, \ref{fig:moriyacorrelations}, and the associated correlation matrices.
}

\subsection{Caveats of the models}
\label{sec:caveats}

We caution that while the models employed to fit the lightcurves provide useful constraints and correlations, they also introduce several assumptions and systematic effects that may influence the derived parameters. We summarize these below.

A key challenge arises from the degeneracy between key parameters such as ejecta mass, explosion energy, and progenitor radius. Fitting models to photometry in only a couple of passbands allows multiple combinations of physical parameters to produce similar light curves. This degeneracy limits our ability to extract unique solutions from lightcurve fitting alone as discussed in \citet{Singh24, Rehemtulla2025}. \citet{Goldberg2019, Goldberg2020} argue that breaking these degeneracies requires external constraints on at least one physical parameter.

\citet{Moriya2023} attempt to mitigate this degeneracy using the progenitor mass–radius relation embedded in the KEPLER stellar grids. Although their lightcurve model grid spans a wide range of progenitor and explosion properties, it is important to note that there are systematic limitations inherited from the underlying stellar evolution framework. For example, they employ a low mixing-length parameter, leading to progenitor radii that are systematically larger than what is inferred from observational constraints. Additionally, the assumption of continuous wind-driven mass loss throughout stellar evolution results in hydrogen-rich envelope masses that are systematically lower at the time of core collapse. These can influence the lightcurves and the physical parameters derived from fitting them \citep{Hsu2024}. While we incorporate velocity information from photospheric-phase spectroscopy to partially constrain kinetic energy, these measurements are not as constraining as those obtained during the brief shock-cooling phase immediately after explosion, which are more diagnostic \citep{Goldberg2019}. We also note that the lowest explosion energy in the model grid is $10^{50}$~erg; lower energies are required to better fit the lightcurves of the faintest SNe in our sample.

The semi-analytical two-component lightcurve models developed by \citet{Nagy2016} assume spherical symmetry and do not accurately capture the early shock-cooling behavior at $t < 20$ days. Accordingly, we restrict our fits to data obtained after 20 days post-explosion. Despite their simplicity, \citet{Nagy2016} and \citet{Jager2020} find that these models yield physical parameters broadly consistent with more sophisticated hydrodynamical simulations for Type IIP SNe. To test this consistency, \citet{Sheng2021} compared fitting results to those from hydrodynamical models such as those in \citet{Martinez2020}, and found that, in several cases, the inferred ejecta masses are in reasonable agreement. We find that the semi-analytical fits return higher explosion energies and ejecta mass for most SNe. This could be because of the two-component density profile and a simplified method for computing bolometric lightcurves. Prior works have also noted that lightcurve modeling can systematically overestimate ZAMS masses compared to direct progenitor imaging or nebular spectroscopy \citep[e.g.,][]{Utrobin2008, Utrobin2009, Maguire2010, Sanders2015}.

The correlations between physical quantities such as peak luminosity, nickel mass, explosion energy, and ejecta mass are preserved in both approaches. Nevertheless, we also observe that, for individual SN, the absolute values derived from the two methods can differ significantly (Figure \ref{fig:comparemodels}). This highlights the presence of methodological systematics and the need for caution when interpreting parameter estimates from any single modeling framework.

\subsection{What are the progenitors of LLIIP SNe?}
\label{sec:progenitor}

\begin{figure*}
    \centering
    \includegraphics[width=0.48\textwidth]{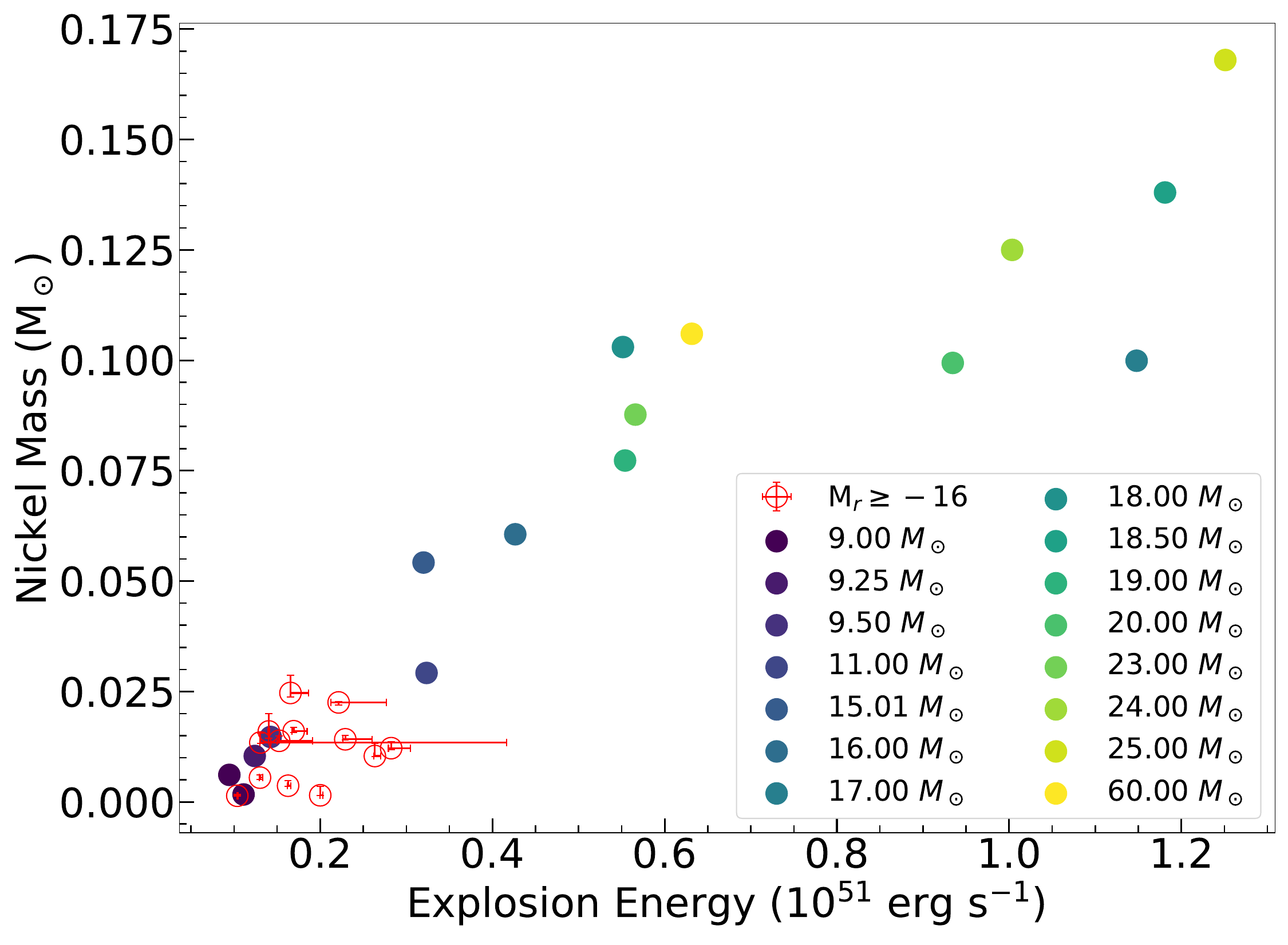}\includegraphics[width=0.48\textwidth]{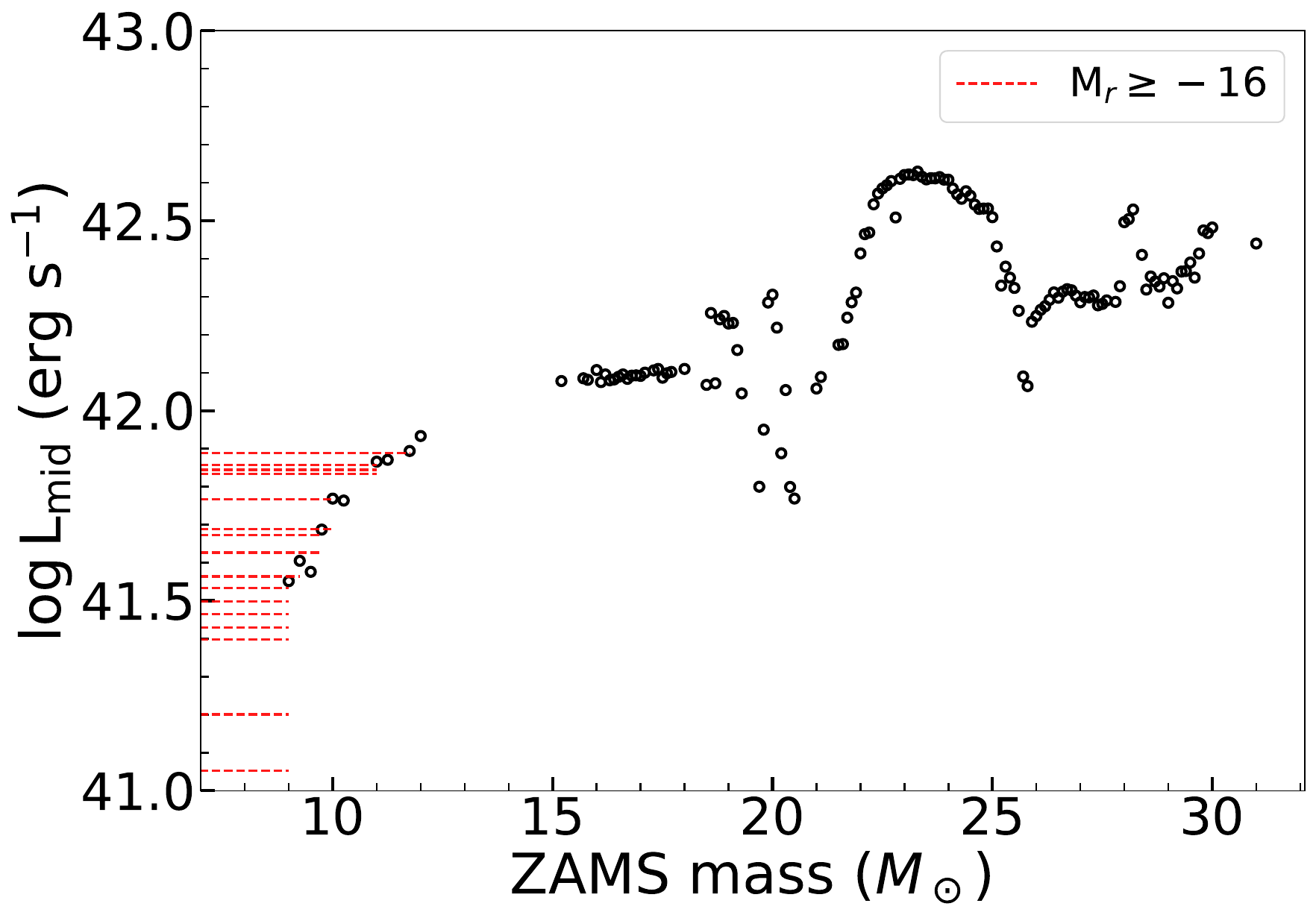}
    
\caption{
Left: Comparison of explosion energies from our SN sample with predictions from recent 3D core-collapse simulations by \citet{Burrows2020}. Right: Comparison of mid-plateau bolometric luminosities for our SN sample (horizontal lines) with theoretical predictions from \citet{Baker2022} (black points), which relate luminosity to iron core mass.
}
\label{fig:CCSNmodels}
\end{figure*}

\subsubsection{7--12 \Msun Red Super Giants that undergo core-collapse SNe}

We compare our derived explosion parameters to theoretical predictions from neutrino-driven explosion models of core-collapse SNe. \citet{Baker2022} show that the properties of a progenitor's core can be estimated from optical photometry during the plateau phase alone. A linear correlation between iron core mass and $L_{50}$ (luminosity 50 days post-explosion) indicates that optical photometry of Type IIP SNe can effectively probe progenitor cores. They find a robust relationship linking progenitor iron core mass to plateau luminosity (see Figure \ref{fig:CCSNmodels}). The bolometric luminosities of our LLIIP SN sample align with ZAMS masses below 12 \Msun. However, we note that these models assume that there is no CSM interaction. Additionally, using best-fit ZAMS mass estimates from \citet{Moriya2023} mode fits, the ZAMS mass for LLIIP SNe are lower than 11 \Msun. Furthermore, ejecta masses inferred from semi-analytical model fitting by \citet{Nagy2016} suggest masses $\sim$8.2 \Msun. Assuming a neutron star mass of 1.4 \Msun, this translates to progenitor masses less than $\sim11$ \Msun.

Also, Recent 3D simulations by \citet{Burrows2020} find explosion energies of approximately 0.1 foe for their lowest ZAMS mass (9 \Msun) progenitor model, compatible with our lowest observed explosion energies from the radiation-hydrodynamical fits (see Figure \ref{fig:CCSNmodels}). Previous 1D simulations \citep[e.g.,][]{Ugliano2012, Ertl2016, Sukhbold2016} yielded explosion energies between 0.1 and 2.0 foe. 

Observations and parameterized 1D explosion models suggest typical nickel yields of $\sim 0.05\,M_\odot$ from core-collapse SNe. However, the low nickel masses we measure imply progenitor initial masses toward the low-mass end of the CCSN spectrum. Stars with low-mass iron cores are known to produce significantly reduced nickel yields. For example, \citet{Stockinger2020} and \citet{Sandoval2021} find nickel masses of $\sim$0.002--0.005 $M_\odot$ for a 9.6 $M_\odot$ progenitor. Recent 3D simulations by \citet{Burrows2024} predict nickel masses in the range 0.002--0.006 $M_\odot$ for a 9 $M_\odot$ progenitor, rising above 0.01 $M_\odot$ for progenitor masses exceeding 9.25 $M_\odot$. These theoretical predictions are consistent with our measured nickel mass for LLIIP SNe.

\subsubsection{sAGB stars that undergo Electron-capture SNe}

\begin{figure*}
    \centering
    \includegraphics[width=0.40\textwidth]{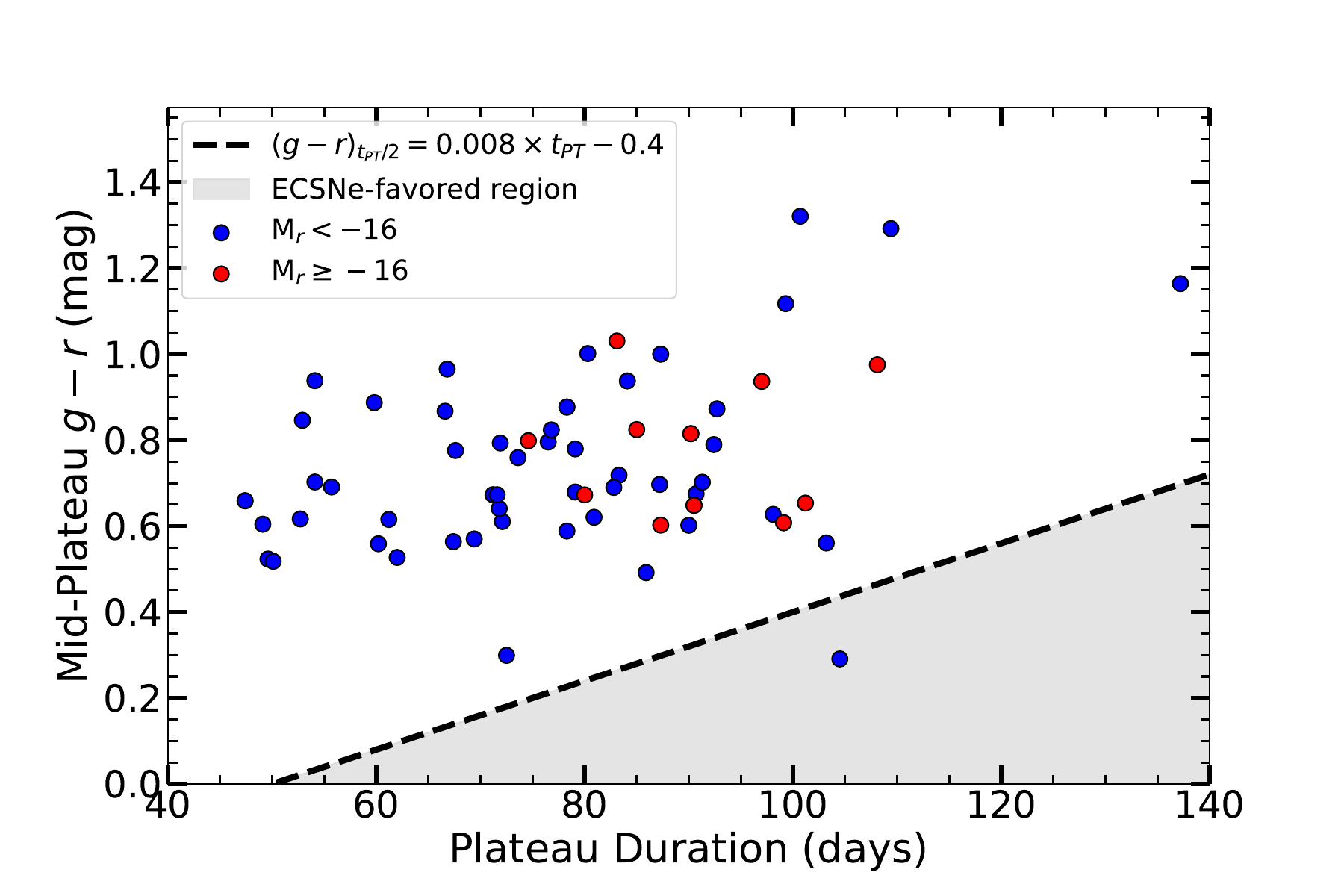}\includegraphics[width=0.33\textwidth]{ZTF21abnlhxs_GP.pdf}\includegraphics[width=0.33\textwidth]{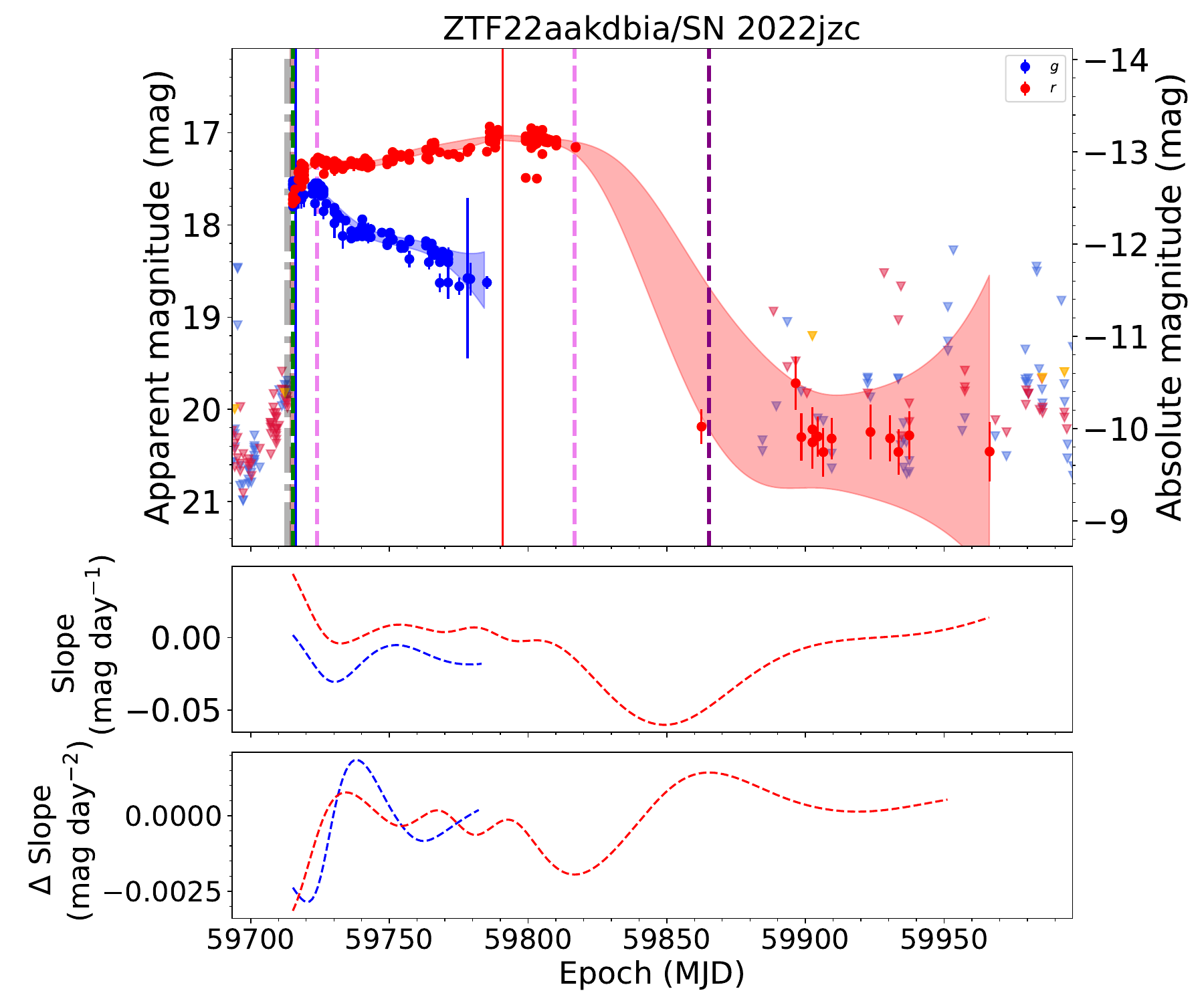}
    \caption{Left: Mid-plateau $(g-r)$ colors compared to the ECSN color criterion from \citet{Sato2024}, distinguishing explosions of sAGB stars from RSGs, shown in dashed black. Light curves of the SNe with the bluest (SN~2021tyw, mid-plateau $(g - r) = 0.2$~mag; center) and reddest (SN~2022jzc, mid-plateau $(g - r) = 1.4$~mag; right) mid-plateau colors.}

    \label{fig:ecsn}
\end{figure*}

Progenitors in the mass range $8-10$ \Msun\ may evolve into super-Asymptotic Giant Branch (sAGB) stars and explode as electron-capture supernovae (ECSNe). ECSN simulations consistently produce low explosion energies and nickel yields significantly lower than typical Fe CCSNe but similar to those observed for the lowest-mass Fe CCSNe. According to \citet{Sato2024}, ECSNe, without significant circumstellar material (CSM) interaction, exhibit distinctively bluer bolometric lightcurve plateaus compared to Fe CCSNe, characterized by mid-plateau $(g-r)$ colors exceeding the relation $0.008 \times t_{PT} - 0.4$. This difference arises from the lower-density envelopes inherent to sAGB progenitors. We plot the mid-plateau colors for our SN sample against the \citet{Sato2024} criterion in Figure~\ref{fig:ecsn}. While SNe~2021tyw, 2022omr, 2021cwe, and 2022prv show bluer mid-plateau colors, their peak luminosities and nickel masses favor an Fe CCSN origin. Among our LLIIP SNe sample, SNe~2023wcr, 2021gmj, 2021zgm, and 2020abcq exhibit notably bluer mid-plateau colors. \textcolor{black}{However, we note that the mid-plateau colors may also be influenced by CSM interaction, as supported by the interaction signatures seen in the spectrum of SN~2021tyw.} The lightcurves of the SNe with the bluest (SN~2021tyw, mid-plateau $(g - r) = 0.2$~mag) and reddest (SN~2022jzc, mid-plateau $(g - r) = 1.4$~mag) colors are shown in Figure~\ref{fig:ecsn}.

We also compare the inferred nickel masses and explosion energies of our LLIIP sample to the ECSN models from \citet{Kozyreva2021}. The nickel masses ($\sim 10^{-3}-10^{-4}\ \Msun$) and explosion energies ($\sim 0.3-1.37 \times 10^{50}\ \mathrm{erg}$) from our sample closely match those predicted by ECSN simulations for their 8.8 and 9.6 \Msun\ progenitor models. \citet{Kozyreva2021} also show that the lightcurves in the $U$ and $B$ bands rises during the first 50 days and then slowly declines, while the light curve in the $V$ and redder bands rises over the same timescale and then settles into a plateau before a steep drop to the radioactive tail. The rising $r$-band light curves seen in some LLIIP SNe, when coupled with their low nickel masses, are consistent with theoretical expectations for explosions from sAGB stars that undergo ECSNe.

Photospheric properties alone do not reliably distinguish between ECSNe from sAGB stars and Fe-core collapse SNe from red supergiants. Nebular-phase spectroscopy, however, offers definitive diagnostic features due to distinct core compositions and nucleosynthetic pathways \citep{Jerkstrand2018}. A comprehensive analysis of nebular spectra for the LLIIP SN sample will be presented in Paper III.

\section{Conclusion}

\label{sec:conclusion}

We present the largest systematic study of lightcurve properties for Type IIP SNe from the volume-limited ZTF CLU survey, analyzing 129 Type IIP events, including 16 LLIIP SNe. We summarize the key takeaways of the paper here:

\begin{enumerate}
\item \textit{Observables:} 

\begin{itemize}

\item The median plateau duration for LLIIP SNe is $89^{+10}_{-20}$ days, compared to $76^{+16}_{-16}$ days for the full Type IIP sample. The median OPTd is $99^{+11}_{-14}$ days for LLIIP SNe and $90^{+16}_{-20}$ days for the full sample. There is no strong correlation with peak magnitude, likely suggesting binary interaction.

\item The plateau slope strongly correlates with peak brightness ($r = -0.68$), with brighter SNe having steeper declines; a significant fraction of LLIIP SNe exhibit positive plateau slopes. This likely implies progenitors for most LLIIP SNe have a smaller radius and different density profile from those of their more luminous counterparts.

\item We observe a strong correlation between the mid-plateau luminosity and nickel mass ($r = -0.89$), extending to the low-luminosity end. This can be explained if a strong core explosion leads to both higher explosion energy and greater nickel yield.

\item The steepest plateau-to-tail drop of $>3.5$ mag was seen in SN~2022aagp. Other objects with sharp drops include SN~2021tyw ($> 3.1$ mag) and SN~2020cxd ($2.5$ mag). These are consistent with ECSNe and failed SNe with very low or zero nickel mass.

\end{itemize}

\item \textit{Explosion and Progenitor Analysis:} 

\begin{itemize}

\item Radiation-hydrodynamical modeling: The ZAMS masses for LLIIP SNe are typically below $12$~M$_\odot$, with a median of $10.5^{+0.4}_{-0.7}$~M$_\odot$, while that of the overall sample is $11.5^{+1.6}_{-1.2}$~M$_\odot$. The kinetic energies are similarly lower for LLIIP SNe, with a median of $0.17^{+0.07}_{-0.03} \times 10^{51}$~erg, compared to $0.76^{+0.89}_{-0.44} \times 10^{51}$~erg for the overall sample. The inferred mass-loss rates just before explosion are relatively low, with a median $\log\ \dot{M} = -3.32^{+0.37}_{-0.26}$ for LLIIP SNe, compared to $-3.01^{+0.83}_{-0.60}$ for the overall sample.

\item Semi-analytical modeling: LLIIP SNe have low nickel masses, with a median of $0.013^{+0.089}_{-0.001}$~M$_\odot$ and a range from $0.0009$ to $0.0420$~M$_\odot$, while that of the overall Type IIP SNe sample is $0.0652^{+0.0680}_{-0.0471}$~M$_\odot$ with a range from $0.0009$ to $0.3098$~M$_\odot$. The median ejecta mass for LLIIP SNe is $8.06^{+0.73}_{-1.71}$~M$_\odot$ compared to $12.8^{+6.2}_{-5.0}$~M$_\odot$ for the overall sample. Their kinetic energies are also lower, with a median of $0.79^{+0.49}_{-0.35} \times 10^{51}$~erg for LLIIP SNe and $4.32^{+4.17}_{-2.91} \times 10^{51}$~erg for the overall sample.

Overall, LLIIP SNe occupy the faint, low-energy, low-mass end of the Type IIP population compared to more luminous events.




\end{itemize}

\item \textit{Correlation of Parameters:} \begin{itemize} 

\item Nickel mass, kinetic and explosion energy, and energy-to-mass ratio strongly correlate with peak absolute magnitude, indicating that fainter SNe have systematically lower nickel mass and explosion energies. This is consistent with theoretical core-collapse SN models, which predict a strong correlation between explosion energy and nickel production.

\item For the semi-analytical fits, nickel mass and kinetic energy correlates with peak brightness ($r = -0.77$ and $r = -0.69$ respectively).  Kinetic energy-to-ejecta mass ratio shows the strongest correlation ($r = -0.77$) with peak brightness. Among physical parameters themselves, kinetic energy correlates strongly with ejecta mass ($r = 0.77$) and with nickel mass ($r = 0.75$).

\item For the radiation-hydrodynamical fits, similar trends are observed: peak brightness correlates tightly with explosion energy ($r = -0.88$), nickel mass ($r = -0.64$) and energy-to-ejecta mass ratio ($r = -0.91$). Explosion energy also correlates with nickel mass ($r = 0.65$), while mass-loss rate shows little dependence on other explosion parameters.

\end{itemize}
\end{enumerate}

The correlations between physical parameters such as peak luminosity, nickel mass, explosion energy, and ejecta mass are preserved across both semi-analytical and radiation-hydrodynamic modeling approaches. Nevertheless, we observe that for individual SNe, the absolute values derived from the two methods can differ significantly. This highlights the presence of methodological systematics and underscores the need for caution when interpreting parameter estimates from any single modeling framework. However, the low explosion energies, nickel masses and progenitor masses inferred for LLIIP SNe are consistent with the explosions of low-mass RSG or sAGB stars with ZAMS masses $\lesssim$12~M$_\odot$. Distinguishing between these progenitor scenarios will require late-time nebular spectroscopy, which will be pursued in Paper III of this series. The models in this paper assumes single-star evolution. Incorporating progenitor models that account for binary interaction and mass transfer will be crucial for constraining the binary fraction and understanding the diversity of envelope stripping among core-collapse progenitors. We also highlight the need for radiation-hydrodynamic model grids that extend to the lowest progenitor masses in the 8–10 M$_\odot$ range and explore explosion energies below $10^{50}$~erg to adequately model the faintest Type IIP SNe. The advent of deep, multi-band, high-cadence surveys such as the Legacy Survey of Space and Time \citep[LSST;][]{Ivezic08} will enable a more complete census of the faintest slow-evolving core-collapse SNe and provide critical insights into the lowest-mass stars that explode. It will also offer deep post-plateau limits, facilitating the discovery of more Type IIP SNe with very low or even zero nickel mass, as expected for electron-capture supernovae or failed explosions.

\section*{Data Availability}

All photometric and spectroscopic data presented in this work will be made publicly available on \href{https://zenodo.org/records/15717884?token=eyJhbGciOiJIUzUxMiJ9.eyJpZCI6IjNhYjE3NjYwLTU2NDEtNDBkZi1iYmI5LTQ1YzQxY2EwYjllNyIsImRhdGEiOnt9LCJyYW5kb20iOiJhNjA2ZTkzNDFjYjU5NGM3ZGIzMmExMTRlOGY1NzU5MCJ9.X0jCmJz6FJbXwmcY_6ZZ1kYCcRHwrABpoZ2epNXLlpIjVIwrFkEEOGD0Z7Cfk7luRPlHvOCwJdWBwB40vX2JoQ}{\texttt{Zenodo}} and \href{https://wiserep.weizmann.ac.il}{\texttt{WISeREP}} upon publication. The codes required to reproduce the lightcurve fitting will be uploaded via \href{https://github.com}{\texttt{GitHub}}. Machine-readable versions of all tables, lightcurve plots, and best-fit model plots will also be available on \href{https://zenodo.org/records/15717884?token=eyJhbGciOiJIUzUxMiJ9.eyJpZCI6IjNhYjE3NjYwLTU2NDEtNDBkZi1iYmI5LTQ1YzQxY2EwYjllNyIsImRhdGEiOnt9LCJyYW5kb20iOiJhNjA2ZTkzNDFjYjU5NGM3ZGIzMmExMTRlOGY1NzU5MCJ9.X0jCmJz6FJbXwmcY_6ZZ1kYCcRHwrABpoZ2epNXLlpIjVIwrFkEEOGD0Z7Cfk7luRPlHvOCwJdWBwB40vX2JoQ}{\texttt{Zenodo}}.

\section{Acknowldegement}

We thank the anonymous referee for their constructive feedback, which helped improve the quality of this manuscript. 

We thank Jared Goldberg, Daichi Hiramatsu and Kishalay De for valuable discussions.

Based on observations obtained with the Samuel Oschin Telescope 48-inch and the 60-inch Telescope at the Palomar Observatory as part of the Zwicky Transient Facility project. ZTF is supported by the National Science Foundation under Grants No. AST-1440341, AST-2034437, and currently Award 2407588. ZTF receives additional funding from the ZTF partnership. Current members include Caltech, USA; Caltech/IPAC, USA; University of Maryland, USA; University of California, Berkeley, USA; University of Wisconsin at Milwaukee, USA; Cornell University, USA; Drexel University, USA; University of North Carolina at Chapel Hill, USA; Institute of Science and Technology, Austria; National Central University, Taiwan, and OKC, University of Stockholm, Sweden. Operations are conducted by Caltech's Optical Observatory (COO), Caltech/IPAC, and the University of Washington at Seattle, USA.

Zwicky Transient Facility access for S.S. was supported by Northwestern University and the Center for Interdisciplinary Exploration and Research in Astrophysics (CIERA).



M.W.C acknowledges support from the National Science Foundation with grant numbers PHY-2117997, PHY-2308862 and PHY-2409481.

D.T. is supported by the Sherman Fairchild Postdoctoral Fellowship at Caltech.

E.O. is supported by the Swedish Research Council (Project No. 2020-00452).

N.Sarin acknowledges support from the Knut and Alice Wallenberg Foundation through the ``Gravity Meets Light" project and by and by the research environment grant ``Gravitational Radiation and Electromagnetic Astrophysical Transients'' (GREAT) funded by the Swedish Research Council (VR) under Dnr 2016-06012.

SED Machine is based upon work supported by the National Science Foundation under
Grant No. 1106171. 

The ZTF forced-photometry service was funded under the Heising-Simons Foundation grant \#12540303 (PI: Graham).

The Gordon and Betty Moore Foundation, through both the Data-Driven Investigator Program and a dedicated grant, provided critical funding for SkyPortal .

This research has made use of the NASA/IPAC Extragalactic Database (NED), which is funded by the National Aeronautics and Space Administration and operated by the California Institute of Technology.

The Liverpool Telescope is operated on the island of La Palma by Liverpool John Moores University in the Spanish Observatorio del Roque de los Muchachos of the Instituto de Astrofisica de Canarias with financial support from the UK Science and Technology Facilities Council.

The W. M. Keck Observatory is operated as a scientific partnership among the California Institute of Technology, the University of California and the National Aeronautics and Space Administration. The Observatory was made possible by the generous financial support of the W. M. Keck Foundation. The authors wish to recognize and acknowledge the very significant cultural role and reverence that the summit of Maunakea has always had within the indigenous Hawaiian community.  We are most fortunate to have the opportunity to conduct observations from this mountain.

\textit{Software:} Global Relay of Observatories Watching Transients Happen Marshal \citep[GROWTH;][]{Kasliwal2019} and the Fritz SkyPortal Marshal  \citep[][]{Duev2019, SkyPortal2019, Coughlin2023}. Astropy \citep{Astropy-Collaboration13}, Matplotlib \citep{Hunter07}, george \citep{Ambikasaran2015}

\bibliography{bibfile}
\bibliographystyle{aasjournal}

\clearpage

\section*{Appendix}
\begin{figure*}
    \centering
    \includegraphics[width=0.333\textwidth]{peakr_vs_ek.pdf}\includegraphics[width=0.33\textwidth]{peakr_vs_ekej.pdf}\includegraphics[width=0.33\textwidth]{peakr_vs_mej.pdf}
    
    \includegraphics[width=0.33\textwidth]{peakr_vs_ni.pdf}\includegraphics[width=0.33\textwidth]{peakr_vs_rad.pdf}\includegraphics[width=0.33\textwidth]{ej_vs_ek.pdf}
    
    \includegraphics[width=0.33\textwidth]{ni_vs_ek.pdf}\includegraphics[width=0.33\textwidth]{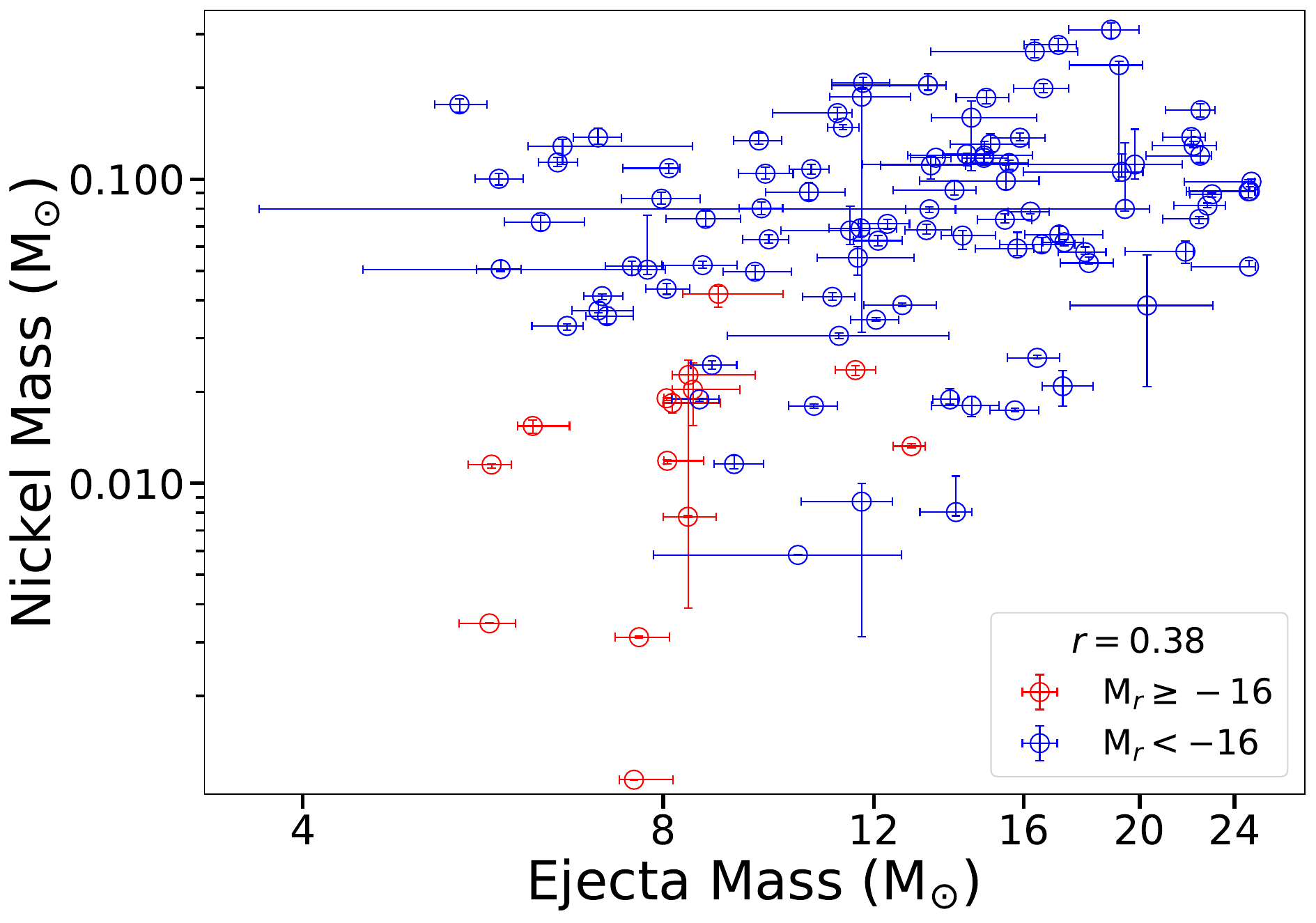}\includegraphics[width=0.33\textwidth]{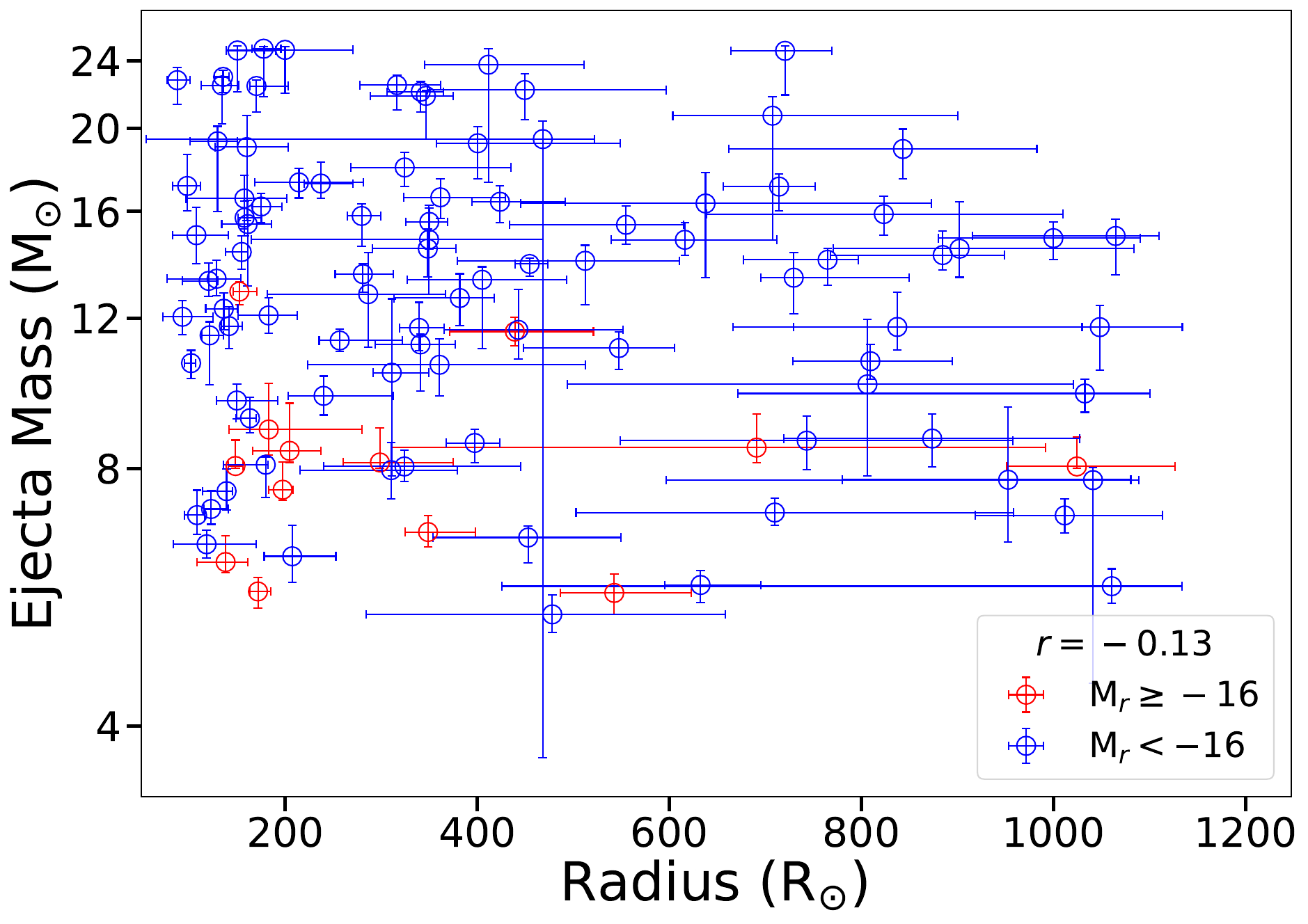}
    
    \includegraphics[width=0.33\textwidth]{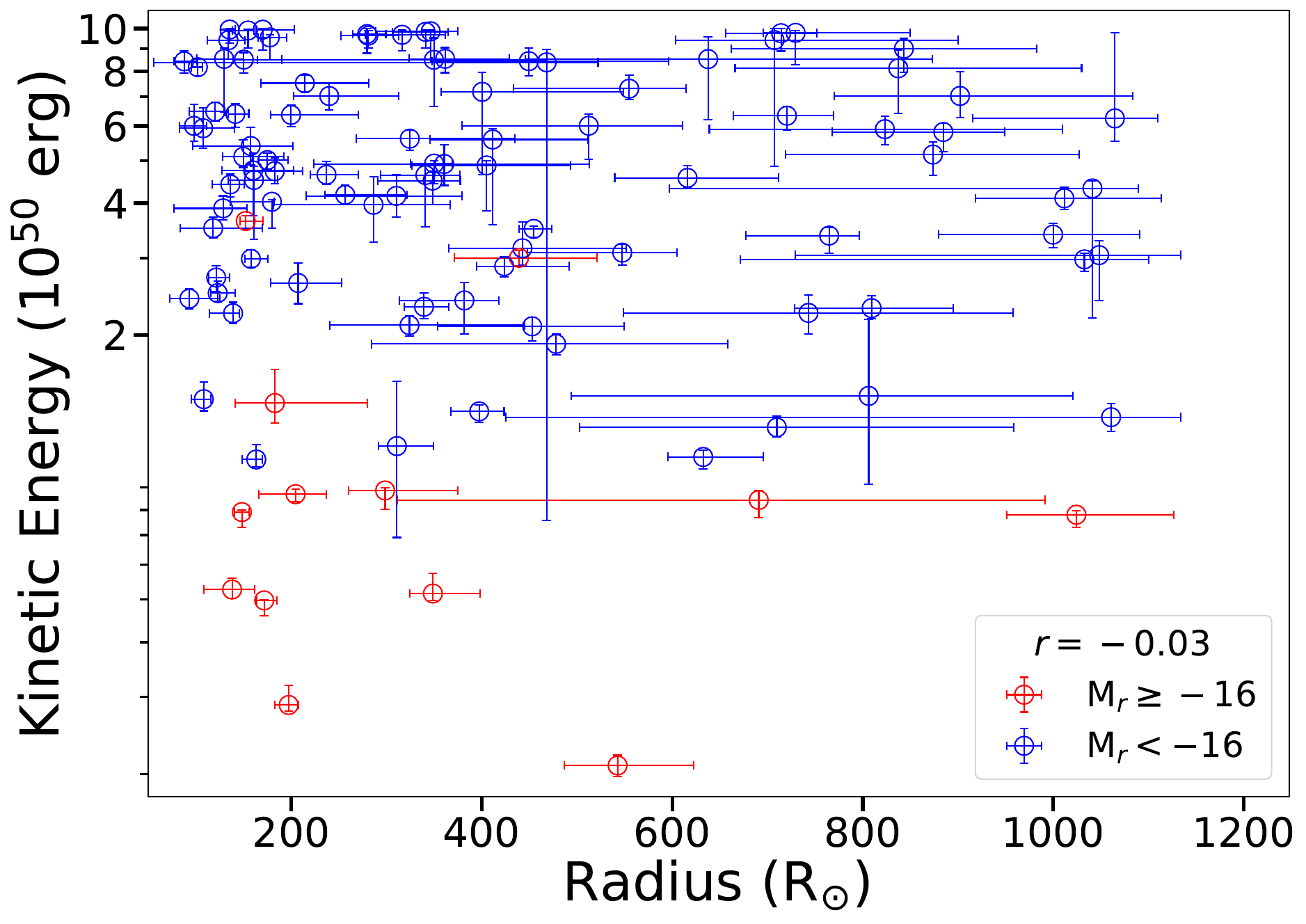}\includegraphics[width=0.33\textwidth]{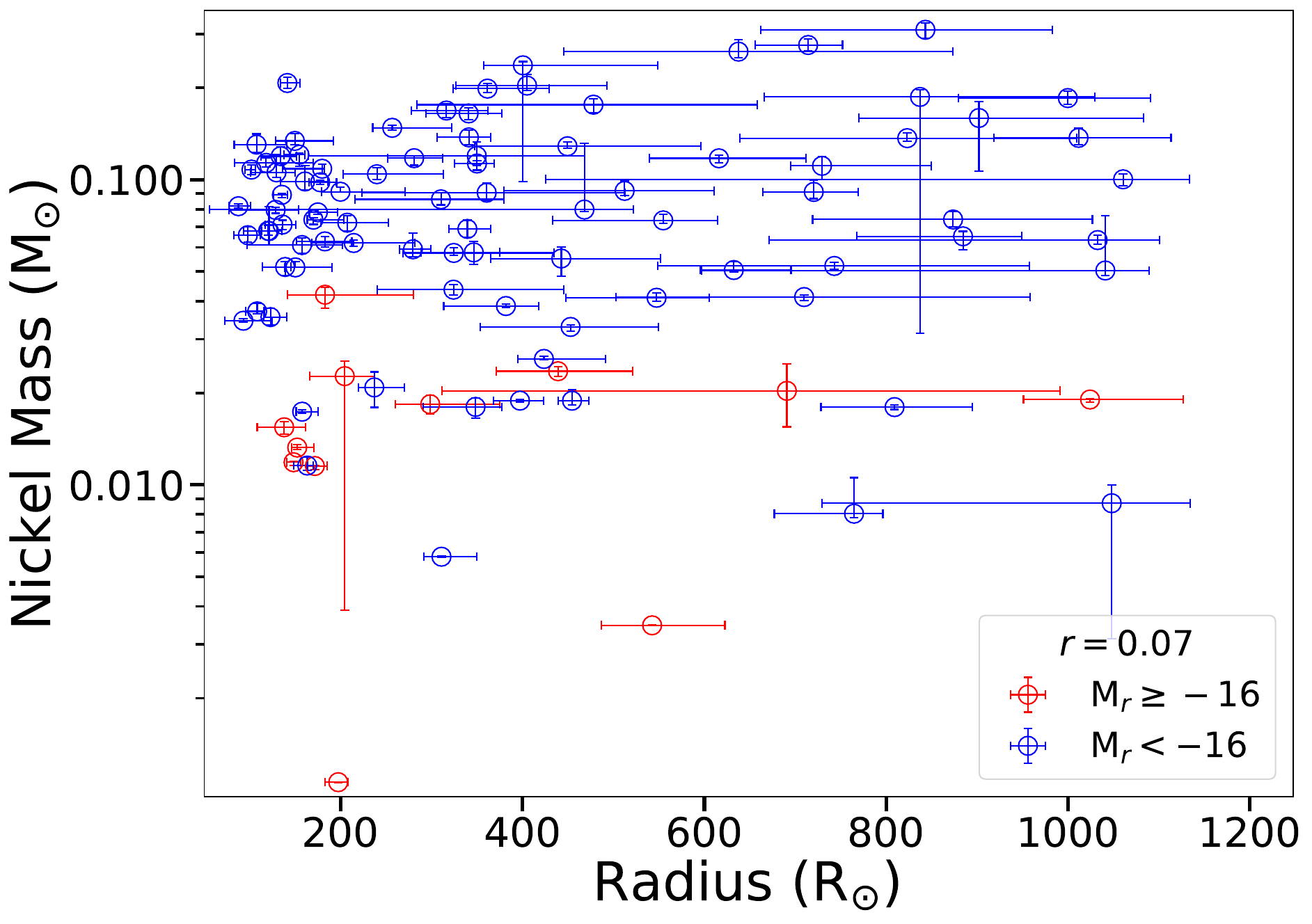}
    
    \caption{All correlations between peak $r$-band magnitude and physical parameters (nickel mass, explosion energy, energy per unit mass, radius, ejecta mass), and among the physical parameters themselves, based on semi-analytical models from \citet{Nagy2016}. LLIIP SNe with $M_r \geq -16$ are shown in red, SNe IIP with $M_r < -16$ in blue.}
    \label{fig:semicorrelationsfull}
\end{figure*}

\begin{figure*}
    \centering
    \includegraphics[width=0.333\textwidth]{moriya_Ni_mass_50_vs_peakr.pdf}\includegraphics[width=0.33\textwidth]{moriya_Energy_50_vs_peakr.pdf}\includegraphics[width=0.33\textwidth]{moriya_ZAMS_50_vs_peakr.pdf}
    
    \includegraphics[width=0.33\textwidth]{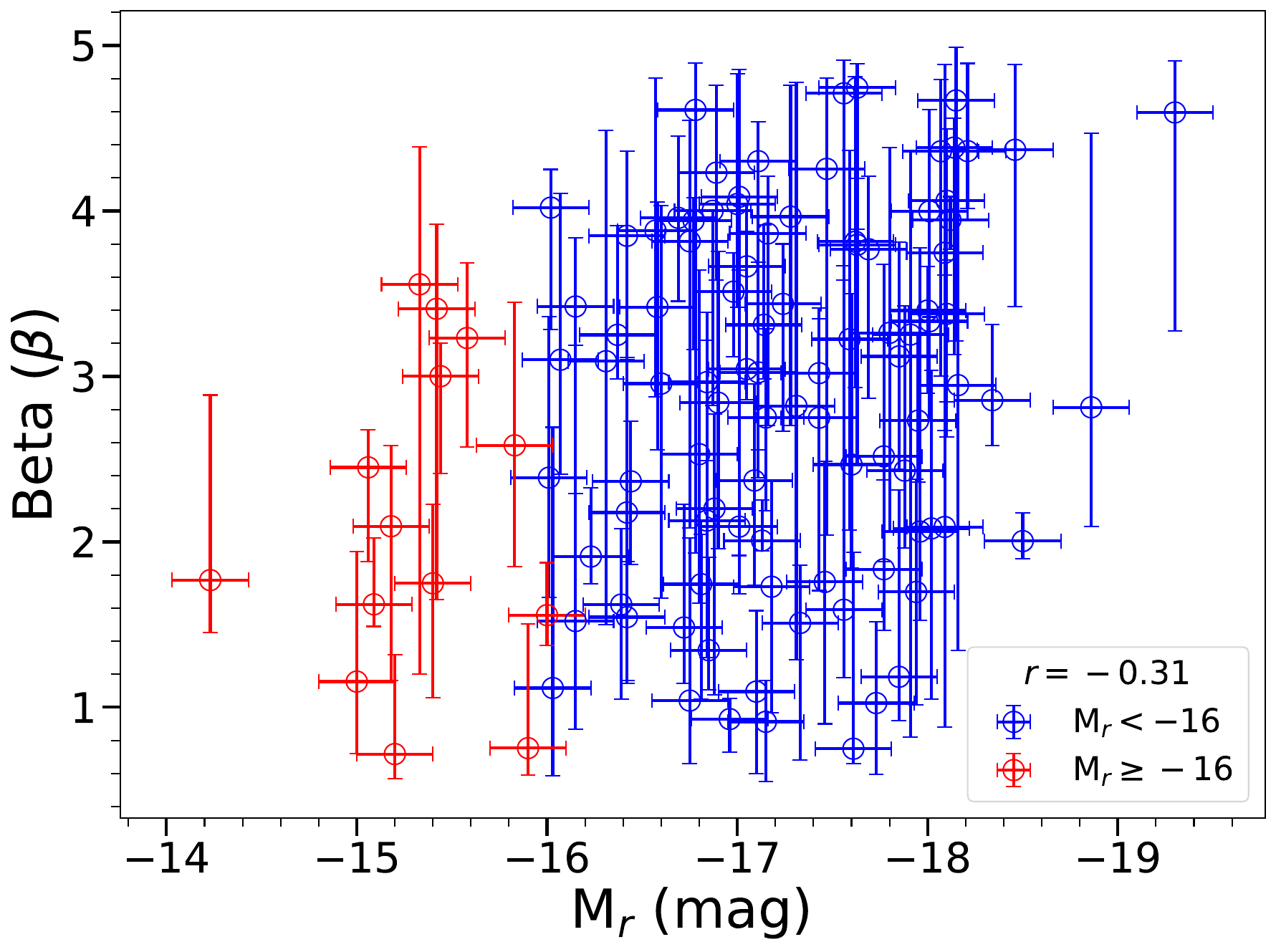}\includegraphics[width=0.33\textwidth]{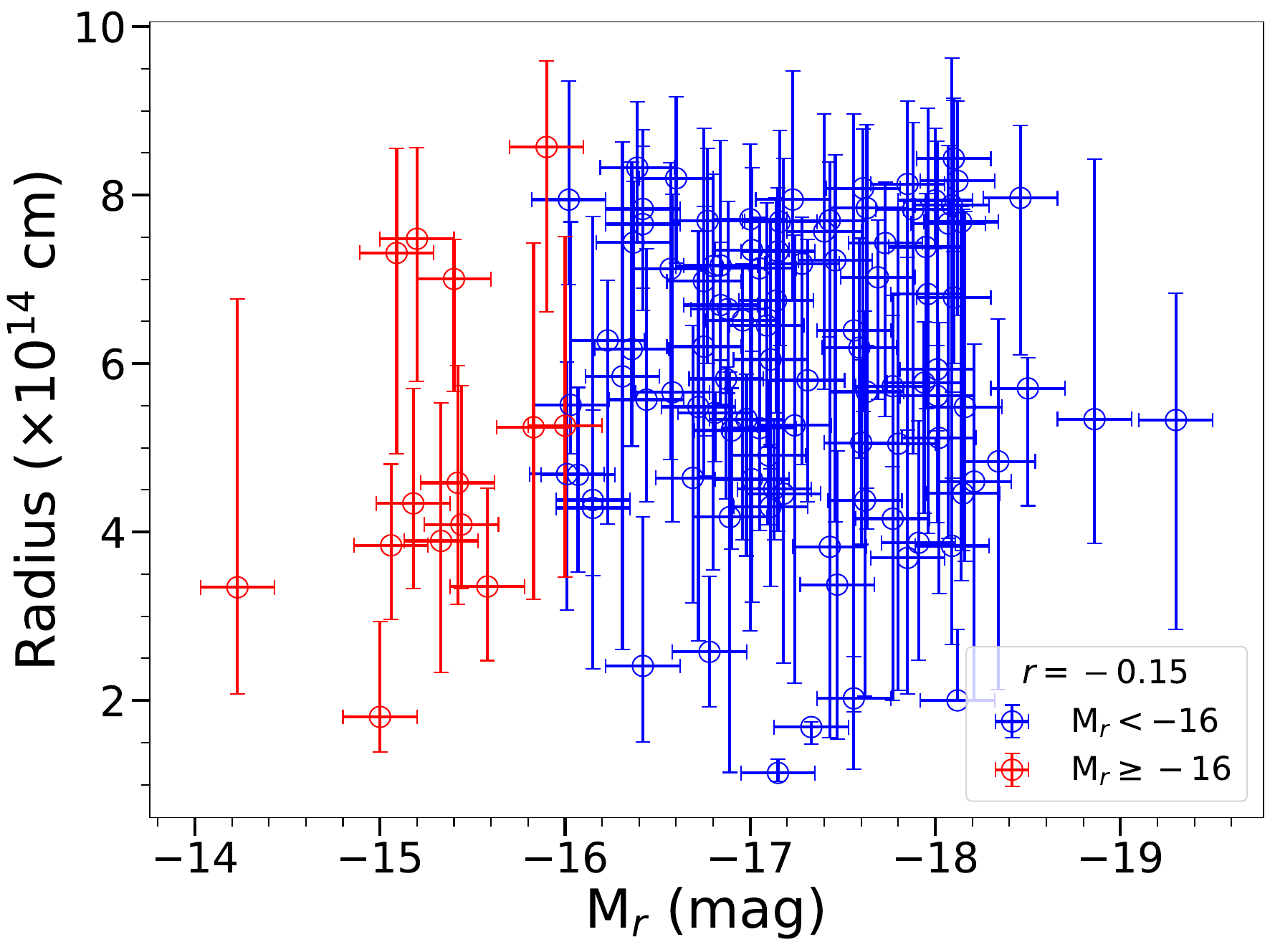}\includegraphics[width=0.33\textwidth]{moriya_Mrate_50_vs_peakr.pdf}

    \includegraphics[width=0.33\textwidth]{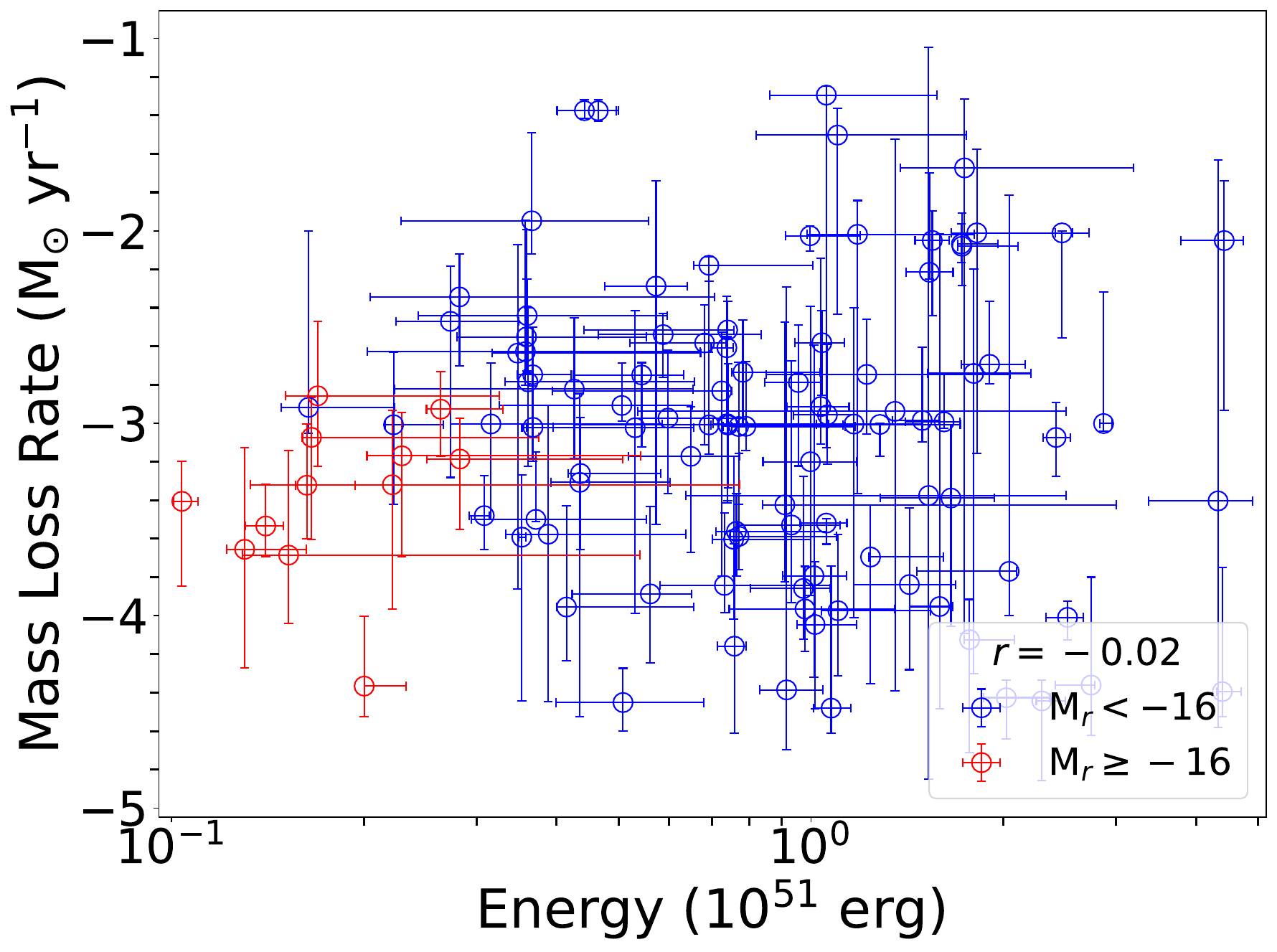}\includegraphics[width=0.33\textwidth]{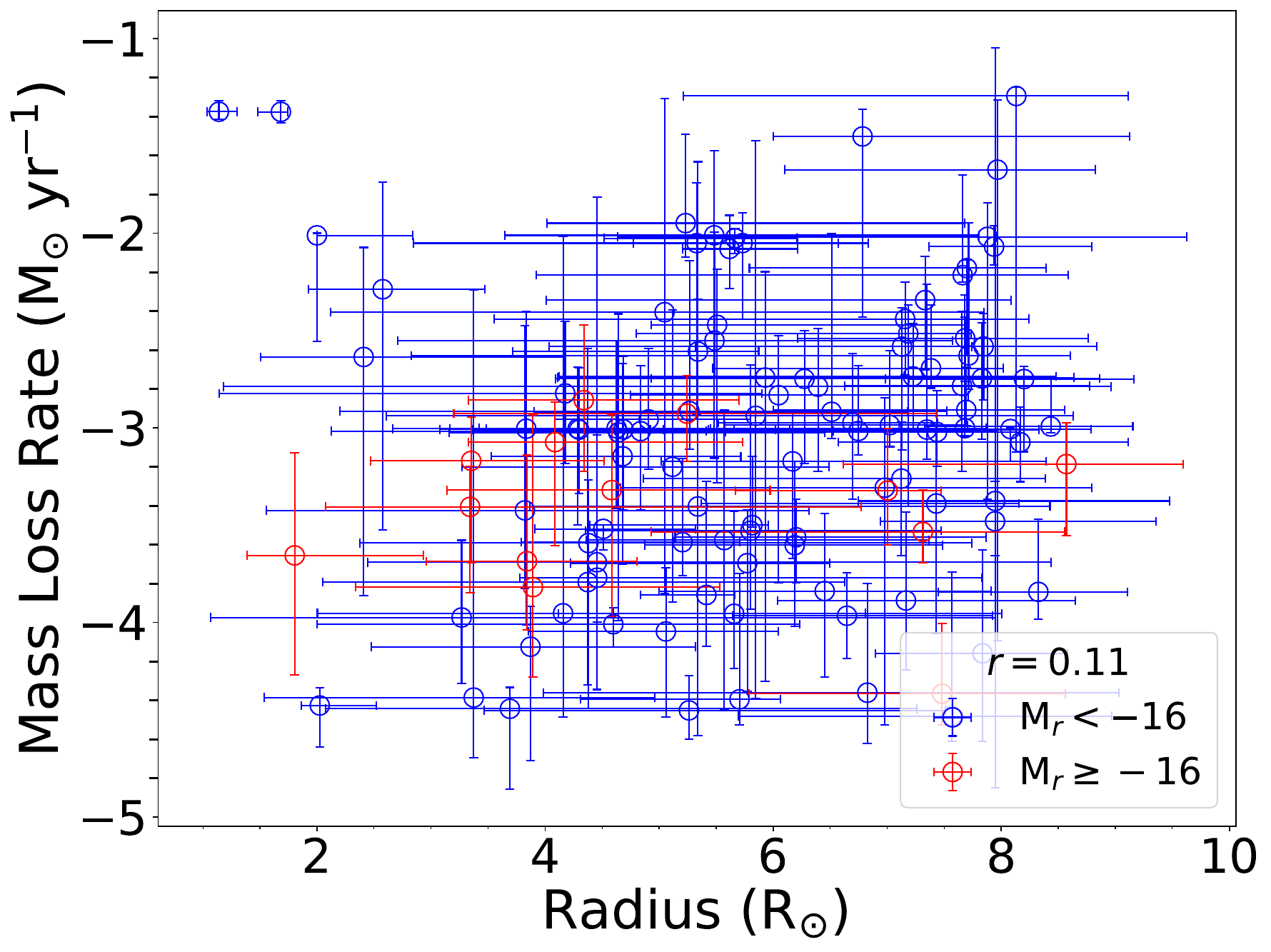}\includegraphics[width=0.33\textwidth]{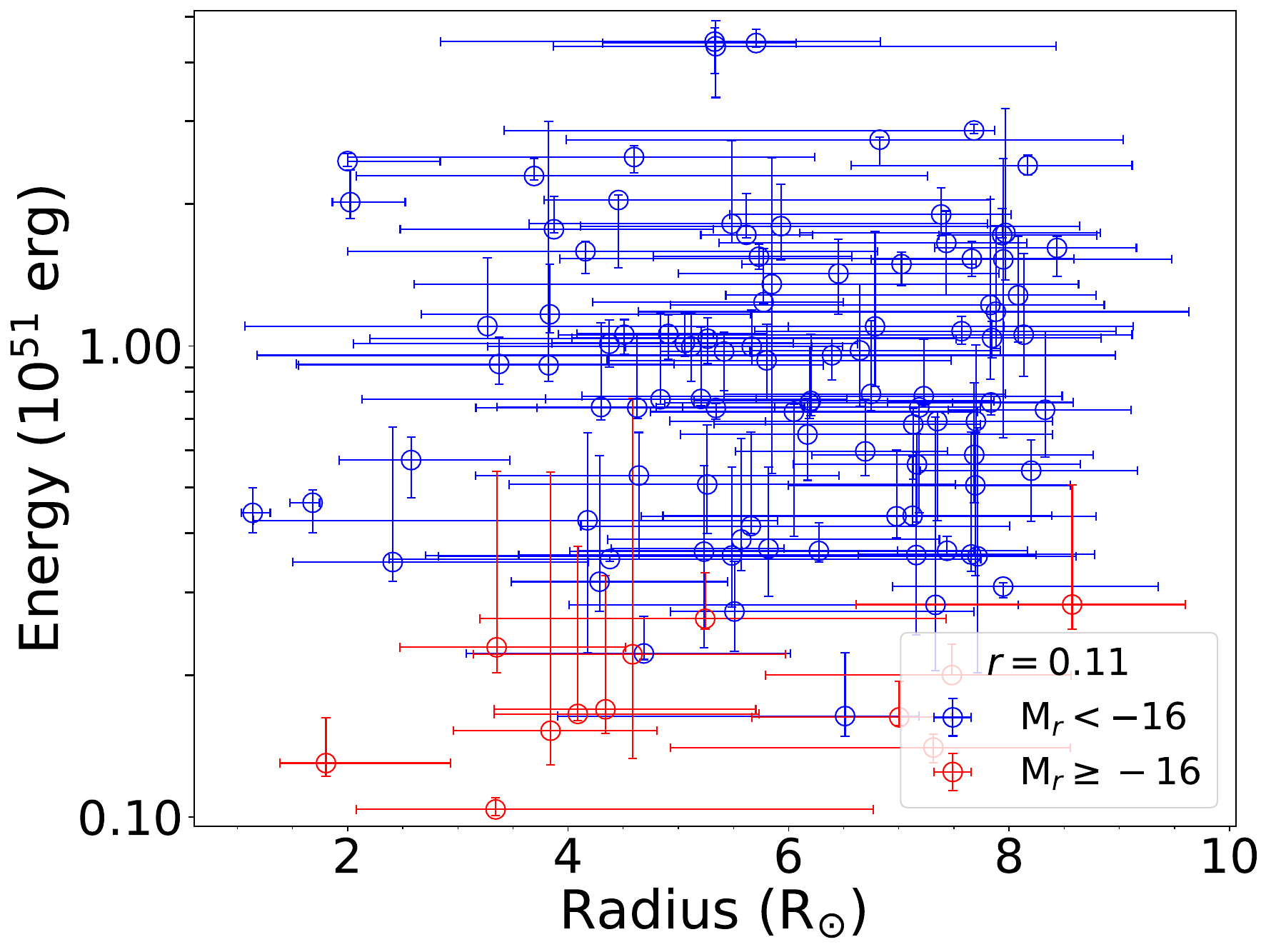}
    
    \includegraphics[width=0.33\textwidth]{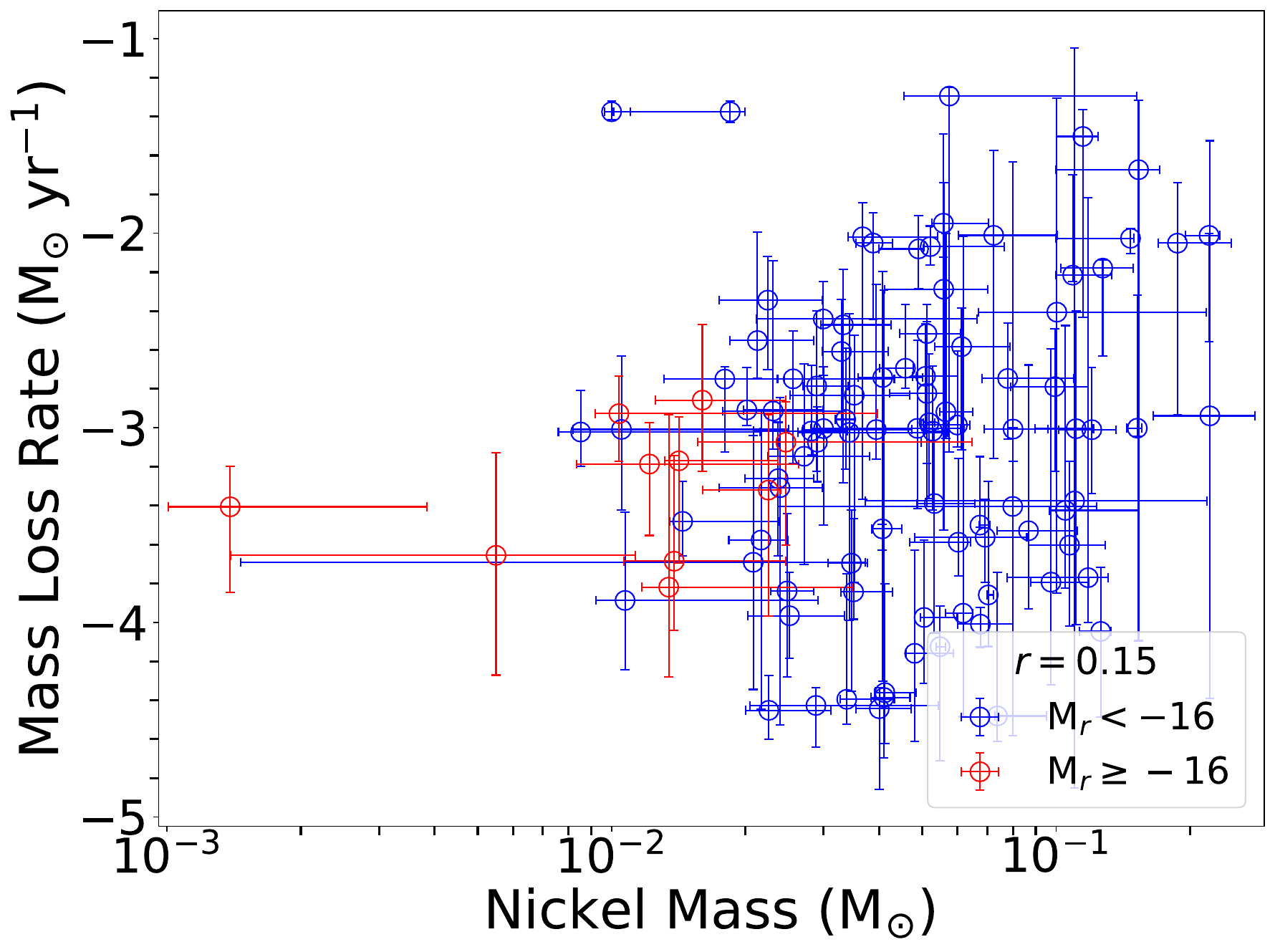}\includegraphics[width=0.33\textwidth]{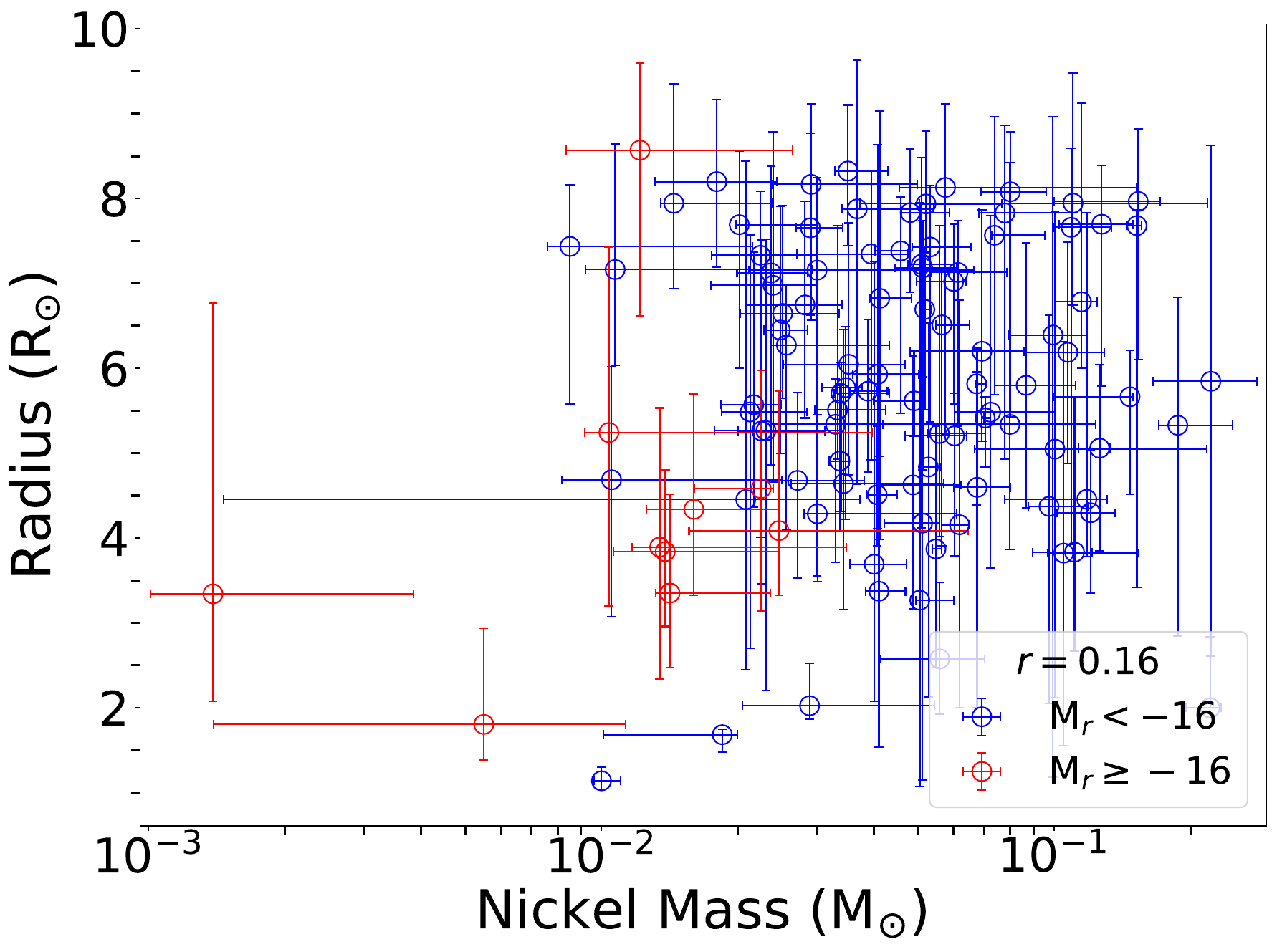}\includegraphics[width=0.33\textwidth]{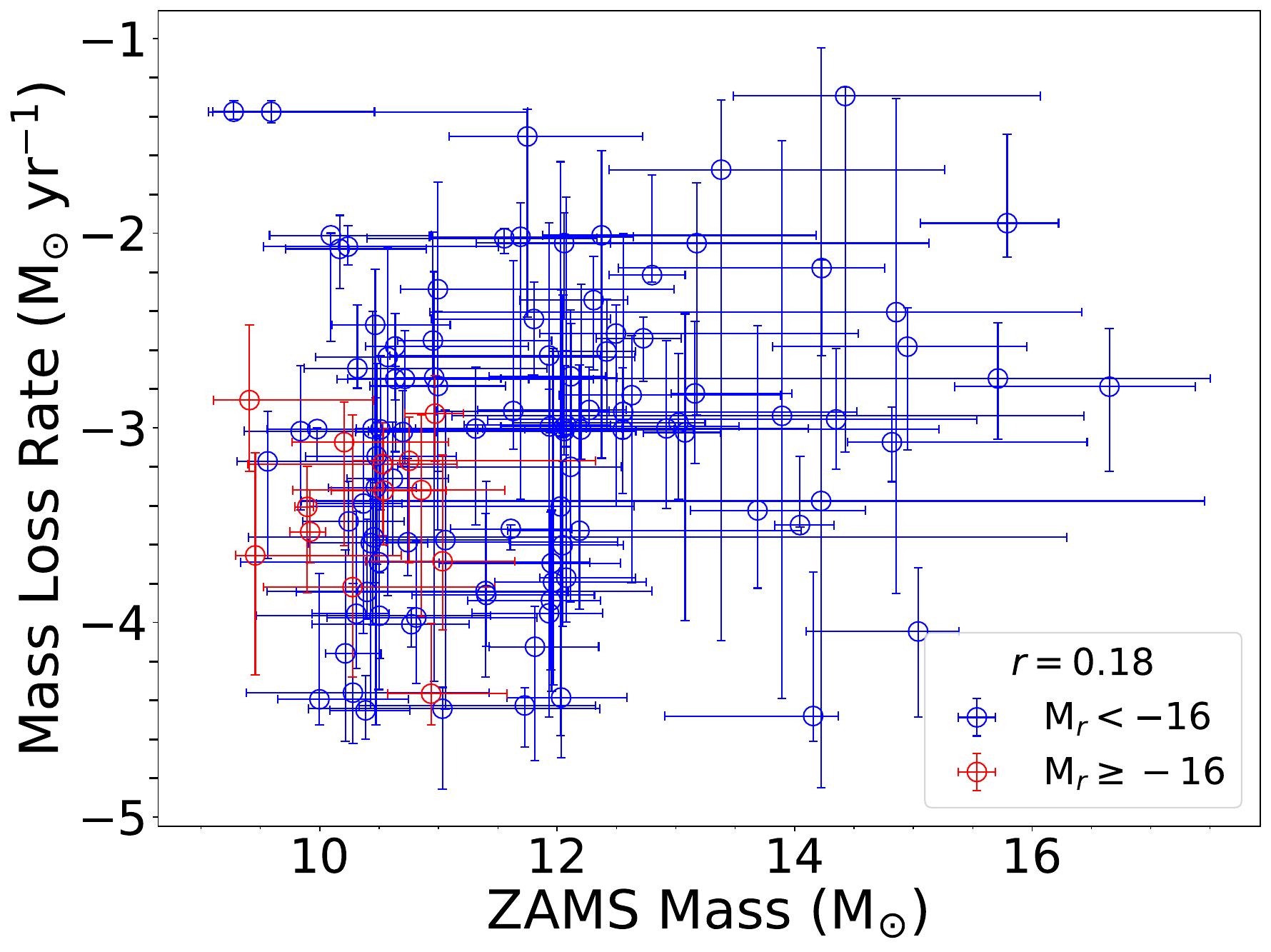}
    
    \includegraphics[width=0.33\textwidth]{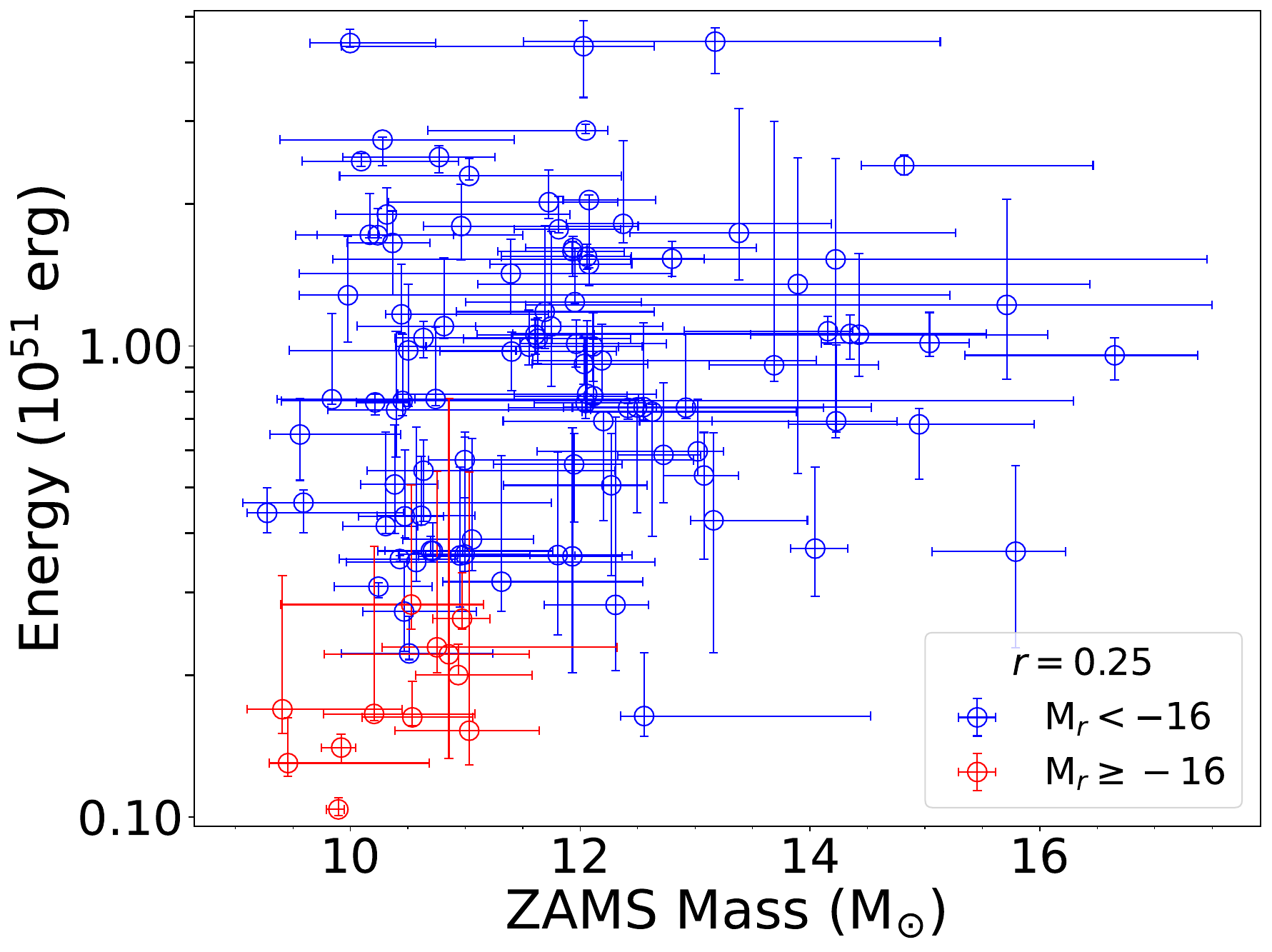}\includegraphics[width=0.33\textwidth]{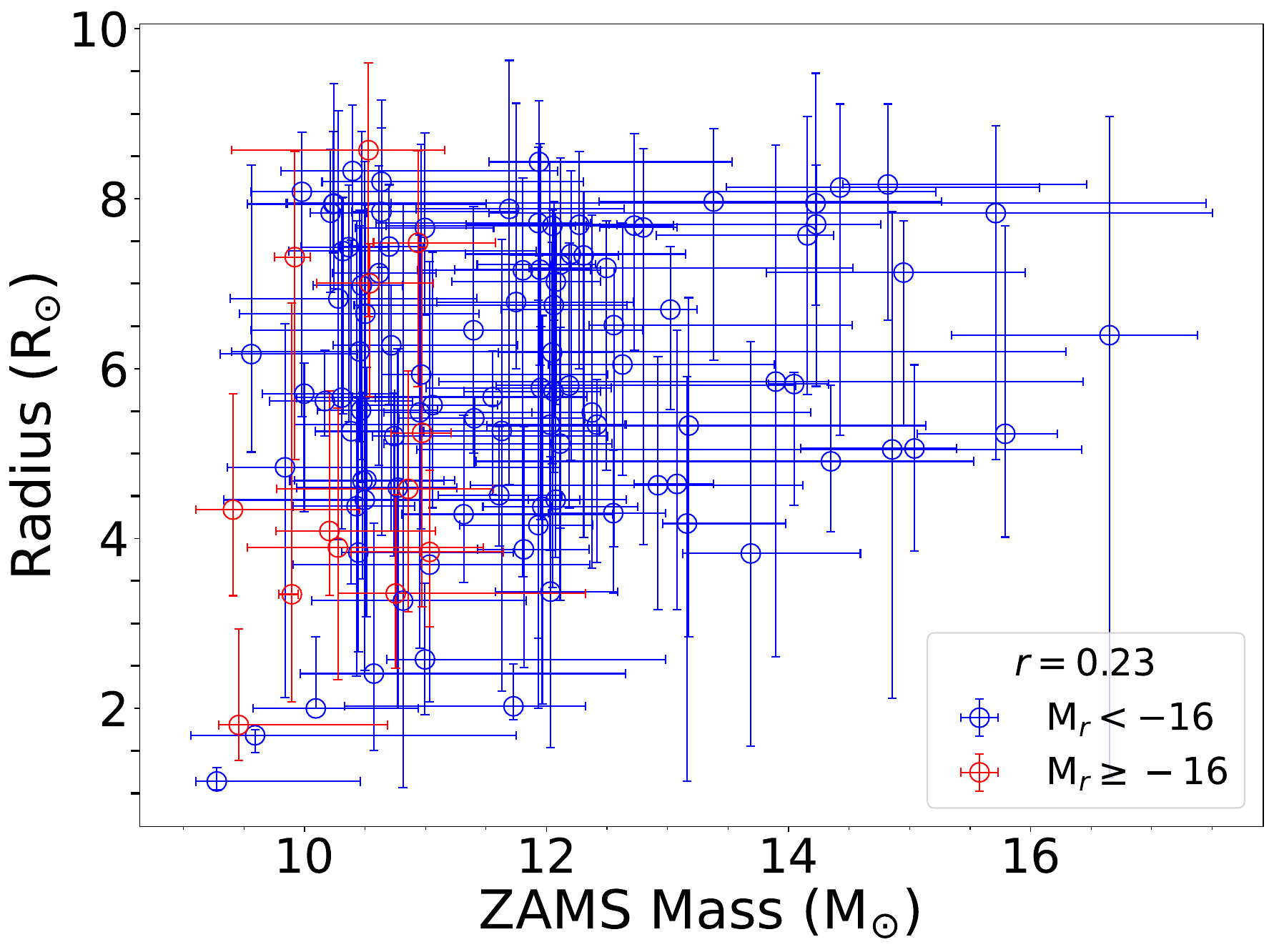}\includegraphics[width=0.33\textwidth]{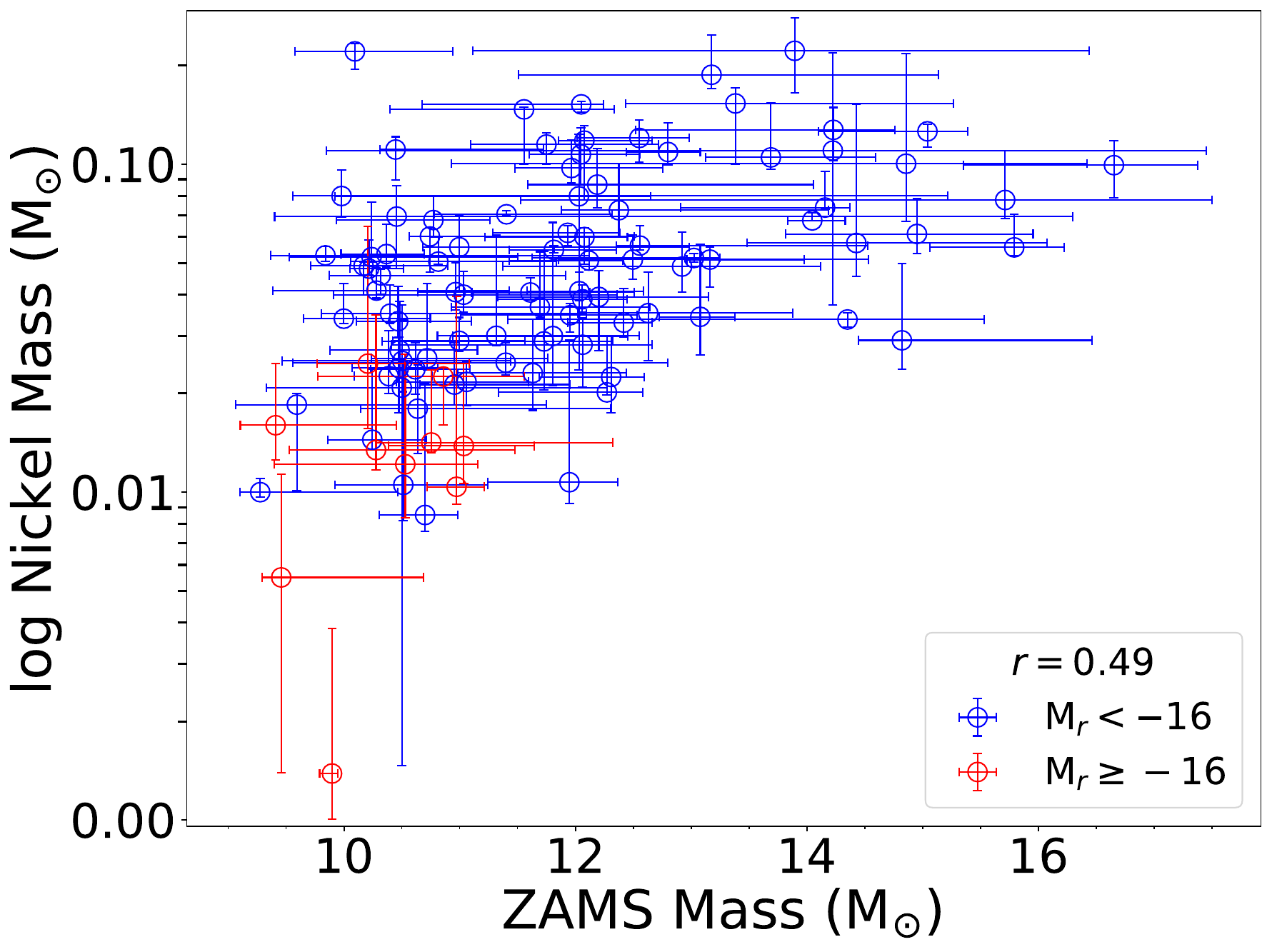}
    
    \caption{Correlations between peak $r$-band magnitude and physical parameters (nickel mass, explosion energy, energy per unit mass, radius, ejecta mass), and among the physical parameters themselves, based on radiation-hydrodynamical model fits from \citet{Moriya2023}. LLIIP SNe with $M_r \geq -16$ are shown in red, those with $M_r < -16$ in blue.}    
\label{fig:moriyacorrelationsfull}
\end{figure*}

\begin{figure*}[ht]
    \centering
    \includegraphics[width=\textwidth]{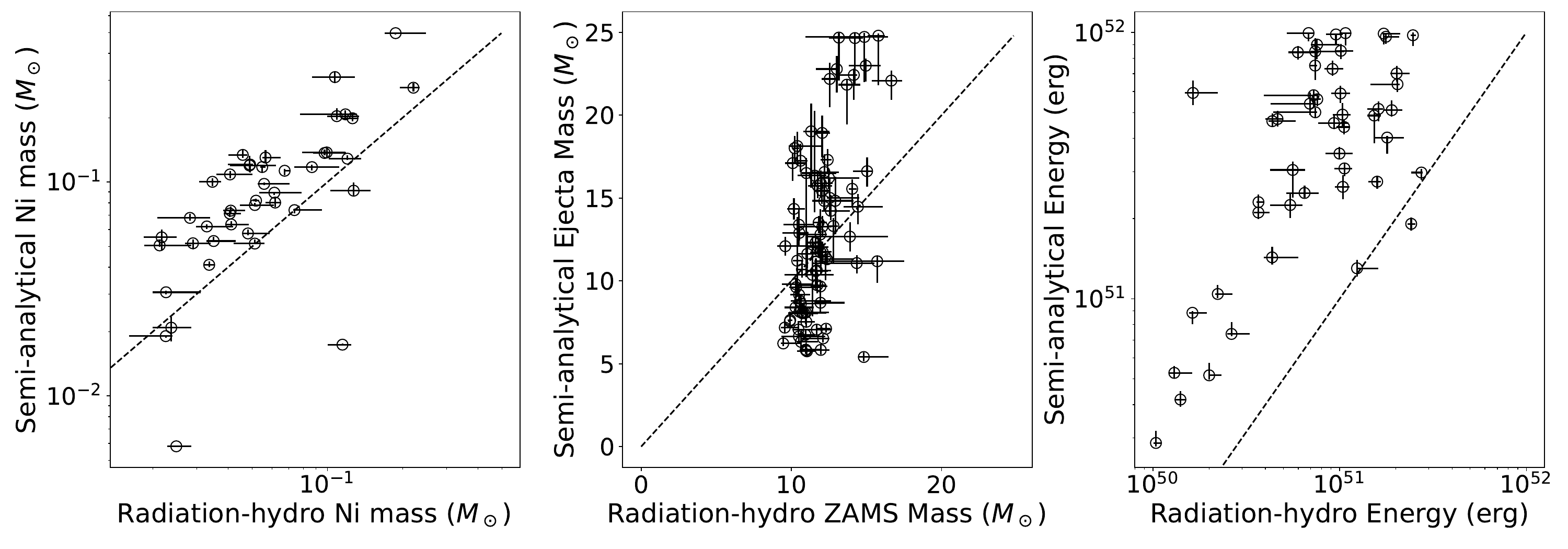}
    \caption{Comparison of physical parameters derived from semi-analytical light curve fits (y-axis) and radiation-hydrodynamical model fits (x-axis). Panels show nickel mass (left), ejecta mass versus ZAMS mass (center), and kinetic energy (right). Dashed lines indicate the one-to-one relation. Error bars reflect the 16th--84th percentile uncertainties from each method.}
    \label{fig:comparemodels}
\end{figure*}

Table~\ref{tab:lightcurveparams} lists the observable lightcurve parameters such as rise time, plateau duration, and decline rate for the entire sample. Tables~\ref{tab:semipriors} and \ref{tab:moriyapriors} summarize the priors used in the semi-analytical and radiation-hydrodynamical model fits, respectively. Tables~\ref{tab:nagyfit} and \ref{moriyafit} present the full posterior estimates of explosion and progenitor properties from the two modeling approaches. Table~\ref{table_vel} provides the spectral observation log and H$\alpha$ velocity measurements. Figure~\ref{fig:semicorrelationsfull} shows all the parameter correlations based on semi-analytical fits, while Figure~\ref{fig:moriyacorrelationsfull} shows the same using radiation-hydrodynamical models. Figure~\ref{fig:comparemodels} compares results between the two modeling methods. 

\FloatBarrier

\renewcommand{\thetable}{\Alph{table}}
\setcounter{table}{0}

\begin{table*}[ht]
\centering
\small
\caption{Light curve parameters for SN~II, sorted by faintest peak $r$-band absolute magnitude ($M_r$). ZTF and IAU names are shown in separate columns. The table includes rise time, plateau duration (Plat. dur.), optically thick duration (OPTd), decline rate during plateau (Slope, in $\times$0.01 mag day$^{-1}$), and the plateau-to-tail drop ($\Delta$mag). The full machine-readable table is available on \href{https://zenodo.org/records/15717884?token=eyJhbGciOiJIUzUxMiJ9.eyJpZCI6IjNhYjE3NjYwLTU2NDEtNDBkZi1iYmI5LTQ1YzQxY2EwYjllNyIsImRhdGEiOnt9LCJyYW5kb20iOiJhNjA2ZTkzNDFjYjU5NGM3ZGIzMmExMTRlOGY1NzU5MCJ9.X0jCmJz6FJbXwmcY_6ZZ1kYCcRHwrABpoZ2epNXLlpIjVIwrFkEEOGD0Z7Cfk7luRPlHvOCwJdWBwB40vX2JoQ}{\texttt{Zenodo}}.}
\label{tab:lightcurveparams}
\begin{tabular}{llrllllrr}
\hline
ZTF & IAU & $M_r$ & Rise & Plat. dur. & OPTd & Slope & $\Delta$mag & Limit? \\
 & & (mag) & (days) & (days) & (days) & ($\times$0.01 mag day$^{-1}$) & (mag) & \\
\hline
ZTF24abtczty & SN~2024abfl & -14.1 & 12.2 ± 0.4 &  99.1 ± 1.5 &  111.7 ± 1.5 & -0.04 ± 0.02 & -2.05 &  no \\
ZTF20aapchqy &  SN~2020cxd & -14.2 &        $-$ & 108.1 ± 1.6 &  112.9 ± 1.6 & -0.42 ± 0.02 & -2.48 &  no \\
ZTF24abmkros &  SN~2024xkd & -14.9 &        $-$ &  90.5 ± 1.3 &   97.1 ± 1.3 & -0.11 ± 0.02 & -1.17 & yes \\
ZTF23aackjhs &   SN~2023bvj & -15.0 &        $-$ &  97.0 ± 2.3 &   99.5 ± 2.3 &  0.14 ± 0.02 &  0.22 & yes \\
ZTF22aakdbia &  SN~2022jzc & -15.1 &  9.7 ± 0.3 &  92.8 ± 5.5 &  111.0 ± 5.5 & -0.18 ± 0.02 & -2.28 &  no \\
ZTF22abtjefa & SN~2022aaad & -15.1 & 16.0 ± 0.3 & 66.5 ± 15.0 &  85.8 ± 15.0 & -0.02 ± 0.03 & -1.38 &  no \\
ZTF22aazmrpx &  SN~2022raj & -15.2 &        $-$ &  80.0 ± 4.2 &   84.7 ± 4.2 &  0.45 ± 0.03 & -0.87 & yes \\
ZTF20acuhren & SN~2020abcq & -15.3 &        $-$ & 101.2 ± 2.8 &  103.2 ± 2.8 & -0.01 ± 0.02 & -2.05 & yes \\
ZTF21aaobkmg &  SN~2021eui & -15.4 &        $-$ &  85.0 ± 3.1 &   88.0 ± 3.1 &  0.09 ± 0.02 & -1.43 & yes \\
ZTF22aaywnyg &  SN~2022pru & -15.4 &        $-$ &  86.1 ± 1.6 &  102.1 ± 1.6 &  0.46 ± 0.02 & -1.09 &  no \\
ZTF24aabppgn &   SN~2024wp & -15.5 &        $-$ &  95.3 ± 2.5 &  109.4 ± 2.4 &   0.1 ± 0.02 & -1.35 & yes \\
ZTF23abnogui &  SN~2023wcr & -15.6 &        $-$ &  87.3 ± 4.3 &  101.1 ± 4.3 &  0.58 ± 0.04 & -1.53 &  no \\
ZTF24aaucrua &  SN~2024nez & -15.8 &        $-$ &  90.2 ± 1.5 &  105.4 ± 1.5 &  0.18 ± 0.02 & -1.58 & yes \\
ZTF22abyivoq & SN~2022acko & -15.8 &        $-$ &  69.8 ± 6.4 &   84.7 ± 6.4 &  0.23 ± 0.04 &  0.02 & yes \\
ZTF20abeohfn &  SN~2020mjm & -15.9 & 10.7 ± 0.1 &  64.3 ± 7.6 &   76.5 ± 7.6 &  0.05 ± 0.03 &  0.05 & yes \\
ZTF22abssiet &  SN~2022zmb & -15.9 &        $-$ & 66.1 ± 11.3 &  76.0 ± 11.3 &  0.18 ± 0.04 & -0.24 & yes \\
ZTF19aamwhat &  SN~2019bzd & -16.0 &        $-$ &  83.1 ± 1.5 &  103.1 ± 1.4 &  0.36 ± 0.02 & -2.00 & yes \\
ZTF22abkhrkd &  SN~2022wol & -16.0 &        $-$ &  79.1 ± 2.6 &   90.8 ± 2.6 &  0.08 ± 0.03 & -1.09 & yes \\
ZTF24aaejecr &  SN~2024btj & -16.0 &        $-$ &  78.3 ± 2.3 &   89.6 ± 2.3 &  0.55 ± 0.03 & -1.66 &  no \\
ZTF24aaplfjd &  SN~2024jxm & -16.0 &        $-$ &  74.6 ± 4.6 &   97.6 ± 4.6 &  0.24 ± 0.03 & -1.30 &  no \\
ZTF19abwztsb &  SN~2019pjs & -16.2 &        $-$ & 68.1 ± 37.2 &  80.0 ± 37.2 &   0.37 ± 0.2 & -1.36 &  no \\
ZTF24aabsmvc &   SN~2024ws & -16.2 &        $-$ & 90.0 ± 23.0 & 100.4 ± 23.0 &  0.26 ± 0.07 & -1.25 & yes \\
ZTF22abyohff & SN~2022acrl & -16.3 &        $-$ &  78.7 ± 6.5 &   88.7 ± 6.4 &  0.33 ± 0.04 & -0.52 & yes \\
ZTF21acgrrnl & SN~2021aayf & -16.3 &        $-$ & 77.0 ± 14.8 &  90.5 ± 14.8 & -0.14 ± 0.04 & -1.44 & yes \\
ZTF21aaeqwov &  AT~2021htp & -16.4 & 13.2 ± 0.6 &  92.4 ± 1.1 &  106.1 ± 1.1 &  0.27 ± 0.02 & -1.30 &  no \\
ZTF23abaxtlq &  SN~2023rix & -16.4 &        $-$ &  75.2 ± 8.3 &   87.3 ± 8.3 &   0.92 ± 0.1 & -2.13 &  no \\
ZTF21abvcxel &  SN~2021wvw & -16.4 &        $-$ &  67.4 ± 1.2 &   69.4 ± 1.2 &  0.47 ± 0.03 & -1.67 & yes \\
ZTF21aantsla &  SN~2021ech & -16.4 &        $-$ &  75.3 ± 8.0 &   86.5 ± 8.0 &  0.25 ± 0.04 & -1.46 & yes \\
ZTF21aafepon &  SN~2021ass & -16.4 &        $-$ &  51.4 ± 9.0 &   67.9 ± 9.0 &  0.16 ± 0.05 & -0.09 & yes \\
ZTF21acpqqgu & SN~2021aewn & -16.4 &        $-$ &  76.3 ± 9.7 &   78.3 ± 9.7 &  0.35 ± 0.05 & -0.77 & yes \\
ZTF22abnujbv &  SN~2022xus & -16.4 &        $-$ & 53.2 ± 20.7 &  62.4 ± 20.7 &  0.58 ± 0.23 & -1.83 &  no \\
ZTF20acmaaan &  SN~2020xyk & -16.4 &        $-$ &  69.4 ± 4.2 &   75.8 ± 4.2 &  0.14 ± 0.03 & -1.96 & yes \\
ZTF22abfxkdm &  SN~2022ubb & -16.6 &        $-$ &  87.3 ± 3.9 &   96.2 ± 3.9 &  0.49 ± 0.03 & -1.66 & yes \\
ZTF22aavbfhz &  AT~2022phi & -16.6 &        $-$ &  78.3 ± 2.0 &   81.3 ± 2.0 &  0.63 ± 0.03 & -1.02 & yes \\
ZTF22abkbjsb &  SN~2022vym & -16.7 &        $-$ &  60.0 ± 9.5 &   66.5 ± 9.5 &   0.62 ± 0.1 & -1.33 &  no \\
ZTF24aabpzuz &   SN~2024vs & -16.7 &        $-$ &  62.0 ± 4.7 &   78.9 ± 4.7 & -0.13 ± 0.03 & -1.33 & yes \\
ZTF21aanzcuj &  SN~2021enz & -16.7 & 18.0 ± 0.8 &  72.1 ± 1.9 &   90.6 ± 1.8 &  0.01 ± 0.03 & -1.54 &  no \\
ZTF24aafqzur &  SN~2024daa & -16.7 &        $-$ &  67.6 ± 1.2 &   86.6 ± 1.2 &  0.15 ± 0.03 & -0.94 & yes \\
ZTF21aagtqna &  SN~2021brb & -16.8 & 21.2 ± 1.4 &  73.6 ± 3.7 &  104.8 ± 3.7 &  0.12 ± 0.03 & -1.26 &  no \\
ZTF18aaszvfn &  SN~2021iaw & -16.8 &        $-$ &  76.8 ± 1.7 &   86.4 ± 1.7 &  0.39 ± 0.03 & -1.23 & yes \\
ZTF19acftfav &  SN~2019ssi & -16.8 & 14.2 ± 0.4 &  61.7 ± 8.0 &   76.4 ± 8.0 &   0.71 ± 0.1 & -1.70 & yes \\
ZTF23aasbvab &  SN~2023ngy & -16.8 &        $-$ &  92.7 ± 1.1 &  102.9 ± 1.1 &  0.83 ± 0.02 & -1.32 & yes \\
ZTF23aasrcyv &  SN~2023nlu & -16.8 & 18.3 ± 1.2 &  71.8 ± 1.8 &   94.0 ± 1.8 &   0.5 ± 0.03 & -0.24 & yes \\
ZTF22abzqwmp & SN~2022adth & -16.8 &        $-$ & 61.0 ± 11.1 &  73.6 ± 11.1 &  0.69 ± 0.13 & -0.81 & yes \\
ZTF22aasojye &  SN~2022omr & -16.9 &        $-$ &  98.1 ± 4.2 &  103.6 ± 4.2 &  0.23 ± 0.02 & -1.76 &  no \\
\hline
\end{tabular}
\end{table*}

\begin{table*}[ht]
\ContinuedFloat
\centering
\small
\caption{Continued.}
\begin{tabular}{llrllllrr}
\hline
ZTF & IAU & $M_r$ & Rise & Plat. dur. & OPTd & Slope & $\Delta$mag & Limit? \\
 & & (mag) & (days) & (days) & (days) & ($\times$0.01 mag day$^{-1}$) & (mag) & \\
\hline
ZTF19aadnxnl &   SN~2019va & -16.9 &        $-$ &  103.2 ± 3.8 &  107.2 ± 3.8 & -0.09 ± 0.02 & -0.80 &  no \\
ZTF19actnyae &  SN~2019vdm & -16.9 &        $-$ &   52.9 ± 1.4 &   61.4 ± 1.4 & -0.02 ± 0.04 & -1.33 &  no \\
ZTF23aaxadel &  SN~2023pbg & -16.9 &        $-$ &   89.8 ± 6.1 &  104.6 ± 6.1 &  0.31 ± 0.03 & -0.82 & yes \\
ZTF21abrluay &  SN~2021vfh & -17.0 &        $-$ &   75.4 ± 8.8 &   88.0 ± 8.8 &  0.08 ± 0.03 & -1.28 & yes \\
ZTF22aafsqud &  SN~2022hql & -17.0 &        $-$ &   61.2 ± 1.6 &   71.7 ± 1.6 &  0.62 ± 0.04 & -1.25 & yes \\
ZTF22abhsxph &  SN~2022vyc & -17.0 &        $-$ &   80.3 ± 1.4 &   94.6 ± 1.4 &  0.51 ± 0.03 & -1.42 & yes \\
ZTF19aazudta &  SN~2019hqm & -17.0 &        $-$ &  73.7 ± 16.5 &  90.0 ± 16.5 &  0.18 ± 0.05 & -1.52 & yes \\
ZTF22abfwxtr &  SN~2022udq & -17.0 &        $-$ &   99.3 ± 2.3 &  111.4 ± 2.3 &  0.41 ± 0.02 & -0.73 & yes \\
ZTF21aafkwtk &  SN~2021apg & -17.0 &        $-$ &   69.3 ± 6.0 &   87.0 ± 6.0 &  0.33 ± 0.04 & -1.65 &  no \\
ZTF19aapafit &  SN~2019cvz & -17.0 &        $-$ &   82.8 ± 2.2 &   98.3 ± 2.2 &  0.12 ± 0.02 & -1.30 &  no \\
ZTF23abbtkrv &  SN~2023rvo & -17.0 & 16.7 ± 1.1 &   90.7 ± 2.2 &  157.1 ± 2.2 & -0.04 ± 0.02 & -0.81 & yes \\
ZTF19acewuwn &  SN~2019ssl & -17.0 &        $-$ &   62.8 ± 8.0 &   83.8 ± 8.0 &  0.19 ± 0.04 & -1.09 & yes \\
ZTF19aailepg &  SN~2019amt & -17.1 &        $-$ & 100.9 ± 10.3 & 109.0 ± 10.3 &   0.3 ± 0.04 & -1.04 & yes \\
ZTF19aazyvub &  SN~2019hnl & -17.1 &        $-$ &   74.6 ± 5.1 &   80.6 ± 5.0 &  0.39 ± 0.04 & -1.44 &  no \\
ZTF21aakvroo &  SN~2021cwe & -17.1 &        $-$ &   72.5 ± 4.2 &   76.5 ± 4.2 & -0.08 ± 0.03 & -0.93 &  no \\
ZTF22abtspsw & SN~2022aagp & -17.1 &  9.0 ± 0.0 &  111.5 ± 6.3 &  122.5 ± 6.3 &  0.99 ± 0.06 & -3.50 & yes \\
ZTF18aaxkqgy &  SN~2018ccb & -17.1 &        $-$ &  70.0 ± 10.6 &  74.2 ± 10.6 &  0.53 ± 0.09 & -1.22 &  no \\
ZTF22abfavpu &  SN~2022tmb & -17.1 &        $-$ &   66.6 ± 1.3 &   80.9 ± 1.3 &  0.77 ± 0.03 & -2.05 &  no \\
ZTF21aapliyn &  SN~2021foj & -17.1 &        $-$ &   91.3 ± 1.8 &   96.4 ± 1.8 &  0.15 ± 0.02 & -1.20 & yes \\
ZTF21acafqtj &  SN~2021yok & -17.1 &        $-$ &  59.9 ± 10.6 &  69.9 ± 10.6 &  0.37 ± 0.07 & -0.20 & yes \\
ZTF21abfiuqf &  SN~2021pla & -17.2 &        $-$ &   52.9 ± 1.2 &   60.6 ± 1.2 &  1.25 ± 0.05 & -1.51 &  no \\
ZTF19aanhhal &  SN~2019cec & -17.2 &        $-$ &   80.9 ± 2.8 &   90.2 ± 2.8 &  0.44 ± 0.03 & -0.96 & yes \\
ZTF22abyokkf & SN~2022acri & -17.2 &        $-$ &   91.7 ± 5.2 &  108.2 ± 5.2 &  0.08 ± 0.02 & -1.14 & yes \\
ZTF18aatyqds &  SN~2018btl & -17.2 & 10.8 ± 0.2 &   76.5 ± 4.4 &   88.8 ± 4.4 &   1.2 ± 0.07 & -1.39 &  no \\
ZTF22aapargp &  SN~2022niw & -17.2 &        $-$ &   54.0 ± 9.6 &   68.7 ± 9.6 &  0.46 ± 0.09 & -1.03 & yes \\
ZTF23aanymcl &  SN~2023kzz & -17.2 &        $-$ &   60.6 ± 2.1 &   64.1 ± 2.1 &  0.68 ± 0.04 & -1.23 & yes \\
ZTF21aavhnpk &  SN~2021jsf & -17.3 &        $-$ &   74.3 ± 9.1 &   78.8 ± 9.0 &   0.7 ± 0.09 & -1.34 & yes \\
ZTF19aaniore &  SN~2019ceg & -17.3 &        $-$ &   87.2 ± 3.0 &  102.0 ± 3.0 &  0.44 ± 0.03 & -1.46 &  no \\
ZTF21aabygea &   SN~2021os & -17.3 & 14.8 ± 0.4 &   83.3 ± 1.8 &  100.1 ± 1.8 &  0.58 ± 0.03 & -1.43 &  no \\
ZTF22absqhkw &  SN~2022zkc & -17.4 &        $-$ &   50.1 ± 4.6 &   52.1 ± 4.6 &  0.62 ± 0.07 & -1.48 & yes \\
ZTF23aaqknaw &  SN~2023lzn & -17.4 &        $-$ &  100.7 ± 1.7 &  104.7 ± 1.6 &  0.55 ± 0.02 & -2.03 & yes \\
ZTF21aaagypx &    SN~2021V & -17.4 &        $-$ &  137.2 ± 1.1 &  146.1 ± 1.1 &  0.56 ± 0.02 & -1.57 & yes \\
ZTF19abbnamr &  SN~2019iex & -17.4 &        $-$ &  94.1 ± 10.0 & 100.6 ± 10.0 &  0.17 ± 0.03 & -1.58 &  no \\
ZTF19aawgxdn &  SN~2019gmh & -17.4 &        $-$ &   95.9 ± 8.0 &  105.5 ± 8.0 &  0.34 ± 0.04 & -1.54 & yes \\
ZTF24aagupsf &  SN~2024egd & -17.5 &        $-$ &   71.0 ± 7.0 &   82.4 ± 7.0 &  0.62 ± 0.07 & -2.46 &  no \\
ZTF18abzrgim &  SN~2018gvt & -17.5 &        $-$ &  73.0 ± 24.1 &  90.4 ± 24.1 & -0.03 ± 0.03 & -0.31 & yes \\
ZTF19acytcsg &  SN~2019wvz & -17.6 &        $-$ &   77.4 ± 9.2 &   93.1 ± 9.2 &  0.25 ± 0.04 & -1.13 &  no \\
ZTF18aawpwlf &  SN~2020hvn & -17.6 &        $-$ &   54.1 ± 3.7 &   62.1 ± 3.7 &  1.76 ± 0.13 & -1.49 & yes \\
ZTF23abhzfww &  SN~2023twg & -17.6 &        $-$ &   90.0 ± 3.0 &   96.0 ± 3.0 &   0.6 ± 0.03 & -1.45 &  no \\
ZTF24aarvbxj &  SN~2024lby & -17.6 &        $-$ &   49.1 ± 1.4 &   60.4 ± 1.4 &  0.73 ± 0.05 & -1.23 &  no \\
ZTF19aclobbu &  SN~2019twk & -17.6 &        $-$ &  73.7 ± 10.3 &  84.5 ± 10.3 &  0.31 ± 0.05 & -0.52 & yes \\
ZTF19actnwtn &  SN~2019vdl & -17.6 &        $-$ &   84.1 ± 1.7 &  102.1 ± 1.7 &  0.39 ± 0.03 & -1.09 & yes \\
ZTF21abouuat &  SN~2021ucg & -17.6 &        $-$ &  81.3 ± 11.4 &  98.7 ± 11.4 &  0.27 ± 0.05 & -1.20 & yes \\
ZTF24aaabbse & SN~2023achj & -17.8 &        $-$ &  103.9 ± 9.5 &  123.8 ± 9.5 &  0.65 ± 0.06 & -1.03 & yes \\
ZTF21acgunkr & SN~2021aaxs & -17.8 &        $-$ &  77.7 ± 16.3 &  89.5 ± 16.3 &  0.99 ± 0.21 & -0.52 & yes \\
\hline
\end{tabular}
\end{table*}

\begin{table*}[ht]
\ContinuedFloat
\centering
\small
\caption{Continued.}
\begin{tabular}{llrllllrr}
\hline
ZTF & IAU & $M_r$ & Rise & Plat. dur. & OPTd & Slope & $\Delta$mag & Limit? \\
 & & (mag) & (days) & (days) & (days) & ($\times$0.01 mag day$^{-1}$) & (mag) & \\
\hline
ZTF21abnlhxs & SN~2021tyw & -17.8 &  8.7 ± 0.2 &  104.5 ± 1.1 &  116.1 ± 1.1 & 0.85 ± 0.02 & -3.10 & yes \\
ZTF19abajxet & SN~2019hyk & -17.8 & 10.9 ± 0.2 &  67.3 ± 10.9 &  79.7 ± 10.9 & 0.95 ± 0.16 & -0.88 & yes \\
ZTF19acbwejj & SN~2019upq & -17.8 & 22.0 ± 7.4 &   87.7 ± 3.3 &  130.3 ± 3.3 & 1.14 ± 0.05 & -1.28 &  no \\
ZTF23abmoxlu & SN~2023vog & -17.8 & 10.9 ± 0.2 &  109.4 ± 2.4 &  123.2 ± 2.4 & 1.19 ± 0.03 & -2.49 &  no \\
ZTF21abhhrpj & SN~2021qiu & -17.9 &        $-$ &   90.7 ± 1.6 &   96.3 ± 1.6 & 0.53 ± 0.02 & -1.93 &  no \\
ZTF22aakdqqg & SN~2022kad & -17.9 & 15.5 ± 0.6 &  63.1 ± 12.8 &  79.6 ± 12.8 & 1.38 ± 0.28 & -1.27 &  no \\
ZTF23aailjjs & SN~2023hcp & -17.9 &        $-$ &   60.2 ± 2.5 &   69.6 ± 2.4 & 0.83 ± 0.05 & -1.57 &  no \\
ZTF19aanrrqu & SN~2019clp & -18.0 &        $-$ &   49.6 ± 4.5 &   53.8 ± 4.5 & 1.17 ± 0.11 & -1.65 &  no \\
ZTF24aadkwni & SN~2024aul & -18.0 &        $-$ &   59.8 ± 1.6 &   68.4 ± 1.6 & 0.84 ± 0.04 & -1.12 &  no \\
ZTF21abjcjmc & SN~2021skn & -18.0 &  7.1 ± 0.0 &   55.7 ± 0.5 &   63.7 ± 0.5 & 2.05 ± 0.04 & -1.41 & yes \\
ZTF23abascqa & SN~2023rbk & -18.0 &        $-$ &   71.9 ± 1.3 &   85.9 ± 1.3 & 1.28 ± 0.04 & -2.02 &  no \\
ZTF24aaemydm & SN~2024chx & -18.0 &        $-$ &   64.3 ± 5.6 &   70.9 ± 5.6 & 0.92 ± 0.09 & -1.40 &  no \\
ZTF24aajxppf & SN~2024grw & -18.0 &        $-$ &   52.7 ± 4.6 &   64.4 ± 4.6 & 1.37 ± 0.13 & -1.32 &  no \\
ZTF21aaqugxm & SN~2021hdt & -18.1 &        $-$ &   49.1 ± 8.9 &   60.9 ± 8.9 & 1.34 ± 0.25 & -1.07 &  no \\
ZTF21ablvzhp & SN~2021tiq & -18.1 &        $-$ &   85.9 ± 4.1 &   98.1 ± 4.1 & 0.56 ± 0.04 & -1.37 & yes \\
ZTF21abgilzj & AT~2021qcr & -18.1 & 16.2 ± 1.0 &   66.8 ± 1.1 &   93.0 ± 1.1 & 1.51 ± 0.04 & -0.92 & yes \\
ZTF21aapkcmr & AT~2021fnj & -18.1 &        $-$ &   54.1 ± 4.6 &   59.1 ± 4.6 & 1.74 ± 0.15 & -1.34 & yes \\
ZTF19abqrhvt & SN~2019nyk & -18.1 &        $-$ &   47.4 ± 2.0 &   60.1 ± 2.0 & 1.54 ± 0.08 & -1.47 & yes \\
ZTF21abnudtb & SN~2021txr & -18.1 &        $-$ &   71.6 ± 1.0 &   80.5 ± 1.0 & 0.63 ± 0.03 & -1.16 &  no \\
ZTF23aaphnyz & SN~2023lkw & -18.1 &        $-$ &  77.5 ± 10.7 &  87.7 ± 10.7 & 0.46 ± 0.07 & -0.95 & yes \\
ZTF22aaolwsd & SN~2022mxv & -18.1 & 12.6 ± 0.2 & 104.8 ± 20.0 & 122.6 ± 20.0 & 1.35 ± 0.26 & -0.87 & yes \\
ZTF19aarykkb & SN~2019dzk & -18.2 &        $-$ &   73.9 ± 2.8 &   82.0 ± 2.8 &  0.3 ± 0.03 & -0.62 & yes \\
ZTF22aativsd & SN~2022ovb & -18.2 & 11.6 ± 0.2 &   71.2 ± 1.9 &   84.3 ± 1.9 & 0.84 ± 0.04 & -0.94 &  no \\
ZTF19abqrhvy & SN~2019odf & -18.2 & 14.0 ± 0.3 &   93.3 ± 8.4 &  107.8 ± 8.4 & 0.71 ± 0.07 & -1.14 & yes \\
ZTF19abbwfgp & SN~2019ikb & -18.2 & 11.9 ± 0.3 &   79.1 ± 2.7 &   91.5 ± 2.7 & 1.05 ± 0.04 & -2.06 & yes \\
ZTF22aavobvq & SN~2022prv & -18.3 & 10.6 ± 0.2 &  62.8 ± 12.0 &  86.5 ± 12.0 &  1.4 ± 0.27 & -2.56 & yes \\
ZTF21aaipypa & SN~2021cgu & -18.3 & 13.2 ± 0.6 &  90.0 ± 13.1 & 106.5 ± 13.1 & 0.98 ± 0.14 & -0.45 & yes \\
ZTF19acrcxri & SN~2019ult & -18.4 &        $-$ &   54.9 ± 1.6 &   65.3 ± 1.6 & 1.07 ± 0.05 & -1.33 & yes \\
ZTF23aaaatjn &  SN~2023cf & -18.5 &        $-$ &   84.6 ± 5.7 &   90.1 ± 5.7 & 0.89 ± 0.06 & -1.92 &  no \\
ZTF22aarycqo & SN~2022ojo & -19.3 &        $-$ &  80.8 ± 11.0 &  96.1 ± 11.0 &  1.42 ± 0.2 & -1.40 &  no \\
\hline
\end{tabular}
\end{table*}

\begin{table}[ht]
\centering
\caption{Priors used in the MCMC fitting with the semi-analytical models from \citet{Nagy2016}.}
\begin{tabular}{ll}
\hline
Parameter & Prior \\
\hline
Initial radius [$R$ (cm)] & $\mathcal{U}[2 \times 10^{12},\ 8 \times 10^{13}]$ \\
Ejected mass [$M_{\rm ej}$ ($M_\odot$)] & $\mathcal{U}[3,\ 25]$ \\
Kinetic energy [$E_{\rm kin}$ ($10^{51}$ erg)] & $\mathcal{U}[0.01,\ 10]$ \\
Thermal energy [$E_{\rm th}$ ($10^{51}$ erg)] & $\mathcal{U}[0.01,\ 10]$ \\
Thomson scattering opacity [$\kappa$ (cm$^2$/g)] & $\mathcal{U}[0.1,\ 0.4]$ \\
Density power-law exponent & $\mathcal{U}[0,\ 3]$ \\
Ionization/recombination temperature [$T_{\rm ion}$ (K)] & $\mathcal{U}[0,\ 20000]$ \\
Nickel mass [$M_{\rm Ni}$ ($M_\odot$)] & $\mathcal{U}[5 \times 10^{-4},\ 0.5]$ \\
Gamma-leakage parameter [$t^2_{\gamma}$ (day$^2$)] & $\mathcal{U}[10^3,\ 10^7]$ \\
\hline
\end{tabular}
\label{tab:semipriors}
\end{table}

\begin{table}[ht]
\centering
\caption{Priors used in the MCMC fitting with the radiation-hydrodynamical models from \citet{Moriya2023}.}
\begin{tabular}{l c}
\hline
Parameter & Prior \\
\hline
$M_\mathrm{ZAMS}$ [\Msun] & $\mathcal{U}(9, 18)$ \\
Explosion Energy [$10^{51}$ erg]           & $\mathcal{U}(0.5, 5.0)$ \\
$M_\mathrm{Ni}$ [\Msun]       & $\mathcal{U}(0.001, 0.3)$ \\
Mass-Loss Rate [$\Msun~\mathrm{yr}^{-1}$]   & $\mathcal{U}(10^{-5.0}, 10^{-1.0})$ \\
Wind Velocity [km~s$^{-1}$]                 & Fixed at 10 \\
R$_\mathrm{CSM}$ [$10^{14}$ cm]                   & $\mathcal{U}(1, 10)$ \\
Wind Structure Parameter ($\beta$)          & $\mathcal{U}(0.5, 5.0)$ \\
\hline
\end{tabular}

\label{tab:moriyapriors}
\end{table}

\begin{table*}
\begin{center}
\caption{Posterior estimates of explosion and progenitor properties based on fits to the semi-analytical models from \citet{Nagy2016}. We list the median values of $^{56}$Ni mass ($M_\mathrm{Ni}$), ejecta mass ($M_\mathrm{ej}$), progenitor radius ($R_0$), and kinetic energy ($E_\mathrm{kin}$). The 16th and 84th percentiles are used to quantify the lower and upper uncertainties around the median. A full machine-readable version of this table is available on \href{https://zenodo.org/records/15717884?token=eyJhbGciOiJIUzUxMiJ9.eyJpZCI6IjNhYjE3NjYwLTU2NDEtNDBkZi1iYmI5LTQ1YzQxY2EwYjllNyIsImRhdGEiOnt9LCJyYW5kb20iOiJhNjA2ZTkzNDFjYjU5NGM3ZGIzMmExMTRlOGY1NzU5MCJ9.X0jCmJz6FJbXwmcY_6ZZ1kYCcRHwrABpoZ2epNXLlpIjVIwrFkEEOGD0Z7Cfk7luRPlHvOCwJdWBwB40vX2JoQ}{\texttt{Zenodo}}.}
\small
\label{tab:nagyfit}
\begin{tabular}{llcccc}
\hline
ZTF name & IAU name & $M_\mathrm{Ni}$ & $M_\mathrm{ej}$ & $R_0$ & $E_\mathrm{kin}$ \\
 &  & $(\times 10^{-2} \, \mathrm{M_\odot})$ & (M$_\odot$) & (R$_\odot$) & $(\times 10^{51}$ erg) \\
\hline
ZTF21abrluay & SN~2021vfh & $11.96^{+1.34}_{-0.84}$ & $14.8^{+1.4}_{-2.0}$ & $350^{+119}_{-185}$ & $8.49^{+1.03}_{-1.85}$ \\
ZTF22abfxkdm & SN~2022ubb & $2.09^{+0.26}_{-0.29}$ & $17.2^{+1.0}_{-0.7}$ & $237^{+33}_{-17}$ & $4.64^{+0.33}_{-0.23}$ \\
ZTF23aasrcyv & SN~2023nlu & $0.56^{+5.65}_{-0.42}$ & $16.5^{+0.9}_{-6.3}$ & $419^{+442}_{-189}$ & $3.16^{+0.44}_{-1.65}$ \\
ZTF22abkhrkd & SN~2022wol & $3.52^{+2.66}_{-1.34}$ & $13.9^{+3.2}_{-5.5}$ & $413^{+326}_{-276}$ & $2.92^{+0.94}_{-1.98}$ \\
ZTF21aaagypx & SN~2021V & $9.12^{+0.82}_{-0.45}$ & $24.7^{+0.3}_{-2.8}$ & $720^{+49}_{-56}$ & $6.33^{+0.31}_{-0.45}$ \\
ZTF18aawpwlf & SN~2020hvn & $7.23^{+0.40}_{-0.44}$ & $6.3^{+0.6}_{-0.4}$ & $207^{+45}_{-29}$ & $2.63^{+0.29}_{-0.27}$ \\
ZTF22aafsqud & SN~2022hql & $2.39^{+1.64}_{-1.34}$ & $12.8^{+1.5}_{-1.7}$ & $286^{+81}_{-105}$ & $3.97^{+0.63}_{-0.70}$ \\
ZTF21acafqtj & SN~2021yok & $6.52^{+0.25}_{-0.64}$ & $14.2^{+0.9}_{-0.6}$ & $885^{+65}_{-117}$ & $5.81^{+0.27}_{-0.57}$ \\
ZTF21acpqqgu & SN~2021aewn & $1.89^{+0.16}_{-0.06}$ & $13.9^{+0.2}_{-0.5}$ & $455^{+19}_{-15}$ & $3.49^{+0.05}_{-0.15}$ \\
ZTF19abqrhvt & SN~2019nyk & $10.79^{+0.40}_{-0.39}$ & $10.6^{+0.4}_{-0.4}$ & $102^{+5}_{-7}$ & $8.17^{+0.36}_{-0.36}$ \\
ZTF21aaipypa & SN~2021cgu & $7.36^{+6.97}_{-2.09}$ & $20.7^{+1.1}_{-5.9}$ & $707^{+193}_{-104}$ & $9.41^{+0.58}_{-4.57}$ \\
ZTF20acuhren & SN~2020abcq & $2.27^{+0.27}_{-1.88}$ & $8.4^{+1.1}_{-0.3}$ & $205^{+32}_{-39}$ & $0.87^{+0.02}_{-0.03}$ \\
ZTF24aaucrua & SN~2024nez & $4.20^{+0.24}_{-0.41}$ & $8.9^{+1.2}_{-0.6}$ & $183^{+97}_{-42}$ & $1.40^{+0.27}_{-0.14}$ \\
ZTF22aapargp & SN~2022niw & $3.84^{+1.78}_{-1.77}$ & $20.3^{+2.7}_{-2.8}$ & $337^{+278}_{-185}$ & $8.66^{+0.90}_{-1.44}$ \\
ZTF23aaqknaw & SN~2023lzn & $5.78^{+0.48}_{-0.49}$ & $21.8^{+0.3}_{-2.4}$ & $346^{+28}_{-58}$ & $9.86^{+0.08}_{-1.51}$ \\
ZTF22abhsxph & SN~2022vyc & $6.11^{+0.33}_{-0.37}$ & $16.6^{+1.1}_{-1.3}$ & $157^{+44}_{-61}$ & $5.40^{+0.55}_{-0.51}$ \\
ZTF22abzqwmp & SN~2022adth & $0.87^{+0.13}_{-0.56}$ & $11.7^{+0.7}_{-1.3}$ & $1048^{+86}_{-319}$ & $3.05^{+0.23}_{-0.65}$ \\
ZTF24aabsmvc & SN~2024ws & $1.94^{+1.68}_{-0.84}$ & $19.0^{+1.7}_{-3.4}$ & $160^{+43}_{-33}$ & $4.75^{+0.45}_{-1.00}$ \\
ZTF21aantsla & SN~2021ech & $5.52^{+0.49}_{-0.69}$ & $11.6^{+1.3}_{-0.9}$ & $443^{+109}_{-77}$ & $3.16^{+0.47}_{-0.27}$ \\
ZTF19aclobbu & SN~2019twk & $21.80^{+2.30}_{-21.34}$ & $16.4^{+3.9}_{-2.2}$ & $89^{+90}_{-18}$ & $6.14^{+2.05}_{-6.01}$ \\
ZTF21abnlhxs & SN~2021tyw & $1.80^{+0.13}_{-0.15}$ & $14.5^{+0.8}_{-1.1}$ & $348^{+29}_{-58}$ & $4.50^{+0.22}_{-0.54}$ \\
ZTF23aanymcl & SN~2023kzz & $9.06^{+0.70}_{-0.53}$ & $10.6^{+0.8}_{-0.8}$ & $361^{+152}_{-137}$ & $4.91^{+0.52}_{-0.52}$ \\
ZTF21abouuat & SN~2021ucg & $13.76^{+0.75}_{-0.72}$ & $22.1^{+0.6}_{-1.2}$ & $341^{+24}_{-35}$ & $9.83^{+0.10}_{-0.79}$ \\
ZTF19acrcxri & SN~2019ult & $18.67^{+1.05}_{-15.54}$ & $11.7^{+1.1}_{-0.7}$ & $837^{+192}_{-171}$ & $8.12^{+0.83}_{-1.71}$ \\
ZTF21aakvroo & SN~2021cwe & $12.83^{+0.71}_{-1.65}$ & $6.6^{+1.9}_{-0.4}$ & $250^{+266}_{-114}$ & $1.97^{+0.89}_{-0.21}$ \\
ZTF22absqhkw & SN~2022zkc & $8.62^{+0.44}_{-0.36}$ & $8.0^{+0.6}_{-0.6}$ & $310^{+69}_{-95}$ & $4.16^{+0.49}_{-0.44}$ \\
ZTF18abzrgim & SN~2018gvt & $26.28^{+2.45}_{-1.20}$ & $16.3^{+1.4}_{-3.0}$ & $638^{+235}_{-192}$ & $8.52^{+1.06}_{-2.32}$ \\
ZTF21aapliyn & SN~2021foj & $10.57^{+1.53}_{-0.42}$ & $19.3^{+0.8}_{-3.3}$ & $129^{+21}_{-29}$ & $8.53^{+0.50}_{-2.08}$ \\
ZTF21abjcjmc & SN~2021skn & $5.04^{+2.59}_{-0.18}$ & $7.8^{+0.3}_{-3.3}$ & $1041^{+48}_{-444}$ & $4.32^{+0.18}_{-2.13}$ \\
ZTF22abyohff & SN~2022acrl & $1.08^{+3.73}_{-0.98}$ & $17.3^{+2.4}_{-8.0}$ & $158^{+95}_{-77}$ & $3.46^{+0.85}_{-1.99}$ \\
ZTF19aazudta & SN~2019hqm & $11.17^{+3.42}_{-1.16}$ & $19.8^{+1.9}_{-8.1}$ & $531^{+274}_{-288}$ & $8.69^{+0.98}_{-5.05}$ \\
ZTF22aarycqo & SN~2022ojo & $16.51^{+20.29}_{-14.52}$ & $7.8^{+1.7}_{-1.2}$ & $953^{+128}_{-173}$ & $1.37^{+1.02}_{-0.58}$ \\
ZTF19aamwhat & SN~2019bzd & $1.16^{+0.08}_{-0.04}$ & $9.2^{+0.5}_{-0.3}$ & $163^{+6}_{-15}$ & $1.04^{+0.09}_{-0.04}$ \\
ZTF22abyokkf & SN~2022acri & $16.85^{+0.86}_{-0.84}$ & $22.5^{+0.6}_{-1.5}$ & $316^{+46}_{-39}$ & $9.68^{+0.24}_{-0.78}$ \\
ZTF24aaabbse & SN~2023achj & $18.51^{+1.00}_{-0.83}$ & $14.9^{+0.7}_{-0.8}$ & $1000^{+91}_{-120}$ & $3.39^{+0.21}_{-0.22}$ \\
ZTF19abqrhvy & SN~2019odf & $23.70^{+0.68}_{-13.83}$ & $19.2^{+0.9}_{-1.8}$ & $400^{+148}_{-43}$ & $7.17^{+0.77}_{-2.52}$ \\
ZTF18aaszvfn & SN~2021iaw & $6.78^{+1.36}_{-0.67}$ & $11.5^{+0.4}_{-1.4}$ & $121^{+14}_{-8}$ & $2.71^{+0.18}_{-0.19}$ \\
ZTF23aaxadel & SN~2023pbg & $9.87^{+1.26}_{-0.60}$ & $15.5^{+1.0}_{-2.4}$ & $161^{+25}_{-27}$ & $4.52^{+0.33}_{-1.21}$ \\
ZTF22abfwxtr & SN~2022udq & $1.56^{+4.38}_{-0.23}$ & $23.8^{+1.0}_{-6.5}$ & $412^{+100}_{-66}$ & $5.59^{+0.32}_{-2.01}$ \\
ZTF21aaobkmg & SN~2021eui & $1.84^{+0.13}_{-0.13}$ & $8.1^{+0.8}_{-0.1}$ & $299^{+76}_{-38}$ & $0.89^{+0.01}_{-0.08}$ \\
ZTF24aabpzuz & SN~2024vs & $9.19^{+0.73}_{-0.30}$ & $14.0^{+0.6}_{-1.6}$ & $512^{+98}_{-133}$ & $6.00^{+0.37}_{-0.98}$ \\
ZTF19aawgxdn & SN~2019gmh & $11.94^{+0.86}_{-0.61}$ & $22.5^{+0.5}_{-2.2}$ & $134^{+18}_{-22}$ & $9.40^{+0.50}_{-0.90}$ \\
ZTF19aarykkb & SN~2019dzk & $0.80^{+0.25}_{-0.02}$ & $14.0^{+0.4}_{-0.9}$ & $765^{+32}_{-88}$ & $3.37^{+0.15}_{-0.29}$ \\
ZTF22aazmrpx & SN~2022raj & $0.09^{+0.24}_{-0.01}$ & $6.7^{+0.3}_{-0.3}$ & $349^{+49}_{-24}$ & $0.52^{+0.06}_{-0.02}$ \\
ZTF21abnudtb & SN~2021txr & $0.74^{+20.24}_{-0.12}$ & $15.0^{+0.7}_{-1.5}$ & $1065^{+45}_{-149}$ & $6.24^{+3.53}_{-0.71}$ \\
ZTF24aafqzur & SN~2024daa & $6.58^{+0.42}_{-0.35}$ & $17.1^{+1.5}_{-1.1}$ & $98^{+14}_{-15}$ & $6.00^{+0.72}_{-0.47}$ \\
ZTF22aavobvq & SN~2022prv & $5.91^{+0.77}_{-0.32}$ & $15.8^{+0.5}_{-1.2}$ & $280^{+20}_{-15}$ & $9.71^{+0.28}_{-0.93}$ \\
ZTF21ablvzhp & SN~2021tiq & $27.63^{+1.35}_{-1.16}$ & $17.1^{+0.6}_{-1.1}$ & $714^{+38}_{-58}$ & $9.76^{+0.24}_{-0.88}$ \\
\hline
\end{tabular}
\end{center}
\end{table*}

\begin{table*}
\ContinuedFloat
\begin{center}
\caption{Continued.}
\small
\begin{tabular}{llcccc}
\hline
ZTF & IAU & $M_\mathrm{Ni}$ & $M_\mathrm{ej}$ & $R_0$ & $E_\mathrm{kin}$ \\
 &  & $(\times 10^{-2} \, \mathrm{M_\odot})$ & (M$_\odot$) & (R$_\odot$) & $(\times 10^{51}$ erg) \\
\hline
ZTF23aaaatjn & SN~2023cf & $15.93^{+2.12}_{-5.28}$ & $14.5^{+1.9}_{-1.1}$ & $902^{+181}_{-132}$ & $7.02^{+0.95}_{-0.75}$ \\
ZTF21acgunkr & SN~2021aaxs & $2.74^{+3.30}_{-1.43}$ & $10.0^{+1.9}_{-2.2}$ & $806^{+214}_{-312}$ & $1.46^{+0.72}_{-0.54}$ \\
ZTF21abgilzj & AT~2021qcr & $6.90^{+0.47}_{-0.41}$ & $11.7^{+0.8}_{-0.7}$ & $339^{+26}_{-20}$ & $2.32^{+0.17}_{-0.14}$ \\
ZTF24aabppgn & SN~2024wp & $2.03^{+0.46}_{-0.48}$ & $8.5^{+0.8}_{-0.3}$ & $691^{+301}_{-379}$ & $0.84^{+0.04}_{-0.07}$ \\
ZTF20acmaaan & SN~2020xyk & $4.96^{+0.25}_{-0.28}$ & $9.5^{+0.7}_{-0.6}$ & $264^{+225}_{-70}$ & $2.78^{+0.29}_{-0.25}$ \\
ZTF23aackjhs & SN~2023bvj & $1.54^{+0.07}_{-0.08}$ & $6.2^{+0.5}_{-0.2}$ & $138^{+23}_{-30}$ & $0.53^{+0.03}_{-0.02}$ \\
ZTF22abtjefa & SN~2022aaad & $1.15^{+0.00}_{-0.03}$ & $5.8^{+0.2}_{-0.2}$ & $172^{+14}_{-8}$ & $0.50^{+0.00}_{-0.04}$ \\
ZTF24aaplfjd & SN~2024jxm & $3.05^{+0.06}_{-0.07}$ & $11.2^{+2.6}_{-2.2}$ & $313^{+288}_{-236}$ & $1.81^{+0.77}_{-0.56}$ \\
ZTF23aaphnyz & SN~2023lkw & $20.34^{+1.83}_{-0.73}$ & $13.3^{+0.5}_{-2.2}$ & $405^{+88}_{-78}$ & $4.88^{+0.12}_{-1.04}$ \\
ZTF23abnogui & SN~2023wcr & $1.19^{+0.01}_{-0.03}$ & $8.1^{+0.6}_{-0.1}$ & $148^{+8}_{-8}$ & $0.79^{+0.01}_{-0.06}$ \\
ZTF24aajxppf & SN~2024grw & $10.86^{+0.40}_{-0.42}$ & $8.1^{+0.2}_{-0.7}$ & $180^{+2}_{-44}$ & $4.03^{+0.05}_{-0.52}$ \\
ZTF24aaejecr & SN~2024btj & $1.89^{+0.02}_{-0.03}$ & $8.6^{+0.3}_{-0.4}$ & $397^{+26}_{-29}$ & $1.34^{+0.05}_{-0.07}$ \\
ZTF19aailepg & SN~2019amt & $12.87^{+0.45}_{-0.18}$ & $22.2^{+1.0}_{-1.7}$ & $449^{+147}_{-102}$ & $8.43^{+0.61}_{-0.62}$ \\
ZTF22aaolwsd & SN~2022mxv & $1.74^{+0.03}_{-0.02}$ & $15.7^{+0.7}_{-0.7}$ & $158^{+18}_{-7}$ & $2.98^{+0.15}_{-0.13}$ \\
ZTF19abqrhvt & SN~2019nyk & $13.71^{+1.02}_{-0.70}$ & $7.1^{+0.3}_{-0.3}$ & $1012^{+102}_{-93}$ & $4.10^{+0.25}_{-0.23}$ \\
ZTF21aapkcmr & AT~2021fnj & $7.41^{+0.46}_{-0.40}$ & $8.7^{+0.6}_{-0.6}$ & $874^{+153}_{-155}$ & $5.16^{+0.35}_{-0.53}$ \\
ZTF21aaqugxm & SN~2021hdt & $17.61^{+0.76}_{-1.01}$ & $5.4^{+0.3}_{-0.3}$ & $478^{+180}_{-194}$ & $1.91^{+0.11}_{-0.11}$ \\
ZTF23abhzfww & SN~2023twg & $13.68^{+0.53}_{-0.31}$ & $15.9^{+0.8}_{-0.9}$ & $823^{+187}_{-184}$ & $5.90^{+0.41}_{-0.46}$ \\
ZTF22abnujbv & SN~2022xus & $2.45^{+0.07}_{-0.08}$ & $8.8^{+0.4}_{-0.4}$ & $259^{+297}_{-97}$ & $1.99^{+0.10}_{-0.10}$ \\
ZTF24abtczty & SN~2024abfl & $0.35^{+0.00}_{-0.00}$ & $5.7^{+0.3}_{-0.3}$ & $543^{+80}_{-56}$ & $0.21^{+0.01}_{-0.01}$ \\
ZTF22aativsd & SN~2022ovb & $20.74^{+0.92}_{-0.78}$ & $11.7^{+0.6}_{-0.7}$ & $141^{+14}_{-8}$ & $6.39^{+0.35}_{-0.43}$ \\
ZTF19acewuwn & SN~2019ssl & $13.03^{+1.09}_{-0.73}$ & $15.0^{+1.2}_{-1.1}$ & $107^{+33}_{-25}$ & $5.93^{+0.66}_{-0.59}$ \\
ZTF23aailjjs & SN~2023hcp & $11.76^{+0.22}_{-0.61}$ & $13.5^{+0.4}_{-0.6}$ & $281^{+32}_{-29}$ & $9.63^{+0.26}_{-0.60}$ \\
ZTF20aapchqy & SN~2020cxd & $0.11^{+0.00}_{-0.00}$ & $7.6^{+0.6}_{-0.2}$ & $197^{+10}_{-14}$ & $0.29^{+0.03}_{-0.01}$ \\
ZTF24aaemydm & SN~2024chx & $12.08^{+0.11}_{-1.02}$ & $14.3^{+0.6}_{-0.6}$ & $155^{+6}_{-17}$ & $9.90^{+0.08}_{-0.87}$ \\
ZTF21acgrrnl & SN~2021aayf & $3.85^{+0.06}_{-0.04}$ & $12.7^{+0.9}_{-0.9}$ & $382^{+36}_{-68}$ & $2.40^{+0.24}_{-0.39}$ \\
ZTF23abaxtlq & SN~2023rix & $1.80^{+0.03}_{-0.03}$ & $10.7^{+0.5}_{-0.5}$ & $809^{+85}_{-81}$ & $2.31^{+0.16}_{-0.13}$ \\
ZTF21aaeqwov & AT~2021htp & $5.30^{+0.25}_{-0.05}$ & $18.1^{+0.9}_{-1.0}$ & $107^{+98}_{-15}$ & $5.42^{+0.35}_{-0.34}$ \\
ZTF24aagupsf & SN~2024egd & $7.36^{+0.34}_{-0.18}$ & $15.4^{+0.8}_{-0.8}$ & $555^{+60}_{-121}$ & $7.29^{+0.53}_{-0.41}$ \\
ZTF22aavbfhz & AT~2022phi & $5.22^{+0.16}_{-0.12}$ & $8.6^{+0.6}_{-0.6}$ & $743^{+215}_{-194}$ & $2.25^{+0.22}_{-0.24}$ \\
ZTF19acytcsg & SN~2019wvz & $19.88^{+0.78}_{-0.61}$ & $16.6^{+0.8}_{-0.9}$ & $362^{+68}_{-38}$ & $8.53^{+0.52}_{-0.60}$ \\
ZTF19aazyvub & SN~2019hnl & $7.13^{+0.24}_{-0.28}$ & $12.3^{+0.5}_{-0.7}$ & $136^{+14}_{-19}$ & $4.41^{+0.26}_{-0.28}$ \\
ZTF19aaniore & SN~2019ceg & $11.75^{+0.31}_{-0.39}$ & $14.8^{+0.7}_{-0.6}$ & $616^{+96}_{-76}$ & $4.57^{+0.30}_{-0.22}$ \\
ZTF22abkbjsb & SN~2022vym & $5.06^{+0.33}_{-0.08}$ & $5.9^{+0.2}_{-0.3}$ & $632^{+63}_{-37}$ & $1.06^{+0.04}_{-0.06}$ \\
ZTF19actnyae & SN~2019vdm & $11.10^{+0.79}_{-1.10}$ & $13.4^{+0.9}_{-1.2}$ & $729^{+120}_{-34}$ & $9.78^{+0.11}_{-1.51}$ \\
ZTF24aadkwni & SN~2024aul & $13.40^{+0.79}_{-0.40}$ & $9.6^{+0.4}_{-0.5}$ & $150^{+42}_{-21}$ & $5.10^{+0.46}_{-0.28}$ \\
ZTF19abajxet & SN~2019hyk & $8.02^{+0.38}_{-0.38}$ & $9.7^{+0.4}_{-0.4}$ & $392^{+366}_{-87}$ & $2.75^{+0.14}_{-0.17}$ \\
ZTF21abfiuqf & SN~2021pla & $3.29^{+0.05}_{-0.10}$ & $6.6^{+0.2}_{-0.4}$ & $453^{+96}_{-99}$ & $2.10^{+0.10}_{-0.15}$ \\
ZTF24abmkros & SN~2024xkd & $0.78^{+0.01}_{-0.01}$ & $8.4^{+0.5}_{-0.4}$ & $309^{+404}_{-166}$ & $0.88^{+0.06}_{-0.05}$ \\
ZTF22aakdqqg & SN~2022kad & $10.03^{+0.41}_{-0.46}$ & $5.8^{+0.3}_{-0.3}$ & $1061^{+73}_{-635}$ & $1.30^{+0.10}_{-0.09}$ \\
ZTF21aafepon & SN~2021ass & $5.17^{+0.20}_{-0.33}$ & $7.5^{+0.5}_{-0.4}$ & $139^{+6}_{-25}$ & $2.24^{+0.13}_{-0.12}$ \\
ZTF21abvcxel & SN~2021wvw & $3.55^{+0.25}_{-0.20}$ & $7.2^{+0.4}_{-0.3}$ & $123^{+18}_{-7}$ & $2.50^{+0.16}_{-0.10}$ \\
ZTF23abbtkrv & SN~2023rvo & $9.81^{+0.18}_{-0.15}$ & $24.8^{+0.1}_{-3.0}$ & $178^{+18}_{-12}$ & $9.54^{+0.14}_{-1.07}$ \\
ZTF19abbnamr & SN~2019iex & $7.40^{+0.12}_{-0.26}$ & $22.4^{+0.4}_{-1.5}$ & $170^{+33}_{-7}$ & $9.94^{+0.05}_{-1.00}$ \\
ZTF20abeohfn & SN~2020mjm & $2.36^{+0.08}_{-0.09}$ & $11.6^{+0.5}_{-0.4}$ & $439^{+82}_{-68}$ & $3.00^{+0.12}_{-0.12}$ \\
ZTF19acbwejj & SN~2019upq & $9.13^{+0.33}_{-0.01}$ & $24.7^{+0.2}_{-2.7}$ & $200^{+71}_{-21}$ & $6.36^{+0.33}_{-0.39}$ \\
ZTF19abbwfgp & SN~2019ikb & $14.79^{+0.30}_{-0.22}$ & $11.3^{+0.4}_{-0.3}$ & $257^{+65}_{-21}$ & $4.18^{+0.23}_{-0.19}$ \\
ZTF21abhhrpj & SN~2021qiu & $16.49^{+0.70}_{-0.75}$ & $11.2^{+0.3}_{-1.3}$ & $341^{+37}_{-47}$ & $4.64^{+0.29}_{-1.10}$ \\
\hline
\end{tabular}
\end{center}
\end{table*}

\begin{table*}
\ContinuedFloat
\begin{center}
\caption{Continued.}
\small
\begin{tabular}{llcccc}
\hline
ZTF & IAU & $M_\mathrm{Ni}$ & $M_\mathrm{ej}$ & $R_0$ & $E_\mathrm{kin}$ \\
 &  & $(\times 10^{-2} \, \mathrm{M_\odot})$ & (M$_\odot$) & (R$_\odot$) & $(\times 10^{51}$ erg) \\
\hline
ZTF22abssiet & SN~2022zmb & $1.32^{+0.03}_{-0.02}$ & $12.9^{+0.3}_{-0.5}$ & $152^{+18}_{-6}$ & $3.64^{+0.10}_{-0.17}$ \\
ZTF22abtspsw & SN~2022aagp & $0.58^{+0.00}_{-0.00}$ & $10.4^{+2.3}_{-2.5}$ & $311^{+39}_{-20}$ & $1.12^{+0.45}_{-0.43}$ \\
ZTF21aavhnpk & SN~2021jsf & $6.28^{+0.24}_{-0.25}$ & $12.1^{+0.6}_{-0.6}$ & $183^{+30}_{-31}$ & $4.73^{+0.34}_{-0.30}$ \\
ZTF23abascqa & SN~2023rbk & $6.34^{+0.23}_{-0.19}$ & $9.8^{+0.4}_{-0.5}$ & $1033^{+68}_{-362}$ & $2.98^{+0.10}_{-0.18}$ \\
ZTF19abwztsb & SN~2019pjs & $4.36^{+0.17}_{-0.18}$ & $8.1^{+0.4}_{-0.3}$ & $324^{+121}_{-84}$ & $2.11^{+0.11}_{-0.12}$ \\
ZTF22aaywnyg & SN~2022pru & $1.90^{+0.01}_{-0.04}$ & $8.1^{+0.7}_{-0.0}$ & $1024^{+102}_{-73}$ & $0.78^{+0.02}_{-0.05}$ \\
ZTF19actnwtn & SN~2019vdl & $30.98^{+1.62}_{-1.92}$ & $18.9^{+1.0}_{-1.5}$ & $843^{+139}_{-181}$ & $9.00^{+0.50}_{-1.06}$ \\
ZTF21aabygea & SN~2021os & $7.81^{+0.18}_{-0.12}$ & $16.2^{+0.6}_{-0.7}$ & $175^{+22}_{-9}$ & $5.01^{+0.20}_{-0.23}$ \\
ZTF23abmoxlu & SN~2023vog & $2.59^{+0.05}_{-0.02}$ & $16.4^{+0.7}_{-0.9}$ & $424^{+68}_{-29}$ & $2.87^{+0.14}_{-0.16}$ \\
ZTF22aakdbia & SN~2022jzc & $0.31^{+0.00}_{-0.00}$ & $7.6^{+0.5}_{-0.3}$ & $363^{+455}_{-172}$ & $0.42^{+0.03}_{-0.03}$ \\
ZTF19aadnxnl & SN~2019va & $11.30^{+0.23}_{-0.55}$ & $15.6^{+0.6}_{-1.1}$ & $350^{+19}_{-25}$ & $4.92^{+0.08}_{-0.49}$ \\
ZTF21aafkwtk & SN~2021apg & $6.20^{+0.14}_{-0.15}$ & $17.3^{+0.7}_{-0.7}$ & $214^{+67}_{-46}$ & $7.51^{+0.33}_{-0.35}$ \\
ZTF19aanrrqu & SN~2019clp & $11.35^{+0.42}_{-0.36}$ & $6.5^{+0.3}_{-0.2}$ & $118^{+52}_{-35}$ & $3.51^{+0.21}_{-0.17}$ \\
ZTF23aasbvab & SN~2023ngy & $3.45^{+0.05}_{-0.04}$ & $12.0^{+0.5}_{-0.6}$ & $93^{+32}_{-20}$ & $2.42^{+0.12}_{-0.13}$ \\
ZTF18aatyqds & SN~2018btl & $4.12^{+0.07}_{-0.10}$ & $7.1^{+0.3}_{-0.2}$ & $710^{+249}_{-207}$ & $1.23^{+0.08}_{-0.06}$ \\
ZTF21aardvtn & AT~2021htp & $5.75^{+0.23}_{-0.10}$ & $18.0^{+0.7}_{-0.9}$ & $324^{+110}_{-56}$ & $5.61^{+0.26}_{-0.34}$ \\
ZTF19acftfav & SN~2019ssi & $3.70^{+0.23}_{-0.07}$ & $7.1^{+0.5}_{-0.4}$ & $108^{+8}_{-13}$ & $1.43^{+0.14}_{-0.08}$ \\
ZTF19aapafit & SN~2019cvz & $8.92^{+0.10}_{-0.20}$ & $23.0^{+0.4}_{-1.0}$ & $135^{+6}_{-9}$ & $9.94^{+0.05}_{-0.69}$ \\
ZTF24aarvbxj & SN~2024lby & $10.42^{+0.54}_{-0.42}$ & $9.7^{+0.5}_{-0.5}$ & $240^{+73}_{-37}$ & $7.02^{+0.48}_{-0.49}$ \\
ZTF21aagtqna & SN~2021brb & $8.19^{+0.16}_{-0.11}$ & $22.8^{+0.8}_{-1.4}$ & $88^{+13}_{-10}$ & $8.42^{+0.47}_{-0.51}$ \\
ZTF22aasojye & SN~2022omr & $5.16^{+0.24}_{-0.00}$ & $24.7^{+0.3}_{-2.6}$ & $150^{+40}_{-12}$ & $8.50^{+0.32}_{-0.57}$ \\
ZTF21aanzcuj & SN~2021enz & $7.98^{+5.16}_{-0.12}$ & $19.4^{+1.0}_{-15.8}$ & $468^{+54}_{-413}$ & $8.37^{+0.59}_{-7.62}$ \\
ZTF22abfavpu & SN~2022tmb & $4.10^{+0.14}_{-0.10}$ & $11.1^{+0.5}_{-0.6}$ & $548^{+58}_{-100}$ & $3.08^{+0.14}_{-0.20}$ \\
ZTF18aaxkqgy & SN~2018ccb & $6.80^{+0.29}_{-0.21}$ & $13.3^{+0.7}_{-0.5}$ & $120^{+10}_{-27}$ & $6.47^{+0.32}_{-0.31}$ \\
ZTF19aanhhal & SN~2019cec & $7.96^{+0.15}_{-0.20}$ & $13.3^{+0.7}_{-0.6}$ & $129^{+25}_{-52}$ & $3.89^{+0.26}_{-0.22}$ \\
\hline
\end{tabular}
\end{center}
\end{table*}

\begin{table*} 
\begin{center} 
\caption{Posterior estimates of explosion and progenitor properties based on fits to the radiation-hydrodynamical models from \citep{Moriya2023}. 
A machine-readable version of this table is available on \href{https://zenodo.org/records/15717884?token=eyJhbGciOiJIUzUxMiJ9.eyJpZCI6IjNhYjE3NjYwLTU2NDEtNDBkZi1iYmI5LTQ1YzQxY2EwYjllNyIsImRhdGEiOnt9LCJyYW5kb20iOiJhNjA2ZTkzNDFjYjU5NGM3ZGIzMmExMTRlOGY1NzU5MCJ9.X0jCmJz6FJbXwmcY_6ZZ1kYCcRHwrABpoZ2epNXLlpIjVIwrFkEEOGD0Z7Cfk7luRPlHvOCwJdWBwB40vX2JoQ}{\texttt{Zenodo}}.}
\small
\label{moriyafit}
\begin{tabular}{llcccccc} 
\hline  
ZTF name & IAU name & ZAMS mass & $M_\mathrm{Ni}$ & $R_\mathrm{CSM}$ & $\beta$ & Energy & log $\dot{M}$  \\ 
 &  & (\Msun) & $(\times 10^{-2}$ \Msun) & ($\times 10^{14}$ cm) & & ($\times 10^{51}$ erg) & (\Msun\,yr$^{-1}$) \\ \hline
ZTF22abkhrkd & SN~2022wol & 10.2 $\pm$ 0.4 & $1.44^{+0.93}_{-0.09}$ & $7.9^{+1.4}_{-1.0}$ & $4.0^{+0.2}_{-0.7}$ & $0.31^{+0.01}_{-0.02}$ & $-0.54$ \\ 
ZTF22abtjefa & SN~2022aaad & 11.0 $\pm$ 0.6 & $1.38^{+1.08}_{-0.32}$ & $3.8^{+1.0}_{-0.9}$ & $2.5^{+0.2}_{-0.6}$ & $0.15^{+0.39}_{-0.02}$ & $-0.57$ \\ 
ZTF19abqrhvt & SN~2019nyk & 11.7 $\pm$ 0.9 & $3.67^{+1.74}_{-0.27}$ & $7.9^{+1.7}_{-3.2}$ & $3.7^{+1.1}_{-1.1}$ & $1.18^{+0.62}_{-0.20}$ & $-0.30$ \\ 
ZTF19acewuwn & SN~2019ssl & 12.6 $\pm$ 1.1 & $5.65^{+0.84}_{-0.17}$ & $6.5^{+0.7}_{-2.6}$ & $0.9^{+0.1}_{-0.2}$ & $0.16^{+0.06}_{-0.02}$ & $-0.47$ \\ 
ZTF21aaqugxm & SN~2021hdt & 14.8 $\pm$ 1.0 & $2.90^{+2.07}_{-0.53}$ & $8.2^{+0.9}_{-1.6}$ & $0.8^{+1.9}_{-0.1}$ & $2.42^{+0.13}_{-0.11}$ & $-0.49$ \\ 
ZTF24aaasazz & SN~2024ov & 10.5 $\pm$ 0.5 & $3.32^{+0.93}_{-0.36}$ & $5.5^{+2.2}_{-0.6}$ & $1.1^{+1.6}_{-0.5}$ & $0.27^{+0.08}_{-0.05}$ & $-0.39$ \\
\hline
\end{tabular} 
\end{center} 
\end{table*}

\begin{table*} 
\ContinuedFloat
\begin{center} 
\caption{Continued.} 
\small
\begin{tabular}{llcccccc} 
\hline  
ZTF name & IAU name & ZAMS mass & $M_\mathrm{Ni}$ & $R_\mathrm{CSM}$ & $\beta$ & Energy & $\dot{M}$  \\ 
 &  & (\Msun) & $(\times 10^{-2}$ \Msun) & ($\times 10^{14}$ cm) & & ($\times 10^{51}$ erg) & (log \Msun\,yr$^{-1}$) \\ \hline
ZTF18aahfqbc & SN~2018aoq & 12.0 $\pm$ 0.8 & $15.23^{+0.34}_{-0.79}$ & $7.7^{+0.2}_{-4.3}$ & $4.4^{+0.2}_{-1.3}$ & $2.87^{+0.09}_{-0.04}$ & $-0.48$ \\ 
ZTF21abhhrpj & SN~2021qiu & 15.7 $\pm$ 3.0 & $7.78^{+3.20}_{-0.97}$ & $7.8^{+1.0}_{-2.9}$ & $2.4^{+1.0}_{-0.5}$ & $1.22^{+0.82}_{-0.37}$ & $-0.44$ \\ 
ZTF20actnuls & SN~2020aawo & 12.1 $\pm$ 0.6 & $3.88^{+0.41}_{-0.34}$ & $5.7^{+0.8}_{-1.0}$ & $1.8^{+0.5}_{-0.4}$ & $1.55^{+0.10}_{-0.09}$ & $-0.31$ \\ 
ZTF22abnujbv & SN~2022xus & 10.6 $\pm$ 1.3 & $1.67^{+4.59}_{-0.61}$ & $2.4^{+1.8}_{-0.9}$ & $3.8^{+0.5}_{-0.7}$ & $0.35^{+0.32}_{-0.03}$ & $-0.42$ \\ 
ZTF23abbtkrv & SN~2023rvo & 15.8 $\pm$ 0.6 & $5.58^{+1.47}_{-0.32}$ & $5.2^{+2.4}_{-1.2}$ & $3.7^{+0.4}_{-0.7}$ & $0.37^{+0.19}_{-0.14}$ & $-0.29$ \\ 
ZTF19aaniore & SN~2019ceg & 12.2 $\pm$ 1.2 & $8.67^{+2.49}_{-1.29}$ & $5.8^{+1.7}_{-1.4}$ & $2.8^{+2.0}_{-1.5}$ & $0.93^{+0.18}_{-0.16}$ & $-0.55$ \\
\hline
\end{tabular} 
\end{center} 
\end{table*}

\begin{table*} 
\ContinuedFloat
\begin{center} 
\caption{Continued.} 
\small
\begin{tabular}{llcccccc} 
\hline  
ZTF name & IAU name & ZAMS mass & $M_\mathrm{Ni}$ & $R_\mathrm{CSM}$ & $\beta$ & Energy & $\dot{M}$  \\ 
 &  & (\Msun) & $(\times 10^{-2}$ \Msun) & ($\times 10^{14}$ cm) & & ($\times 10^{51}$ erg) & (log \Msun\,yr$^{-1}$) \\ \hline
ZTF22abfxkdm & SN~2022ubb & 10.6 $\pm$ 0.4 & $2.37^{+0.48}_{-0.38}$ & $7.1^{+1.3}_{-2.3}$ & $3.9^{+0.9}_{-1.0}$ & $0.44^{+0.15}_{-0.02}$ & $-0.51$ \\ 
ZTF22aavbfhz & AT~2022phi & 10.6 $\pm$ 1.1 & $1.80^{+0.65}_{-0.48}$ & $8.2^{+1.0}_{-1.0}$ & $3.0^{+1.1}_{-1.3}$ & $0.54^{+0.09}_{-0.12}$ & $-0.44$ \\ 
ZTF22aakdbia & SN~2022jzc & 9.9 $\pm$ 0.1 & $1.60^{+4.04}_{-1.49}$ & $7.3^{+1.2}_{-2.4}$ & $1.6^{+0.4}_{-0.1}$ & $0.14^{+0.01}_{-0.01}$ & $-0.55$ \\ 
ZTF22aazmrpx & SN~2022raj & 10.9 $\pm$ 0.5 & $0.15^{+2.03}_{-0.05}$ & $7.5^{+1.1}_{-1.7}$ & $0.7^{+0.6}_{-0.2}$ & $0.20^{+0.03}_{-0.00}$ & $-0.64$ \\ 
ZTF20acfkzcg & AT~2021htp & 11.4 $\pm$ 0.8 & $7.04^{+0.19}_{-0.05}$ & $5.4^{+0.2}_{-0.6}$ & $1.7^{+0.3}_{-0.7}$ & $0.97^{+0.10}_{-0.17}$ & $-0.59$ \\ 
ZTF21abrluay & SN~2021vfh & 12.9 $\pm$ 1.4 & $4.88^{+1.34}_{-0.81}$ & $4.6^{+1.5}_{-1.5}$ & $2.1^{+1.6}_{-0.4}$ & $0.74^{+0.32}_{-0.04}$ & $-0.48$ \\
\hline
\end{tabular} 
\end{center} 
\end{table*}

\begin{table*}[ht]
\caption{Spectral log and H I $\lambda$6563 velocity measurements.}
\centering
\footnotesize
\begin{minipage}[t]{0.48\textwidth}
\centering
\begin{tabular}{ccccc}
\hline
Source & Date & Phase & Inst. & H I $\lambda$6563 \\
& & (days) & & (km~s$^{-1}$) \\
\hline
SN~2019env & 2019-06-25 & 51 & P60 & $7980 \pm 460$ \\
SN~2018bjo & 2018-06-12 & 56 & P200 & $6720 \pm 40$ \\
SN~2018cbb & 2018-05-17 & 45 & P200 & $6390 \pm 110$ \\
SN~2018bdv & 2018-05-11 & 16 & P60 & $10780 \pm 640$ \\
SN~2024iwm & 2024-06-01 & 18 & SEDM & $-$ \\
SN~2024iwm & 2024-06-06 & 23 & Deveny+LMI & $-$ \\
\hline
\end{tabular}
\end{minipage}%
\hfill
\begin{minipage}[t]{0.48\textwidth}
\centering
\begin{tabular}{ccccc}
\hline
Source & Date & Phase & Inst. & H I $\lambda$6563 \\
& & (days) & & (km~s$^{-1}$) \\
\hline
SN~2019dvw & 2019-04-24 & 62 & P200 & $-$ \\
SN~2019dvd & 2019-05-02 & 12 & P60 & $-$ \\
AT~2019esn & 2019-05-13 & 13 & P200 & $-$ \\
SN~2019gss & 2019-06-26 & 29 & P200 & $9130 \pm 550$ \\
SN~2019gmh & 2019-07-03 & 32 & P60 & $8880 \pm 560$ \\
SN~2019gmh & 2019-08-15 & 73 & P60 & $-$ \\
\hline
\end{tabular}
\end{minipage}%
\label{table_vel}
\end{table*}

\begin{table*}[ht]
\ContinuedFloat
\caption{Continued.}
\centering
\footnotesize
\begin{minipage}[t]{0.48\textwidth}
\centering
\begin{tabular}{ccccc}
\hline
Source & Date & Phase & Inst. & H I $\lambda$6563 \\
& & (days) & & (km~s$^{-1}$) \\
\hline
SN~2019ust & 2019-11-24 & 13 & P200 & $-$ \\
SN~2019ust & 2019-11-26 & 14 & P60 & $-$ \\
SN~2019ust & 2019-12-10 & 28 & P60 & $-$ \\
SN~2019ust & 2020-01-02 & 51 & P60 & $-$ \\
SN~2019vdl & 2019-12-16 & 29 & P60 & $-$ \\
SN~2019vdl & 2019-12-18 & 31 & P60 & $-$ \\
\hline
\end{tabular}
\end{minipage}%
\hfill
\begin{minipage}[t]{0.48\textwidth}
\centering
\begin{tabular}{ccccc}
\hline
Source & Date & Phase & Inst. & H I $\lambda$6563 \\
& & (days) & & (km~s$^{-1}$) \\
\hline
SN~2020eaf & 2021-01-12 & 311 & LRIS & $-$ \\
SN~2020fqv & 2020-04-15 & 16 & P60 & $12630 \pm 870$ \\
SN~2020gpe & 2020-05-29 & 48 & P60 & $8520 \pm 460$ \\
SN~2020ghv & 2020-04-23 & 12 & P60 & $-$ \\
SN~2020hyb & 2020-05-18 & 27 & LT & $-$ \\
SN~2020jfo & 2020-05-15 & 11 & P60 & $10810 \pm 0$ \\
\hline
\end{tabular}
\end{minipage}%
\label{table_vel}
\end{table*}

\begin{table*}[ht]
\ContinuedFloat
\caption{Continued.}
\centering
\footnotesize
\begin{minipage}[t]{0.48\textwidth}
\centering
\begin{tabular}{ccccc}
\hline
Source & Date & Phase & Inst. & H I $\lambda$6563 \\
& & (days) & & (km~s$^{-1}$) \\
\hline
SN~2020sje & 2020-10-11 & 42 & P60 & $-$ \\
SN~2020svn & 2020-09-19 & 12 & P60 & $-$ \\
SN~2020svn & 2020-09-20 & 13 & P60 & $-$ \\
SN~2020svn & 2020-09-21 & 14 & P60 & $-$ \\
SN~2020svn & 2020-09-22 & 15 & P60 & $-$ \\
SN~2020szs & 2020-10-27 & 45 & P60 & $7590 \pm 460$ \\
\hline
\end{tabular}
\end{minipage}%
\hfill
\begin{minipage}[t]{0.48\textwidth}
\centering
\begin{tabular}{ccccc}
\hline
Source & Date & Phase & Inst. & H I $\lambda$6563 \\
& & (days) & & (km~s$^{-1}$) \\
\hline
SN~2020acjg & 2021-11-06 & 140 & DBSP & $-$ \\
SN~2020acjg & 2022-02-05 & 228 & LRIS & $-$ \\
SN~2020acjg & 2022-03-04 & 255 & LRIS & $-$ \\
SN~2020adhh & 2021-01-09 & 21 & DIS & $-$ \\
SN~2020aeqx & 2021-01-12 & 15 & LRIS & $8560 \pm 470$ \\
SN~2021V & 2021-01-12 & 12 & SEDM & $-$ \\
\hline
\end{tabular}
\end{minipage}%
\label{table_vel}
\end{table*}

\begin{table*}[ht]
\ContinuedFloat
\caption{Continued.}
\centering
\footnotesize
\begin{minipage}[t]{0.48\textwidth}
\centering
\begin{tabular}{ccccc}
\hline
Source & Date & Phase & Inst. & H I $\lambda$6563 \\
& & (days) & & (km~s$^{-1}$) \\
\hline
SN~2021tyw & 2021-08-12 & 20 & EFOSC2 & $-$ \\
SN~2021tyw & 2021-08-17 & 25 & SEDM & $6110 \pm 570$ \\
SN~2021tyw & 2021-08-21 & 29 & SEDM & $-$ \\
SN~2021tyw & 2021-08-30 & 38 & SEDM & $-$ \\
SN~2021tyw & 2021-09-13 & 51 & EFOSC2 & $-$ \\
SN~2021tyw & 2021-09-13 & 51 & EFOSC2 & $-$ \\
\hline
\end{tabular}
\end{minipage}%
\hfill
\begin{minipage}[t]{0.48\textwidth}
\centering
\begin{tabular}{ccccc}
\hline
Source & Date & Phase & Inst. & H I $\lambda$6563 \\
& & (days) & & (km~s$^{-1}$) \\
\hline
SN~2021afud & 2022-01-12 & 44 & DBSP & $-$ \\
SN~2022cru & 2022-04-27 & 118 & LRIS & $4370 \pm 60$ \\
SN~2022ces & 2022-03-02 & 25 & SEDM & $9320 \pm 460$ \\
SN~2022gwg & 2022-04-19 & 12 & SEDM & $-$ \\
SN~2022hql & 2022-04-26 & 12 & SEDM & $12530 \pm 460$ \\
SN~2022hss & 2022-05-02 & 16 & SEDM & $13000 \pm 460$ \\
\hline
\end{tabular}
\end{minipage}%
\label{table_vel}
\end{table*}

\begin{table*}[ht]
\ContinuedFloat
\caption{Continued.}
\centering
\footnotesize
\begin{minipage}[t]{0.48\textwidth}
\centering
\begin{tabular}{ccccc}
\hline
Source & Date & Phase & Inst. & H I $\lambda$6563 \\
& & (days) & & (km~s$^{-1}$) \\
\hline
SN~2022xav & 2022-11-18 & 43 & DBSP & $8580 \pm 570$ \\
SN~2022yma & 2022-11-26 & 36 & DBSP & $7670 \pm 510$ \\
SN~2022zkc & 2022-11-17 & 14 & SEDM & $12570 \pm 850$ \\
SN~2022zmb & 2022-11-18 & 15 & DBSP & $9550 \pm 460$ \\
SN~2022zmb & 2023-03-20 & 136 & LRIS & $-$ \\
SN~2022zmb & 2023-04-26 & 172 & LRIS & $3090 \pm 260$ \\
\hline
\end{tabular}
\end{minipage}%
\hfill
\begin{minipage}[t]{0.48\textwidth}
\centering
\begin{tabular}{ccccc}
\hline
Source & Date & Phase & Inst. & H I $\lambda$6563 \\
& & (days) & & (km~s$^{-1}$) \\
\hline
SN~2022acko & 2023-11-08 & 336 & LRIS & $3730 \pm 210$ \\
SN~2022acko & 2023-12-07 & 365 & LRIS & $3580 \pm 240$ \\
SN~2022acko & 2024-02-15 & 435 & LRIS & $3700 \pm 50$ \\
SN~2022acrl & 2022-12-19 & 18 & SEDM & $10800 \pm 460$ \\
SN~2022acri & 2022-12-18 & 16 & SEDM & $12050 \pm 910$ \\
SN~2022adtt & 2023-01-17 & 30 & LRIS & $7930 \pm 600$ \\
\hline
\end{tabular}
\end{minipage}%
\label{table_vel}
\end{table*}

\begin{table*}[ht]
\ContinuedFloat
\caption{Continued.}
\centering
\footnotesize
\begin{minipage}[t]{0.48\textwidth}
\centering
\begin{tabular}{ccccc}
\hline
Source & Date & Phase & Inst. & H I $\lambda$6563 \\
& & (days) & & (km~s$^{-1}$) \\
\hline
SN~2023udb & 2023-10-17 & 15 & LRIS & $-$ \\
SN~2023twg & 2023-10-17 & 19 & SEDM & $13220 \pm 560$ \\
SN~2023vcj & 2023-12-09 & 57 & SEDM & $-$ \\
SN~2023vcj & 2023-12-10 & 58 & ALFOSC & $6380 \pm 260$ \\
SN~2023vcj & 2023-12-28 & 76 & SEDM & $-$ \\
SN~2023vcj & 2023-12-29 & 77 & SEDM & $-$ \\
\hline
\end{tabular}
\end{minipage}%
\hfill
\begin{minipage}[t]{0.48\textwidth}
\centering
\begin{tabular}{ccccc}
\hline
Source & Date & Phase & Inst. & H I $\lambda$6563 \\
& & (days) & & (km~s$^{-1}$) \\
\hline
SN~2024wp & 2024-07-09 & 177 & LRIS & $4700 \pm 120$ \\
SN~2024aul & 2024-02-27 & 36 & SPRAT & $-$ \\
AT~2024bzq & 2024-02-23 & 16 & SEDM & $-$ \\
SN~2024btx & 2024-05-08 & 90 & LRIS & $4340 \pm 160$ \\
SN~2024atk & 2024-04-10 & 66 & SEDM & $7450 \pm 560$ \\
SN~2024chx & 2024-04-03 & 50 & SPRAT & $6280 \pm 250$ \\
\hline
\end{tabular}
\end{minipage}%
\label{table_vel}
\end{table*}

\pagebreak

\end{document}